М. А. СУХОРОЛЬСЬКИЙ

# ФУНКЦІОНАЛЬНІ ПОСЛІДОВНОСТІ ТА РЯДИ

М. А. СУХОРОЛЬСЬКИЙ

# ФУНКЦІОНАЛЬНІ ПОСЛІДОВНОСТІ ТА РЯДИ

ВИДАННЯ ДРУГЕ, ВИПРАВЛЕНЕ

Викладено теорію функціональних послідовностей та рядів; розглянуто рівномірно збіжні, збіжні в середньому і слабко збіжні послідовності та ряди. Розвинуто послідовнісний підхід до побудови узагальнених методів підсумовування рядів та узагальнених розв'язків задач математичної фізики.

Для широкого кола науковців, аспірантів та студентів.

Рис. 11. Бібліогр. 24 назв.

The theory of the functional sequences and series is presented; uniformly convergent, convergent in the sense of a mean square and weakly convergent sequences and series are considered. Sequential approach to constructing generalized methods of series summarizing and generalized solutions of the problems of mathematical physics is developed.

For the broad sections of scientist, graduate and post-graduate students.

Fig. 11. Ref. 24.

# ЗМІСТ

# ПЕРЕДМОВА

Функціональні послідовності та ряди широко використовуються для зображення та обчислення значень функцій; вони є теоретичною основою побудови та дослідження математичних структур інших розділів математики, а також покладені в основу побудови ефективних методів розв'язування задач прикладної математики та фізики. Водночас теорія послідовностей і рядів систематично поповнюється новими математичними поняттями, ідеями та методами. Тому важливою є задача достатньо повного та системного викладу основних положень цієї теорії на доступному для широкого кола науковців математичному рівні.

У книзі послідовності та ряди розглядаються як окремі взаємно зв'язані математичні структури. Висвітлено три рівні їх збіжності – рівномірну збіжність, збіжність в середньому та слабку збіжність. Спочатку викладено рівномірну збіжність, збіжність в середньому та слабку збіжність функціональних послідовностей. Потім ґрунтуючись на цих дослідженнях розглянуто різні рівні збіжності функціональних рядів, їх застосування до розвинення функцій у ряди та побудови узагальнених розв'язків крайових задач математичної фізики.

Виклад основних тем книги ґрунтується на матеріалі класичних підручників з математичного аналізу. Для викладу підрозділів, які стосуються слабкої збіжності послідовностей та рядів, узагальнених методів підсумовування рядів та узагальнених розв'язків крайових задач, використано підручники та літературу навчального спрямування з поглибленим викладом цих тем.

У першому розділі наведені класичні результати з теорії функціональних послідовностей. У перших двох (допоміжних) підрозділах викладені основні властивості неперервних та диференційовних функцій, а також властивості невласних інтегралів з нескінченними межами. Наступні підрозділи присвячені викладу основних відомостей з теорії рівномірно збіжних, збіжних в середньому та слабко збіжних функціональних послідовностей. Детально вивчаються властивості слабко збіжних дельтоподібних послідовностей функцій.

Другий розділ присвячений дослідженню властивостей рівномірно збіжних функціональних рядів. Розглядаються ряди за довільними системами функцій, степеневі ряди і тригонометричні ряди. Встановлено достатні умови рівномірної збіжності степеневих рядів та рядів Фур'є за тригонометричними системами функцій. Наведено також достатні умови рівномірної збіжності подвійних тригонометричних рядів.

У третьому розділі розглядаються збіжні в середньому функціональні ряди. Викладено властивості ортогональних систем функцій та рядів за цими системами. Досліджено повноту систем тригонометричних функцій та систем алгебраїчних многочленів у просторі інтегровних з квадратом функцій. Розглядається збіжність в середньому рядів за цими системами, а також повнота і збіжність рядів за системами функцій від двох змінних.

Четвертий розділ присвячений викладу теорії узагальненого підсумовування рядів, що ґрунтується на математичному апараті інтегральних операторів згладжування з використанням дельтоподібних послідовностей функцій. Сформульовано загальний підхід до побудови методів підсумовування тригонометричних рядів. Наведено основні найчастіше використовувані у аналізі узагальнені методи підсумовування рядів.

У п'ятому розділі викладено побудову розв'язків основних задач математичної фізики методом Фур'є; розглянуто достатні умови існування класичних розв'язків рівнянь коливань струни та мембрани. Інтегральні оператори згладжування функцій використано для побудови узагальнених розв'язків некоректно поставлених крайових задач. В основі узагальнених розв'язків задач лежить концепція послідовнісного підходу до побудови узагальнених функцій. Узагальнені розв'язки зображено у вигляді границь послідовностей функцій – сум рівномірно збіжних рядів Фур'є.

Теоретичний матеріал доповнюється значною кількістю прикладів, які ілюструють відповідні поняття і твердження. Кожний підрозділ першого розділу і всі наступні розділи доповнені достатньою для закріплення теоретичних знань кількістю завдань для самостійного розв'язування.

Рівень строгості викладу теоретичного матеріалу вибрано в залежності від складності досліджуваних об'єктів та важливості відповідних тверджень та понять для подальших досліджень. Так, дослідження збіжності функціональних послідовностей ґрунтується на неперервності їх членів, що не є принциповим, однак суттєво спрощує доведення відповідних тверджень теорії послідовностей і безпосередньо переноситься на доведення тверджень теорії рядів. Дослідження розвинень функцій від двох змінних ґрунтуються (з метою спрощення викладок) на послабленні вимог стосовно гладкості функцій.

# Р О З Д І Л  I

# ФУНКЦІОНАЛЬНІ ПОСЛІДОВНОСТІ
______________________________________________

## 1.1. Неперервні функції

**1.1.1. Неперервність функції в точці.** Розглянемо найважливіші властивості функцій $[12, 23]$, які лежать в основі понять, що характеризують функціональні послідовності.

Нехай $y = f(x)$ – функція, визначена на проміжку $(a,b) \subset R$, $a < b$.

***О з н а ч е н н я  1 (Коші).*** *Функція $y = f(x)$ називається неперервною в точці $x_0 \in (a,b)$, якщо для будь-якого числа $\varepsilon > 0$ існує таке число $\delta > 0$, що для всіх $x \in (a,b)$, які задовольняють умову $|x - x_0| < \delta$, виконується нерівність*

$$|f(x) - f(x_0)| < \varepsilon.$$

***О з н а ч е н н я  1' (Гейне).*** *Функція $y = f(x)$ називається неперервною в точці $x_0 \in (a,b)$, якщо якою б не була послідовність точок $\{x_n\} \subset (a,b)$, що збігається до точки $x_0$, $x_0 \neq x_n$, відповідна послідовність $\{f(x_n)\}$ збігається до значення функції $f(x_0)$ в цій точці.*

З цих означень випливає, що визначена в інтервалі $(a,b)$ функція $y = f(x)$ неперервна в точці $x_0 \in (a,b)$, якщо

$$\lim_{x \to x_0} f(x) = f(x_0). \qquad (1.1)$$

При дослідженні функцій на неперервність можна скористатися одним з означень неперервності з використанням поняття приросту функції. Візьмемо дві довільні точки деякого проміжку $x_0, x_0 + \Delta x \in (a,b)$. Тоді приросту аргумента $\Delta x$ в точці $x_0$ відповідає приріст функції

$$\Delta y = f(x_0 + \Delta x) - f(x_0).$$

***О з н а ч е н н я   1"***. *Функція $y = f(x)$ неперервна в точці $x_0 \in (a, b)$, якщо нескінченно малому приросту аргумента в цій точці відповідає нескінченно малий приріст функції, тобто*
$$\lim_{\Delta x \to 0} \Delta y = 0.$$

Легко переконатись, що наведені означення неперервності функції в точці еквівалентні між собою, тобто, коли функція неперервна в точці за яким-небудь одним означенням, то вона неперервна і за іншим означенням.

***О з н а ч е н н я   2***. *Функція $y = f(x)$ називається неперервною в точці $x_0 \in (a, b)$ справа (зліва), якщо існує число $f(x_0)$, існує правостороння (лівостороння) границя функції в $x_0$,*
$$\lim_{\Delta x \to +0} f(x_0 + \Delta x) = f(x_0 + 0) \ \left( \lim_{\Delta x \to -0} f(x_0 + \Delta x) = f(x_0 - 0) \right),$$
*і правостороння (лівостороння) границя дорівнює значенню функції в цій точці,*
$$f(x_0 + 0) = f(x_0) \ \ (f(x_0 - 0) = f(x_0)).$$

Очевидно, що необхідною і достатньою умовою неперервності функції в точці є її неперервність справа і зліва.

***О з н а ч е н н я   3***. *Якщо функція $y = f(x)$ в точці $x_0$ не є неперервною, то точка $x_0$ називається точкою розриву функції, а функція $y = f(x)$ називається розривною в точці $x_0$.*

*Точка розриву $x_0$ функції $y = f(x)$ називається точкою розриву першого роду, якщо в цій точці існують скінченні лівостороння і правостороння границі.*

*Якщо ці границі рівні між собою, то точка $x_0$ називається точкою усувного розриву.*

*Точка розриву $x_0$ функції $y = f(x)$ називається точкою розриву другого роду, якщо в цій точці не існує або дорівнює нескінченності хоча б одна з односторонніх границь.*

У математичному аналізі чільне місце займають елементарні функції, які утворені з використанням суперпозиції та раціональних операцій (додавання, віднімання, множення та ділення) над основними елементарними функціями. Тому важливими для дослідження неперервності елементарних функцій є наступні твердження.

***Т е о р е м а   1 .*** *Нехай визначені на проміжку $(a,b)$ функції $y = f(x)$ і $y = g(x)$ неперервні в точці $x_0 \in (a,b)$.*

*Тоді неперервними в цій точці є функції: а) $cf(x)$, де $c$ – стала; б) $f(x)+g(x)$; в) $f(x)g(x)$; г) $\dfrac{f(x)}{g(x)}$, якщо, крім того, $g(x_0) \neq 0$.*

*Д о в е д е н н я .* Відповідні твердження випливають безпосередньо з означення неперервності і властивостей границь функцій.

Доведемо, наприклад, неперервність функції $f(x)g(x)$. Оскільки функції $f(x)$ і $g(x)$ неперервні в точці $x_0$ і для кожної з них виконується рівність (1.1), справедлива рівність
$$\lim_{x \to x_0} f(x)g(x) = \lim_{x \to x_0} f(x) \lim_{x \to x_0} g(x) = f(x_0)g(x_0).$$
Виконання цієї рівності підтверджує неперервність функції $f(x)g(x)$ в точці $x_0$.

Теорему доведено.

Нехай маємо складну функцію, яка є суперпозицією двох функцій $y = f(t)$ і $t = \varphi(x)$, що визначені, відповідно, в областях $(\alpha, \beta)$ і $(a, b)$. Для дослідження неперервності цієї функції скористаємося наступною теоремою.

***Т е о р е м а   2 .*** *Якщо функція $t = \varphi(x)$ неперервна в точці $x_0 \in (a,b)$, а функція $y = f(t)$ неперервна в точці $t_0 = \varphi(x_0)$, $t_0 \in (\alpha, \beta)$, то складна функція $y = f(\varphi(x))$ неперервна в точці $x_0$.*

*Д о в е д е н н я .* Скористаємося означенням неперервності функції за Коші. Потрібно довести, що для будь-якого як завгодно малого числа $\varepsilon > 0$ існує таке число $\delta > 0$, що для всіх $x \in (a,b)$, які задовольняють нерівність $|x - x_0| < \delta$, виконується нерівність $|f(\varphi(x)) - f(\varphi(x_0))| < \varepsilon$.

Справді, функція $y = f(t)$ неперервна в точці $t_0 \in (\alpha, \beta)$. Тому для довільного $\varepsilon > 0$ знайдеться число $\delta_1 > 0$, що з нерівності $|t - t_0| < \delta_1$ випливає нерівність $|f(t) - f(t_0)| < \varepsilon$. Функція $t = \varphi(x)$ також неперервна в точці $x_0 \in (a, b)$. Тому для будь-якого наперед заданого числа, зокрема $\delta_1$, існує число $\delta > 0$, що з нерівності $|x - x_0| < \delta$ випливає нерівність $|\varphi(x) - \varphi(x_0)| < \delta_1$. Звідси випливає, що для довільного $\varepsilon > 0$ існує $\delta > 0$, що з нерівності $|x - x_0| < \delta$ випливає нерівність $|f(t) - f(t_0)| < \varepsilon$ або

$$|f(\varphi(x)) - f(\varphi(x_0))| < \varepsilon.$$

Теорему доведено.

**1.1.2. Неперервність функції на проміжку.**

*О з н а ч е н н я  4 .* Якщо функція $y = f(x)$ *неперервна в кожній точці інтервалу* $(a, b)$, *то вона називається неперервною на інтервалі* $(a, b)$ *і, крім того, якщо функція* $y = f(x)$ *неперервна на кінцях інтервалу, відповідно, справа і зліва, то вона називається неперервною на сегменті* $[a, b]$.

Функція $y = f(x)$, визначена на проміжку $(a, b)$, називається *обмеженою зверху (знизу),* якщо множина її значень обмежена зверху (знизу), тобто існує стала $M$ така, що для кожного $x \in (a, b)$ виконується нерівність $f(x) \leq M$ (відповідно $f(x) \geq M$).

Функція $y = f(x)$, обмежена на проміжку $(a, b)$ як зверху, так і знизу, називається *обмеженою* на цьому проміжку.

Верхня (нижня) межа множини значень функції $y = f(x)$, заданої на проміжку $(a, b)$, називається *верхньою (нижньою) межею або гранню* функції на цьому проміжку і позначається

$$\sup_{x \in (a,b)} f(x) \left( \inf_{x \in (a,b)} f(x) \right).$$

За такого означення верхня (нижня) грань функції може бути як скінченною, так і нескінченною величиною. Крім того [8], якщо функція монотонно зростає (спадає) на інтервалі $(a, b)$, то існують скінченні або нескінченні односторонні границі

$$\lim_{x \to b-0} f(x) = \sup_{x \in (a,b)} f(x), \quad \lim_{x \to a+0} f(x) = \inf_{x \in (a,b)} f(x)$$

(відповідно $\lim\limits_{x \to b-0} f(x) = \inf\limits_{x \in (a,b)} f(x)$, $\lim\limits_{x \to a+0} f(x) = \sup\limits_{x \in (a,b)} f(x)$).

Функція $y = f(x)$, визначена на проміжку $(a, b)$, називається *монотонно зростаючою (спадною)*, якщо для будь-яких двох точок $x_1, x_2 \in (a, b)$ таких, що $x_1 < x_2$, виконується нерівність $f(x_1) \leq f(x_2)$ (відповідно $f(x_1) \geq f(x_2)$).

Функція $y = f(x)$, задана на проміжку $(a, b)$, приймає в точці $x_0 \in (a, b)$ найбільше (найменше) значення, якщо $f(x_0) \geq f(x)$ (відповідно $f(x_0) \leq f(x)$) для кожної точки $x \in (a, b)$ і пишуть

$$f(x_0) = \max_{x \in (a,b)} f(x) \left( f(x_0) = \min_{x \in (a,b)} f(x) \right).$$

Очевидно, якщо функція $y = f(x)$ приймає в точці $x_0 \in (a, b)$ найбільше (найменше) значення, то $f(x_0) = \sup_{x \in (a,b)} f(x)$ (відповідно $f(x_0) = \inf_{x \in (a,b)} f(x)$) і, зокрема, якщо функція $y = f(x)$ неперервна на сегменті $[a, b]$, то $\sup_{x \in [a,b]} f(x)$ ($\inf_{x \in [a,b]} f(x)$) можна замінити на $\max_{x \in [a,b]} f(x)$ ($\min_{x \in [a,b]} f(x)$).

Розглядаючи функцію $y = f(x)$, неперервну на проміжку $(a, b)$, згідно з означенням Коші, число $\delta = \delta(\varepsilon, x_0)$ залежить від $\varepsilon$ і від точки $x_0$. Важливий випадок неперервної функції на проміжку, коли можна вказати число $\delta = \delta(\varepsilon)$, яке не залежить від розглядуваної точки.

*О з н а ч е н н я  5*. *Функція $y = f(x)$, неперервна на проміжку $(a, b)$, називається рівномірно неперервною на цьому проміжку, якщо для будь-якого числа $\varepsilon > 0$ існує число $\delta = \delta(\varepsilon) > 0$ таке, що для будь-яких двох точок $x_1, x_2 \in (a, b)$, які задовольняють нерівність $|x_1 - x_2| < \delta$, справджується така нерівність:*

$$|f(x_1) - f(x_2)| < \varepsilon.$$

*П р и к л а д  1*. Покажемо, що функція $y = x$ рівномірно неперервна в інтервалі $(-\infty, \infty)$.

Візьмемо довільне число $\varepsilon > 0$. Виберемо $\delta = \varepsilon$. Тоді з

нерівності $|x_1 - x_2| < \delta$ випливає нерівність $|f(x_1) - f(x_2)| = |x_1 - x_2| < \delta$. Оскільки число $\delta$ залежить тільки від $\varepsilon$ і не залежить від точок розглядуваного проміжку, функція $y = x$ є рівномірно неперервною на цьому проміжку.

*П р и к л а д  2*. Покажемо, що функція $y = \dfrac{1}{x}$ не є рівномірно неперервною на проміжку $(0, 1]$.

Як частка від ділення двох неперервних функцій вона неперервна функція. Візьмемо дві точки $x_1 = \dfrac{1}{2n}$, $x_2 = \dfrac{1}{n}$ $(n = 2, 3, ...)$. Оскільки $|x_1 - x_2| = \dfrac{1}{2n}$, відстань між цими точками може бути вибрана як завгодно малою.

Для будь-якого числа $\varepsilon \in \left(0; \dfrac{1}{2}\right)$ не існує числа $\delta = \delta(\varepsilon) > 0$ такого, що з нерівності $|x_1 - x_2| < \delta$ випливала б нерівність $|f(x_1) - f(x_2)| < \varepsilon$. Тому, що при $n > \dfrac{1}{2\delta}$ маємо $|x_1 - x_2| < \delta$ і $|f(x_1) - f(x_2)| = \dfrac{1}{x_1} - \dfrac{1}{x_2} = n > \varepsilon$.

Отже, не кожна неперервна на проміжку функція рівномірно неперервна на ньому.

Для неперервної на відрізку $[a, b]$ функції $y = f(x)$ характерні наступні твердження.

***Т е о р е м а  3.** (Вейєрштрасс). Якщо функція $y = f(x)$ неперервна на відрізку $[a, b]$, то вона на ньому обмежена.*

*Д о в е д е н н я*. Припустимо протилежне – існує неперервна функція $f(x)$ на відрізку $[a, b]$, яка необмежена на цьому відрізку. З цього випливає, що для будь-якого числа $N > 0$ існує така точка $x_N \in [a, b]$, що $|f(x_N)| > N$. Вибираючи послідовність $\{N\} = \{1, 2, ..., n, ...\}$, одержимо послідовність точок $\{x_n\} \in [a, b]$, для яких $|f(x_n)| > n$. За теоремою Больцано – Вейєрштрасса з обмеженої послідовності $\{x_n\}$ можна виділити збіжну підпослідовність $\{x_{n_k}\}$, $\lim\limits_{k \to \infty} x_{n_k} = x_0$. З умови $a \leq x_{n_k} \leq b$ випливає, що $a \leq x_0 \leq b$. Зазначимо, що $\left|f(x_{n_k})\right| > n_k$ і, відповідно, $\lim\limits_{k \to \infty} n_k = +\infty$, а отже, $\lim\limits_{k \to \infty} f(x_{n_k}) = \infty$.

З іншого боку, внаслідок неперервності функції $f(x)$ в точці $x_0 \in [a, b]$, маємо $-\infty < \lim\limits_{k \to \infty} f(x_{n_k}) = f(x_0) < +\infty$. Одержане протиріччя доводить теорему.

**Т е о р е м а 4. (Вейєрштрасс).** *Якщо функція $y = f(x)$ неперервна на відрізку $[a, b]$, то вона приймає на цьому відрізку як найменше, так і найбільше значення.*

Д о в е д е н н я. Якщо функція $y = f(x)$ неперервна на відрізку $[a, b]$, то за теоремою 3 вона обмежена, а отже, $\sup\limits_{x \in [a, b]} f(x) = M < +\infty$.

Припустимо, що функція $y = f(x)$ не досягає своєї верхньої грані, тобто $f(x) \neq M$ для всіх $x \in [a, b]$. Тоді функція $\varphi(x) = \dfrac{1}{M - f(x)}$ неперервна на відрізку $[a, b]$.

За означенням верхньої грані функції $y = f(x)$ на відрізку $[a, b]$ величина $M - f(x)$ може бути вибрана (за рахунок $x \in [a, b]$) як завгодно малою, тобто: для довільного $\varepsilon > 0$ існує точка $x_\varepsilon \in [a, b]$ така, що $0 \leq M - f(x_\varepsilon) < \varepsilon$, а отже, $\varphi(x_\varepsilon) = \dfrac{1}{M - f(x_\varepsilon)} > \dfrac{1}{\varepsilon}$. Внаслідок того, що $\varepsilon$ – довільне мале число і, відповідно, $\dfrac{1}{\varepsilon}$ – велике число, функція $\varphi(x)$ необмежена на відрізку $[a, b]$, що суперечить теоремі 3. Отже, існує хоч би одна точка $\xi \in [a, b]$ така, що $f(\xi) = M$.

Якщо $\inf\limits_{x \in [a, b]} f(x) = m$, то $\sup\limits_{x \in [a, b]} [-f(x)] = -m$ і, ґрунтуючись

на вже доведеному твердженні, існує точка $\eta \in [a, b]$, що $-f(\eta) = -m$, тобто $f(\eta) = m$.

Теорему доведено.

*З а у в а ж е н н я*. Якщо функція неперервна на інтервалі або пів-інтервалі і, крім того, обмежена на ньому, то вона може не мати найбільшого чи найменшого значення. Наприклад, функція $y = x$ на інтервалі $(0, 1)$.

***Т е о р е м а 5***. *(Коші). Якщо функція $y = f(x)$ неперервна на відрізку $[a, b]$ і $f(a) = A$, $f(b) = B$, то для довільного числа $C$, що лежить між $A$ і $B$, існує точка $\xi \in [a, b]$, така що $f(\xi) = C$.*

*Д о в е д е н н я*. Нехай для визначеності $A < C < B$. Розділимо відрізок $[a, b]$ на дві рівні частини точкою $x_0$ і знайдемо значення функції в цій точці. Якщо $f(x_0) = C$, то точка $\xi = x_0$ знайдена. Якщо $f(x_0) = C_1 \neq C$, то вибираємо ту половину відрізка $[a, b]$, на кінцях якої значення функції лежать з різних боків числа $C$, і позначимо її через $[a_1, b_1]$.

Розділимо відрізок $[a_1, b_1]$ знову на дві частини і т. д. В результаті такого поділу прийдемо до шуканої точки $\xi$, в якій $f(\xi) = C$ або одержимо послідовність вкладених відрізків $[a_n, b_n]$, довжина яких прямує до нуля і таких, що

$$f(a_n) < C < f(b_n). \quad (1.2)$$

Нехай $\xi$ – спільна точка системи відрізків $[a_n, b_n]$, $n = 1, 2, \ldots$. Тоді

$$\xi = \lim_{n \to \infty} a_n = \lim_{n \to \infty} b_n.$$

Оскільки функція $y = f(x)$ неперервна, знайдемо

$$f(\xi) = \lim_{n \to \infty} f(a_n) = \lim_{n \to \infty} f(b_n),$$

а, переходячи в нерівності (1.2) до границі, одержимо

$$\lim_{n \to \infty} f(a_n) \leq C \leq \lim_{n \to \infty} f(b_n).$$

З останніх двох співвідношень випливає $f(\xi) = C$.

Теорему доведено.

***Н а с л і д о к .*** *Якщо функція* $y = f(x)$ *неперервна на відрізку* $[a, b]$, *то множина значень цієї функції є відрізок* $[m, M]$, *де* $m$ *і* $M$ – *найменше і найбільше значення функції відповідно.*

*Д о в е д е н н я .* Якщо $m = \min\limits_{x \in [a,b]} f(x)$ і $M = \max\limits_{x \in [a,b]} f(x)$, то за теоремою 4 існують точки $\alpha \in [a, b]$ і $\beta \in [a, b]$, що $f(\alpha) = m$ і $f(\beta) = M$. Тепер твердження наслідку випливає безпосередньо з теореми 5, застосованої до відрізка $[\alpha, \beta]$.

***Т е о р е м а  6 (Больцано – Коші).*** *Якщо на кінцях відрізка значення неперервної функції протилежні за знаком, то існує хоча б одна точка* $x_0 \in (a, b)$ *така, що* $f(x_0) = 0$.

*Д о в е д е н н я .* Сформульована теорема є наслідком теореми 5, оскільки значення $C = 0$ лежить між протилежними за знаком значеннями функції на кінцях відрізка.

***Т е о р е м а  7 (Кантор).*** *Якщо функція* $y = f(x)$ *неперервна на відрізку* $[a, b]$, *то вона рівномірно неперервна на цьому відрізку.*

*Д о в е д е н н я .* Припустимо, що функція $y = f(x)$ неперервна на відрізку $[a, b]$, однак не є рівномірно неперервною на цьому відрізку. Тоді існує число $\varepsilon > 0$, що для достатньо малого $\delta > 0$ знайдуться дві точки $x', x'' \in [a, b]$ такі, що $|x' - x''| < \delta$, однак $|f(x') - f(x'')| \geq \varepsilon$.

Виберемо нескінченно малу послідовність $\delta_n = \dfrac{1}{n}$, $n = 1, 2, \ldots$. Тоді можна стверджувати, що для розглянутого $\varepsilon$ і для будь-якого номера $n$ знайдуться дві точки $x'_n, x''_n \in [a, b]$ такі, що

$$|x'_n - x''_n| < \delta_n, \text{ однак } |f(x'_n) - f(x''_n)| \geq \varepsilon. \qquad (1.3)$$

Оскільки послідовність $\{x'_n\} \subset [a, b]$ обмежена, за теоремою Больцано – Вейєрштрасса з неї можна виділити збіжну підпослідовність $\{x'_{n_k}\} \subset [a, b]$, $k = 1, 2, \ldots$, таку, що $\lim\limits_{k \to \infty} x'_{n_k} = \xi$ і $\xi \in [a, b]$. Внаслідок першої нерівності (1.3) підпослідовність $\{x''_{n_k}\} \subset [a, b]$ також збігається до точки $\xi$.

Оскільки функція $y = f(x)$ неперервна на відрізку $[a,b]$, вона також неперервна в точці $\xi$. Тоді за означенням неперервності функції за Гейне обидві послідовності функцій $\{f(x'_{n_k})\}$ і $\{f(x''_{n_k})\}$ повинні збігатися до $f(\xi)$ і, відповідно, $\{f(x'_{n_k}) - f(x''_{n_k})\}$ – нескінченно мала величина. Цей результат суперечить другій нерівності (1.3).

Отже, припущення про те, що неперервна на відрізку $[a,b]$ функція не є рівномірно неперервною не правильне.

Теорему доведено.

**1.1.3. Диференційовні функції.** Нехай функція $y = f(x)$ визначена на проміжку $(a,b)$ і $x_0, x_0 + \Delta x$ – дві точки цього проміжку.

*О з н а ч е н н я  6 . Якщо існує границя відношення приросту функції $\Delta y = f(x_0 + \Delta x) - f(x_0)$ в точці $x_0$ до приросту аргумента $\Delta x$ в цій точці, то ця границя називається похідною від функції $y = f(x_0)$ в точці $x_0$ і позначається*

$$f'(x_0) = \lim_{\Delta x \to 0} \frac{\Delta y}{\Delta x}.$$

*Якщо функція $y = f(x)$ визначена справа (зліва) в точці $x_0$ і існує скінчена або нескінченна границя*

$$\lim_{\Delta x \to +0} \frac{f(x_0 + \Delta x) - f(x_0)}{\Delta x} \quad \left(\lim_{\Delta x \to -0} \frac{f(x_0 + \Delta x) - f(x_0)}{\Delta x}\right),$$

*то ця границя називається, відповідно, скінченною або нескінченною похідною справа (зліва) від функції в точці $x_0$.*

Під виразом "функція має похідну" розуміємо, що вона має скінчену похідну.

Функція $y = f(x)$ називається *диференційовною в точці* $x_0$, якщо її приріст в цій точці можна зобразити у вигляді

$$\Delta y = A(x_0)\Delta x + \varepsilon(\Delta x)\Delta x, \qquad (1.4)$$

де $A(x_0)$ – скінченна величина (число); $\varepsilon(\Delta x)$ – нескінченно мала величина.

Лінійна частина приросту функції $y = f(x)$ в точці $x_0$

називається *диференціалом функції* в цій точці, $dy = A(x_0)\Delta x$, або з урахуванням того, що приріст незалежної змінної дорівнює її диференціалу, $\Delta x = dx$, записують так
$$dy = A(x_0)dx. \qquad (1.5)$$

**Т е о р е м а  8 .** *Необхідною і достатньою умовою диференційовності функції в точці $x_0$ є існування похідної в цій точці, при цьому для диференціала справедлива формула*
$$dy = f'(x_0)dx. \qquad (1.6)$$

*Н е о б х і д н і с т ь .* Нехай функція $y = f(x)$ диференційовна в точці $x_0$, тобто справедлива формула (1.4). Тоді
$$\lim_{\Delta x \to 0} \frac{\Delta y}{\Delta x} = A(x_0) + \lim_{\Delta x \to 0} \varepsilon(\Delta x) = A(x_0).$$
Тому $f'(x_0) = \lim_{\Delta x \to 0} \frac{\Delta y}{\Delta x} = A(x_0)$. Звідси з урахуванням формули (1.5), одержимо формулу (1.6).

*Д о с т а т н і с т ь .* Нехай існує похідна, тобто існує границя $f'(x_0) = \lim_{\Delta x \to 0} \frac{\Delta y}{\Delta x}$. Тоді $\frac{\Delta y}{\Delta x} = f'(x_0) + \varepsilon(\Delta x)$, де $\lim_{\Delta x \to 0} \varepsilon(\Delta x) = 0$. Для $\Delta x \neq 0$ маємо $\Delta y = A(x_0)\Delta x + \varepsilon(\Delta x)\Delta x$ і $\varepsilon(\Delta x)$ – нескінченно мала величина. Це підтверджує диференційовність функції $y = f(x)$ в точці $x_0$.

Теорему доведено.

Якщо функція $y = f(x)$ диференційовна в кожній точці інтервалу $(a, b)$, то вона називається *диференційовною на інтервалі* і, крім цього, якщо існують скінченні похідні справа в точці $a$ і зліва в точці $b$, то функція називається *диференційовною на сегменті* $[a, b]$.

Наступне твердження встановлює зв'язок між диференційовністю і неперервністю функції в точці.

**Т е о р е м а  9 .** *Якщо функція диференційовна в точці, то вона неперервна в цій точці.*

*Д о в е д е н н я .* З означення диференційовності функції випливає рівність (1.4), яку можна записати у вигляді $\Delta y = f'(x_0)\Delta x + \varepsilon(\Delta x)\Delta x$. Переходячи тут до границі, одержимо

$$\lim_{\Delta x \to 0} \Delta y = \lim_{\Delta x \to 0} A(x_0)\Delta x + \lim_{\Delta x \to 0} \varepsilon(\Delta x)\Delta x = 0.$$

Границя приросту функції в точці $x_0$ дорівнює нулю. Тому за означенням функція $y = f(x)$ неперервна в точці $x_0$.

Теорему доведено.

*З а у в а ж е н н я .* З неперервності функції в точці не випливає її диференційовність в цій точці.

Наступні властивості похідних визначають правила диференціювання функцій.

***Т е о р е м а 10 .*** *Якщо функції $f(x)$ і $g(x)$ в точці $x$ мають похідні, то в цій точці похідні також мають:*

а) *функція $y = f(x) \pm g(x)$ і справедлива формула*
$$y' = f'(x) \pm g'(x);$$

б) *функція $y = f(x)g(x)$ і справедлива формула*
$$y' = f'(x)g(x) + f(x)g'(x);$$

в) *функція $y = \dfrac{f(x)}{g(x)}$ за умови $g(x) \neq 0$ і справедлива формула*
$$y' = \frac{f'(x)g(x) - f(x)g'(x)}{[g(x)]^2}.$$

*Д о в е д е н н я .* Відповідні твердження випливають безпосередньо з означення похідної і властивостей границь функцій.

Доведемо, наприклад, формулу для обчислення похідної від функції $y = \dfrac{f(x)}{g(x)}$. Надамо аргументу $x$ приріст $\Delta x$. Тоді функції матимуть прирости $\Delta f = f(x + \Delta x) - f(x)$ і $\Delta g = g(x + \Delta x) - g(x)$. Звідси маємо $f(x + \Delta x) = f(x) + \Delta f$, $g(x + \Delta x) = g(x) + \Delta g$ і знайдемо приріст частки

$$\Delta y = \frac{f(x + \Delta x)}{g(x + \Delta x)} - \frac{f(x)}{g(x)} = \frac{f(x) + \Delta f}{g(x) + \Delta g} - \frac{f(x)}{g(x)} = \frac{g(x)\Delta f - f(x)\Delta g}{[g(x) + \Delta g]g(x)}.$$

Знайдемо відношення

$$\frac{\Delta y}{\Delta x} = \frac{g(x)\dfrac{\Delta f}{\Delta x} - f(x)\dfrac{\Delta g}{\Delta x}}{[g(x) + \Delta g]\,g(x)}.$$

За умовою теореми функції $f(x)$ і $g(x)$ мають похідні в точці $x$, $f'(x) = \lim\limits_{\Delta x \to 0} \dfrac{\Delta f}{\Delta x}$, $g'(x) = \lim\limits_{\Delta x \to 0} \dfrac{\Delta g}{\Delta x}$, $\lim\limits_{\Delta x \to 0} \Delta g = 0$ і в цій точці $g(x) \neq 0$. Тому

$$\lim_{\Delta x \to 0} \frac{\Delta y}{\Delta x} = \frac{g(x) \lim\limits_{\Delta x \to 0} \dfrac{\Delta f}{\Delta x} - f(x) \lim\limits_{\Delta x \to 0} \dfrac{\Delta g}{\Delta x}}{\left[g(x) + \lim\limits_{\Delta x \to 0} \Delta g\right] g(x)} = \frac{g(x)f'(x) - f(x)g'(x)}{[g(x)]^2}.$$

З цієї рівності й випливає відповідна формула пункту в).
Теорему доведено.

**Т е о р е м а  11.** *Якщо функції $y = f(u)$ і $u = g(x)$ мають похідні, відповідно, в точках $u_0 = g(x_0)$ і $x_0$, то складна функція $y = f(g(x))$ також має похідну і справедлива формула*

$$y' = f'_u(u_0) g'(x_0).$$

*Д о в е д е н н я.* Оскільки функція $y = f(u)$ в точці $u_0 = g(x_0)$ має похідну $y'_u(u_0)$ і, відповідно, диференційовна в цій точці, її приріст згідно (1.4) можна записати у вигляді

$$\Delta y = y'_u(u_0) \Delta u + \varepsilon(\Delta u) \Delta u.$$

Розділимо цей вираз на $\Delta x$ і перейдемо до границі при $\Delta x \to 0$

$$\lim_{\Delta x \to 0} \frac{\Delta y}{\Delta x} = y'_u(u_0) \lim_{\Delta x \to 0} \frac{\Delta u}{\Delta x} + \lim_{\Delta x \to 0} \varepsilon(\Delta u) \frac{\Delta u}{\Delta x}.$$

Враховуючи, що існує похідна від функції $u = g(x)$ в точці $x_0$, $\lim\limits_{\Delta x \to 0} \dfrac{\Delta u}{\Delta x} = g'(x_0) = u'(x_0)$ і $\lim\limits_{\Delta x \to 0} \alpha(\Delta u) = 0$, оскільки $\lim\limits_{\Delta x \to 0} \Delta u = 0$, одержимо

$$\lim_{\Delta x \to 0} \frac{\Delta y}{\Delta x} = y'_u(u_0) u'_x(x_0) \text{ або } y' = f'(u_0) g'(x_0).$$

Теорему доведено.

Одне з центральних місць у математичному аналізі займають теореми про середнє значення.

**Т е о р е м а  12 (Ролль).** *Нехай функція $f(x)$ задовольняє умови: а) визначена і неперервна на відрізку $[a, b]$; б) диференційовна в інтервалі $(a, b)$; в) на кінцях відрізка набуває* однакових значень $f(a) = f(b)$.

*Тоді всередині інтервалу $(a, b)$ знайдеться хоча б одна точка $c \in (a, b)$, в якій $f'(c) = 0$.*

Д о в е д е н н я. Якщо функція $f(x)$ є сталою на відрізку $[a, b]$, то $f'(x) = 0$ і за точку $c$ можна взяти будь-яку точку інтервалу $(a, b)$.

Довільна неперервна функція за теоремою 4 на відрізку $[a, b]$ набуває найбільшого $M$ і найменшого $m$ значень, $m < M$. Через те, що $f(a) = f(b)$, то хоча б одне з чисел $M$ і $m$ досягається функцією всередині інтервалу $(a, b)$. Нехай, наприклад, число $m$ досягається в точці $c \in (a, b)$, $f(c) = m$.

Покажемо, що $f'(c) = 0$. Очевидно існує окіл $(c - \delta, c + \delta) \subset (a, b)$, що для всіх $x$ з цього околу $f(x) \geq f(c)$ і, відповідно, $f(x) - f(c) \geq 0$, якщо $x > c$, і $f(x) - f(c) \geq 0$, якщо $x < c$. Тоді

$$\frac{f(x) - f(c)}{x - c} \geq 0, \text{ якщо } x > c, \quad \frac{f(x) - f(c)}{x - c} \leq 0, \text{ якщо } x < c.$$

Перейдемо у цих нерівностях до границі при $x \to c$. Границя лівих частин нерівностей існує, оскільки існує похідна, і дорівнює похідній $f'(c)$. Одержимо

$$f'(c) \geq 0 \text{ і } f'(c) \leq 0.$$

Звідси випливає рівність $f'(c) = 0$.

Теорему доведено.

З а у в а ж е н н я. Теорема Ролля має геометричну інтерпретацію. Якщо функція задовольняє умови теореми, то її графік суцільна гладка лінія, крайні точки якої знаходяться на однаковій віддалі від осі $Ox$ (крива називається гладкою, якщо в кожній її точці можна провести дотичну). Тоді на графіку знайдеться хоч би одна точка, в якій дотична паралельна до осі $Ox$.

**Т е о р е м а 13 (Лагранж).** *Нехай функція $f(x)$ задовольняє умови: а) визначена і неперервна на відрізку $[a, b]$; б) диференційовна в інтервалі $(a, b)$.*

*Тоді в інтервалі* $(a,b)$ *знайдеться хоча б одна точка* $c \in (a,b)$, *в якій справджується рівність*

$$\frac{f(b)-f(a)}{b-a} = f'(c). \qquad (1.7)$$

*Д о в е д е н н я*. Розглянемо допоміжну функцію

$$F(x) = f(x) - \frac{f(b)-f(a)}{b-a}x.$$

Функція $F(x)$ задовольняє умови теореми Ролля. Справді,

$$F(a) = F(b) = \frac{bf(a)-af(b)}{b-a}$$

і функція $F(x)$ на відрізку $[a,b]$ неперервна як різниця двох неперервних функцій. Всередині інтервалу $(a,b)$ функція $F(x)$ має похідну

$$F'(x) = f'(x) - \frac{f(b)-f(a)}{b-a},$$

оскільки існує похідна $f'(x)$.

Отже, за теоремою Ролля існує точка $c \in (a,b)$, що $F'(c) = 0$, звідси випливає формула (1.7).

Теорему доведено.

*З а у в а ж е н н я  1*. Теоремі Лагранжа можна дати геометричну інтерпретацію. Очевидно, що точки $A(a, f(a))$ і $B(b, f(b))$ є кінцевими точками дуги $AB$ графіка функції $y = f(x)$. Розглянемо також хорду $AB$, кутовий коефіцієнт якої

$$tg\,\alpha = \frac{f(b)-f(a)}{b-a},$$

де $\alpha$ – кут, що утворює хорда з віссю $Ox$.

З теореми Лагранжа випливає, що існує значення $x = c$, таке що дотична (з кутовим коефіцієнтом $k = f'(c)$) до графіка функції $y = f(x)$, проведена в точці $C(c, f(c))$, паралельна до хорди $AB$,

$$f'(c) = tg\,\alpha.$$

*З а у в а ж е н н я  2*. Якщо у формулі (1.7) прийняти $a = x_0$, $b-a = \Delta x$ і, відповідно, $b = x_0 + \Delta x$, то

$$f(x_0 + \Delta x) = f(x_0) + f'(c)\Delta x,$$

яка називається *формулою скінченних приростів*. Її можна записати ще у такому вигляді

$$f(x_0 + \Delta x) = f(x_0) + f'(x_0 + \theta \Delta x)\Delta x, \quad 0 < \theta < 1.$$

Зазначимо, що наближена формула

$$f(x_0 + \Delta x) \approx f(x_0) + f'(x_0)\Delta x$$

називається формулою *нескінченно малих приростів*.

***Т е о р е м а  14 (Коші).*** *Нехай функції* $f(x)$, $g(x)$ *задовольняють умови: а) визначені і неперервні на відрізку* $[a, b]$; *б) диференційовні в інтервалі* $(a, b)$; *в) похідна* $g'(x)$ *всередині інтервалу* $(a, b)$ *не дорівнює нулю.*

*Тоді в інтервалі* $(a, b)$ *знайдеться хоча б одна точка* $c \in (a, b)$, *в якій виконується рівність*

$$\frac{f(b) - f(a)}{g(b) - g(a)} = \frac{f'(c)}{g'(c)}. \qquad (1.8)$$

*Д о в е д е н н я .* З умов теореми випливає, що $g(a) \neq g(b)$. Дійсно, якщо $g(a) = g(b)$, то функція $g(x)$ задовольняє умови теореми Ролля і знайдеться точка $\xi \in (a, b)$, в якій $g'(\xi) = 0$, а це суперечить третій умові теореми.

Розглянемо допоміжну функцію

$$F(x) = f(x) - \frac{f(b) - f(a)}{g(b) - g(a)} g(x).$$

Легко бачити, що функція $F(x)$ задовольняє умови теореми Ролля. Тому існує точка $c \in (a, b)$, що $F'(c) = 0$ або, що те саме

$$f'(c) - \frac{f(b) - f(a)}{g(b) - g(a)} g'(c) = 0.$$

Звідси випливає формула (1.8).

Теорему доведено.

***Т е о р е м а  15 .*** *Нехай функція* $f(x)$ *визначена в інтервалі* $(a, b)$ *і має в точці* $x_0 \in (a, b)$ *похідні до* $(n + 1)$-*го порядку включно.*

*Тоді справедлива формула Тейлора*

$$f(x) = \sum_{k=0}^{n} \frac{f^{(k)}(x_0)}{k!}(x-x_0)^k + R_n(x), \qquad (1.9)$$

*де*

$$R_n(x) = \frac{f^{(n+1)}(c)}{(n+1)!}(x-x_0)^{n+1}, \; c = x_0 + \theta(x-x_0), \; 0 < \theta < 1, \; (1.10)$$

*– додатковий член у формі Лагранжа.*

*Д о в е д е н н я .* Фіксуємо довільне значення $x$ з відрізку $(a, b)$. Для визначеності вважаємо, що $x > x_0$ (для значень $x < x_0$ доводимо аналогічно). Позначимо через $t$ змінну величину, що приймає значення з відрізка $[x_0, x]$ і введемо функцію

$$\varphi(t) = f(x) - \sum_{k=0}^{n} \frac{f^{(k)}(t)}{k!}(x-t)^k.$$

Розглянемо також функцію $\psi(t) = (x-t)^{n+1}$.

Легко бачити, що функція $\varphi(t)$ має на відрізку $[x_0, x]$ похідну

$$\varphi'(t) = -\left[ f'(t) + f''(t)(x-t) + \frac{f'''(t)}{2!}(x-t)^2 + \ldots + \frac{f^{(n+1)}(t)}{n!}(x-t)^n \right] +$$

$$+ f'(t) + f''(t)(x-t) + \frac{f'''(t)}{2!}(x-t)^2 + \ldots + \frac{f^{(n)}(t)}{(n+1)!}(x-t)^{n-1} =$$

$$= -\frac{f^{(n+1)}(t)}{n!}(x-t)^n.$$

До функцій $\varphi(t)$ і $\psi(t)$ застосуємо на відрізку $[x_0, x]$ теорему Коші

$$\frac{\varphi(x) - \varphi(x_0)}{\psi(x) - \psi(x_0)} = \frac{\varphi'(c)}{\psi'(c)}, \; c = x_0 + \theta(x-x_0), \; 0 < \theta < 1.$$

Підставивши сюди величини

$$\varphi(x) = 0, \; \varphi(x_0) = R_n(x), \; \psi(x_0) = (x-x_0)^{n+1},$$

$$\varphi'(c) = -\frac{f^{(n+1)}(c)}{n!}(x-c)^n, \; \psi'(c) = -(n+1)(x-c)^n,$$

одержимо

$$\frac{-R_n(x)}{-(x-x_0)^{n+1}} = \frac{-f^{(n+1)}(c)(x-c)^n}{-n!(n+1)(x-c)^n}.$$

Звідси $R_n(x) = \frac{f^{(n+1)}(c)}{(n+1)!}(x-x_0)^{n+1}$, $c = x_0 + \theta(x-x_0)$, $0 < \theta < 1$.

Теорему доведено.

**1.1.4. Модуль неперервності функції.** Нехай функція $f(x)$ визначена і неперервна на відрізку $[a, b]$. Множину всіх неперервних на відрізку $[a, b]$ функцій позначають через $C[a,b]$.

*О з н а ч е н н я  7 .* Для кожного $\delta > 0$ модулем неперервності функції $f(x)$ на відрізку $[a, b]$ називається точна верхня грань модуля різниці $|f(x') - f(x'')|$ на множині всіх точок $x', x'' \in [a, b]$, що задовольняють умову $|x' - x''| < \delta$, і записуємо

$$\omega(\delta, f) = \sup_{\substack{|x'-x''|<\delta \\ x', x'' \in [a, b]}} |f(x') - f(x'')|. \qquad (1.11)$$

У термінах модуля неперервності рівномірна неперервність виражається наступним чином [8].

*Т е о р е м а  16 .* Для того, щоби функції $f(x)$, неперервна на відрізку $[a, b]$, була рівномірно неперервною на цьому відрізку, необхідно і достатньо, щоби

$$\lim_{\delta \to 0} \omega(\delta, f) = 0. \qquad (1.12)$$

*Н е о б х і д н і с т ь .* Нехай функція $f(x)$ рівномірно неперервна на відрізку $[a, b]$. Тоді для будь-якого числа $\varepsilon > 0$ існує число $\delta(\varepsilon) > 0$ таке, що для будь-яких двох точок $x', x' \in [a, b]$, які задовольняють нерівність $|x' - x''| < \delta(\varepsilon)$, справджується нерівність

$$|f(x') - f(x'')| < \varepsilon.$$

Звідси випливає, що для будь-якого $\delta < \delta(\varepsilon)$ виконується нерівність

$$\omega(\delta, f) = \sup_{\substack{|x'-x''|<\delta \\ x', x'' \in [a, b]}} |f(x') - f(x'')| < \varepsilon,$$

тобто $\lim_{\delta \to 0} \omega(\delta, f) = 0$. Необхідність умови (1.12) доведено.

*Д о с т а т н і с т ь .* Нехай виконується умова (1.12). Тоді для будь-якого числа $\varepsilon > 0$ існує число $\delta(\varepsilon) > 0$ таке, що як тільки $0 < \delta < \delta(\varepsilon)$, то

$$\omega(\delta, f) = \sup_{\substack{|x'-x''|<\delta \\ x', x'' \in [a,b]}} |f(x') - f(x'')| < \varepsilon.$$

З виконання цієї нерівності випливає, що для $\varepsilon > 0$ існує $\delta > 0$ таке, що для всіх $x', x' \in [a, b]$, що задовольняють умову $|x' - x''| < \delta$, виконається нерівність $|f(x') - f(x'')| < \varepsilon$, тобто функція $f(x)$ рівномірно неперервна на відрізку $[a, b]$.

Теорему доведено.

***Т е о р е м а 17 .*** *Якщо функція $f(x)$ неперервна на відрізку $[a, b]$ і її похідна обмежена на цьому відрізку, то модуль неперервності функції $f(x)$ має оцінку*

$$\omega(\delta, f) \leq M \delta, \qquad (1.13)$$

*де $M$ – стала.*

*Д о в е д е н н я .* З теореми Лагранжа випливає, що для будь-яких двох точок $x', x' \in [a, b]$ знайдеться точка $\xi$, яка лежить між ними і така, що

$$|f(x') - f(x'')| = f'(\xi)|x' - x''|.$$

Оскільки похідна $f'(x)$ обмежена на $[a, b]$, то знайдеться стала $M$ така, що $|f'(x)| \leq M$ для всіх $x \in [a, b]$ і з врахуванням записаної рівності маємо

$$|f(x') - f(x'')| \leq M |x' - x''| < M \delta$$

для всіх $x', x' \in [a, b]$, що задовольняють умову $|x' - x''| < \delta$. Ця рівність стверджує оцінку (1.13).

Теорему доведено.

***О з н а ч е н н я  8 .*** *Неперервна функція $f(x)$ на відрізку $[a, b]$ належить класу Ліпшіца $C_1[a, b]$, якщо існує стала $M$, $0 < M < +\infty$ , така, що для будь яких двох точок $x_1, x_2 \in [a, b]$ виконується нерівність*

$$|f(x_1) - f(x_2)| \le M|x_1 - x_2|. \qquad (1.14)$$

З теореми 17 випливає, що неперервна на сигменті $[a, b]$ функція $f(x)$ з обмеженою похідною належить класу Ліпшіца.

*П р и к л а д  3*. Знайти модуль неперервності функції $y = x^2$ на проміжку $[1, +\infty)$ і дослідити її на рівномірну неперервність.

Для довільного $\delta > 0$ і будь-яких фіксованих точок $x'$ і $x''$ таких, що $1 \le x' - \delta \le x'' \le x' < +\infty$ маємо

$$\omega(\delta, x^2) = \sup_{|x'-x''| \le \delta} |x'^2 - x''^2| \ge |x'^2 - (x' - \delta)^2| = |2x'\delta - \delta^2|.$$

Якщо $x \in [1, +\infty)$, то $\omega(\delta, x^2) = +\infty$. Оскільки рівність (1.12) не виконується, функція $y = x^2$ не є рівномірно неперервною на проміжку $[1, +\infty)$.

Зауважимо, що на будь-якому скінченному проміжку функція $y = x^2$ рівномірно неперервна.

*П р и к л а д  4*. Знайти модуль неперервності і дослідити функцію $y = \sin\dfrac{1}{x}$ на рівномірну неперервність на проміжку $(0, \infty)$.

Розглянемо довільні точки $x'$ і $x''$ з цього проміжку, що задовольняють умову $|x' - x''| < \delta$. Тоді

$$\omega\left(\delta, \sin\frac{1}{x}\right) = \sup_{|x'-x''| \le \delta} \left|\sin\frac{1}{x'} - \sin\frac{1}{x''}\right| \le \sup_{|x'-x''| \le \delta}\left[\left|\sin\frac{1}{x'}\right| + \left|\sin\frac{1}{x''}\right|\right] \le$$
$$\le \sup_{|x'-x''| \le \delta} 2 = 2.$$

Отже, функція $y = \sin\dfrac{1}{x}$ не є рівномірно неперервною на інтервалі $(0, \infty)$.

### 1.1.5. Завдання до першого параграфа

1. Показати, що $x = 0$ є точкою розриву першого роду функції
$$y = \begin{cases} \dfrac{|x|}{x}, & x \neq 0, \\ 0, & x = 0. \end{cases}$$

2. Показати, що $x = 0$ є точкою усувного розриву функції
$$y = e^{-\frac{1}{x^2}}.$$

3. Показати, що $x = 0$ є точкою розриву другого роду функції $y = e^{-\frac{1}{x}}$.

4. Знайти модуль неперервності функції $y = x^2$ і дослідити її на рівномірну неперервність на проміжку $[0, 1]$.

5. Довести, що функція $y = |x|$ у точці $x = 0$ похідної не має.

6. Неперервні похідні якого порядку мають функції: а) $y = x\sqrt[3]{x}$, $x \in (-1, 1)$; б) $y = \left|x^3\right|$, $x \in (-1, 1)$.

### 1.2. Інтеграл Рімана

**1.2.1. Властивості інтеграла.** Нехай функція $f(x)$ визначена на проміжку $[a, b]$. Розіб'ємо довільно цей проміжок на $n$ частин точками
$$x_0 = a < x_1 < x_2 < \ldots < x_{i-1} < x_i < \ldots < x_n = b.$$
Найбільшу з довжин $\Delta x_i = x_i - x_{i-1}$, $i = 1, 2, \ldots n$, позначимо через $\lambda$. Виберемо в кожному з проміжків $[x_{i-1}, x_i]$ точку $x = \xi_i$ і складемо суму
$$\sigma = \sum_{i=0}^{n} f(\xi_i) \Delta x_i.$$

*О з н а ч е н н я  1 . Скінченна границя суми $\sigma$ при $\lambda \to 0$ називається визначеним інтегралом (Рімана) від функції $f(x)$ на проміжку $[a, b]$,*

$$\int\limits_a^b f(x)dx = \lim_{\lambda \to 0} \sigma.$$

Як допоміжний засіб дослідження інтегралів виступають суми Дарбу $[6, 12, 23]$. Позначимо через $m_i$ і $M_i$, відповідно, нижню і верхню межі функції $f(x)$ на проміжку $[x_{i-1}, x_i]$,

$$m_i = \inf_{x \in [x_{i-1}, x_i]} f(x), \ M_i = \sup_{x \in [x_{i-1}, x_i]} f(x), \ i = 1, 2, ..., n,$$

і складемо суми

$$s = \sum_{i=0}^n m_i \Delta x_i, \ S = \sum_{i=0}^n M_i \Delta x_i, \qquad (1.15)$$

які називаються *нижньою і верхньою інтегральними сумами* або *сумами Дарбу*.

Суми Дарбу для інтегровної функції мають такі властивості:

1) якщо до існуючих точок поділу відрізка додати нові точки, то нижня сума Дарбу може тільки зрости, а верхня сума може тільки зменшитися;

2) кожна верхня сума Дарбу не менша від кожної нижньої суми Дарбу, хоч би для іншого розбиття відрізка.

3) яке б не було розбиття справедлива нерівність $s \le \sigma \le S$.

Виходячи із цих властивостей випливає, що множина $\{s\}$ всіх нижніх сум обмежена зверху, наприклад, будь-якою верхньою сумою $S$. Аналогічно множина $\{S\}$ всіх верхніх сум обмежена знизу, наприклад, будь-якою нижньою сумою $s$. Тоді ці множини мають, відповідно, скінченні точну верхню грань і точну нижню грань

$$I_* = \sup\{s\}, \ I^* = \inf\{S\},$$

які для будь-яких сум Дарбу справджують умову

$$s \le I_* \le I^* \le S. \qquad (1.16)$$

Якщо позначити через $\omega_i = M_i - m_i$ коливання функції на $i$-му проміжку розбиття, то можна записати

$$S - s = \sum_{i=1}^n \omega_i \Delta x_i.$$

Тоді умову існування визначеного інтеграла можна записати

у вигляді

$$\lim_{\lambda \to 0} \sum_{i=1}^{n} \omega_i \Delta x_i = 0 . \qquad (1.17)$$

**Т е о р е м а 1 .** *Якщо функція* $y = f(x)$ *інтегровна на відрізку* $[a, b]$, *то вона обмежена на цьому відрізку.*

*Д о в е д е н н я .* Доведемо методом від супротивного. Нехай функція $f(x)$ необмежена на відрізку $[a, b]$ і нехай зафіксовано деяке розбиття відрізка. Тоді функція необмежена хоч би на одному з відрізків, наприклад, на відрізку $[x_0, x_1]$. На цьому відрізку існує послідовність $\{\xi_1^{(k)}\} \subset [x_0, x_1]$, $k = 1, 2, \ldots$, така, що

$$\lim_{k \to \infty} f(\xi_1^{(k)}) = +\infty . \qquad (1.18)$$

Зафіксуємо точки $\xi_i \in [x_{i-1}, x_i]$, $i = 2, 3, \ldots, n$, і побудуємо інтегральну суму

$$\sigma(f; \xi_1^{(k)}, \xi_2, \ldots, \xi_n) = f(\xi_1^{(k)}) \Delta x_1 + \sum_{i=2}^{n} f(\xi_i) \Delta x_i .$$

Перейдемо тут до границі при $k \to \infty$ і врахуємо граничну рівність (1.18),

$$\lim_{k \to \infty} \sigma(f; \xi_1^{(k)}, \xi_2, \ldots, \xi_n) = \lim_{k \to \infty} \left[ f(\xi_1^{(k)}) \Delta x_1 + \sum_{i=2}^{n} f(\xi_i) \Delta x_i \right] = +\infty .$$

Одержана суперечність доводить обмеженість функції. Теорему доведено.

*З а у в а ж е н н я .* Умова обмеженості функції є тільки необхідною умовою інтегровності функції. Прикладом обмеженої, однак не інтегровної функції, є функція Діріхле $[23]$

$$f(x) = \begin{cases} 1, & \text{якщо } x - \textit{раціональне число}, \\ 0, & \text{якщо } x - \textit{ірраціональне число}. \end{cases}$$

Розглянемо інтегральну суму для цієї функції на відрізку $[0, 1]$, зафіксувавши певне розбиття. Якщо точки $\xi_i \in [x_{i-1}, x_i]$ є раціональними числами, то одержимо

$$\sigma = \sum_{i=0}^{n} f(\xi_i) \Delta x_i = \sum_{i=0}^{n} \Delta x_i = 1 .$$

Якщо ж $\xi_i$ – ірраціональні числа, то $\sigma = \sum_{i=0}^{n} f(\xi_i)\Delta x_i = 0$. Ці співвідношення справедливі для будь-якого розбиття і, відповідно, інтегральна сума не має границі.

*Т е о р е м а 2 (критерій інтегровності).* *Для того щоби функція $f(x)$ була інтегровною на відрізку $[a, b]$, необхідно і достатньо, щоб виконувалася рівність*

$$\lim_{\lambda \to 0}(S - s) = 0 . \qquad (1.19)$$

*Н е о б х і д н і с т ь* . Нехай обмежена на відрізку $[a, b]$ функція $f(x)$ інтегровна і існує інтеграл

$$I = \int_{a}^{b} f(x)dx .$$

Тоді існує границя

$$\lim_{\lambda \to 0} \sigma = I .$$

Звідси для будь-якого $\varepsilon > 0$ існує $\delta = \delta(\varepsilon)$ таке, що якщо $\lambda < \delta$, то $|\sigma - I| < \varepsilon$ або $I - \varepsilon < \sigma < I + \varepsilon$. За властивістю 3 сум Дарбу і рівностей $\lim_{\lambda \to 0} S = \lim_{\lambda \to 0} s = I$, якщо тільки $\lambda < \delta$, одержимо

$$I - \varepsilon \leq s \leq S \leq I + \varepsilon .$$

Отже, якщо $\lambda < \delta$, то $0 \leq S - s \leq 2\varepsilon$. Звідси випливає рівність (1.19).

*Д о с т а т н і с т ь* . Нехай функція $f(x)$ обмежена і справедлива рівність (1.19). Тоді з умови (1.16) безпосередньо випливає, що $I_* = I^*$ і, позначивши їх значення через $I$, маємо

$$s \leq I \leq S .$$

Якщо $\sigma$ інтегральна сума, що відповідає тому ж розбиттю проміжка, що і суми $s$ і $S$, то за властивістю 3 маємо

$$s \leq \sigma \leq S .$$

Умова (1.19) стверджує, що для достатньо малого $\lambda$ суми $s$ і $S$ відрізняються менше, ніж на довільно вибране $\varepsilon > 0$. Тоді з цих нерівностей випливає

$$|\sigma - I| < \varepsilon ,$$

що підтверджує існування границі інтегральної суми
$$I = \lim_{\lambda \to 0} \sigma$$
і, відповідно, інтегровності функції $f(x)$ на відрізку $[a, b]$.

Теорему доведено.

Відмітимо деякі класи інтегровних функцій [21].

***Т е о р е м а  3 .*** *Якщо функція $f(x)$ неперервна на проміжку $[a, b]$, то вона інтегровна.*

*Д о в е д е н н я* . Оскільки функція $f(x)$ неперервна на проміжку $[a, b]$, то за теоремою Кантора вона рівномірно неперервна і за заданим $\varepsilon > 0$ знайдеться таке $\delta > 0$, що якщо тільки проміжок $[a, b]$ розбити на відрізки $\Delta x_i < \delta$, $i = 1, 2, ..., n$, то $\omega_i < \dfrac{\varepsilon}{b-a}$, де $\omega_i = M_i - m_i$ – коливання функції на $i$-му проміжку розбиття. Звідси

$$S - s = \sum_{i=1}^{n} \omega_i \Delta x_i < \frac{\varepsilon}{b-a} \sum_{i=1}^{n} \Delta x_i = \varepsilon.$$

Оскільки $\varepsilon$ – довільне мале число, то умова (1.17) виконується і, відповідно, функція $f(x)$ інтегровна на проміжку $[a, b]$.

Теорему доведено.

***Т е о р е м а  4 .*** *Якщо функція $f(x)$ обмежена на відрізку $[a, b]$ і має скінченне число точок розриву першого роду, то вона інтегровна.*

*Д о в е д е н н я* . Нехай $\xi_1, \xi_2, ..., \xi_k$ – точки розриву функції $f(x)$, $M$ і $m$ – точна верхня і нижня грані функції $f(x)$ на відрізку $[a, b]$.

Виберемо довільне число $\varepsilon > 0$. Існує число $\delta' < \dfrac{\varepsilon}{2k(M-m)}$ таке, що система околів з центрами в точках розриву

$$\left(\xi_1 - \frac{\delta'}{2},\ \xi_1 + \frac{\delta'}{2}\right), \left(\xi_2 - \frac{\delta'}{2},\ \xi_2 + \frac{\delta'}{2}\right), ..., \left(\xi_k - \frac{\delta'}{2},\ \xi_k + \frac{\delta'}{2}\right),$$

має сумарну довжину, меншу $\dfrac{\varepsilon}{2(M-m)}$.

На решту (замкнутих) проміжках, які будемо називати додатковими проміжками, функція неперервна і задовольняє умови теореми Кантора. Розіб'ємо додаткові проміжки на частини з довжинами $\Delta x_i < \delta$ так, щоб коливання функції на кожному з них задовольняло умову $\omega_i < \dfrac{\varepsilon}{2(b-a)}$, де $\omega_i = M_i - m_i$.

Об'єднуючи розбиття додаткових проміжків і системи околів з приєднаними до них кінцевими точками, одержимо розбиття відрізка $[a, b]$. Для цього розбиття маємо

$$S - s = \sum_{i=1}^{n} \omega_i \, \Delta x_i = {\sum_{i=1}^{n}}' \omega_i \, \Delta x_i + {\sum_{i=1}^{n}}'' \omega_i \, \Delta x_i.$$

В суму з одним штрихом віднесені складові, які відповідають відрізкам розбиття, що покривають точки розриву. З урахуванням оцінки для *коливання функції* на цих відрізках $\omega_i \le M - m$ і сумарної довжини відрізків, маємо оцінку

$${\sum_{i=1}^{n}}' \omega_i \, \Delta x_i < (M - m){\sum_{i=1}^{n}}' \Delta x_i < \dfrac{\varepsilon}{2}.$$

В суму з двома штрихами віднесені складові, що відповідають додатковим проміжкам. Враховуючи оцінку для коливань функції на цих відрізках, одержимо

$${\sum_{i=1}^{n}}'' \omega_i \, \Delta x_i < \dfrac{\varepsilon}{2(b-a)} {\sum_{i=1}^{n}}'' \Delta x_i < \dfrac{\varepsilon}{2}.$$

Отже, існує $\delta > 0$, що якщо $\Delta x_i < \delta$, то $S - s = {\sum_{i=1}^{n}}' \omega_i \, \Delta x_i + {\sum_{i=1}^{n}}'' \omega_i \, \Delta x_i < \varepsilon$. Ця нерівність стверджує, що $f(x)$ інтегровна на відрізку $[a, b]$.

Теорему доведено.

***Т е о р е м а  5 .*** *Монотонна і обмежена на проміжку $[a, b]$ функція $f(x)$ інтегровна.*

*Д о в е д е н н я .* Для визначеності приймемо, що $f(x)$ –

монотонно зростаюча функція. Тоді її коливання на проміжках $[x_{i-1}, x_i]$, будуть
$$\omega_i = f(x_i) - f(x_{i-1}).$$

Задамося $\varepsilon > 0$ і виберемо $\delta = \dfrac{\varepsilon}{f(b) - f(a)}$. Якщо тільки $\Delta x_i < \delta$, то
$$S - s = \sum_{i=1}^{n} \omega_i \Delta x_i < \delta \sum_{i=1}^{n} [f(x_i) - f(x_{i-1})] = \delta [f(b) - f(a)] = \varepsilon.$$
Звідси випливає інтегровність функції.

Теорему доведено.

Відмітимо також основні властивості визначених інтегралів.

**Т е о р е м а   6 .** *Нехай функції $f(x)$, $g(x)$ інтегровні на відрізку $[a, b]$ і $c$, $k$ – дійсні числа, $a \leq c \leq b$.*

*Тоді справедливі рівності:*

*а)* $\int\limits_a^a f(x)dx = 0$; *б)* $\int\limits_a^b f(x)dx = -\int\limits_b^a f(x)dx$;

*в)* $\int\limits_a^b f(x)dx = \int\limits_a^c f(x)dx + \int\limits_c^b f(x)dx$;

*г)* $\int\limits_a^b [f(x) \pm g(x)]dx = \int\limits_a^b f(x)dx \pm \int\limits_a^b g(x)dx$; *д)* $\int\limits_a^b kf(x)dx = k\int\limits_a^b f(x)dx$;

*е)* $\int\limits_a^b f(x)dx \leq \int\limits_a^b g(x)dx$, *якщо* $f(x) \leq g(x)$ *для всіх* $x \in [a, b]$.

*Д о в е д е н н я* ґрунтується на властивостях інтегральних сум. Доведемо, наприклад, твердження г). Запишемо інтегральну суму для інтеграла у лівій частині відповідної рівності
$$\sigma = \sum_{i=0}^{n} [f(\xi_i) \pm g(\xi_i)] \Delta x_i = \sum_{i=0}^{n} f(\xi_i) \Delta x_i \pm \sum_{i=0}^{n} g(\xi_i) \Delta x_i.$$

Перейшовши тут до границі при $\lambda \to 0$ і врахувавши, що у правій частині цієї рівності маємо інтегральні суми для інтегрових функцій $f(x)$ і $g(x)$, дістанемо

$$\lim_{\lambda \to 0} \sigma = \lim_{\lambda \to 0}\left[\sum_{i=0}^{n} f(\xi_i)\Delta x_i \pm \sum_{i=0}^{n} g(\xi_i)\Delta x_i\right] = \lim_{\lambda \to 0}\sum_{i=0}^{n} f(\xi_i)\Delta x_i \pm$$

$$\pm \lim_{\lambda \to 0}\sum_{i=0}^{n} g(\xi_i)\Delta x_i = \int_{a}^{b} f(x)dx \pm \int_{a}^{b} g(x)dx.$$

Твердження доведено.

*З а у в а ж е н н я* . Властивість в) справедлива також, коли точка $c$ знаходиться поза відрізком $[a,b]$ і функція $f(x)$ інтегрована на відрізку $[c,b]$ або $[a,c]$.

***Т е о р е м а*** *7 . Якщо функція $f(x)$ інтегровна на проміжку $[a,b]$, то на цьому проміжку інтегровною є функція $|f(x)|$ і виконується рівність*

$$\left|\int_{a}^{b} f(x)dx\right| \leq \int_{a}^{b} |f(x)|\,dx. \qquad (1.20)$$

*Д о в е д е н н я* . Спочатку доведемо, що за умови теореми інтегровною є функція $|f(x)|$ на відрізку $[a,b]$. Використаємо критерій інтегровності. Нехай $\omega_i = M_i - m_i$ . коливання функції $f(x)$ на відрізку $[x_{i-1}, x_i]$. Тоді

$$S - s = \sum_{i=1}^{n} \omega_i \Delta x_i$$

Покажемо, що для коливання $\omega'$ функції $|f(x)|$ на відрізку $[x_{i-1}, x_i]$ справедлива нерівність $\omega_i' \leq \omega_i$. Справді для будь-яких точок $x', x'' \in [x_{i-1}, x_i]$ маємо нерівність

$$\big||f(x')| - |f(x'')|\big| \leq |f(x') - f(x'')| \leq \omega_i,$$

тому $\omega_i' \leq \omega_i$.

Тоді $\sum_{i=1}^{n} \omega_i' \Delta x_i \leq \sum_{i=1}^{n} \omega_i \Delta x_i$.

Проте інтегральна сума у правій частині цієї нерівності за умовою прямує до нуля при $\lambda \to 0$. Тоді і ліва частина цієї нерівності прямує до нуля при $\lambda \to 0$.

Отже, з інтегровності за Ріманом функції $f(x)$ випливає інтегровність функції $|f(x)|$.

Оскільки для будь-якого $x \in [a,b]$ справедливі нерівності $f(x) \le |f(x)|$, $f(x) \ge -|f(x)|$, то

$$\int\limits_a^b f(x)dx \le \int\limits_a^b |f(x)|dx, \ \int\limits_a^b f(x)dx \ge -\int\limits_a^b |f(x)|\,dx.$$

Об'єднуючи ці нерівності, одержимо нерівність (1.20).
Теорему доведено.

***Т е о р е м а  8 (узагальнена теорема про середнє).*** *Нехай: а) функції $f(x)$ і $g(x)$ інтегровні на відрізку $[a,b]$; б) функція $f(x)$ обмежена на $[a,b]$, $m \le f(x) \le M$; в) функція $g(x)$ не змінює знак на $[a,b]$, $g(x) \ge 0$ або $g(x) \le 0$.*

*Тоді виконується рівність*

$$\int\limits_a^b f(x)g(x)dx = \mu \int\limits_a^b g(x)dx, \qquad (1.21)$$

*де $m \le \mu \le M$; $m, M$ – скінченні числа.*

*Д о в е д е н н я*. Нехай $g(x) \ge 0$ і $a < b$. Тоді розглянемо нерівність

$$m\,g(x) \le f(x)g(x) \le M\,g(x)$$

і проінтегруємо її з урахуванням властивостей інтеграла

$$m\int\limits_a^b g(x)dx \le \int\limits_a^b f(x)g(x)dx \le M\int\limits_a^b g(x)dx.$$

Оскільки функція $g(x) \ge 0$ невід'ємна, то $\int\limits_a^b g(x)dx \ge 0$. Якщо цей інтеграл дорівнює нулю, то формула (1.21) очевидна. Нехай інтеграл більший від нуля. Розділимо на нього одержану подвійну нерівність,

$$m \le \int\limits_a^b f(x)g(x)dx \left[\int\limits_a^b g(x)dx\right]^{-1} \le M.$$

Ввівши позначення $\mu = \int\limits_a^b f(x)g(x)dx \left[\int\limits_a^b g(x)dx\right]^{-1}$, матимемо формулу (1.21).

Зміна знаку функції $g(x)$ не порушує схеми доведення теореми.

Теорему доведено.

**Н а с л і д о к  1 .** *Якщо справджуються умови теореми і, крім того, функція $f(x)$ неперервна на відрізку $[a, b]$, то*

$$\int\limits_a^b f(x)g(x)dx = f(c)\int\limits_a^b g(x)dx, \text{ де } c \in [a, b].$$

Цей результат випливає з (1.21) з урахуванням теореми 5 (п. 1.1).

**Н а с л і д о к  2 (теорема про середнє).** *Нехай функції $f(x)$ інтегровна на відрізку $[a, b]$ і $m \le f(x) \le M$.*

*Тоді справедлива формула*

$$\int\limits_a^b f(x)dx = \mu(b - a).$$

Цю формулу одержимо з формули (1.21), якщо приймемо $g(x) = 1$.

Наслідок 2 має назву *інтегральної теореми про середнє*.

Нехай функція $f(x)$ інтегровна на відрізку $[a, b]$, тоді вона інтегровна на будь-якому відрізку $[a, x]$, $a \le x \le b$, тобто має сенс інтеграл

$$F(x) = \int\limits_a^x f(t)dt. \qquad (1.22)$$

Функція $F(x)$, визначена на відрізку $[a, b]$, називається *інтегралом зі змінною верхньою межею*.

**Т е о р е м а  9 .** *Якщо функція $f(x)$ інтегровна на відрізку $[a, b]$, то функція* (1.22) *неперервна на цьому відрізку.*

*Д о в е д е н н я* . Нехай точки $x$ і $x + \Delta x$ належать відрізку $[a, b]$. Знайдемо приріст функції $F(x)$ в точці $x$

$$\Delta F = \int_a^{x+\Delta x} f(t)dt - \int_a^x f(t)dt = \int_a^x f(t)dt +$$
$$= \int_x^{x+\Delta x} f(t)dt - \int_a^x f(t)dt = \int_x^{x+\Delta x} f(t)dt .$$

Оскільки функція $f(x)$ інтегровна на відрізку $[a, b]$, вона обмежена на цьому відрізку, тобто $|f(x)| \le M < \infty$ для всіх $x \in [a, b]$. Застосовуючи цю нерівність для оцінки приросту $\Delta F$, одержимо

$$|\Delta F| = \left| \int_x^{x+\Delta x} f(t)dt \right| \le \left| \int_x^{x+\Delta x} |f(t)|dt \right| \le \left| \int_x^{x+\Delta x} M\, dt \right| \le M |\Delta x|.$$

З цієї оцінки випливає гранична рівність $\lim_{\Delta x \to 0} \Delta F = 0$ для всіх $x \in [a, b]$, що стверджує неперервність функції $F(x)$ на відрізку $[a, b]$.

Теорему доведено.

***Т е о р е м а 10 .*** *Якщо функція $f(x)$ інтегровна на відрізку $[a, b]$ і неперервна в точці $x_0 \in [a, b]$, то функція (1.22) диференційовна в точці $x_0$ і справедлива формула*

$$\frac{d\Phi(x_0)}{dx} = f(x_0).$$

*Д о в е д е н н я* . Для функції $F(x)$ в будь-яких точках $x, x + \Delta x \in [a, b]$ справедлива залежність

$$\Phi(x + \Delta x) - \Phi(x) = \int_x^{x+\Delta x} f(t)dt = \mu\, \Delta x ,$$

де $m' \le \mu \le M'$, $m', M'$ – точні верхня і нижня грані функції $f(x)$ на відрізку $[x, x + \Delta x]$. Звідси

$$\frac{\Phi(x + \Delta x) - \Phi(x)}{\Delta x} = \mu . \qquad (1.23)$$

Оскільки функція $f(x)$ неперервна в точці $x_0$, для довільного $\varepsilon > 0$ знайдеться $\delta > 0$, що при $|\Delta x| < \delta$ і для всіх значень $t \in [x_0, x_0 + \Delta x]$ виконується нерівність
$$f(x_0) - \varepsilon < f(t) < f(x_0) + \varepsilon$$
і, відповідно,
$$f(x_0) - \varepsilon \leq m' \leq \mu \leq M' \leq f(x_0) + \varepsilon.$$

Звідси одержимо нерівність
$$|\mu - f(x_0)| \leq \varepsilon,$$
яку запишемо з урахуванням (1.23) у вигляді
$$\left| \frac{\Phi(x_0 + \Delta x) - \Phi(x_0)}{\Delta x} - f(x_0) \right| \leq \varepsilon.$$

Ця нерівність стверджує, що
$$\lim_{\Delta x \to 0} \frac{\Phi(x_0 + \Delta x) - \Phi(x_0)}{\Delta x} = f(x_0).$$

Теорему доведено.

*Н а с л і д о к .* *Для кожної неперервної на відрізку $[a, b]$ функції $f(x)$ існує первісна і однією з первісних є визначений інтеграл (1.22) зі змінною верхньою межею.*

Дійсно за теоремою 10 такою первісною є функція (1.22), оскільки в кожній точці відрізка $[a, b]$ справедлива формула $\frac{d\Phi(x)}{dx} = f(x)$.

Розглянемо неперервну на відрізку $[a, b]$ функцію $f(x)$. Функція (1.22) є первісною для $f(x)$. Якщо $F(x)$ будь-яка інша первісна для функції $f(x)$, то
$$\Phi(x) = F(x) + C.$$
Сталу $C$ визначимо, прийнявши $x = a$. Оскільки $\Phi(a) = 0$, одержимо $C = -F(a)$ і
$$\Phi(x) = F(x) - F(a).$$

Ця формула при $x = b$ набуде вигляду

$$\int_a^b f(t)dt = F(b) - F(a) \equiv F(x)\big|_a^b. \qquad (1.24)$$

Формула (1.24) називається *формулою Ньютона – Лейбніца*.

***Т е о р е м а  11 (заміна змінної).*** *Нехай: а) функція $f(x)$ неперервна на відрізку $[a,b]$; б) функція $x = \varphi(t)$ неперервна разом зі своєю похідною $\varphi'(t)$ на відрізку $[\alpha, \beta]$, при цьому $a = \varphi(\alpha) < \varphi(t) < \varphi(\beta) = b$, $\alpha \leq t \leq \beta$.*

*Тоді*

$$\int_a^b f(x)dx = \int_\alpha^\beta f(\varphi(t))\varphi'(t)dt \qquad (1.25)$$

*або*

$$\int_a^b f(x)dx = \int_\alpha^\beta f(\varphi(t))\, d\varphi(t).$$

*Д о в е д е н н я*. За умовою складна функція $f(\varphi(t))$ неперервна, тому інтеграли у обох частинах формули (1.25) існують.

Нехай $\Phi(x)$ – яка-небудь первісна на відрізку $[a,b]$ для функції $f(x)$. Тоді має сенс складна функція $\Phi(\varphi(t))$, яка є первісною для функції $f(\varphi(t))\varphi'(t)$. За формулою Ньютона – Лейбніца маємо

$$\int_a^b f(x)dx = \Phi(b) - \Phi(a),$$

$$\int_\alpha^\beta f(\varphi(t))\varphi'(t)dt = \Phi(\varphi(\beta)) - \Phi(\varphi(\alpha)) = \Phi(b) - \Phi(a).$$

З цих рівностей випливає формула (1.25).

Теорему доведено.

*З а у в а ж е н н я*. Важливим випадком формули (1.25) є інтеграл $\int_{-a}^{a} f(x)\, dx$. Розбивши цей інтеграл на два інтеграли і зробивши в другому інтегралі заміну $x = -t$, одержимо

$$\int_{-a}^{a} f(x)dx = \int_{0}^{a} f(x)dx + \int_{-a}^{0} f(x)dx = \int_{0}^{a} f(x)dx + \int_{0}^{a} f(-t)dt =$$

$$= \int_{0}^{a} [f(x) + f(-x)]dx .$$

Якщо $f(x)$ – парна функція, $f(-x) = f(x)$, то $\int_{-a}^{a} f(x)dx = 2\int_{0}^{a} f(x)dx$, якщо ж $f(x)$ – непарна функція, то $f(-x) = -f(x)$, то $\int_{-a}^{a} f(x)dx = 0$.

**Т е о р е м а  12 (інтегрування частинами).** *Якщо функції $u = u(x)$ і $v = v(x)$ неперервні разом зі своїми похідними на відрізку $[a, b]$, то справедлива формула*

$$\int_{a}^{b} u\,dv = [uv]\Big|_{a}^{b} - \int_{a}^{b} v\,du , \qquad (1.26)$$

*яка називається формулою інтегрування частинами для визначеного інтеграла.*

*Д о в е д е н н я .* Маємо

$$\int_{a}^{b} (uv)' dx = \int_{a}^{b} (u'v + uv')dx = \int_{a}^{b} u\,dv + \int_{a}^{b} v\,du .$$

Всі ці інтеграли існують, оскільки підінтегральні функції неперервні. За формулою Ньютона – Лейбніца

$$\int_{a}^{b} (uv)' dx = [uv]\Big|_{a}^{b} .$$

З цих формул випливає формула (1.26).
Теорему доведено.

*П р и к л а д  1 .* Обчислити інтеграл $I = \int_{0}^{\pi^2/4} \sin\sqrt{x}\,dx$.

Зробивши заміну $t = \sqrt{x}$, звідки $x = t^2$, $dx = 2t\,dt$, $\alpha = 0$, $\beta = \dfrac{\pi}{2}$, знайдемо
$$I = 2\int\limits_0^{\pi/2} t\sin t\,dt.$$
До останнього інтегралу застосуємо формулу інтегрування частинами (1.26), $u = t$, $dv = \sin t\,dt$, $du = dt$, $v = -\cos t$. Отже,
$$I = -2(t\cos t)\Big|_0^{\pi/2} + 2\int\limits_0^{\pi/2}\cos t\,dt = 2.$$

*П р и к л а д  2.* Обчислити інтеграл $I_n = \int\limits_0^{\pi/2}\sin^n x\,dx$.

Запишемо цей інтеграл у вигляді $I_n = \int\limits_0^{\pi/2}\sin^{n-1}x\sin x\,dx$ і використаємо формулу інтегрування частинами $u = \sin^{n-1}x$, $dv = \sin x\,dx$, $du = (n-1)\sin^{n-2}x\cos x\,dx$, $v = -\cos x$. Одержимо
$$I_n = -\left(\sin^{n-1}x\cos x\right)\Big|_0^{\pi/2} + (n-1)\int\limits_0^{\pi/2}\sin^{n-2}x\cos^2 x\,dx =$$
$$= (n-1)\int\limits_0^{\pi/2}\sin^{n-2}x\,dx - (n-1)\int\limits_0^{\pi/2}\sin^n x\,dx$$
або
$$I_n = (n-1)I_{n-2} - (n-1)I_n.$$
Звідси одержимо рекурентну формулу
$$I_n = \frac{n-1}{n}I_{n-2}.$$

Розглянемо випадок парного $n = 2m$, $m \geq 0$.
$I_0 = \int\limits_0^{\pi/2}dx = \dfrac{\pi}{2}$, $I_2 = \dfrac{1}{2}I_0 = \dfrac{1}{2}\cdot\dfrac{\pi}{2}$, $I_4 = \dfrac{3}{4}I_2 = \dfrac{1\cdot 3}{2\cdot 4}\cdot\dfrac{\pi}{2}$,

$$I_6 = \frac{5}{6}I_4 = \frac{1 \cdot 3 \cdot 5}{2 \cdot 4 \cdot 6} \cdot \frac{\pi}{2}.$$

За методом індукції

$$I_{2m} = \frac{1 \cdot 3 \cdot 5 \cdot \ldots \cdot (2m-1)}{2 \cdot 4 \cdot 6 \cdot \ldots \cdot 2m} \frac{\pi}{2} = \frac{(2m)!}{2^{2m}(m!)^2} \frac{\pi}{2}.$$

Якщо $n$ – непарне, $n = 2m+1$, то знайдемо $I_1 = 1$, $I_3 = \frac{2}{3}$, …,

$$I_{2m+1} = \frac{2 \cdot 4 \cdot 6 \cdot \ldots \cdot 2m}{3 \cdot 5 \cdot \ldots \cdot (2m+1)} = \frac{2^{2m}(m!)^2}{(2m+1)!}.$$

*З а у в а ж е н н я*. Нехай виконуються умови теореми 15 (п. 1.1). З використанням формули інтегрування частинами одержимо додатковий член формули Тейлора в інтегральній формі

$$R_n(x) = \frac{1}{n!} \int_{x_0}^{x} f^{(n+1)}(t)(x-t)^n \, dt. \qquad (1.27)$$

Дійсно, застосуємо $n$ раз формулу інтегрування частинами до функції $f(x)$

$$f(x) = f(x_0) + \int_{x_0}^{x} f'(t)dt = f(x_0) + f'(x_0)(x-x_0) + \int_{x_0}^{x} f''(t)(x-t)dt =$$

$$= f(x_0) + f'(x_0)(x-x_0) + f''(x_0)\frac{(x-x_0)^2}{2!} + \int_{x_0}^{x} f'''(t)\frac{(x-t)^2}{2!} dt = \ldots =$$

$$= \sum_{k=0}^{n} \frac{f^{(k)}(x_0)}{k!}(x-x_0)^k + \int_{x_0}^{x} f^{(n+1)}(t)\frac{(x-t)^n}{n!} dt.$$

Звідси $R_n(x) = \frac{1}{n!} \int_{x_0}^{x} f^{(n+1)}(t)(x-t)^n \, dt$.

Оскільки множник $(x-t)^{n+1}$ підінтегральної функції не змінює знак, до інтеграла у формулі додаткового члена можна застосувати узагальнену теорему про середнє

$$R_n(x) = \frac{1}{n!}\int_{x_0}^{x} f^{(n+1)}(t)(x-t)^n dt = \frac{f^{(n+1)}(c)}{n!}\int_{x_0}^{x}(x-t)^n dt =$$

$$= \frac{f^{(n+1)}(c)}{(n+1)!}(x-t)^{n+1},$$

де точка $c$ належить проміжку $[x_0, x]$.

Отже, ми одержали лагранжову форму додаткового члена (1.10).

**1.2.2. Невласні інтеграли з нескінченними межами.** Поняття інтеграла для скінченного проміжку узагальнюється на випадок нескінченного проміжку [9, 21]. Нехай функція $f(x)$ визначена на проміжку $[a, +\infty)$ і інтегровна на кожному скінченному відрізку $[a, \eta]$, $\eta < +\infty$,

$$F(\eta) = \int_{a}^{\eta} f(x) dx. \qquad (1.28)$$

***О з н а ч е н н я  2***. *Границя (скінченна або нескінченна) інтеграла $F(\eta)$ при $\eta \to +\infty$ називається невласним інтегралом від функції $f(x)$ на проміжку $[a, +\infty)$*

$$\int_{a}^{+\infty} f(x) dx = \lim_{\eta \to +\infty} \int_{a}^{\eta} f(x) dx.$$

*При цьому, якщо границя інтеграла у цієї рівності існує, то невласний інтеграл називається збіжним, в протилежному випадку називається розбіжним.*

Аналогічно визначаються невласні інтеграли на проміжку $(-\infty, b]$.

***О з н а ч е н н я  3***. *Скінченна або нескінченна границя*

$$\lim_{\substack{\zeta \to -\infty \\ \eta \to +\infty}} \int_{\zeta}^{\eta} f(x) dx = \int_{-\infty}^{+\infty} f(x) dx,$$

*називається невласним інтегралом від функції $f(x)$ на проміжку $(-\infty, +\infty)$.*

*Якщо ця границя існує, то невласний інтеграл називається збіжним, в протилежному випадку називається розбіжним.*

**О з н а ч е н н я  4** . Якщо інтеграл $\int\limits_{-\infty}^{+\infty} f(x)dx$ розбігається, але існує границя

$$\lim_{\eta \to +\infty} \int\limits_{-\eta}^{\eta} f(x)dx = V.P. \int\limits_{-\infty}^{+\infty} f(x)dx ,$$

*то вона називається головним значенням за Коші розбіжного інтеграла від функції $f(x)$ на проміжку $(-\infty, +\infty)$.*

*З а у в а ж е н н я.* Головне значення за Коші збіжного інтеграла дорівнює самому інтегралу, головне значення розбіжного інтеграла від непарної функції завжди дорівнює нулю, а головне значення парної функції існує тоді і тільки тоді, коли існує невласний інтеграл $\int\limits_{0}^{+\infty} f(x)dx$,

$$V.P. \int\limits_{-\infty}^{+\infty} f(x)dx = 2\int\limits_{0}^{+\infty} f(x)dx .$$

*П р и к л а д  3* . Знайти головне значення за Коші інтеграла від функції $f(x) = \sin x$ на проміжку $(-\infty, +\infty)$.

Дійсно,

$$V.P. \int\limits_{-\infty}^{+\infty} \sin x\, dx = \lim_{\eta \to +\infty} \int\limits_{-\eta}^{\eta} \sin x\, dx = 0 .$$

*П р и к л а д  4* . Покажемо, що інтеграл $\int\limits_{1}^{+\infty} \dfrac{dx}{x^\alpha}$ збігається, якщо $\alpha > 1$, і розбігається, якщо $\alpha \le 1$.

Дійсно, якщо $\alpha \ne 1$, то

$$\int\limits_{1}^{+\infty} \frac{dx}{x^\alpha} = \lim_{\eta \to +\infty} \int\limits_{1}^{\eta} \frac{dx}{x^\alpha} = \lim_{\eta \to +\infty} \left( \frac{\eta^{1-\alpha}}{1-\alpha} - \frac{1}{1-\alpha} \right) = \begin{cases} +\infty, & \alpha < 1, \\ -\dfrac{1}{1-\alpha}, & \alpha > 1. \end{cases}$$

Якщо $\alpha = 1$, то

$$\int\limits_{1}^{+\infty} \frac{dx}{x} = \lim_{\eta \to +\infty} \int\limits_{1}^{\eta} \frac{dx}{x} = \lim_{\eta \to +\infty} \ln \eta = +\infty.$$

**Т е о р е м а   12 (критерій Коші).** *Для того щоби інтеграл $\int\limits_{a}^{+\infty} f(x)dx$ збігався, необхідно і достатньо, щоб для будь-якого $\varepsilon > 0$ можна було вказати число $A = A(\varepsilon) > a$ таке, що для будь-яких чисел $\eta'$ і $\eta''$, що задовольняють умову $\eta' > A$ і $\eta'' > A$, виконувалася нерівність*

$$\left| \int\limits_{\eta'}^{\eta''} f(x)dx \right| < \varepsilon. \qquad (1.29)$$

*Д о в е д е н н я .* Збіжність інтеграла (1.28) еквівалентна існуванню при $\eta \to \infty$ границі функції $F(\eta)$. За критерієм Коші (існування границі) маємо: для того щоб існувала границя $\lim\limits_{\eta \to \infty} F(\eta)$, необхідно і достатньо, щоб для довільного числа $\varepsilon > 0$ існувало число $A = A(\varepsilon) > a$ таке, що для будь-яких двох значень аргумента $\eta = \eta'$ і $\eta = \eta''$, що справджують умову $\eta' > A$ і $\eta'' > A$, виконувалася нерівність

$$\left| F(\eta'') - F(\eta') \right| < \varepsilon.$$

Звідси, оскільки

$$\left| F(\eta'') - F(\eta') \right| = \left| \int\limits_{a}^{\eta''} f(x)dx - \int\limits_{a}^{\eta'} f(x)dx \right| = \left| \int\limits_{\eta'}^{\eta''} f(x)dx \right|,$$

одержимо нерівність (1.29).

Теорему доведено.

*Н а с л і д о к   1 .* *Якщо збігається невласний інтеграл від абсолютної величини функції $f(x)$, $\int\limits_{a}^{+\infty} |f(x)|dx$, то збігається* *також невласний інтеграл від цієї функції, $\int\limits_{a}^{+\infty} f(x)dx$. При цьому останній інтеграл називається абсолютно збіжним* (детально див. п. 1.4.6).

Дійсно, застосовуючи критерій Коші до інтеграла $\int\limits_{a}^{+\infty} |f(x)|dx$, який за умовою збігається, одержимо $\int\limits_{\eta'}^{\eta''} |f(x)|dx < \varepsilon$. Однак $\left| \int\limits_{\eta'}^{\eta''} f(x)dx \right| \leq \int\limits_{\eta'}^{\eta''} |f(x)|dx < \varepsilon$, що є умовою збіжності інтеграла $\int\limits_{a}^{+\infty} f(x)dx$.

**Н а с л і д о к  2 .** *Якщо функція $g(x)$ обмежена на проміжку $[a, +\infty)$, $|g(x)| \leq L < \infty$, а функція $f(x)$ абсолютно інтегровна на цьому проміжку, то і добуток цих функцій $f(x)g(x)$ – абсолютно інтегровна функція на $[a, +\infty)$.*

Доведення цього твердження випливає з нерівності $|f(x)g(x)| \leq L|f(x)|$ для всіх $x \in [a, +\infty)$.

*Функція $f(x)$ називається абсолютно інтегровною на проміжку $[a, +\infty)$, якщо разом з інтегралом $\int\limits_{a}^{+\infty} f(x)dx$ збігається інтеграл $\int\limits_{a}^{+\infty} |f(x)|\, dx$.*

Питання збіжності невласних інтегралів не завжди можна вирішити з використанням первісної для підінтегральної функції. Ефективними для дослідження збіжності невласних інтегралів є ознаки збіжності (достатні ознаки).

***Т е о р е м а   13 (ознака порівняння).*** *Якщо для $x \geq A$, де $A \geq a$, справедлива нерівність $|f(x)| \leq g(x)$, то із збіжності інтеграла $\int\limits_{a}^{+\infty} g(x)dx$ випливає збіжність інтеграла $\int\limits_{a}^{+\infty} f(x)dx$.*

*Д о в е д е н н я*. Якщо інтеграл $\int\limits_{a}^{+\infty} g(x)dx$ збігається, то за критерієм збіжності для довільного числа $\varepsilon > 0$ знайдеться число $A = A(\varepsilon) > a$ таке, що для будь-яких двох чисел $\eta' > A$ і $\eta'' > A$, виконується нерівність

$$\left| \int\limits_{\eta'}^{\eta''} g(x)\,dx \right| < \varepsilon. \qquad (1.30)$$

Тоді відповідно до умови теореми і властивостей інтеграла знайдемо

$$\left| \int\limits_{\eta'}^{\eta''} f(x)dx \right| \leq \int\limits_{\eta'}^{\eta''} |f(x)|dx \leq \int\limits_{\eta'}^{\eta''} g(x)dx.$$

Звідси з урахуванням нерівності (1.30) випливає, що для будь-яких $\eta' > A$ і $\eta'' > A$ справедлива нерівність

$$\left| \int\limits_{\eta'}^{\eta''} f(x)dx \right| < \varepsilon.$$

Отже, інтеграл $\int\limits_{a}^{+\infty} f(x)\,dx$ збігається.

Теорему доведено.

***Н а с л і д о к   1*** *. Нехай на проміжку $0 < a \leq x < +\infty$ справджується нерівність $|f(x)| \leq \dfrac{c}{x^\lambda}$, де $c, \lambda$ – довільні сталі і $\lambda > 1$.*

*Тоді інтеграл $\int\limits_{a}^{+\infty} f(x)dx$ збігається.*

*Якщо ж на цьому проміжку існують сталі $c, \lambda$, $\lambda \leq 1$, і справедлива нерівність $f(x) \geq \dfrac{c}{x^\lambda}$, то інтеграл розбігається.*

Перша частина цього твердження випливає з наслідку 1, якщо взяти $g(x) = \dfrac{c}{x^\lambda}$, оскільки при $\lambda > 1$ інтеграл $\int\limits_a^{+\infty} g(x)dx$ збігається. Другу частину доведемо, ґрунтуючись на тому, що невласний інтеграл від функції $g(x) = \dfrac{c}{x^\lambda}$ при $\lambda \leq 1$ розбігається. Якщо би інтеграл $\int\limits_a^{+\infty} f(x)dx$ збігався, то це б суперечило першій частині твердження.

**Н а с л і д о к  2.** *Якщо при $\lambda > 1$ існує скінченна границя $\lim\limits_{x \to +\infty} f(x)x^\lambda = c$, $0 \leq c < +\infty$, то інтеграл $\int\limits_a^{+\infty} f(x)dx$ збігається. Якщо ж при $\lambda \leq 1$ існує додатна границя $\lim\limits_{x \to +\infty} f(x)x^\lambda = c > 0$, то інтеграл $\int\limits_a^{+\infty} f(x)dx$ розбігається.*

Дійсно, з існування границі при $x \to +\infty$ випливає обмеженість функції $|f(x)|x^\lambda$, тобто $|f(x)| \leq \dfrac{c}{x^\lambda}$. Застосовуючи першу частину наслідку 1, одержимо першу частину твердження.

Справедливість другої частини твердження випливає з таких міркувань. Оскільки $c > 0$, можна вказати число $\varepsilon > 0$, що $c - \varepsilon > 0$. За означенням границі цьому $\varepsilon$ відповідає таке число $A > 0$, що якщо тільки $x > A$, то виконується нерівність $f(x) > \dfrac{c - \varepsilon}{x^\lambda}$, $\lambda \leq 1$. Тоді виконуються умови другої частини наслідку 1.

*П р и к л а д  5*. Дослідити на збіжність інтеграл $\int\limits_{1}^{+\infty} \frac{\ln x \, dx}{\sqrt{x^3+1}}$.

Виберемо $\lambda = \frac{3}{2} - \varepsilon$, де $\varepsilon > 0$ таке, що виконується умова $\frac{3}{2} - \varepsilon > 1$. Тоді інтеграл $\int\limits_{1}^{+\infty} \frac{dx}{x^{3/2-\varepsilon}}$ збігається і

$$\lim_{x \to +\infty} \frac{\ln x}{\sqrt{x^3+1}} x^{3/2-\varepsilon} = \lim_{x \to +\infty} \frac{\ln x}{x^{\varepsilon}} = 0.$$

Отже, за наслідком 2 інтеграл збігається.

**Т е о р е м а   14 (ознака Діріхле).** *Нехай: а) функція $f(x)$ неперервна і має обмежену первісну,*

$$|F(x)| \leq K, \text{ де } F(x) = \int\limits_{a}^{x} f(t)dt, \ K = const, \ a \leq x < +\infty;$$

*б) функція $g(x)$ монотонно прямує до нуля при $x \to +\infty$, $\lim\limits_{x \to +\infty} g(x) = 0$.*

*Тоді збігається інтеграл*

$$\int\limits_{a}^{+\infty} f(x)g(x)dx. \tag{1.31}$$

*Д о в е д е н н я* проведемо за додаткової умови, що похідна $g'(x)$ неперервна на проміжку $[a, +\infty)$. Скористаємося критерієм Коші збіжності невласного інтеграла. Розглянемо інтеграл $\int\limits_{\eta'}^{\eta''} f(x)g(x)dx$, де $\eta' < \eta''$, $\eta' > a$ $\eta'' > a$, і проінтегруємо його частинами

$$\int\limits_{\eta'}^{\eta''} f(x)g(x)dx = \left[F(x)g(x)\right]\Big|_{\eta'}^{\eta''} - \int\limits_{\eta'}^{\eta''} F(x)g'(x)dx.$$

Знайдемо оцінку інтеграла з урахуванням того, що за умовою $|F(x)| \leq K$, а функція $g(x)$ не зростає, має неперервну похідну і,

відповідно, $g(x) \geq 0$  $g'(x) \leq 0$

$$\left|\int_{\eta'}^{\eta''} f(x)g(x)dx\right| \leq K[g(\eta') + g(\eta'')] + K\int_{\eta'}^{\eta''}[-g'(x)]dx = 2K g(\eta'). \quad (1.32)$$

Нехай $\varepsilon > 0$ - довільне число. Оскільки $\lim\limits_{x \to \infty} g(x) = 0$, то можна вибрати достатньо велике число $A > 0$ таке, що якщо $\eta' > A$, то виконується нерівність $g(\eta') < \dfrac{\varepsilon}{2K}$. Враховуючи її в (1.32), одержимо

$$\left|\int_{\eta'}^{\eta''} f(x)g(x)dx\right| < \varepsilon. \quad (1.33)$$

Звідси випливає, що яке б не було число $\varepsilon > 0$ існує число $A > 0$ таке, що для будь-яких чисел $\eta' > A$ і $\eta'' > A$ виконується нерівність (1.33). За критерієм Коші інтеграл (1.31) збігається.

Теорему доведено.

*П р и к л а д  6*. Показати, що збігаються інтеграли $\int\limits_1^{+\infty} \dfrac{\sin x}{x^\lambda} dx$, $\int\limits_1^{+\infty} \dfrac{\cos x}{x^\lambda} dx$, $\lambda > 0$.

Легко переконатися, що якщо $f(x) = \sin x$ (або $f(x) = \cos x$) і $g(x) = \dfrac{1}{x^\lambda}$, то виконуються умови теореми 14.

*П р и к л а д  7*. Показати, що збігається інтеграл Френеля $\int\limits_0^{+\infty} \sin x^\alpha \, dx$, $\alpha > 1$.

Запишемо цей інтеграл у вигляді $\int\limits_0^{+\infty} \sin x^\alpha dx =$ $= \int\limits_0^1 \sin x^\alpha dx + \int\limits_1^{+\infty} \sin x^\alpha dx$. Перший інтеграл існує, а другий інтеграл запишемо у вигляді $\int\limits_{1}^{+\infty} \frac{x^{\alpha-1}\sin x^{\alpha}}{x^{\alpha-1}}dx$. Якщо прийняти

$f(x) = x^{\alpha-1}\sin x^{\alpha}$, $g(x) = \frac{1}{x^{\alpha-1}}$ і, відповідно, $F(x) = -\frac{\cos x^{\alpha}}{\alpha}$, то легко переконатися, що виконуються умови теореми 14.

Отже, інтеграл Френеля збігається.

**1.2.3. Невласні інтеграли, залежні від параметра.** У багатьох дослідженнях функцію від двох змінних $f(x, y)$ розглядають як функцію за змінною $x \in [a, +\infty)$ на множині значень $y \in [c, d]$. Тоді змінна $y$ є параметром, а функція $f(x, y)$ визначає сім'ю функцій від змінної $x$.

Сформулюємо поняття рівномірної збіжності за параметром. Нехай функція $f(x, y)$ визначена для всіх $x \in [a, +\infty)$ і всіх $y \in [c, d]$, а функція $\varphi(x)$ визначена в області $[a, +\infty)$.

*О з н а ч е н н я  5 . Функція $f(x, y)$ рівномірно прямує на множині $[a, +\infty)$ до функції $\varphi(x)$ при $y \to y_0 \in [c, d]$, якщо для будь-якого $\varepsilon > 0$ існує такий окіл $O(y_0)$ точки $y_0$, що для всіх $x \in [a, +\infty)$ і всіх $y \in O(y_0)$ виконується нерівність*

$$|f(x, y) - \varphi(x)| < \varepsilon.$$

Наведемо важливі у подальших дослідженнях властивості невласних інтегралів, залежних від параметра [9].

*О з н а ч е н н я  6 . Інтеграл*

$$F(y) = \int\limits_{a}^{+\infty} f(x, y)dx, \ a < +\infty, \tag{1.34}$$

*називається збіжним на проміжку $[c, d]$, $-\infty < c < d < +\infty$, якщо він збігається для кожного $y_0 \in [c, d]$, тобто, для кожного фіксованого $y_0 \in [c, d]$ і для довільного $\varepsilon > 0$ можна вказати число $\eta = \eta(y_0, \varepsilon) < +\infty$ таке, що для будь-якого $\eta'$, $\eta < \eta' < +\infty$, виконується нерівність*

$$\left|\int\limits_{\eta'}^{+\infty} f(x, y_0)dx\right| < \varepsilon.$$

**О з н а ч е н н я  7 .** *Збіжний на проміжку* $[c, d]$ *інтеграл* (1.34) *називається рівномірно збіжним на цій множині, якщо для довільного* $\varepsilon > 0$ *існує число* $\eta = \eta(\varepsilon) < +\infty$ *таке, що для всіх* $y \in [c, d]$ *і будь-якого* $\eta'$, $\eta < \eta' < \infty$, *виконується нерівність*

$$\left|\int\limits_{\eta'}^{+\infty} f(x, y)dx\right| < \varepsilon.$$

Сформулюємо (виходячи з критерію рівномірної збіжності функції) критерій рівномірної збіжності інтеграла (1.34).

**Т е о р е м а  15 (критерій Коші) .** *Для того, щоби інтеграл* (1.34) *збігався рівномірно відносно* $y \in [c, d]$, *необхідно і достатньо, щоб для довільного числа* $\varepsilon > 0$ *знайшлося таке число* $\eta(\varepsilon)$, *не залежне від* $y$, *щоби нерівність*

$$\left|\int\limits_{\eta'}^{\eta''} f(x, y)dx\right| < \varepsilon$$

*виконувалася для всіх* $y \in [c, d]$, *якщо тільки* $\eta' > \eta(\varepsilon)$ *і* $\eta'' > \eta(\varepsilon)$.

Розглянемо ознаки збіжності і основні властивості інтеграла (1.34)

**Т е о р е м а  16 (ознака Вейєрштрасса).** *Нехай існує функція* $\varphi(x)$, *визначена на проміжку* $[a, \infty)$ *і інтегровна на будь-якому відрізку* $[a, \eta]$, *де* $a < \eta < \infty$, *така, що:* а) $|f(x, y)| \leq \varphi(x)$ *для всіх* $x \in [a, +\infty)$ *і* $y \in [c, d]$; б) *інтеграл* $\int\limits_{a}^{+\infty} \varphi(x)dx$ *збігається.*

*Тоді інтеграл* (1.34) *рівномірно збігається.*

*Д о в е д е н н я .* Інтеграл $\int\limits_{a}^{+\infty} \varphi(x)dx$ збігається, тому за означенням для довільного $\varepsilon > 0$ існує число $\eta(\varepsilon) < \infty$ таке, що якщо $\eta(\varepsilon) \leq \eta < +\infty$, то $\int\limits_{\eta}^{+\infty} \varphi(x)dx < \varepsilon$. Тоді, ґрунтуючись на умовах а) і б), одержимо нерівність

$$\left| \int\limits_{\eta}^{+\infty} f(x,y)dx \right| \leq \int\limits_{\eta}^{+\infty} |f(x,y)|dx \leq \int\limits_{\eta}^{+\infty} \varphi(x)dx < \varepsilon,$$

яка справедлива для всіх $y \in [c,d]$ і $\eta(\varepsilon) \leq \eta < +\infty$.

Отже, інтеграл (1.34) збігається рівномірно на множині $[c,d]$.
Теорему доведено.

*П р и к л а д  8*. З використанням ознаки Вейєрштрасса встановимо, що інтеграл $\int\limits_{0}^{+\infty} \dfrac{\sin x \, dx}{1+x^2+y^2}$ рівномірно збігається на всій дійсній осі $-\infty < y < +\infty$.

Оскільки виконуються умови теореми 16, $\dfrac{1}{1+x^2+y^2} \leq \dfrac{1}{1+x^2}$ для всіх значень $x$ і $y$, а також збігається інтеграл $\int\limits_{0}^{+\infty} \dfrac{dx}{1+x^2} = \dfrac{\pi}{2}$, то інтеграл $\int\limits_{0}^{+\infty} \dfrac{\sin x \, dx}{1+x^2+y^2}$ рівномірно збігається.

***Т е о р е м а  17***. *Нехай: а) функція $f(x,y)$ визначена для всіх $x \in [a,+\infty)$ і $y \in [c,d]$, а також неперервна за змінною $x$; б) функція $f(x,y)$ рівномірно прямує на відрізку $[a,\eta]$, $a < \eta < \infty$, до функції $\varphi(x)$ при $y \to y_0$; в) інтеграл $\int\limits_{a}^{+\infty} f(x,y)dx$ збігається рівномірно для всіх $y \in [c,d]$.*

*Тоді справедлива рівність*

$$\lim_{y \to y_0} \int\limits_{a}^{+\infty} f(x,y)dx = \int\limits_{a}^{+\infty} \lim_{y \to y_0} f(x,y)dx = \int\limits_{a}^{+\infty} \varphi(x)dx. \qquad (1.35)$$

*Д о в е д е н н я .* Зафіксуємо $\eta$ таке, що $a < \eta < \infty$, і розглянемо граничний перехід при $y \to y_0$ під знаком інтеграла $\int\limits_a^\eta f(x, y)dx$. Оскільки виконуються умови а) і б), за відповідною теоремою [20] для звичайного інтегралу одержимо

$$\lim_{y \to y_0} \int\limits_a^\eta f(x, y)dx = \int\limits_a^\eta \lim_{y \to y_0} f(x, y)dx = \int\limits_a^\eta \varphi(x)dx. \qquad (1.36)$$

Тому за означенням невласного інтеграла рівність (1.35) можна записати у вигляді

$$\lim_{y \to y_0} \lim_{\eta \to +\infty} \int\limits_a^\eta f(x, y)dx = \lim_{\eta \to +\infty} \lim_{y \to y_0} \int\limits_a^\eta f(x, y)dx.$$

Отже, доведення теореми зводиться до доведення можливості зміни порядку граничних переходів для функції

$$F(y, \eta) = \int\limits_a^\eta f(x, y)dx.$$

Дійсно, з формули (1.36) випливає існування границі
$$\lim_{y \to y_0} F(y, \eta) = \phi(\eta). \qquad (1.37)$$

За умовою в) існує границя

$$\lim_{\eta \to +\infty} F(y, \eta) = \lim_{\eta \to +\infty} \int\limits_a^\eta f(x, y)dx = \int\limits_a^{+\infty} f(x, y)dx = \psi(y)$$

і при цьому прямування до границі відбувається рівномірно на множині $[c, d]$.

Отже, функція $F(y, \eta)$ рівномірно на $[c, d]$ прямує до $\psi(y)$ при $\eta \to +\infty$. Тоді для довільного $\varepsilon > 0$ існує $\eta_\varepsilon$, таке що для довільного $\eta > \eta_\varepsilon$ і для всіх $y \in [c, d]$ виконується нерівність

$$|F(y, \eta) - \psi(y)| < \frac{\varepsilon}{2}. \qquad (1.38)$$

Якщо $\eta_1 > \eta_\varepsilon$ і $\eta_2 > \eta_\varepsilon$, то
$$|F(y, \eta_1) - F(y, \eta_2)| \leq |F(y, \eta_1) - \psi(y)| + |F(y, \eta_2) - \psi(y)| < \varepsilon.$$

Переходячи тут до границі при $y \to y_0$, одержимо
$$|\phi(\eta_1) - \phi(\eta_2)| < \varepsilon. \qquad (1.39)$$
З цієї нерівності за критерієм Коші існує границя $\lim_{\eta \to +\infty} \phi(\eta) = A$.

Отже, показано, що існує границя
$$\lim_{\eta \to +\infty} \lim_{y \to y_0} F(y, \eta) = A.$$

Зафіксуємо $\eta_1 > a$. Оскільки існує границя (1.37), для заданого $\varepsilon > 0$ знайдеться $\delta > 0$ таке, що для всіх $y \in (y_0 - \delta, y_0 + \delta)$, справджується нерівність
$$|F(y, \eta_1) - \phi(\eta_1)| < \varepsilon. \qquad (1.40)$$
Перепишемо нерівність (1.38) при $\eta = \eta_1$ і нерівність (1.39) при $\eta \to +\infty$ у вигляді
$$|F(y, \eta_1) - \psi(y)| < \frac{\varepsilon}{2}, \quad |\phi(\eta_1) - A| < \varepsilon. \qquad (1.41)$$
З нерівностей (1.40) і (1.41) для всіх $y \in (y_0 - \delta, y_0 + \delta)$ маємо
$$|\psi(y) - A| = |[\psi(y) - F(y, \eta_1)] + [F(y, \eta_1) - \phi(\eta_1)] + [\phi(\eta_1) - A]| \leq$$
$$\leq |\psi(y) - F(y, \eta_1)| + |F(y, \eta_1) - \phi(\eta_1)| + |\phi(\eta_1) - A| < 3\varepsilon.$$

Отже, $\lim_{y \to y_0} \lim_{\eta \to \infty} F(y, \eta) = A$, тобто справедлива зміна порядку граничних переходів для функції $F(y, \eta)$.

Теорему доведено.

**Т е о р е м а  18 .** *Нехай: а) функція $f(x, y)$ визначена і неперервна для всіх $x \in [a, \infty)$ і $y \in [c, d]$; б) інтеграл*
$$F(y) = \int_a^{+\infty} f(x, y) dx \qquad (1.42)$$
*збігається рівномірно для всіх $y \in [c, d]$.*

*Тоді функція $F(y)$ неперервна на проміжку $[c, d]$.*

*Д о в е д е н н я .* Яке б не було $y_0 \in [c, d]$, функція $f(x, y)$ при $y \to y_0$ рівномірно збігається відносно $x$ на відрізку $[a, \eta]$, $a \leq \eta < +\infty$, до функції $f(x, y_0)$. Тому за теоремою 17 в інтегралі (1.42) можна перейти до границі під знаком інтеграла

$$\lim_{y \to y_0} F(y) = \lim_{y \to y_0} \int_a^{+\infty} f(x, y) dx = \int_a^{+\infty} \lim_{y \to y_0} f(x, y) dx = F(y_0).$$

Остання рівність показує, що функція $F(y)$ неперервна на проміжку $[c, d]$.

Теорему доведено.

***Т е о р е м а   19***. *Нехай: а) функція $f(x, y)$ визначена і неперервна для всіх $x \in [a, +\infty)$ і $y \in [c, d]$; б) інтеграл* (1.42) *збігається рівномірно для всіх $y \in [c, d]$.*

*Тоді виконується рівність*

$$\int_c^d dy \int_a^{+\infty} f(x, y) dx = \int_a^{+\infty} dx \int_c^d f(x, y) dy. \qquad (1.43)$$

*Д о в е д е н н я .* Якщо $a < \eta < +\infty$, то за теоремою про зведення подвійного інтеграла до повторних маємо

$$\int_c^d dy \int_a^{\eta} f(x, y) dx = \int_a^{\eta} dx \int_c^d f(x, y) dy. \qquad (1.44)$$

За умовою функція

$$F(y, \eta) = \int_a^{\eta} f(x, y) dx$$

неперервна за змінною $y$ і при $\eta \to +\infty$ рівномірно на відрізку $[c, d]$ прямує до граничної функції $F(y)$. Тому за теоремою для звичайного інтеграла, залежного від параметра, можна перейти до границі під знаком інтеграла

$$\lim_{\eta \to +\infty} \int_c^d F(y, \eta) dy = \int_c^d \lim_{\eta \to +\infty} F(y, \eta) dy = \int_c^d F(y) dy = \int_c^d dy \int_a^{\infty} f(x, y) dx.$$

Граничний перехід при $\eta \to +\infty$ у правій частині рівності (1.44) також можливий, внаслідок означення невласного інтеграла,

$$\lim_{\eta \to +\infty} \int_a^{\eta} dx \int_c^d f(x, y) dy = \int_a^{+\infty} dx \int_c^d f(x, y) dy.$$

Теорему доведено.

**Т е о р е м а  20.** *Нехай: а) функції $f(x,y)$ і $\dfrac{\partial f(x,y)}{\partial y}$ визначені і неперервні для всіх $x \in [a, +\infty)$ і $y \in [c,d]$; б) інтеграл $F(y) = \int\limits_a^{+\infty} f(x,y)dx$ збігається на відрізку $[c,d]$; в) інтеграл $\int\limits_a^{+\infty} \dfrac{\partial f(x,y)}{\partial y}dx$ рівномірно збігається на $[c,d]$.*

*Тоді функція $F(y) = \int\limits_a^{+\infty} f(x,y)dx$ неперервно диференційована на цьому відрізку і справедлива формула*

$$\frac{d}{dy}\int\limits_a^{+\infty} f(x,y)dx = \int\limits_a^{+\infty} \frac{\partial f(x,y)}{\partial y}dx. \qquad (1.45)$$

*Д о в е д е н н я.* Розглянемо вираз

$$\frac{F(y_0 + \Delta y) - F(y_0)}{\Delta y} = \int\limits_a^{\infty} \frac{f(x, y_0 + \Delta y) - f(x, y_0)}{\Delta y}dx, \qquad (1.46)$$

де $y_0, y_0 + \Delta y \in [c,d]$, і покажемо, що тут можливий граничний перехід при $\Delta y \to 0$.

Внаслідок рівномірної збіжності інтеграла $\int\limits_a^{\infty} \dfrac{\partial f(x,y)}{\partial y}dx$ (за критерієм збіжності) для довільного $\varepsilon > 0$ знайдеться таке $\eta_\varepsilon > a$, що якщо тільки $\eta' > \eta_\varepsilon$ і $\eta'' > \eta_\varepsilon$, то для всіх $y \in [c,d]$ виконується нерівність

$$\left| \int\limits_{\eta'}^{\eta''} \frac{\partial f(x,y)}{\partial y}dx \right| < \varepsilon. \qquad (1.47)$$

Покажемо, що справедлива також нерівність

$$\left| \int\limits_{\eta'}^{\eta''} \frac{f(x, y_0 + \Delta y) - f(x, y_0)}{\Delta y}dx \right| < \varepsilon \qquad (1.48)$$

для всіх можливих $\Delta y$. Для цього розглянемо функцію

$$\Phi(y) = \int\limits_{\eta'}^{\eta''} f(x, y) dx.$$

За теоремою для інтеграла Рімана, залежного від параметра [21], оскільки функції $f(x, y)$, $\dfrac{\partial f(x, y)}{\partial y}$ визначені і неперервні для всіх $x \in [\eta', \eta'']$ і $y \in [c, d]$, справедлива формула

$$\frac{d\Phi(y)}{dy} = \int\limits_{\eta'}^{\eta''} \frac{\partial f(x, y)}{\partial y} dx.$$

Абсолютна величина цієї похідної, внаслідок (1.47), задовольняє нерівність $\left|\dfrac{d\Phi(y)}{dy}\right| < \varepsilon$, $y \in [c, d]$. Тоді, застосувавши формулу Лагранжа до відношення $\dfrac{\Phi(y_0 - \Delta y) - \Phi(y_0)}{\Delta y} = \dfrac{d\Phi(y_0 + \theta \Delta y)}{dy}$, маємо оцінку $\left|\dfrac{d\Phi(y_0 + \theta \Delta y)}{dy}\right| < \varepsilon$ і, відповідно, виконується нерівність (1.48).

Теорему доведено.

*П р и к л а д  9*. Покажемо, що функція $\Phi(y) = \int\limits_{0}^{+\infty} \dfrac{\sin x \, dx}{1 + x^2 + y^2}$ неперервно диференційовна на будь-якому відрізку. Виконуються умови (теореми 20): функції $f(x, y) = \dfrac{\sin x}{1 + x^2 + y^2}$, $\dfrac{\partial f(x, y)}{\partial y} = \dfrac{-2y \sin x}{(1 + x^2 + y^2)^2}$ неперервні для будь-яких значень змінних $x$ і $y$; інтеграл $\int\limits_{0}^{+\infty} \dfrac{\sin x \, dx}{1 + x^2 + y^2}$ (приклад 8) збігається; інтеграл

$$\frac{d\Phi(y)}{dy} = -2y \int\limits_0^{+\infty} \frac{\sin x \, dx}{\left(1+x^2+y^2\right)^2}$$ збігається рівномірно (виконуються умови теореми Вейєрштрасса).

Отже, функція $\Phi(y)$ неперервно диференційована.

*П р и к л а д 10*. Обчислити інтеграл Діріхле $[5, 9]$

$$I(\alpha) = \int\limits_0^{+\infty} \frac{\sin \alpha x}{x} \, dx \, .$$

Цей інтеграл збігається для усіх значень $\alpha$. Дійсно, якщо $\alpha = 0$, то очевидно $I(0)=0$. Якщо $\alpha \neq 0$, то, провівши заміну $t = \alpha x$, одержимо

$$I(\alpha) = \int\limits_0^{+\infty} \frac{\sin t}{t} \, dt = I(1) \, , \text{ якщо } \alpha > 0 \, ,$$

$$I(\alpha) = -\int\limits_0^{+\infty} \frac{\sin t}{t} \, dt = -I(1) \, , \text{ якщо } \alpha < 0 \, .$$

Інтеграл $I(1)$ збігається за теоремою 14, тому інтеграл $I(\alpha)$ також збігається.

Розглянемо загальніший інтеграл

$$I(\alpha, \beta) = \int\limits_0^{+\infty} e^{-\beta x} \frac{\sin \alpha x}{x} \, dx \, , \qquad (1.49)$$

який для будь-якого $\beta > 0$ рівномірно збігається відносно параметра $\alpha$, $-\infty < \alpha < \infty$. Знайдемо похідну і обчислимо інтеграл

$$\frac{\partial I(\alpha, \beta)}{\partial \alpha} = \int\limits_0^{+\infty} e^{-\beta x} \cos \alpha x \, dx = \frac{\beta}{\alpha^2 + \beta^2} \, .$$

Звідси

$$I(\alpha, \beta) = \int\limits_0^{\alpha} \frac{\beta \, d\alpha}{\alpha^2 + \beta^2} = \operatorname{arctg} \frac{\alpha}{\beta} \, , \, \beta > 0 \, .$$

Обґрунтуємо можливість переходу до границі в цьому інтегралі при $\beta \to +0$.

Покажемо, що інтеграл (1.49) при будь-якому фіксованому

$\alpha \neq 0$ рівномірно збігається відносно $\beta$ на деякому відрізку $[0, b]$. За означенням 7 інтеграл $I(\alpha, \beta)$ на проміжку $[0, b]$ збігається рівномірно, якщо для довільного $\varepsilon > 0$ існує $\eta(\varepsilon)$ таке, що для всіх $\beta \in [0, b]$ і будь-яких $\eta$, $\eta(\varepsilon) < \eta < \infty$, виконується нерівність

$$\left| \int\limits_{\eta}^{+\infty} e^{-\beta x} \frac{\sin \alpha x}{x} dx \right| < \varepsilon. \qquad (1.50)$$

Дійсно, інтегруючи частинами, одержимо

$$\int\limits_{\eta}^{+\infty} e^{-\beta x} \frac{\sin \alpha x}{x} dx = \frac{e^{-\beta x}}{x} \frac{\alpha \cos \alpha x + \beta \sin \alpha x}{\alpha^2 + \beta^2} \bigg|_{\eta}^{+\infty} +$$

$$+ \int\limits_{\eta}^{+\infty} e^{-\beta x} \frac{\alpha \cos \alpha x + \beta \sin \alpha x}{\alpha^2 + \beta^2} \frac{dx}{x^2}.$$

Виберемо $\eta_\varepsilon$ таким, щоби при $\eta > \eta_\varepsilon$ виконувалися нерівності

$$\left| \frac{e^{-\beta \eta}}{\eta} \frac{\alpha \cos \alpha \eta + \beta \sin \alpha \eta}{\alpha^2 + \beta^2} \right| \leq \frac{\alpha + b}{\alpha^2} \frac{1}{\eta} < \frac{\varepsilon}{2},$$

$$\left| \int\limits_{\eta}^{+\infty} e^{-\beta x} \frac{\alpha \cos \alpha x + \beta \sin \alpha x}{\alpha^2 + \beta^2} \frac{dx}{x^2} \right| \leq \frac{\alpha + b}{\alpha^2} \int\limits_{\eta}^{+\infty} \frac{dx}{x^2} < \frac{\varepsilon}{2}.$$

Тоді при $\eta > \eta_\varepsilon$ справедлива нерівність (1.48), а отже, інтеграл $I(\alpha, \beta)$ рівномірно збігається за параметром $\beta$ на довільному проміжку $[0, b]$. За теоремою 17 знайдемо

$$I(\alpha) = I(\alpha, 0) = \lim_{\beta \to +0} I(\alpha, \beta) = \lim_{\beta \to +0} \operatorname{arctg} \frac{\alpha}{\beta} = \frac{\pi}{2} \operatorname{sign} \alpha.$$

Отже,

$$I(\alpha) = \int\limits_{0}^{+\infty} \frac{\sin \alpha x}{x} dx = \begin{cases} \pi/2, & \alpha > 0, \\ 0, & \alpha = 0, \\ -\pi/2, & \alpha < 0. \end{cases}$$

Значення цього інтеграла використовується для обчислення інших важливих інтегралів (див. завдання 7).

### 1.2.4. Завдання до другого параграфа

1. Показати, що $\int_0^1 (1-x^2)^n dx = \dfrac{2^{2n}(n!)^2}{(2n+1)^2}$ (використати приклад 2).

2. Дослідити на збіжність інтеграли $\int_0^{+\infty} \dfrac{\sin ax\, dx}{k^2+x^2}$, $\int_0^{+\infty} \dfrac{\cos ax\, dx}{k^2+x^2}$, $k>0, a>0$.

3. Показати, що невласний інтеграл $\int_1^{+\infty} \dfrac{\sin^2 x}{x} dx$ розбігається (використати рівність $\sin^2 x = \dfrac{1-\cos 2x}{2}$).

4. Показати, що невласний інтеграл $\int_1^{+\infty} \dfrac{|\sin x|}{x} dx$ розбігається (використати нерівність $|\sin x| \geq \sin^2 x$).

5. Показати, що невласний інтеграл $\int_1^{+\infty} \dfrac{\operatorname{arctg} x}{x^\alpha} dx$ збігається при $\alpha > 1$.

6. Обчислити: а) $V.P. \int_{-\infty}^{+\infty} \operatorname{arctg} x\, dx$; б) $V.P. \int_{-\infty}^{+\infty} \dfrac{x^{2k+1}}{\sqrt{x^{2k}+1}} dx$, де $k$ – натуральне число.

7. Показати, що інтеграли а) $\int_0^{+\infty} e^{-\alpha x^2} dx = \dfrac{1}{2}\sqrt{\dfrac{\pi}{\alpha}}$, б) $\int_0^{+\infty} \dfrac{dx}{\alpha+x^2} = \dfrac{\pi}{2\sqrt{\alpha}}$ можна диференціювати за параметром (при $\alpha > 0$) і вивести формули
$$\int_0^{\infty} e^{-\alpha x^2} x^{2n} dx = \dfrac{(2n-1)!!}{2^{n+1}\alpha^n}\sqrt{\dfrac{\pi}{\alpha}}, \quad \int_0^{\infty} \dfrac{dx}{(\alpha+x^2)^{n+1}} = \dfrac{(2n-1)!!}{2(2n)!!\alpha^n}\dfrac{\pi}{\sqrt{\alpha}}.$$

8. Вивести формулу
$$\int_{-\infty}^{\infty} \dfrac{1-\cos \alpha x}{x^2} dx = |\alpha|\pi,$$
використовуючи формулу інтегрування частинами і результат прикладу 10.

### 1.3. Рівномірно збіжні функціональні послідовності

**1.3.1. Функціональні послідовності.** Дослідження збіжності функціональних послідовностей безпосередньо стосується різних питань аналізу, зокрема питань збіжності невласних інтегралів та функціональних рядів. Кожному невласному інтегралу ставиться у відповідність послідовність звичайних інтегралів, а кожному функціональному ряду відповідає функціональна послідовність його частинних сум [5, 8, 12].

Розглянемо функціональну послідовність $\{f_n(x)\}_{n=1}^{\infty}$, де $f_1(x)$, $f_2(x)$, ... — функції дійсної змінної (члени послідовності), неперервні на скінченному або нескінченному проміжку $(a,b) \subset R$.

Фіксуємо довільну точку $x_0 \in (a,b)$ і розглянемо значення членів послідовності у цій точці. Тоді одержимо числову послідовність. Якщо ця послідовність збігається, то функціональна послідовність збігається в точці $x_0$.

*Множина всіх точок збіжності функціональної послідовності є областю її збіжності.*

Хоч у конкретних випадках область збіжності функціональної послідовності може складати лише частину області визначення функцій $f_n(x)$, вважаємо, що область збіжності послідовності – проміжок $(a,b)$.

Сукупність граничних значень послідовності $\{f_n(x)\}$ на множині її збіжності $(a,b)$ є значення деякої функції $f(x)$ на цій множині. Вважаємо, що функції $f_n(x)$ неперервні на проміжку $(a,b)$.

*О з н а ч е н н я  1 . Послідовність $\{f_n(x)\}$ визначених на проміжку $(a,b)$ функцій збігається до функції $f(x)$ в точці $x_0 \in (a,b)$, якщо для будь-якого числа $\varepsilon > 0$ можна вказати номер $N = N(\varepsilon, x_0)$ такий, що для всіх $n > N$ виконується нерівність $|f_n(x_0) - f(x_0)| < \varepsilon$ і пишуть*

$$\lim_{n \to \infty} f_n(x_0) = f(x_0).$$

*П р и к л а д  1*. Розглянемо послідовність функцій $\{f_n(x)\}$, кожна з яких неперервна функція на сегменті $[0, 1]$ і має вигляд

$$f_n(x) = \begin{cases} 1 - nx, & 0 \le x \le 1/n, \\ 0, & 1/n < x \le 1. \end{cases} \quad (1.51)$$

Графік функції $y = f_n(x)$ наведено на рис. 1.1. Послідовність $\{f_n(x)\}$ збігається в усіх точках сегменту $[0, 1]$. В точці $x = 0$ послідовність збігається до одиниці, оскільки $f_n(0) = 1$ для всіх номерів $n$. Якщо фіксувати будь-яку точку $x \in (0, 1]$, то всі значення $f_n(x)$, починаючи з деякого номера (залежного від $x$) дорівнюють нулеві.

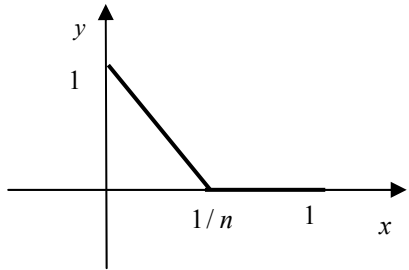

Рис. 1.1.

Отже, послідовність збігається на сегменті $[0, 1]$ до граничної функції

$$f(x) = \begin{cases} 1, & x = 0, \\ 0, & 0 < x \le 1, \end{cases}$$

яка розривна в точці $x = 0$.

*П р и к л а д  2*. Розглянемо послідовність функцій $\{f_n(x)\}$, члени якої нескінченно диференційовані функції на інтервалі $(-\infty, \infty)$ і мають вигляд

$$f_n(x) = e^{-nx^2}.$$

Графік функції $y = f_n(x)$ наведено на рис. 1.2.

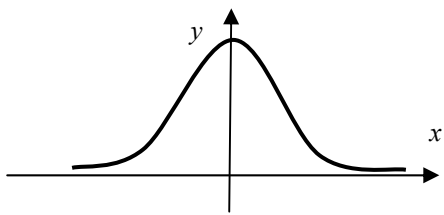

Рис. 1.2.

Послідовність збігається у кожній точці вказаного інтервалу. В точці $x = 0$ послідовність збігається до одиниці, оскільки $f_n(0) = 1$ для всіх номерів $n$. В будь-якій іншій точці $x$, $|x| \in (0, \infty)$, всі значення $f_n(x)$, починаючи з деякого номера (залежного від $x$), будуть менші як-завгодно малого $\varepsilon > 0$.

Гранична функція цієї послідовності – розривна функція

$$f(x) = \begin{cases} 1, & x = 0, \\ 0, & |x| > 0. \end{cases}$$

Розглянуті приклади показують, що умови неперервності чи навіть диференційовності членів послідовності не забезпечують неперервності граничної функції.

*О з н а ч е н н я  2*. *Послідовність $\{f_n(x)\}$, визначених на проміжку $(a, b)$ функцій, збігається до функції $f(x)$ рівномірно на цьому проміжку, якщо для як завгодно малого числа $\varepsilon > 0$ існує номер $N = N(\varepsilon)$ такий, що для всіх $n > N$ і всіх $x \in (a, b)$ виконується нерівність*

$$|f_n(x) - f(x)| < \varepsilon. \qquad (1.52)$$

*П р и к л а д  3*. Розглянемо послідовність функцій (1.51) на сегменті $[\delta, 1]$, де $0 < \delta < 1$,

$$f_n(x) = \begin{cases} 1 - nx, & \delta \leq x \leq 1/n, \\ 0, & 1/n < x \leq 1. \end{cases}$$

Яке б не було число $\delta$ знайдеться такий номер $n$, починаючи з якого всі елементи $f_n(x)$ дорівнюють нулеві на сегменті $[\delta, 1]$.

Гранична функція цієї послідовності також дорівнює нулеві, $f(x) = 0$, $x \in [\delta, 1]$, і тому нерівність $|f_n(x) - f(x)| < \varepsilon$ виконується для будь-якого $\varepsilon > 0$, починаючи з цього номера.

Отже, розглянута послідовність рівномірно збігається на сегменті $[\delta, 1]$.

*З а у в а ж е н н я   1* . У визначенні рівномірної збіжності послідовності суттєвим є те, що існує номер $N = N(\varepsilon)$, який залежить тільки від $\varepsilon$ і не залежить від $x$.

*З а у в а ж е н н я   2* . З означення 2 безпосередньо випливає, що якщо послідовність $\{f_n(x)\}$ збігається до функції $f(x)$ рівномірно на проміжку $(a, b)$, то вона рівномірно збігається до $f(x)$ і на будь-якій частині цього проміжку.

*З а у в а ж е н н я   3* . Із збіжності послідовності $\{f_n(x)\}$ на сегменті $[a, b]$ не випливає її рівномірна збіжність на цьому сегменті.

*П р и к л а д   4* . Легко переконатися, що послідовність (1.51) не збігається рівномірно.

Розглянемо послідовність точок $x_n = \dfrac{1}{2n}$ $(n = 1, 2, ...)$, що належать сегменту $[0, 1]$. Для кожного номера $n$ справедливі співвідношення $f_n(x_n) = \dfrac{1}{2}$, $f(x_n) = 0$ і, відповідно, $|f_n(x_n) - f(x_n)| = \dfrac{1}{2}$. Отже, якщо вибрати $\varepsilon < \dfrac{1}{2}$, то нерівність $|f_n(x) - f(x)| < \varepsilon$ не можна задовольнити одночасно для всіх $x \in [0, 1]$ ні для якого номера $n$.

Справедливе і друге (еквівалентне означенню 2) означення рівномірної збіжності послідовності функцій.

*О з н а ч е н н я   2'* . *Послідовність $\{f_n(x)\}$ визначених на проміжку $(a, b)$ функцій збігається до функції $f(x)$ рівномірно на цьому проміжку, якщо верхня грань модулів відхилень функцій $f_n(x)$ від граничної функції $f(x)$ для всіх $x \in (a, b)$ прямує до нуля, коли $n \to \infty$, тобто*

$$\rho_n = \sup_{x \in (a,\,b)} \left| f(x) - f_n(x) \right| \underset{n \to \infty}{\to} 0.$$

Якщо виконується друге означення, то для будь-якого $\varepsilon > 0$ знайдеться номер $N$ такий, що $\rho_n < \varepsilon$ для всіх $n > N$. Тоді $\left| f(x) - f_n(x) \right| \leq \rho_n < \varepsilon$ для всіх $x \in (a,\,b)$ і $n > N$, тобто виконується перше означення.

Якщо тепер виконується перше означення, то для будь-якого $\varepsilon > 0$ знайдеться номер $N$ такий, що для всіх $n > N$ виконується нерівність (1.52) для всіх $x \in (a,\,b)$. Візьмемо верхню межу лівої частини цієї нерівності за змінною $x \in (a,\,b)$, одержимо $\rho_n < \varepsilon$ для всіх $n > N$, звідки $\rho_n \underset{n \to \infty}{\to} 0$, тобто виконується друге означення.

**1.3.2. Рівномірна фундаментальність.** Поняття фундаментальної числової послідовності може бути перенесене на функціональні послідовності.

*О з н а ч е н н я  3 . Послідовність $\{f_n(x)\}$ визначених на проміжку $(a,\,b)$ функцій називається рівномірно фундаментальною на цьому проміжку, якщо для будь-якого малого числа $\varepsilon > 0$ існує номер $N = N(\varepsilon)$ такий, що для всіх $n > N$, $m > N$ і всіх $x \in (a,\,b)$ виконується нерівність*

$$\left| f_n(x) - f_m(x) \right| < \varepsilon. \qquad (1.53)$$

Зауважимо, що числова послідовність є окремим випадком послідовності сталих функцій. У цьому випадку поняття фундаментальної послідовності і рівномірно фундаментальної послідовності співпадають.

*Т е о р е м а  1 (критерій Коші).* Для того, щоби послідовність $\{f_n(x)\}$ визначених на проміжку $(a,\,b)$ функцій збігалася рівномірно на цьому проміжку до деякої функції $f(x)$, необхідно і достатньо, щоби вона була рівномірно фундаментальною на ньому.

*Н е о б х і д н і с т ь .* Нехай послідовність $\{f_n(x)\}$ збігається рівномірно на множині $(a,\,b)$ до функції $f(x)$. Фіксуємо довільне число $\varepsilon > 0$. Для цього числа знайдеться номер $N$ такий, що для всіх $m > N$ і всіх $x \in (a,\,b)$ справджується нерівність $|f_m(x) - f(x)| < \dfrac{\varepsilon}{2}$. Якщо $n \geq m$, то така нерівність тим більше виконується $|f_n(x) - f(x)| < \dfrac{\varepsilon}{2}$. Оскільки модуль суми не перевищує суми модулів, то одержимо

$$|f_n(x) - f_m(x)| = |[f_n(x) - f(x)] + [f(x) - f_m(x)]| \leq$$
$$\leq |f_n(x) - f(x)| + |f(x) - f_m(x)| < \dfrac{\varepsilon}{2} + \dfrac{\varepsilon}{2} = \varepsilon.$$

Отже, нерівність (1.53) виконується для всіх $n \geq N$, $m \geq N$ і всіх $x \in (a, b)$.

*Д о с т а т н і с т ь*. Нехай послідовність $\{f_n(x)\}$ рівномірно фундаментальна на множині $(a, b)$. Тоді справджується нерівність (1.53) і за критерієм Коші для числової послідовності існують граничні значення відповідних числових послідовностей, які є значеннями деякої функції $f(x)$. Переходячи в цій нерівності до границі при $n \to \infty$, одержимо нерівність $|f(x) - f_m(x)| < \varepsilon$, яка справедлива для всіх $m > N$ і всіх $x \in (a, b)$. Внаслідок довільності числа $\varepsilon > 0$, послідовність $\{f_n(x)\}$ збігається рівномірно до граничної функції на множині $(a, b)$.

Теорему доведено.

***Т е о р е м а  2 (ознака Діні).*** *Нехай послідовність $\{f_n(x)\}$ неперервних на проміжку $(a, b)$ функцій не спадає (або не зростає) в кожній точці сегменту $[a, b]$ і збігається до граничної функції $f(x)$ на цьому сегменті.*

*Тоді, якщо функція $f(x)$ неперервна на $[a, b]$, то збіжність послідовності $\{f_n(x)\}$ є рівномірною на цьому сегменті.*

*Д о в е д е н н я*. Для визначеності приймемо, що послідовність $\{f_n(x)\}$ не спадає на сегменті $[a, b]$.

Розглянемо послідовність $\{r_n(x)\}$, де $r_n(x) = f(x) - f_n(x)$. Характерним для неї є те що: 1) вона не зростає на сегменті $[a, b]$; 2) функції $r_n(x)$ невід'ємні і неперервні на сегменті $[a, b]$; 3) в кожній точці $x \in [a, b]$ існує границя $\lim\limits_{n \to \infty} r_n(x) = 0$. Потрібно довести, що послідовність $\{r_n(x)\}$ збігається до нуля рівномірно на сегменті $[a, b]$.

Припустимо, що для деякого $\varepsilon > 0$ не знайдеться ні одного номера $n$ такого, що $r_n(x) < \varepsilon$ зразу для всіх $x \in [a, b]$. Тоді для будь-якого номера $n$ знайдеться точка $x_n \in [a, b]$ така, що
$$r_n(x_n) \geq \varepsilon. \qquad (1.54)$$

Із послідовності $\{x_n\}$ можна виділити згідно теореми Больцано – Вейєрштрасса підпослідовність $\{x_{n_k}\}$, що збігається до деякої точки $x_0 \in [a, b]$.

Всі функції $r_m(x)$ для будь-якого номера $m$ неперервні в точці $x_0$ і тому для будь-якого номера справедлива рівність
$$\lim\limits_{k \to \infty} r_m(x_{n_k}) = r_m(x_0). \qquad (1.55)$$

З іншого боку, вибравши для будь-якого фіксованого номера $m$ більший від нього номер $n_k$, одержимо (внаслідок не зростання послідовності) $r_m(x_{n_k}) \geq r_{n_k}(x_{n_k})$. Порівнюючи цю нерівність з нерівністю (1.54), одержимо таку нерівність
$$r_m(x_{n_k}) \geq \varepsilon, \qquad (1.56)$$
яка справедлива для будь-якого фіксованого номера $m$ і більшого від нього номера $n_k$. Остаточно, порівнюючи нерівності (1.55) і (1.56), прийдемо до такої нерівності $r_m(x_0) \geq \varepsilon$ для будь-якого номера $m$. Остання нерівність суперечить збіжності послідовності $\{r_n(x)\}_{n=1}^{\infty}$ до нуля в точці $x_0$.

Одержане протиріччя доводить теорему.

*П р и к л а д 5*. В ознаці Діні важливою є умова монотонності послідовності функцій у кожній точці сегмента. Кожна з функцій послідовності $\{f_n(x)\}$,
$$f_n(x) = \begin{cases} \sin nx, & 0 \leq x \leq \pi/n, \\ 0, & \pi/n < x \leq \pi, \end{cases}$$
неперервна на сегменті $[0, \pi]$, рис. 1.3.

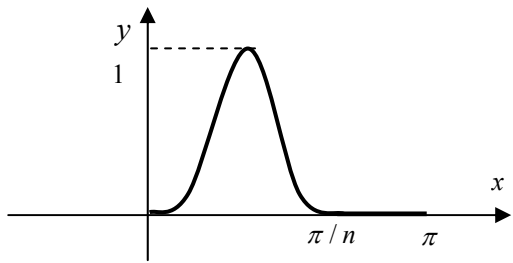

Рис. 1.3

Ця послідовність збігається до функції $f(x) \equiv 0$ в кожній точці $x \in [0, \pi]$, однак вона не збігається рівномірно на цьому сегменті, оскільки $|f_n(x_n) - f(x_n)| = 1$ для всіх $n$, якщо $x_n = \dfrac{\pi}{2n}$.

*П р и к л а д 6*. Послідовність $\{x^n\}$ збігається не рівномірно на відрізку $[0, 1]$.

Гранична функція цієї послідовності -
$$f(x) = \lim_{n \to \infty} x^n = \begin{cases} 0, & 0 \leq x < 1, \\ 1, & x = 1. \end{cases}$$

Тому $\rho_n = \sup\limits_{x \in [0,1]} |f(x) - x^n| = \sup\limits_{x \in [0,1]} |x^n| = 1$ і $\rho_n$ не прямує до нуля, коли $n \to \infty$.

*П р и к л а д 7*. Послідовність $\{x^n\}$ збігається рівномірно до функції $f(x) \equiv 0$ на сегменті $[0, q]$, $0 < q < 1$, оскільки всі умови теореми Діні виконані.

Дійсно: 1) для будь-якого $x \in [0, q]$ послідовність збігається до нуля; 2) всі функції $x^n$ і гранична функція $f(x) \equiv 0$ неперервні на $[0, q]$; 3) послідовність $\{x^n\}$ не зростає на $[0, q]$.

*З а у в а ж е н н я*. Рівномірна збіжність послідовності функцій на проміжку $(a, b)$ еквівалентна збіжності числової послідовності $\{\varepsilon_n\}$, члени якої $\varepsilon_n$ – точні верхні межі функції $|f_n(x) - f(x)|$ на цьому проміжку.

***Т е о р е м а 3 (ознака Вейєрштрасса).*** *Якщо існує числова*

послідовність $\{a_n\}$, $a_n \geq 0$, така, що
$$\lim_{n \to \infty} a_n = 0, \quad |f(x) - f_n(x)| \leq a_n \qquad (1.57)$$
для всіх $n$ і всіх $x \in (a, b)$, то послідовність $\{f_n(x)\}$ рівномірно збігається до функції $f(x)$ на проміжку $(a, b)$.

*Доведення*. З першої умови (1.57) випливає, що для будь-якого числа $\varepsilon > 0$ існує номер $N = N(\varepsilon)$ такий, що для всіх $n \geq N$ виконується нерівність $a_n < \varepsilon$. Тоді в силу другої умови (1.57) маємо $|f(x) - f_n(x)| < \varepsilon$ для всіх $n \geq N$ і всіх $x \in (a, b)$, а це умова рівномірної збіжності послідовності до функції $f(x)$ на множині $(a, b)$.

Теорему доведено.

***Теорема 4***. *Якщо функціональна послідовність $\{f_n(x)\}$ неперервних на проміжку $(a, b)$ функцій збігається рівномірно на проміжку $(a, b)$ (інтервалі, сегменті, пів-сегменті) до граничної функції $f(x)$, то гранична функція $f(x)$ неперервна на $(a, b)$. При цьому в довільній точці $x_0 \in (a, b)$ справджується рівність*

$$\lim_{x \to x_0} \left[ \lim_{n \to \infty} f_n(x) \right] = \lim_{n \to \infty} \left[ \lim_{x \to x_0} f_n(x) \right]. \qquad (1.58)$$

*Доведення*. Послідовність $\{f_n(x)\}$ збігається до функції $f(x)$ у будь-якій точці проміжку $(a, b)$. Розглянемо точку $x_0 \in (a, b)$ і оцінимо різницю $f(x) - f(x_0)$ для всіх $x$ з достатньо малого околу точки $x_0$. Для будь-якого номера $n$ і будь-якого $x \in (a, b)$ справедлива тотожність
$$f(x) - f(x_0) \equiv [f_n(x) - f_n(x_0)] +$$
$$+ [f(x) - f_n(x)] - [f(x_0) - f_n(x_0)].$$

З цієї тотожності одержимо нерівність
$$|f(x) - f(x_0)| \leq |f_n(x) - f_n(x_0)| +$$
$$+ |f(x) - f_n(x)| + |f(x_0) - f_n(x_0)| \qquad (1.59)$$

Фіксуємо довільне число $\varepsilon > 0$. Оскільки послідовність $\{f_n(x)\}$ збігається рівномірно на множині $(a, b)$ до функції $f(x)$ і, відповідно, послідовність $\{f_n(x_0)\}$ збігається до числа $f(x_0)$, знайдеться номер $n$, починаючи з якого

$$|f(x) - f_n(x)| < \frac{\varepsilon}{3}, \quad x \in (a, b),$$

$$|f(x_0) - f_n(x_0)| < \frac{\varepsilon}{3}. \qquad (1.60)$$

Для фіксованого $\varepsilon > 0$ і вибраного $n$ можна вказати число $\delta > 0$ таке, що

$$|f_n(x) - f_n(x_0)| < \frac{\varepsilon}{3} \qquad (1.61)$$

для всіх $x \in (a, b)$, що справджують умову $0 < |x - x_0| < \delta$.

Врахувавши нерівності (1.60) і (1.61) в (1.59), знайдемо $|f(x) - f(x_0)| < \varepsilon$. Отже, функція $f(x)$ неперервна в точці $x_0$, а також неперервна на проміжку $(a, b)$. Внаслідок цього

$$\lim_{x \to x_0} \left[ \lim_{n \to \infty} f_n(x) \right] = \lim_{x \to x_0} f(x) = f(x_0) =$$

$$= \lim_{n \to \infty} f_n(x_0) = \lim_{n \to \infty} \left[ \lim_{x \to x_0} f_n(x) \right],$$

а отже, справедлива формула (1.58).

Теорему доведено.

**1.3.3. Почленне інтегрування та диференціювання послідовностей.**

***Т е о р е м а  5***. *Якщо послідовність $\{f_n(x)\}$ неперервних на сегменті $[a, b]$ функцій збігається рівномірно на $[a, b]$, то гранична функція $f(x) = \lim\limits_{n \to \infty} f_n(x)$ інтегровна (існують інтеграли Рімана) на $[a, b]$ і справедлива формула*

$$\lim_{n \to \infty} \int_a^x f_n(t)\,dt = \int_a^x f(t)\,dt. \qquad (1.62)$$

*Д о в е д е н н я*. Фіксуємо довільне число $\varepsilon > 0$. Внаслідок рівномірної збіжності послідовності $\{f_n(x)\}$ до функції $f(x)$ на сегменті $[a, b]$ знайдеться номер $N = N(\varepsilon)$ такий, що для всіх

$n > N$ і всіх $x \in [a, b]$ виконується нерівність

$$|f_n(x) - f(x)| < \frac{\varepsilon}{b-a}. \qquad (1.63)$$

Оскільки функції $f_n(x)$ і $f(x)$ неперервні і, відповідно, інтегровані на сегменті $[a, b]$, то оцінюючи інтеграли від цих функцій з урахуванням властивостей інтегровних функцій і нерівності (1.63), одержимо

$$\left| \int_a^x f_n(t)dt - \int_a^x f(t)dt \right| = \left| \int_a^x [f_n(t) - f(t)]dt \right| \leq$$

$$\leq \int_a^x |f_n(t) - f(t)|dt \leq \frac{\varepsilon}{b-a} \int_a^x dt \leq \varepsilon.$$

Ця нерівність виконується для всіх $n > N$, а отже, справедлива формула (1.62).

Теорему доведено.

***Т е о р е м а  6.*** *Нехай кожна функція послідовності $\{f_n(x)\}$ має неперервну похідну на сегменті $[a, b]$, послідовність похідних $\{f_n'(x)\}$ збігається рівномірно на цьому сегменті і послідовність $\{f_n(x)\}$ збігається (хоч би в одній точці $x_0 \in [a, b]$).*

*Тоді послідовність $\{f_n(x)\}$ збігається до граничної функції $f(x)$ рівномірно на сегменті $[a, b]$ і гранична функція має неперервну похідну $f'(x)$ на $[a, b]$, яка є граничною функцією послідовності $\{f_n'(x)\}$, тобто*

$$\lim_{n \to \infty} f_n'(x) = f'(x).$$

*Д о в е д е н н я.* Функція має похідну на сегменті $[a, b]$, якщо існують її похідні в точках інтервалу $(a, b)$, правостороння похідна в точці $x = a$ і лівостороння похідна в точці $x = b$.

Спочатку покажемо, що послідовність $\{f_n(x)\}$ рівномірно збігається на сегменті $[a, b]$. Функції $f_n(x)$ неперервні на сегменті $[a, b]$, оскільки їх похідні неперервні на цьому сегменті. Ґрунтуючись на збіжності числової послідовності $\{f_n(x_0)\}$ і на рівномірній збіжності послідовності $\{f'_n(x)\}$ на сегменті $[a, b]$ встановимо, що для довільного $\varepsilon > 0$ знайдеться номер $N = N(\varepsilon)$ такий, що

$$|f_n(x_0) - f_m(x_0)| < \frac{\varepsilon}{2}, \; |f'_n(x) - f'_m(x)| < \frac{\varepsilon}{2(b-a)}. \quad (1.64)$$

для всіх $n > N$, $m > N$, а також для будь-яких $x \in [a, b]$.

Перетворимо вираз $f_n(x) - f_m(x)$, $x \in [a, b]$, зі застосуванням формули Лагранжа (функції $f_n(x)$ неперервні на сегменті $[x_0, x]$ і диференційовані всередині цього сегменту)

$$f_n(x) - f_m(x) = [f_n(x) - f_m(x)] - [f_n(x_0) - f_m(x_0)] +$$
$$+ [f_n(x_0) - f_m(x_0)] = [f'_n(\xi) - f'_m(\xi)](x - x_0) + [f_n(x_0) - f_m(x_0)].$$

Точка $\xi$ лежить між точками $x$ і $x_0$. Оцінюючи одержану залежність з урахуванням нерівностей (1.64) і $|x - x_0| \leq b - a$, знайдемо

$$|f_n(x) - f_m(x)| \leq |f'_n(\xi) - f'_m(\xi)||x - x_0| + |f_n(x_0) - f_m(x_0)| < \varepsilon.$$

Ця нерівність виконується для всіх $n > N$, $m > N$ і будь-яких $x \in [a, b]$.

Отже, згідно з теоремою 1, послідовність $\{f_n(x)\}$ рівномірно збігається на сегменті $[a, b]$ до деякої неперервної функції $f(x)$.

Доведемо тепер, що в будь-якій точці $x_0 \in [a, b]$ гранична функція $f(x)$ має похідну і ця похідна є граничною функцією послідовності $\{f'_n(x)\}$.

Розглянемо сегмент $[x_0, x]$, що належить сегменту $[a, b]$ ($x - x_0 > 0$, якщо $x_0 = a$ і $x - x_0 < 0$, якщо $x_0 = b$). Застосовуючи формулу Лагранжа до функції $f_n(t) - f_m(t)$ на сегменті $[x_0, x]$, одержимо формулу

$$\frac{f_n(x) - f_m(x) - [f_n(x_0) - f_m(x_0)]}{x - x_0} = f'_n(\xi) - f'_m(\xi),$$

яку перепишемо ще так

$$\frac{f_n(x) - f_n(x_0)}{x - x_0} - \frac{f_m(x) - f_m(x_0)}{x - x_0} = f'_n(\xi) - f'_m(\xi). \quad (1.65)$$

Точка $\xi$ лежить між точками $x$ і $x_0$.

Внаслідок рівномірної збіжності послідовності $\{f_n'(x)\}$ до функції $f'(x)$ на сегменті $[a, b]$, для довільного $\varepsilon > 0$ знайдеться номер $N = N(\varepsilon)$ такий, що для всіх $n > N$, $m > N$ і $x \in [a, b]$ справджується нерівність $|f_n'(x) - f_m'(x)| < \varepsilon$. Враховуючи її у формулі (1.65), одержимо

$$\left| \frac{f_n(x) - f_n(x_0)}{x - x_0} - \frac{f_m(x) - f_m(x_0)}{x - x_0} \right| < \varepsilon .$$

Отже, послідовність $\left\{ \dfrac{f_n(x) - f_n(x_0)}{x - x_0} \right\}$ збігається (за критерієм Коші) рівномірно на сегменті $[x_0, x]$. Згідно теореми 4 функція $\dfrac{f(x) - f(x_0)}{x - x_0}$ є граничною для цієї послідовності і вона має границю при $x \to x_0$, при цьому можлива перестановка граничних переходів

$$f'(x) = \lim_{x \to x_0} \frac{f(x) - f(x_0)}{x - x_0} = \lim_{x \to x_0} \left[ \lim_{n \to \infty} \frac{f_n(x) - f_n(x_0)}{x - x_0} \right] =$$
$$= \lim_{n \to \infty} \left[ \lim_{x \to x_0} \frac{f_n(x) - f_n(x_0)}{x - x_0} \right] = \lim_{n \to \infty} f_n'(x_0) .$$

Цим підтверджено, що функція $f(x)$ має похідну в точці $x_0$ і вона дорівнює $\lim\limits_{n \to \infty} f_n'(x_0)$.

Неперервність функції $f'(x)$ на сегменті $[a, b]$ випливає з умов рівномірної збіжності послідовності неперервних функцій $f_n'(x)$ на цьому сегменті (теорема 4).

Теорему доведено.

*З а у в а ж е н н я .* Стосовно рівномірно збіжних послідовностей неперервних функцій можна стверджувати, що рівномірна збіжність залишає граничну функцію в класі неперервних функцій, в класі інтегровних функцій і в класі диференційовних функцій (у випадку рівномірної збіжності похідних).

## 1.3.4. Завдання до третього параграфа

1. Дослідити на рівномірну збіжність послідовність $\{f_n(x)\}$, якщо $f_n(x) = \dfrac{\sin nx}{n}$, $x \in R$.

(Для $x \in R$ маємо $\lim\limits_{n \to \infty} f_n(x) = 0$ і $f(x) = 0$; $\forall \varepsilon > 0$ $\exists N = N(\varepsilon) > \dfrac{1}{\varepsilon}$ $\forall n \geq N$ $\forall x \in R$ $\Rightarrow \left| \dfrac{\sin x}{n} \right| < \dfrac{1}{n} < \varepsilon$. Отже, послідовність рівномірно збігається).

2. Дослідити на рівномірну збіжність послідовність, загальний член якої визначається за формулою $f_n(x) = x^n$, $x \in [0, 1)$.

(Якщо $x \in [0, 1)$, то $\lim\limits_{n \to \infty} x^n = 0$ і $f(x) = 0$. Розглянемо послідовність точок $x_n = 2^{-1/n}$ $(n = 1, 2, ...)$, $x_n \in [0, 1)$. Маємо $f_n(x_n) = \dfrac{1}{2}$, $f(x_n) = 0$ і, відповідно, $|f_n(x_n) - f(x_n)| = \dfrac{1}{2}$. Отже, якщо вибрати $\varepsilon < \dfrac{1}{2}$, то нерівність $|f_n(x) - f(x)| < \varepsilon$ не можна задовольнити одночасно для всіх $x \in [0, 1)$ ні для якого номера $n$. Отже, послідовність збігається не рівномірно).

3. Дослідити на рівномірну збіжність послідовність, загальний член якої визначається за формулою $f_n(x) = \dfrac{x}{n}$, $x \in [0, 1]$.

(Послідовність рівномірно збігається, доведення аналогічне завданню 1).

4. Показати, що послідовність $\{f_n(x)\}$, де $f_n(x) = \dfrac{x}{1 + (nx)^2}$, збігається рівномірно на сегменті $[0, 1]$.

(Справедлива оцінка $0 \leq f_n(x) = \dfrac{1}{n} \dfrac{nx}{1 + (nx)^2} \leq \dfrac{1}{n}$. Достатньо вибрати $n > \dfrac{1}{\varepsilon}$, щоби для довільного $\varepsilon > 0$ виконувалося $f_n(x) < \varepsilon$)

5. Показати, що послідовність $\{f_n(x)\}$, де $f_n(x) = \dfrac{nx}{1 + (nx)^2}$, не збігається рівномірно на сегменті $[0, 1]$.

6. Показати, що послідовність $\{f_n(x)\}$, де $f_n(x) = \dfrac{1}{n + x^2}$, збігається рівномірно на проміжку $(-\infty, +\infty)$.

## 1.4. Збіжні в середньому функціональні послідовності

**1.4.1. Поняття нормованого простору.** Множини дійсних чисел, множини векторів площини чи тривимірного простору, а також множини функцій можна охарактеризувати спільними властивостями. Важливими з них ті, що приводять до поняття лінійного простору та метрики в ньому. Поняття модуля дійсного числа чи модуля вектора дозволяє ввести метрику на числовій осі або у просторі векторів. Введення поняття метрики у функціональних просторах дозволяє ширше розглянути питання збіжності послідовностей функцій та їх граничних елементів $[6, 9, 11, 12, 20]$.

*О з н а ч е н н я   1* . *Множина E елементів $f, g, \varphi, \ldots$, називається лінійним простором, якщо в ній визначені операція додавання $f + g \in E$ і операція множення на скаляр $\lambda f \in E$ так, що для будь-яких елементів цієї множини і довільних скалярів справджуються наступні аксіоми:*

*1)* $f + g = g + f$ ;
*2)* $f + (g + \varphi) = (f + g) + \varphi$ ;
*3) існує нульовий елемент $o \in E$ такий, що $f + o = f$ ;*
*4)* $\lambda(\mu f) = (\lambda \mu) f$ ;
*5)* $1 \cdot f = f$ ;
*6) поряд з $f \in E$ існує протилежний елемент $-f \in E$ такий, що $f + (-f) = o$ ;*
*7)* $(\lambda + \mu) f = \lambda f + \mu f$ ;
*8)* $\lambda(f + g) = \lambda f + \lambda g$ .

Числові множники (скаляри) $\lambda, \mu, \ldots$ у лінійному просторі дійсні або комплексні числа. У першому випадку $E$ – дійсний простір, в другому випадку $E$ – комплексний простір.

*П р и к л а д   1* . Множина $R$ дійсних чисел є лінійним простором, оскільки сума двох дійсних чисел є дійсним числом, добуток дійсного числа (множника) на дійсне число також є дійсним числом. Аксіоми лінійного простору справджуються тривіально.

Подібним способом можна ввести лінійний простір комплексних чисел з множенням на комплексні числа.

*П р и к л а д 2*. Розглянемо множину $R^m$ елементів, якими є упорядковані стовпчики з $m$ дійсних чисел $u = \{x_i\}_{i=1}^{m}$, $v = \{y_i\}_{i=1}^{m}$, де $x_i, y_i$ – координати. Під сумою елементів множини розуміємо стовпчик, координати якого – суми відповідних координат $u + v = \{x_i + y_i\}_{i=1}^{m}$. Під стовпчиком $\lambda u$, де $\lambda$ – дійсне число, розуміємо стовпчик $\lambda u = \{\lambda x_i\}_{i=1}^{m}$. Нульовим елементом множини є стовпчик $o = \{0\}_{i=1}^{m}$. Оскільки операції над стовпчиками зводяться до операцій над координатами – дійсними числами, для яких аксіоми виконуються, то ці аксіоми виконуються і для стовпчиків.

Отже, $R^m$ – лінійний простір.

Зазначимо, що лінійним простором є також множина $M_{m \times n}$ матриць розміру $m \times n$.

*П р и к л а д 3*. Лінійним простором є множина $P_n$ многочленів $P_n(t) = \sum_{i=0}^{n} a_i t^{n-i}$ степеня не більшого, ніж $n$, де $a_i$ – довільні дійсні числа, $t \in (-\infty, +\infty)$. Оскільки добуток многочленів на дійсне число і сума двох многочленів є многочленами, множина $P_n$ – лінійний простір.

Аналогічно можна розглянути комплексний лінійний простір многочленів степеня не більшого, ніж $n$. Його елементи – многочлени комплексної змінної з комплексними коефіцієнтами.

*П р и к л а д 4*. Розглянемо простір $C[a, b]$ неперервних на відрізку $[a, b]$ функцій. Якщо функції $f(x)$ і $g(x)$ неперервні на $[a, b]$, а $\lambda$ – дійсне число, то $f(x) + g(x)$ і $\lambda f(x)$ також неперервні функції на цьому відрізку. Нуль-елементом простору є функція, що тотожно дорівнює нулю на $[a, b]$.

Отже, $C[a, b]$ – лінійний простір.

Можливі також варіанти лінійних просторів функцій комплексної змінної.

***О з н а ч е н н я 2 .*** *Лінійний простір E називається нормованим простором, якщо кожному елементу* $f \in E$

поставлено у відповідність невід'ємне число $\|f\|_E$ (норма $f$) таке, що справджуються наступні аксіоми:

*1)* $\|f\|_E \geq 0$ *і* $\|f\|_E = 0$ *лише в тому випадку, коли* $f = o$;

*2)* $\|\lambda f\|_E = |\lambda| \cdot \|f\|_E$;

*3)* $\|f + g\|_E \leq \|f\|_E + \|g\|_E$, *де* $f, g \in E$.

Зауважимо, що норма – це визначена на множині $E$ функція з невід'ємними значеннями, аксіома 1) має назву умови *невиродженності*, аксіома 2) – умови *однорідності* і аксіома 3) – *нерівності трикутника*.

***Т е о р е м а  1 .*** *Якщо $E$ – нормований простір і $f, g \in E$, то справедлива нерівність* $\|f - g\|_E \geq \|f\|_E - \|g\|_E$.

*Д о в е д е н н я .* За нерівністю трикутника маємо $\|f\|_E = \|(f - g) + g\|_E \leq \|f - g\|_E + \|g\|_E$. Звідси $\|f - g\|_E \geq \|f\|_E - \|g\|_E$.

Теорему доведено.

У нормованому просторі між елементами $f, g \in E$ можна ввести метрику (віддаль)
$$\rho(f, g) = \|f - g\|_E.$$

***О з н а ч е н н я  3 .*** *Лінійний простір називається метричним простором, якщо кожній парі його елементів $f, g \in E$ поставлено у відповідність дійсне число $\rho(f, g)$, що справджує аксіоми:*

*1)* $\rho(f, g) \geq 0$ *і* $\rho(f, g) = 0$ *лише в тому випадку, коли* $f = g$;

*2)* $\rho(f, g) = \rho(g, f)$;

*3)* $\rho(f, g) \leq \rho(f, u) + \rho(u, g)$, *де* $u \in E$.

Отже, будь-який нормований простір є метричним простором з метрикою $\rho(f, g) = \|f - g\|_E$.

У нормованому просторі важливим є питання збіжності послідовності його елементів.

***О з н а ч е н н я  4 .*** *Послідовність елементів $\{f_n\} \subset E$ називається збіжною до елемента $f \in E$, якщо для будь-якого $\varepsilon > 0$ можна вказати номер $N = N(\varepsilon)$ такий, що для всіх $n > N$* *виконується нерівність* $\|f - f_n\|_E < \varepsilon$ *і пишуть*

$$f = \lim_{n \to \infty}^{E} f_n \text{ або } \lim_{n \to \infty} \|f - f_n\|_E = 0.$$

Визначена таким чином збіжність у просторі $E$ називається *збіжністю за нормою* (простору $E$).

Збіжна за нормою послідовність має звичні для числової послідовності властивості.

***Т е о р е м а  2 .*** *Якщо послідовність* $\{f_n\}$ *збіжна у просторі* $E$, *то:*

*1) вона має тільки одну границю;*

*2) вона обмежена за нормою;*

*3) якщо* $\lim\limits_{n\to\infty}^{E} f_n = f$, $\lim\limits_{n\to\infty}^{E} g_n = g$, *то* $\lim\limits_{n\to\infty}^{E} (f_n + g_n) = f + g$;

*4) в будь-якому околі точки $f$ лежать всі члени послідовності* $\{f_n\}$, *за виключенням хіба-що скінченного їх числа;*

*5) норма є неперервною функцією на лінійному нормованому просторі;*

*6) якщо* $\lim\limits_{n\to\infty} \lambda_n = \lambda$ *і* $\lim\limits_{n\to\infty}^{E} f_n = f$, *то* $\lim\limits_{n\to\infty}^{E} \lambda_n f_n = \lambda f$.

*Д о в е д е н н я .* 1) Оскільки послідовність $\{f_n\} \subset E$ має границю, візьмемо довільне число $\varepsilon > 0$ і виберемо номер $N$ такий, що для всіх $n > N$ виконується нерівність $\|f - f_n\| < \dfrac{\varepsilon}{2}$.

Нехай послідовність має дві границі $\lim\limits_{n\to\infty}^{E} f_n = f^1$, $\lim\limits_{n\to\infty}^{E} f_n = f^2$ і $f^1 \neq f^2$. Тоді $\|f^1 - f_n\|_E < \dfrac{\varepsilon}{2}$, $\|f^2 - f_n\|_E < \dfrac{\varepsilon}{2}$ для $\varepsilon > 0$ і всіх $n > N$. За нерівністю трикутника знайдемо $\|f^1 - f^2\|_E =$
$= \|(f^1 - f_n) + (f_n - f^2)\|_E \leq \|f^1 - f_n\|_E + \|f_n - f^2\|_E < \varepsilon$, що заперечує прийняте припущення. Отже, $\|f^1\|_E = \|f^2\|_E$.

2) Нехай $\lim\limits_{n\to\infty}^{E} f_n = f$. Тоді існує натуральне число $N$ таке, що

якщо $n \geq N$, то $\|f_n - f\|_E \leq 1$ і, відповідно, $\|f_n\|_E \leq \|(f_n - f) + f\|_E \leq \|f_n - f\|_E + \|f\|_E \leq \|f\|_E + 1$ Позначимо через $M = \max\{\|f_1\|_E, \|f_2\|_E, ..., \|f_{N-1}\|_E, \|f\|_E + 1\}$. Тоді, очевидно, $\|f_n\|_E \leq M$ для всіх $n = 1, 2, ...$, що доводить обмеженість послідовності за нормою.

3) Оскільки послідовності $\{f_n\} \subset E$ і $\{g_n\} \subset E$ мають границі, для довільного числа $\varepsilon > 0$ можна вибрати номер $N = N(\varepsilon)$ такий, що для всіх $n > N$ виконуються нерівності $\|f - f_n\|_E < \dfrac{\varepsilon}{2}$, $\|g - g_n\|_E < \dfrac{\varepsilon}{2}$. Перетворимо норму виразу $(f + g) - (f_n + g_n)$ з урахуванням цих нерівностей $\|(f + g) - (f_n + g_n)\|_E =$
$= \|(f - f_n) + (g - g_n)\|_E \leq \|f - f_n\|_E + \|g - g_n\|_E \leq \varepsilon$.

Одержана нерівність виконується для всіх $n > N$, а отже, $\lim\limits_{n \to \infty} (f_n + g_n) \overset{E}{=} f + g$.

4) Нехай $S_\delta(f, f_0) = \{f : \rho(f, f_0) < \delta\}$ – довільний окіл точки $f_0$ (відкрита куля) і $\lim\limits_{n \to \infty} f_n \overset{E}{=} f_0$. Тоді для вибраного $\delta > 0$ існує номер $N$ такий, що для всіх $n \geq N$ виконується нерівність $\|f - f_n\|_E \leq \delta$. Очевидно, якщо $N > 1$, то зовні околу можуть лежати хіба що деякі з членів $f_1, f_2, ..., f_{N-1}$ послідовності. В протилежному випадку всі члени послідовності лежать в цьому околі.

5) Покажемо, що $\lim\limits_{n \to \infty} \|f_n\|_E = \|f\|_E$. За означенням $\lim\limits_{n \to \infty} \|f_n - f\|_E = 0$, тобто для будь-якого числа $\varepsilon > 0$ існує номер $N = N(\varepsilon)$ такий, що для всіх $n > N$ виконується нерівність $\|f - f_n\|_E < \varepsilon$. Звідси з урахуванням нерівності $\|f - f_n\|_E \geq \|f\|_E - \|f_n\|_E$ (теорема 1) отримаємо $\|f\|_E - \|f_n\|_E < \varepsilon$.

6) З урахуванням третьої аксіоми норми і властивості

лінійності норми оцінимо вираз

$$\|\lambda_n f_n - \lambda f\|_E = \|(f_n - f)\lambda + (\lambda_n - \lambda)f_n\|_E \le \|f_n - f\|_E |\lambda| +$$
$$+ |\lambda_n - \lambda| \|f\|_E.$$ Звідси маємо

$$\lim_{n \to \infty} \|\lambda_n f_n - \lambda f\|_E \le |\lambda| \lim_{n \to \infty} \|f_n - f\|_E + \|f\|_E \lim_{n \to \infty} |\lambda_n - \lambda| = 0.$$

Отже, $\lim\limits_{n \to \infty} \lambda_n f_n \overset{E}{=} \lambda f$.

Теорему доведено.

В одному лінійному просторі можна визначити декілька норм.

*О з н а ч е н н я  5 .* *Дві норми простору E називаються еквівалентними, якщо послідовність $\{f_n\}$ збігається до елемента $f \in E$ за одною нормою, то вона збігається до цього елемента за іншою нормою.*

*П р и к л а д  5 .* Розглянемо простір $R^m$. Для елемента $u = \{x_i\}_{i=1}^m \in R^m$ визначають, зокрема, норми

$$\|u\|_1 = \sum_{i=1}^m |x_i|, \quad \|u\|_2 = \left(\sum_{i=1}^m x_i^2\right)^{1/2}, \quad \|u\|_\infty = \max_{i=1,m} |x_i|. \quad (1.66)$$

Кожна з цих норм задовольняє аксіоми норми. Норма $\|u\|_1$ називається *октаедричною*, норма $\|u\|_2$ – *евклідовою* (або сферичною) і норма $\|u\|_\infty$ – *кубічною*.

Легко переконатися, що норми (1.66) еквівалентні. Вони справджують нерівності

$$\max_{i=1,m} |x_i| \le \sum_{i=1}^m |x_i| \le m^{1/2} \left(\sum_{i=1}^m x_i^2\right)^{1/2} \le m \max_{i=1,m} |x_i|. \quad (1.67)$$

Перша і третя нерівності очевидні. Друга нерівність випливає з нерівності Шварца для елементів простору $R^m$

$$\sum_{i=1}^m |x_i y_i| \le \left(\sum_{i=1}^m |x_i|^2\right)^{1/2} \left(\sum_{i=1}^m |y_i|^2\right)^{1/2},$$

де $\{x_i\}_{i=1}^m$, $\{y_i\}_{i=1}^m \in R^m$, якщо прийняти $y_i = 1$.

Покажемо, наприклад, що із збіжності за третьою нормою (1.66) випливає збіжність за другою нормою (1.66). Якщо послідовність $\{u_n\}_{n=1}^{\infty} \in R^m$ збігається до елемента $u \in R^m$ за третьою нормою (1.66), то для будь-якого $\varepsilon > 0$ можна вказати номер $N = N(\varepsilon)$ такий, що для всіх $k > N$ і $n > N$ виконується нерівність $\|u_n - u_k\|_{\infty} < m^{-1/2}\varepsilon$. Враховуючи тут другу нерівність (1.67),

$$\|u_n - u_k\|_2 = \left(\sum_{i=1}^{m}(x_{n,i} - x_{k,i})^2\right)^{1/2} \le m^{1/2} \max_{i=\overline{1,m}} |x_{n,i} - x_{k,i}| =$$
$$= m^{1/2}\|u_n - u_k\|_{\infty} < \varepsilon,$$

отримаємо, що для вибраного $\varepsilon > 0$, знайденого $N = N(\varepsilon)$ і всіх $k > N$, $n > N$ випливає нерівність $\|u_n - u_k\|_2 < \varepsilon$.

Отже, послідовність $\{u_n\}_{n=1}^{\infty}$ збігається в $R^m$ за другою нормою (1.66).

Важливим поняттям, пов'язаним зі збіжністю, є поняття повноти простору.

***О з н а ч е н н я  6 .*** *Послідовність $\{f_n\} \subset E$ називається фундаментальною послідовністю в просторі $E$, якщо для будь-якого $\varepsilon > 0$ знайдеться номер $N = N(\varepsilon)$ такий, що як тільки $n > N$ і $m > N$, то*

$$\|f_n - f_m\|_E < \varepsilon.$$

***Т е о р е м а  3 .*** *Якщо послідовність $\{f_n\}$ збігається в $E$, то вона фундаментальна.*

*Д о в е д е н н я .* Якщо послідовність $\{f_n\}$ збігається до $f \in E$, то для довільного $\varepsilon > 0$ знайдеться номер $N = N(\varepsilon)$ такий, що для всіх $n > N$ виконується нерівність $\|f - f_n\|_E < \dfrac{\varepsilon}{2}$. Якщо $m > N$, то також справедлива нерівність $\|f - f_m\|_E < \dfrac{\varepsilon}{2}$. За нерівністю трикутника маємо $\|f_m - f_n\|_E \le \|(f_m - f) + (f - f_n)\|_E \le$
$\le \|f_m - f\|_E + \|f - f_n\|_E < \varepsilon.$

Отже, послідовність $\{f_n\}$ фундаментальна.

Теорему доведено.

***О з н а ч е н н я  7 .*** *Нормований простір E називається повним простором, якщо будь-яка фундаментальна послідовність $\{f_n\} \subset E$ збігається до елемента $f$ цього простору.*

*Повний нормований простір називається простором Банаха.*

*П р и к л а д  6 .* Нехай $Q$ – множина раціональних чисел і $R$ – множина дійсних чисел. Можна вважати, що $Q$ і $R$ – лінійні простори, елементами яких, відповідно, раціональні і дійсні числа, а скалярами є також, відповідно, раціональні та дійсні числа $[11\,c.118]$. Ці простори нормовані, якщо за норми елементів вибрати їх модулі. При цьому $R$ – повний простір (простір Банаха), а $Q$ – неповний простір.

Дійсно, фундаментальна послідовність $x_n = \left(1+\dfrac{1}{n}\right)^n$ належить простору $Q$, а її граничний елемент $\lim\limits_{n\to +\infty}\left(1+\dfrac{1}{n}\right)^n = e$ не належить цьому простору, $e = 2{,}718281...$ – ірраціональне число.

**1.4.2. Простір неперервних функцій.** Розглянемо лінійний простір $C[a,b]$ всіх неперервних на відрізку $[a,b]$ функцій. Якщо $f(x)$ – елемент цього простору, то його норма –
$$\|f\|_{C[a,b]} = \sup_{x\in[a,b]} |f(x)|.$$

Перша і друга аксіоми норми (означення 2) виконуються. Виконання третьої аксіоми випливає з властивостей модуля числа. Для будь-яких $x_1, x_2 \in [a,b]$ маємо нерівність $|f(x_1)+g(x_2)| \le$
$\le |f(x_1)| + |g(x_2)| \le \sup\limits_{x\in[a,b]} |f(x)| + \sup\limits_{x\in[a,b]} |g(x)|$. Ця нерівність виконується також для максимальних значень функцій, тобто справедлива нерівність трикутника
$$\|f+g\|_{C[a,b]} \le \|f\|_{C[a,b]} + \|g\|_{C[a,b]}.$$

Отже, $C[a,b]$ – нормований простір.

***Т е о р е м а  4 .*** *Збіжність функціональної послідовності*

$\{f_n(x)\}$ *за нормою в просторі* $C[a,b]$ *є рівномірною збіжністю цієї послідовності на сегменті* $[a,b]$.

*Доведення*. Із збіжності послідовності $\{f_n(x)\} \subset C[a,b]$ до функції $f(x)$ випливає (за означенням 4), що для будь-якого $\varepsilon > 0$ знайдеться номер $N = N(\varepsilon)$ такий, що для всіх $n > N$ виконується нерівність $\|f - f_n\|_{C[a,b]} \equiv \sup\limits_{x \in [a,b]} |f(x) - f_n(x)| < \varepsilon$. Однак, якщо $\sup\limits_{x \in [a,b]} |f(x) - f_n(x)| < \varepsilon$, то $|f(x) - f_n(x)| < \varepsilon$ для всіх $x \in [a,b]$.

Отже, збіжність в $C[a,b]$ за нормою є рівномірною збіжністю неперервних функцій на $[a,b]$.

Теорему доведено.

***Теорема 5.*** *Функціональний простір* $C[a,b]$ – *повний простір (простір Банаха).*

*Доведення*. Розглянемо фундаментальну послідовність $\{f_n(x)\} \subset C[a,b]$ і покажемо, що вона має границю $f(x) \in C[a,b]$. Для будь-якого $\varepsilon > 0$ знайдеться номер $N = N(\varepsilon)$ такий, що для всіх $n > N$ і $m > N$ виконується нерівність $\|f_n - f_m\|_{C[a,b]} =$
$= \max\limits_{x \in [a,b]} |f_n(x) - f_m(x)| < \varepsilon$, а отже,

$$|f_n(x) - f_m(x)| < \varepsilon$$

для всіх $x \in [a,b]$ і $n > N$, $m > N$. Тому функціональна послідовність $\{f_n(x)\}$ фундаментальна (в сенсі поточкової збіжності) і кожна з функцій $f_n(x)$ неперервна на сегменті $[a,b]$, а отже (за теоремами 1 і 4 п. 1.3), гранична функція $f(x)$ неперервна на цьому сегменті і, відповідно, $f(x) \in C[a,b]$.

Теорему доведено.

**Важливими в аналізі є простори** $C_\alpha[a,b]$ **неперервних функцій, що справджують умову Гельдера** [5]**.** Неперервна на відрізку $[a,b]$ функція $f(x)$ задовольняє на цьому відрізку умову Гельдера (належить класу Гельдера $C_\alpha[a,b]$) з показником $\alpha$,

$0 < \alpha \leq 1$, якщо існує стала $M > 0$ така, що для всіх $x_1, x_2 \in [a, b]$ виконується нерівність
$$|f(x_1) - f(x_2)| \leq M |x_1 - x_2|^{\alpha}.$$

Точну верхню межу відношення $\dfrac{|f(x_1) - f(x_2)|}{|x_1 - x_2|^{\alpha}}$ на множині всіх точок $x_1, x_2 \in [a, b]$, $x_1 \neq x_2$, називають константою Гельдера функції $f(x)$ на відрізку $[a, b]$
$$K_{\alpha}(f) = \sup_{x_1, x_2 \in [a,b]} \frac{|f(x_1) - f(x_2)|}{|x_1 - x_2|^{\alpha}}.$$

Нормою Гельдера функції $f(x)$ на відрізку $[a, b]$ називають суму верхньої межі $|f(x)|$ на цьому відрізку і константи Гельдера
$$\|f\|_{C_{\alpha}[a, b]} = \sup_{x \in [a, b]} |f(x)| + K_{\alpha}(f).$$

Перша і друга аксіоми норми випливають з властивостей модуля числа. Аналогічно одержимо і третю аксіому. Для будь-яких точок $x_1', x_2', x_1'', x_2'' \in [a, b]$, $x_1' \neq x_2'$, $x_1'' \neq x_2''$ маємо нерівність
$$\left| f(x_1') + g(x_1'') + \frac{f(x_1') - f(x_2')}{(x_1' - x_2')^{\alpha}} + \frac{g(x_1'') - g(x_2'')}{(x_1'' - x''')^{\alpha}} \right| \leq$$
$$\leq |f(x_1')| + |g(x_1'')| + \frac{|f(x_1') - f(x_2')|}{|x_1' - x_2'|^{\alpha}} + \frac{|g(x_1'') - g(x_2'')|}{|x_1'' - x_2''|^{\alpha}} \leq$$
$$\leq \sup_{x \in [a, b]} |f(x)| + \sup_{x \in [a, b]} |g(x)| + K_{\alpha}(f) + K_{\alpha}(g) =$$
$$= \|f\|_{C_{\alpha}[a, b]} + \|g\|_{C_{\alpha}[a, b]}.$$

Оскільки ця нерівність виконується для будь яких значень аргумента, вона виконується і для тих значень, в яких відповідні функції приймають найбільші значення, тобто
$$\|f + g\|_{C_{\alpha}[a, b]} \leq \|f\|_{C_{\alpha}[a, b]} + \|g\|_{C_{\alpha}[a, b]}.$$

*П р и к л а д  7*. Функція $f(x) = \sqrt{x}$ належить класу $C_{1/2}[0, b]$, $0 < b < \infty$. Для будь-яких точок $x_1, x_2 \in [a, b]$, $x_1 > x_2$, виконується нерівність $\sqrt{x_1} - \sqrt{x_2} \leq \sqrt{x_1} + \sqrt{x_2}$. Домножимо ліву і праву частини цієї нерівності на додатну величину $\sqrt{x_1} - \sqrt{x_2}$, одержимо $\left(\sqrt{x_1} - \sqrt{x_2}\right)^2 \leq x_1 - x_2$ або $|f(x_1) - f(x_1)| \leq |x_1 - x_2|^{1/2}$. При цьому константа Гельдера $K_{1/2}(f) = 1$, а норма Гельдера $\|f\|_{C_{1/2}[0,\,b]} = \sqrt{b} + 1$.

**1.4.3. Простір неперервно диференційовних функцій.** Лінійний простір $k$ раз неперервно диференційовних функцій $f(x)$ на сегменті $[a,b]$ з нормою

$$\|f\|_{C^k[a,\,b]} = \sum_{i=0}^{k} \sup_{x \in [a,\,b]} \left| f^{(i)}(x) \right|$$

називається простором $C^k[a,b]$, де $f^{(i)}(x)$ – похідна $i$-го порядку від функції $f(x)$.

Функції неперервні на сегменті $[a,b]$ в тому розумінні, що: а) функція $f(x)$ і всі її похідні до $k$-го порядку (включно) неперервні в інтервалі $(a,b)$; б) функція $f(x)$ і всі її похідні мають граничні значення справа в точці $x = a$ і зліва в точці $x = b$.

Покажемо, що аксіоми норми виконуються. Перша і друга аксіоми норми виконуються тривіально. Третю аксіому – нерівність трикутника одержимо з оцінки виразу

$$\sum_{i=1}^{k} \left| f^{(i)}(x_1) + g^{(i)}(x_2) \right| \leq \sum_{i=1}^{k} \left[ \left| f^{(i)}(x_1) \right| + \left| g^{(i)}(x_2) \right| \right] =$$
$$= \sum_{i=1}^{k} \left| f^{(i)}(x_1) \right| + \sum_{i=1}^{k} \left| g^{(i)}(x_2) \right| \leq \sum_{i=1}^{k} \sup_{x \in [a,\,b]} \left| f^{(i)}(x) \right| + \sum_{i=1}^{k} \sup_{x \in [a,\,b]} \left| g^{(i)}(x) \right|$$

у довільних точках $x_1, x_2 \in [a,b]$. Ця нерівність виконується і для максимальних значень функцій, тобто

$$\|f + g\|_{C^k[a,\,b]} \leq \|f\|_{C^k[a,\,b]} + \|g\|_{C^k[a,\,b]}.$$

У просторі $C^k[a,b]$ справедливі аналогічні з простором неперервних функцій теореми про рівномірну збіжність та повноту простору.

*Збіжність функціональної послідовності $\{f_n(x)\}$ за нормою в просторі $C^k[a,b]$ є рівномірною збіжністю послідовностей*

$\{f_n^{(i)}(x)\}$, $i = \overline{0, k}$, на сегменті $[a, b]$.

*Функціональний простір $C^{(k)}[a, b]$ – повний простір (простір Банаха).*

**Важливими є також простори неперервно диференційовних функцій, що справджують умову Гельдера.** Лінійний простір $k$ раз неперервно диференційовних функцій $f(x)$ на сегменті $[a, b]$ з похідною $k$-го порядку, що задовольняє умову Гельдера

$$\left| f^{(k)}(x_1) - f^{(k)}(x_2) \right| \le M \left| x_1 - x_2 \right|^\alpha, \; 0 < \alpha \le 1, \; M = const,$$

називається простором $C_\alpha^k[a, b]$. Норма простору задається виразом

$$\| f \|_{C_\alpha^k[a, b]} = \sum_{i=0}^{k} \sup_{x \in [a, b]} \left| f^{(i)}(x) \right| + K_\alpha\left(f^{(k)}\right),$$

де $K_\alpha\left(f^{(k)}\right) = \sup\limits_{x_1, x_2 \in [a, b]} \dfrac{\left| f^{(k)}(x_1) - f^{(k)}(x_2) \right|}{\left| x_1 - x_2 \right|^\alpha}$, $x_1, x_2 \in [a, b]$, $x_1 \ne x_2$.

*З а у в а ж е н н я .* Якщо функція $f^{(k)}(x)$ кусково-гладка на відрізку $[a, b]$ і, відповідно, функція $f^{(k+1)}(x)$ обмежена на цьому відрізку, то $f(x) \in C_1^k[a, b]$.

**1.4.4. Простори неперервних функцій, інтегровних в середньому.** Лінійний простір функцій $f(x)$, неперервних на сегменті $[a, b]$, в якому введено норму

$$\| f \|_{L_p} = \left( \int_a^b | f(x) |^p dx \right)^{1/p} \quad 1 \le p < \infty, \tag{1.68}$$

називається простором $CL_p[a, b]$.

Покажемо, що виконуються аксіоми норми. Спочатку доведемо допоміжне твердження.

*Л е м а 1 .* Нехай функції $f(x)$ і $g(x)$ неперервні на відрізку $[a, b]$, $1 < p < \infty$, $\dfrac{1}{p} + \dfrac{1}{q} = 1$.

*Тоді справедлива нерівність Гельдера*

$$\left|\int\limits_a^b f(x)g(x)dx\right| \leq \left(\int\limits_a^b |f(x)|^p dx\right)^{1/p} \left(\int\limits_a^b |g(x)|^q dx\right)^{1/q}. \quad (1.69)$$

*Доведення.* Використаємо арифметичну нерівність

$$xy \leq \frac{x^p}{p} + \frac{y^q}{q}, \quad 0 \leq x < \infty, \ 0 \leq y < \infty, \quad (1.70)$$

доведення якої ґрунтується на дослідженні функції $\varphi(x) = \frac{x^p}{p} + \frac{y_0^q}{q} - xy_0$ за фіксованого $y = y_0 > 0$. Оскільки мінімальне значення цієї функції $\varphi(x_0) = 0$ досягається в точці $x_0 = y_0^{1/(p-1)} = y_0^{q/p}$ і $y_0 > 0$ – довільне число, одержимо нерівність $\varphi(x) \geq 0$. Звідки випливає нерівність (1.70).

Приймаючи в нерівності (1.70) $x(t) = \frac{f(t)}{\|f\|_{L_p}}$, $y(t) = \frac{g(t)}{\|g\|_{L_q}}$ і інтегруючи її на відрізку $[a, b]$ (функції $f(t)$ і $g(t)$ неперервні на цьому відрізку), одержимо

$$\frac{1}{\|f\|_{L_p} \|g\|_{L_q}} \int\limits_a^b |f(x)g(x)|dx \leq \frac{1}{p\|f\|_{L_p}^p} \int\limits_a^b |f(x)|^p dx +$$

$$+ \frac{1}{q\|f\|_{L_q}^q} \int\limits_a^b |g(x)|^q dx = \frac{1}{p} + \frac{1}{q} = 1.$$

Звідси, одержимо (з урахуванням виразів норм функцій $f(t)$ і $g(t)$) нерівність (1.69).

Лему доведено.

*Зауваження.* Для випадку $p = 2$ нерівність (1.69) називають *нерівністю Коші – Буняковського*

$$\left|\int\limits_a^b f(x)g(x)dx\right| \leq \left(\int\limits_a^b |f(x)|^2 dx\right)^{1/2} \left(\int\limits_a^b |g(x)|^2 dx\right)^{1/2}.$$

З властивостей інтеграла безпосередньо випливають перші дві аксіоми. Третя аксіома трикутника є нерівністю Мінковського

$$\left(\int_a^b |f(x)+g(x)|^p dx\right)^{1/p} \leq \left(\int_a^b |f(x)|^p dx\right)^{1/p} + \left(\int_a^b |g(x)|^p dx\right)^{1/p} \quad (1.71)$$

або у скороченій формі

$$\|f+g\|_{L_p} \leq \|f\|_{L_p} + \|g\|_{L_p}.$$

Доведемо цю нерівність.

I. Для випадку простору $CL_1[a,b]$ нерівність трикутника випливає безпосередньо з оцінки інтеграла

$$\int_a^b |f(x)+g(x)|dx \leq \int_a^b [|f(x)|+|g(x)|]dx \leq \int_a^b |f(x)|dx + \int_a^b |g(x)|dx.$$

Звідси з урахуванням формули (1.65) для $p=1$ знайдемо

$$\|f+g\|_{L_1} \leq \|f\|_{L_1} + \|g\|_{L_1}.$$

II. Нерівність (1.71) для випадку $1 < p < \infty$ одержимо з використанням нерівності Гельдера. Перетворимо норму суми функцій з урахуванням нерівності (1.69)

$$\|f+g\|_{L_p}^p = \int_a^b |f(x)+g(x)|^p dx \leq \int_a^b |f(x)+g(x)|^{p-1}|f(x)|dx +$$

$$+ \int_a^b |f(x)+g(x)|^{p-1}|g(x)|dx \leq$$

$$\leq \left(\int_a^b |f(x)+g(x)|^{(p-1)q}dx\right)^{1/q} \left(\int_a^b |f(x)|^p dx\right)^{1/p} +$$

$$+ \left(\int_a^b |f(x)+g(x)|^{(p-1)q}dx\right)^{1/q} \left(\int_a^b |g(x)|^p dx\right)^{1/p} \leq$$

$$\leq \left(\int_a^b |f(x)+g(x)|^p dx\right)^{1/q} \left(\int_a^b |f(x)|^p dx\right)^{1/p} +$$

$$+ \left(\int_a^b |f(x)+g(x)|^p dx\right)^{1/q} \left(\int_a^b |g(x)|^p dx\right)^{1/p} =$$

$$= \| f + g \|_{L_p}^{p/q} \left( \| f \|_{L_p} + \| g \|_{L_p} \right).$$

Звідси, скорочуючи на вираз $\| f + g \|_{L_p}^{p/q}$, одержимо нерівність (1.71).

*З а у в а ж е н н я 1*. Норма (1.68) узагальнює норми, які виникають у задачах визначення середнього відхилення між кривими $y = f(x)$ і $y = g(x)$. В деяких випадках вигідно використовувати усереднене відхилення

$$\frac{1}{b-a}\int_a^b |f(x) - g(x)| dx = \rho_1(f,g)$$

в інших – середньоквадратичне відхилення

$$\left( \frac{1}{b-a}\int_a^b |f(x) - g(x)|^2 dx \right)^{1/2} = \rho_2(f,g).$$

Введені інтеграли характеризують віддаль між елементами $f$ і $g$, породжені, відповідно, нормами

$$\| f \|_{L_1} = \int_a^b |f(x)| dx, \ \ \| f \|_{L_2} = \left( \int_a^b |f(x)|^2 dx \right)^{1/2}.$$

*З а у в а ж е н н я 2*. Виконується граничне співвідношення

$$\lim_{p \to \infty} \left( \int_a^b |f(x)|^p dx \right)^{1/p} = \sup_{x \in [a,b]} |f(x)|, \quad f \in C[a,b],$$

яке випливає з подання інтегралу через границю інтегральної суми з урахуванням критерія інтегровності (1.19).

Тому, якщо вважати $p = \infty$, то можна прийняти позначення $\| f \|_{L_\infty}$ для норми функції, розуміючи під $CL_\infty[a,b]$ простір неперервних функцій на сегменті $[a, b]$.

Збіжні послідовності функцій $\{f_n(x)\} \subset CL_p[a, b]$ прийнято називати *збіжними в p-середньому*. Для збіжних послідовностей справедливі твердження теореми 2.

**Т е о р е м а 6**. *Якщо послідовність* $\{f_n(x)\}$ *збігається в p-*

*середньому на сегменті* $[a,b]$ *до функції* $f(x)$, *то цю послідовність можна почленно інтегрувати на сегменті* $[a,b]$ *і справедлива формула*

$$\lim_{n\to\infty}\int_a^b f_n(x)\,dx = \int_a^b f(x)\,dx.$$

Д о в е д е н н я . Фіксуємо довільне число $\varepsilon > 0$. Для додатного числа $\dfrac{\varepsilon}{b-a}$, внаслідок збіжності послідовності за нормою, можна вказати номер $N = N(\varepsilon)$ такий, що для всіх $n > N$ і всіх $x \in [a, b]$ справедлива нерівність

$$\left(\int_a^b |f_n(x) - f(x)|^p\,dx\right)^{1/p} < \frac{\varepsilon}{(b-a)^{1/q}} \quad \left(\frac{1}{p} + \frac{1}{q} = 1\right). \qquad (1.72)$$

Використовуючи нерівність (1.69) при $g(x) \equiv 1$, одержимо

$$\left|\int_a^b f_n(x)dx - \int_a^b f(x)dx\right| = \left|\int_a^b [f_n(x) - f(x)]dx\right| \le$$

$$\le \left(\int_a^b |f_n(x) - f(x)|^p\,dx\right)^{1/p} \left(\int_a^b dx\right)^{1/q} =$$

$$\le \left(\int_a^b |f_n(x) - f(x)|^p\,dx\right)^{1/p} (b-a)^{1/q}.$$

Врахувавши тут нерівність (1.72), одержимо таку нерівність

$$\left|\int_a^b f_n(x)dx - \int_a^b f(x)dx\right| \le \varepsilon.$$

Звідси випливає інтегровність послідовності.
Теорему доведено.

**Т е о р е м а   7 .** *Якщо послідовність* $\{f_n(x)\}$ *збігається рівномірно до функції* $f(x)$ *на сегменті* $[a,b]$, *то вона збігається в* $p$-*середньому до* $f(x)$ *на* $[a,b]$.

Д о в е д е н н я . Фіксуємо довільне число $\varepsilon > 0$. Для додатного числа $\dfrac{\varepsilon}{(b-a)^{1/p}}$, внаслідок рівномірної збіжності послідовності, можна вказати номер $N = N(\varepsilon)$ такий, що для всіх $n > N$ і всіх $x \in [a, b]$ справедлива нерівність

$$|f_n(x) - f(x)| < \frac{\varepsilon}{(b-a)^{1/p}}.$$

Звідси для всіх $n > N$ маємо

$$\left( \int\limits_a^b |f_n(x) - f(x)|^p\, dx \right)^{1/p} \le \left( \frac{\varepsilon^p}{b-a} \int\limits_a^b dx \right)^{1/p} = \varepsilon.$$

Отже, послідовність $\{f_n(x)\}$ збігається в $p$-середньому до функції $f(x)$ на $[a, b]$.

Теорему доведено.

*З а у в а ж е н н я*. Обернене твердження не виконується – не кожна послідовність, що збігається в $p$-середньому, збігається рівномірно.

Зауважимо також, що простори функцій $CL_p[a, b]$, $1 \le p < \infty$, є неповними просторами.

*П р и к л а д  8*. Розглянемо послідовність функцій $\{f_n(x)\} \subset CL_p[-1, 1]$, $1 \le p < \infty$, члени якої неперервні функції. Графік функції $y = f(x)$ зображено на рис. 1.4.

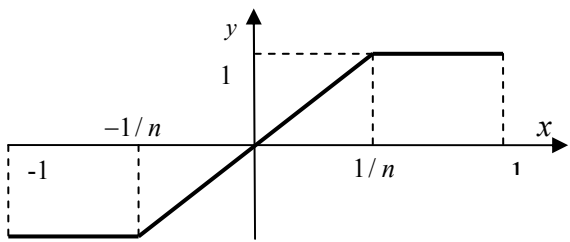

Рис. 1.4.

Аналітичний вираз $n$-го члена послідовності наступний

$$f_n(x) = \begin{cases} -1, & -1 \leq x \leq -1/n, \\ nx, & -1/n \leq x \leq 1/n, \\ 1, & 1/n \leq x \leq 1. \end{cases}$$

Покажемо, що послідовність $\{f_n(x)\}$ – фундаментальна за нормою. Враховуючи нерівність $|f_n(x)| \leq 1$, одержимо для $m > n$

$$\|f_n(x) - f_m(x)\|_{L_p} = \left(\int_{-1}^{1} |f_n(x) - f_m(x)|^p dx\right)^{1/p} \leq$$

$$\leq \left(\int_{-1}^{1} |f_n(x)|^p dx\right)^{1/p} + \left(\int_{-1}^{1} |f_m(x)|^p dx\right)^{1/p} =$$

$$= \left(\int_{-1/n}^{1/n} |f_n(x)|^p dx\right)^{1/p} + \left(\int_{-1/n}^{1/n} |f_m(x)|^p dx\right)^{1/p} \leq 2\left(\int_{-1/n}^{1/n} dx\right)^{1/p} = 2\left(\frac{2}{n}\right)^{1/p}.$$

Отже, $\|f_n(x) - f_m(x)\|_{L_p} \to 0$, коли $n \to \infty$. Тому послідовність $\{f_n(x)\}$ фундаментальна в просторі $CL_p[-1, 1]$, $1 \leq p < \infty$.

У кожній точці сегменту $[-1, 1]$ послідовність функцій $\{f_n(x)\}$ збігається до граничної функції $f(x) = \mathrm{sgn}\, x$,

$$f(x) = \begin{cases} -1, & -1 \leq x < 0, \\ 0, & x = 0, \\ 1, & 0 < x \leq 1. \end{cases}$$

Покажемо також, що послідовність $\{f_n(x)\}$ збігається до функції $f(x)$ за нормою простору $CL_p[-1, 1]$. Дійсно,

$$\|f(x) - f_n(x)\|_{L_p} = \left(\int_{-1}^{1} |f(x) - f_n(x)|^p dx\right)^{1/p} =$$

$$= \left(\int_{-1/n}^{1/n} |f(x) - f_n(x)|^p dx\right)^{1/p} \leq 2\left(\frac{2}{n}\right)^{1/p}$$

і, відповідно, $\|f(x) - f_n(x)\|_{L_p} \to 0$, коли $n \to \infty$.

Водночас гранична функція $f(x)$ не належить простору $CL_p[-1, 1]$, $1 \le p < \infty$, тобто $CL_p[-1, 1]$ – неповний простір.

**1.4.5. Простори кусково-неперервних функцій, інтегровних в середньому.** У розглянутому прикладі 8 фундаментальна послідовність неперервних функцій збігається до розривної функції, що має розрив першого роду. Введемо відповідний клас функцій [11].

***О з н а ч е н н я  8***. *Функція $f(x)$, визначена на відрізку $[a, b]$, називається кусково-неперервною на $[a, b]$, якщо відрізок $[a, b]$ можна розбити на скінченне число відрізків так, що всередині кожного з них $f(x)$ неперервна і має скінченні границі при наближенні аргумента до кінців цих відрізків.*

Якщо на множині кусково-неперервних функцій на $[a, b]$, ввести (за аналогією з операціями на множині неперервних функцій) операції додавання функцій і множення функції на число, то одержимо лінійний простір. Позначимо його через $Q[a, b]$. Нуль-елементом простору $Q[a, b]$ вважається функція, яка тотожно дорівнює нулеві на $[a, b]$.

Очевидно в просторі $Q[a, b]$ має сенс вираз (1.68), тобто існує інтеграл Рімана, як інтеграл від кусково-неперервної функції. Вивчимо питання щодо можливості використання цього виразу як норми у просторі $Q[a, b]$. Розглянемо спочатку допоміжне твердження.

***Т е о р е м а  8***. *Кожна функція $f(x) \in Q[a, b]$ зображається як границя в розумінні збіжності в $p$-середньому послідовності неперервних функцій з $C[a, b]$, тобто справедлива рівність*

$$\lim_{n \to \infty} \int_a^b |f(x) - f_n(x)|^p dx = 0 \text{ або } \lim_{n \to \infty} f_n(x) \stackrel{L_p}{=} f(x).$$

*Д о в е д е н н я*. Нехай функція $f(x) \in Q[a, b]$ має точки розриву $x_i$ $(i = 0, 1, ..., m)$, серед яких є і кінцеві точки відрізка, $a = x_0 < x_1 < ... < x_m = b$. Для кожної з цих точок побудуємо $\delta$-окіл $|x - x_i| < \delta$, $2\delta < \min\limits_{i=1,m} |x_i - x_{i-1}|$, і викинемо їх з відрізка $[a, b]$. У проміжках, що залишилися, функція $f(x)$ неперервна. Довизначимо її до неперервної на $[a, b]$ функції $\bar{f}_\delta(x)$, вважаючи, що $\bar{f}_\delta(x_i) = f(x_i)$ і $\bar{f}_\delta(x)$ лінійна на викинутих околах. Очевидно (за побудовою), справедлива поточкова збіжність

$$\lim_{\delta \to 0} \bar{f}_\delta(x) = f(x).$$

Введемо позначення $M = \max\limits_{x \in [a,b]} f(x)$ і знайдемо оцінку

$$\int\limits_a^b \left| f(x) - \bar{f}_\delta(x) \right|^p dx = \int\limits_a^{a+\delta} \left| f(x) - \bar{f}_\delta(x) \right|^p dx + \sum_{i=1}^{m-1} \int\limits_{x_i - \delta}^{x_i + \delta} \left| f(x) - \bar{f}_\delta(x) \right|^p dx \le$$

$$\le (2M)^p \delta + m(2M)^p 2\delta + (2M)^p \delta = (2M)^p 2(m+1)\delta.$$

Звідси

$$\lim_{\delta \to 0} \bar{f}_\delta(x) \stackrel{L_p}{=} f(x).$$

Легко показати, що послідовність $\{\bar{f}_{\delta(n)}(x)\}$, де $\delta(n) = \dfrac{1}{n}$, фундаментальна в $CL_p[a, b]$. Прийнявши $\bar{f}_{\delta(n)} = f_n(x)$, одержимо твердження теореми.

Теорему доведено.

За результатами теореми 8 можна стверджувати, що для двох кусково-неперервних на $[a, b]$ функцій $f_1(x)$ і $f_2(x)$, що приймають різні значення у скінченній кількості точок, вираз (1.68) приймає однакове значення, тобто будь-яка функція $f(x)$, що приймає ненульові значення у скінченій кількості точок, обертає вираз (1.68) в нуль.

Отже, якщо ввести норму (1.68) у просторі $Q[a, b]$, то умова *невиродженості* не виконується, тобто з умови $\|f(x)\|_{L_p} = 0$ не випливає рівність $f(x) = 0$. Однак цю невідповідність можна виправити наступним чином.

*Дві функції $f(x)$ і $g(x)$ з простору $Q[a, b]$ називаються еквівалентними і записуємо $f(x) \sim g(x)$, якщо вони відрізняються не більше, ніж в скінченній кількості точок.*

Відзначимо, що нерівності Гельдера (1.69) і нерівність Мінковського (1.71) виконуються і для кусково-неперервних функцій, оскільки в обох випадках розглядаються інтеграли Рімана.

Тоді величина (1.68) для випадку простору $Q[a, b]$ має такі властивості:

1) $\|f(x)\|_{L_p} \geq 0$ і $\|f(x)\|_{L_p} = 0$ лише тоді, коли $f(x) \sim 0$;
2) $\|\lambda f(x)\|_{L_p} = |\lambda| \|f(x)\|_{L_p}$;
3) $\|f(x) + g(x)\|_{L_p} \leq \|f(x)\|_{L_p} + \|g(x)\|_{L_p}$.

*Величина $\|f(x)\|_{L_p}$ на лінійному просторі $Q[a, b]$ називається нормою (з точністю до еквівалентності), а простір $QL_p[a, b]$ нормованим (з точністю до еквівалентності).*

Отже, ми прийшли до нормованого простору $QL_p[a, b]$, який є розширенням простору $CL_p[a, b]$, $CL_p[a, b] \subset QL_p[a, b]$. Однак і простір $QL_p[a, b]$ не є повним. Зокрема, він не містить еквівалентних функцій, що відрізняються у нескінченній (зліченній) кількості точок.

**1.4.6. Простори абсолютно інтегровних функцій.** Елементи просторів $CL_p[a, b]$, $QL_p[a, b]$, $1 \leq p < \infty$, є обмеженими функціями. Поповнимо ці простори необмеженими функціями [5].

Розглянемо лінійний простір $C(a, b)$, що складається з неперервних на інтервалі $(a, b)$ функцій $f(x)$.

Сукупність функцій $f(x) \in C(a, b)$ зі скінченною нормою

$$\|f\|_{L_p} = \left(\int_a^b |f(x)|^p dx\right)^{1/p}, \quad 1 \leq p < \infty, \qquad (1.73)$$

позначимо через $CL_p(a, b)$. При цьому справедливе вкладення
$$CL_p[a, b] \subset CL_p(a, b).$$

Легко навести приклад функції $f(x) \in CL_p(a,b)$, що не належить $CL_p[a,b]$. Функція $f(x) = \dfrac{1}{x^\alpha}$, $x \in (0,1)$, належить простору $CL_p(0,1)$ при $\alpha < \dfrac{1}{p}$ і не належить простору $CL_p[0,1]$.

Введемо простір $L'_p(a,b)$ абсолютно інтегровних у степені $p$ функцій, який є деяким аналогом простору кусково-неперервних функцій.

*О з н а ч е н н я   9 . Функція $f(x)$ називається абсолютно інтегровною у степені $p$ на проміжку $(a,b)$, тобто $f(x) \in L'_p(a,b)$, якщо цей проміжок можна розбити на скінченне число частин точками $x_i$, $i = 0, 1, ..., n$,*

$$a = x_0 < x_1 < ... < x_n = b,$$

*таким чином, що всередині кожного проміжку $[x_{i-1}, x_i]$, $i = 1, 2, ..., n$, існує інтеграл (за Ріманом)*

$$\int\limits_{\lambda_{i-1}}^{\lambda_i} |f(x)|^p dx$$

*для будь-яких $\lambda_i$, $x_{i-1} < \lambda_{i-1} < \lambda_i < x_i$, а також існує (хоч би як невласний) інтеграл*

$$\int\limits_{x_{i-1}}^{x_i} |f(x)|^p dx.$$

Очевидно, для абсолютно інтегровної у степені $p$ функції справедлива формула

$$\int\limits_a^b |f(x)|^p dx = \sum_{i=1}^n \int\limits_{x_{i-1}}^{x_i} |f(x)|^p dx.$$

Норма абсолютно інтегрованої функції $f(x)$ задається формулою (1.73) і тоді простір $L'_p(a,b)$ стає нормованим простором. При цьому аксіома невиродженності виконується з точністю до еквівалентності.

Розширенням простору $L'_p(a,b)$ є простір $L_p(a,b)$, який складаються з неперервних функцій, кусково-неперервних функцій, абсолютно інтегровних функцій і деяких «абстрактних елементів», які можна розглядати, як класи еквівалентних функцій.

Справедливі відношення між просторами
$$CL_p[a,b] \subset QL_p[a,b] \subset L'_p(a,b) \subset L_p(a,b).$$

При цьому норма в просторі $L_p(a,b)$ визначається за формулою (1.68), однак інтеграл у ній розуміємо як інтеграл Лебега.

Проте функції з простору $L_p(a,b)$ можна охарактеризувати не звертаючись до інтеграла Лебега $[9,11]$.

*Функція $f(x)$ належить простору $L_p(a,b)$, якщо існує інтеграл Рімана $\int_a^b |f(x)|^p dx < \infty$ (хоч би в розумінні невласного інтеграла), або вона є граничним елементом фундаментальної послідовності функцій $\{f_n(x)\}$, $f_n(x) \in CL_p[a,b]$, і границя інтегралів Рімана від функцій $f_n(x)$ існує, при цьому вона ототожнюється з інтегралом від $p$-го степеня модуля цієї функції,*

$$\lim_{n \to \infty} \int_a^b |f_n(x)|^p dx = \int_a^b |f(x)|^p dx.$$

Ми не розглядаємо спосіб, за яким фундаментальній послідовності $\{f_n(x)\} \subset CL[a,b]$ ставиться у відповідність функція $f(x) \in L_p(a,b)$. Однак можна показати, що послідовність $\{f_n(x)\}$ однозначно визначає норму (1.68).

Дійсно, якщо $\{f_n(x)\} \subset CL_p[a,b]$ – фундаментальна послідовність, то з нерівності трикутника одержимо нерівність

$$\left| \left( \int_a^b |f_n(x)|^p dx \right)^{1/p} - \left( \int_a^b |f_m(x)|^p dx \right)^{1/p} \right| \leq \left( \int_a^b |f_n(x) - f_m(x)|^p dx \right)^{1/p}.$$

Звідси (внаслідок фундаментальності послідовності) випливає існування границі

$$\lim_{n \to \infty} \int_a^b |f_n(x)|^p dx,$$

яку ототожнюємо з величиною

$$\int_a^b |f(x)|^p dx.$$

Зроблені викладки дають орієнтовне поняття про простір $L_p[a,b]$, яким і обмежимось.

*З а у в а ж е н н я  1*. Необхідність конструктивного описання інтеграла Лебега виникає за бажання описати інтеграл $\int_a^b |f(x)|^p dx$ без звертання до фундаментальної послідовності.

*З а у в а ж е н н я  2*. Важливим у наступних застосуваннях є простір абсолютно інтегровних функцій $L'_1[a,b]$.

*Функцій $f(x)$ називається абсолютно інтегровною на проміжку $[a,b]$, якщо проміжок $[a,b]$ можна розбити на частини точками $x_i$, $i = 0, 1, ..., n$,*

$$a = x_0 < x_1 < ... < x_n = b,$$

*таким чином, що всередині кожного проміжка $[x_{i-1}, x_i]$, $i = 1, 2, ..., n$, існує інтеграл (за Ріманом)*

$$\int_{\lambda_{i-1}}^{\lambda_i} |f(x)| dx$$

*для будь-яких $\lambda_i$, $x_{i-1} < \lambda_{i-1} < \lambda_i < x_i$, а також існує (хоч би як невласний) інтеграл*

$$\int_{x_{i-1}}^{x_i} |f(x)| dx.$$

*З а у в а ж е н н я  3*. Наведені в попередніх пунктах простори функцій узагальнюються на випадок нескінченого проміжку. Нехай функція $f(x)$ визначена на проміжку $a \le x < +\infty$.

*Будемо вважати, що* $f(x)$ *належить класу* $CL_p[a, +\infty)$, *якщо вона неперервна на проміжку* $a \leq x < +\infty$, *інтегровна (існує інтеграл Рімана) на будь-якому скінченому проміжку і збігається невласний інтеграл*

$$\int_a^{+\infty} |f(x)|^p dx,$$

*де* $1 \leq p < +\infty$.

Тоді норма в цьому просторі визначається наступним чином

$$\|f\|_{L_p[a,+\infty)} = \left(\int_a^{+\infty} |f(x)|^p dx\right)^{1/p}.$$

Можна показати, виходячи з властивостей інтеграла, що $CL_p[a,+\infty)$ – лінійний, але не повний простір. Повним простором є простір $L_p[a,+\infty)$ і справедливе вкладення $CL_p[a,+\infty) \subset QL_p[a,+\infty) \subset L'_p[a,+\infty) \subset L_p[a,+\infty)$.

### 1.4.7. Завдання до четвертого параграфа

1. Вивести нерівність Гельдера і нерівність трикутника в $CL_2[a, b]$, грунтуючись на властивостях квадратного тричлена $\int_a^b [f(x) - \lambda g(x)]^2 dx > 0$.

2. Знайти норму функцій в $L_2[a, b]$:

а) $f(x) = x^2$, $x \in [0, 1]$;
б) $f(x) = \cos x$ $x \in [0, \pi]$;
в) $f(x) = 1 - x$, $x \in [0, 1]$.

3. Нехай задана послідовність функцій

$$f_n(x) = \begin{cases} n^\alpha - n^{1+\alpha} x, & 0 \leq x \leq \dfrac{1}{n}, \\ 0, & \dfrac{1}{n} < x \leq 1. \end{cases}$$

За якого значення $\alpha \geq 0$ послідовність збігається до нуля в розумінні середнього квадратичного? ($0 \leq \alpha < \dfrac{1}{2}$).

4. Дослідити на рівномірну і середньоквадратичну збіжність послідовність

$$f_n(x) = \begin{cases} \sin \pi n^\alpha x, & 0 \le x \le n^{-\alpha}, \\ 0, & n^{-\alpha} < x < \pi. \end{cases}$$

(Для будь-якого $\alpha > 0$ збігається до нуля нерівномірно, $\max_{x \in [0,\,\pi]} f_n(x) = 1$. Для всіх $\alpha > 0$ збігається до нуля в розумінні середнього квадратичного).

5. Показати, що послідовність $\{f_n(x)\}$, де $f_n(x) = n\sqrt{x}\,e^{-nx}$, яка збігається до нуля, в кожній фіксованій точці $x \in [0, 1]$, не збігається в просторі $CL_2[0, 1]$.

(Оскільки $\int\limits_0^1 [f_n(x) - 0]^2 dx = n^2 \int\limits_0^1 x e^{-nx} dx \to 1$ при $n \to \infty$, послідовність не збігається в $CL_2[0, 1]$).

6. Дослідити на збіжність послідовність $\{f_n(x)\}$, де $f_n(x) = x^n$, в просторі $CL_\infty[0, 1)$.

(Якщо $x \in [0, 1)$, то $\lim\limits_{n \to \infty} f_n = 0$ і, відповідно, $f(x) = 0$. Однак, оскільки $\|f_n(x) - 0\| = \|f_n(x)\| = \max\limits_{x \in (-1, 1)} |f_n(x)| = 1$, то послідовність не збігається в $CL_\infty[0, 1)$ (збігається до нуля не рівномірно)).

7. Дослідити на збіжність послідовність $\{f_n(x)\}$, де $f_n(x) = \dfrac{x}{1 + (nx)^2}$, в просторі $CL_1[0, 1]$.

8. Дослідити на збіжність послідовність $\{f_n(x)\}$, де $f_n(x) = \dfrac{1}{x^2 + n^2}$, в просторі $CL_p[0, 1]$.

9. Довести справедливість вкладення
$CL_{p_2}(a, b) \subset CL_{p_1}(a, b)$, $1 \le p_1 < p_2 < \infty$,
де $(a, b)$ – скінченний проміжок (показати нерівність $\|f\|_{L_{p_2}} \le C \|f\|_{L_{p_1}}$, грунтуючись на нерівності Гельдера)

10. Довести, що функція $f(x) = \dfrac{1}{\sqrt{x}}$ не належить простору $L_2[-\pi, \pi]$.

### 1.5. Слабко збіжні функціональні послідовності

**1.5.1. Слабко збіжні послідовності функціоналів.** Розглянемо важливий клас функціональних послідовностей, збіжність яких пов'язується зі збіжністю відповідних послідовностей функціоналів [4, 9, 14, 20].

*О з н а ч е н н я 1*. *Простір $D$ називається лінійним простором зі збіжністю, якщо в ньому визначено поняття збіжності послідовності його елементів, такі що операції додавання елементів простору і множення їх на число є неперервними.*

З цього означення випливає, що для будь-яких послідовностей $\{f_n\}$ і $\{g_n\}$ з множини $D$, що збігаються відповідно до граничних елементів $f \in D$ і $g \in D$, та довільних чисел $\lambda$, $\mu$, послідовність $\{\lambda f_n + \mu g_n\}$ також збігається і

$$\lim_{n \to \infty} (\lambda f_n + \mu g_n) \overset{D}{=} \lambda f + \mu g,$$

а також, якщо $\{\lambda_n\}$ – числова послідовність і $\lim_{n \to \infty} \lambda_n = \lambda$, то $\lim_{n \to \infty} \lambda_n f = \lambda f$ для будь-якого $f \in D$.

Прикладами лінійних просторів зі збіжністю є нормовані простори, однак існують простори зі збіжністю, в яких не можна ввести норму, що визначає задану збіжність.

*О з н а ч е н н я 2*. *Закон, за яким кожному елементу $f \in D$ ставиться у відповідність число, називається функціоналом над цим простором.*

*Значення функціонала $l$ на елементі $f \in D$ позначається $(l, f)$.*

Функціонал лінійний, якщо з того, що $f \in D$, $g \in D$ і $\lambda$, $\mu$ – довільні числа, випливає рівність

$$(l, \lambda f + \mu g) = \lambda (l, f) + \mu (l, g).$$

Функціонал обмежений, якщо існує стала $C > 0$ така, що для будь-якого $f \in D$ справджується нерівність

$$|(l, f)| < C \|f\|_D.$$

*О з н а ч е н н я 3*. *Лінійний функціонал $l$, визначений на лінійному просторі $D$ зі збіжністю, називається неперервним, якщо для будь-якої збіжної послідовності $\{f_n\} \subset D$ до $f \in D$, $\lim_{n \to \infty} f_n \overset{D}{=} f$, випливає збіжність послідовності $\{(l, f_n)\}$,*

$$\lim_{n\to\infty}(l, f_n) = (l, f).$$

Означення неперервності функціоналу також можна сформулювати у термінах околів.

***О з н а ч е н н я  3'.*** *Лінійний функціонал $l$, визначений на лінійному просторі $D$, називається неперервним в цьому просторі, якщо в будь-якій точці $f_0 \in D$ і для будь-якого $\varepsilon > 0$ існує такий окіл точки $f_0$, що для всіх $f$ з цього околу*

$$|(l, f) - (l, f_0)| < \varepsilon.$$

***Л е м а  1***. *Лінійні неперервні функціонали, визначені на лінійному просторі $D$ зі збіжністю, утворюють лінійний простір $D'$.*

***Д о в е д е н н я***. Покажемо, якщо $l_1$ і $l_2$ – лінійні функціонали, $\alpha$, $\beta$ – довільні числа і $\lambda f + \mu g \in D$, то $\alpha l_1 + \beta l_2$ – також лінійний функціонал. Дійсно

$$(\alpha l_1 + \beta l_2, \lambda f + \mu g) = \alpha(l_1, \lambda f + \mu g) + \beta(l_2, \lambda f + \mu u) =$$
$$= \alpha[\lambda(l_1, f) + \mu(l_1, g)] + \beta[\lambda(l_2, f) + \mu(l_2, g)] =$$
$$= \lambda[\alpha(l_1, f) + \beta(l_2, f)] + \mu[\alpha(l_1, g) + \beta(l_2, g)] =$$
$$= \lambda(\alpha l_1 + \beta l_2, f) + \mu(\alpha l_1 + \beta l_2, g).$$

Покажемо, що якщо $l_1$ і $l_2$ – неперервні функціонали і $\lim\limits_{n\to\infty}^{D} f_n = f$, $\{f_n\} \subset D$, $f \in D$, то $\alpha l_1 + \beta l_2$ – неперервний функціонал,

$$\lim_{n\to\infty}(\alpha l_1 + \beta l_2, f_n) =$$
$$= \lim_{n\to\infty}[\alpha(l_1, f_n) + \beta(l_2, f_n)] = \alpha \lim_{n\to\infty}(l_1, f_n) + \beta \lim_{n\to\infty}(l_2, f_n) =$$
$$= \alpha(l_1, f) + \beta(l_2, f) = (\alpha l_1 + \beta l_2, f).$$

Отже, за означенням 3 функціонал $\alpha l_1 + \beta l_2$ неперервний.
Лему доведено.

***Л е м а  2***. *Якщо лінійний функціонал $l$ визначений у просторі $D$ і неперервний в деякій точці $f_0 \in D$, то він неперервний у просторі $D$.*

***Д о в е д е н н я***. Оскільки функціонал $l$ неперервний в точці

$f_0$, для будь-якої збіжної послідовності $\{f_n\} \subset D$, $\lim\limits_{n \to \infty} f_n \stackrel{D}{=} f_0$, справедлива рівність $\lim\limits_{n \to \infty} (l, f_n) = (l, f_0)$.

Розглянемо послідовність $\{u_n = f_n + g_0 - f_0\} \subset D$, яка внаслідок лінійності простору $D$, збігається, $\lim\limits_{n \to \infty} u_n \stackrel{D}{=} g_0$, де $g_0$ – довільна точка в $D$ і $(l, g_0)$ – значення функціоналу в цій точці.

Знайдемо границю послідовності $\{(l, u_n)\}$. Оскільки функціонал $l$ лінійний, маємо
$$\lim_{n \to \infty}(l, u_n) = \lim_{n \to \infty}[(l, f_n) + (l, g_0) - (l, f_0)] =$$
$$= \lim_{n \to \infty}(l, f_n) + (l, g_0) - (l, f_0) = (l, g_0).$$

Отже, $\lim\limits_{n \to \infty}(l, u_n) = (l, g_0)$, тобто функціонал неперервний в довільній точці $g_0$.

Лему доведено.

*Т е о р е м а   1 .* *Для того щоби лінійний функціонал $l$, визначений в просторі $D$, був неперервний, необхідно і достатньо, щоби для будь-якої послідовності $\{f_n\} \subset D$, такої що $\lim\limits_{n \to \infty} f_n \stackrel{D}{=} 0$, відповідна послідовність $\{(l, f_n)\}$ збігалася до нуля,*
$$\lim_{n \to \infty}(l, f_n) = 0.$$

*Н е о б х і д н і с т ь .* Якщо функціонал $l$ неперервний в $D$, то він неперервний в точці $f_0 = 0$, тобто для будь-якої послідовності $\{f_n\} \subset D$, такої що $\lim\limits_{n \to \infty} f_n \stackrel{D}{=} 0$, справджується рівність $\lim\limits_{n \to \infty}(l, f_n) = 0$.

*Д о с т а т н і с т ь .* Нехай для будь-якої послідовності $\{f_n\} \subset D$, такої що $\lim\limits_{n \to \infty} f_n \stackrel{D}{=} 0$, справедлива для всіх $n$ нерівність $\lim\limits_{n \to \infty}(l, f_n) = 0$. Тоді за лемою 2 функціонал неперервний в $D$.

Теорему доведено.

***Т е о р е м а   2 .*** *Якщо лінійний функціонал* $l$, *визначений в просторі* $D$ *і обмежений в околі нульової точки, то він неперервний в* $D$.

*Д о в е д е н н я .* Нехай $\{f_n\} \subset D$ – довільна послідовність така, що $\lim\limits_{n \to \infty} f_n \stackrel{D}{=} 0$. Оскільки функціонал $l$ обмежений, то $|(l, f_n)| < C \|f\|_D$ і, відповідно, $\lim\limits_{n \to \infty}(l, f_n) = 0$, а отже, за теоремою 1 функціонал неперервний в $D$.

Теорему доведено.

*З а у в а ж е н н я .* Теорема 2 справедлива також за умови обмеженості функціонала в деякому околі нульової точки.

*П р и к л а д   1 .* Неперервним функціоналом $l$ в просторі $D = L_2[a, b]$ є скалярний добуток функцій

$$(l, u) = \int_a^b u(x) g(x)\, dx, \qquad (1.74)$$

де $g(x)$ – фіксована функція з $L_2[a, b]$.

Лінійність функціоналу випливає з лінійності скалярного добутку. Обмеженість функціоналу випливає з нерівності Гельдера (1.69),

$$|(l, u)| = \left| \int_a^b u(x) g(x) dx \right| \leq \left( \int_a^b |u(x)|^2 dx \int_a^b |g(x)|^2 dx \right)^{1/2} < +\infty . \quad (1.75)$$

Обмежений лінійний функціонал завжди неперервний, оскільки для будь-якого числа $\varepsilon > 0$ можна підібрати число $\delta > 0$ таке, що, які б не були функції $u, v \in L_2[a, b]$, з нерівності $\|u - v\|_{L_2[a, b]} < \delta$ випливає нерівність $|(l, u - v)| < \varepsilon$.

Дійсно, фіксуємо довільне число $\varepsilon > 0$. Оцінимо вираз $|(l, u - v)|$ з урахуванням нерівності (1.75),

$$|(l, u - v)| = \left| \int_a^b [u(x) - v(x)] g(x) dx \right| \leq$$

$$\leq \left( \int\limits_a^b |u(x)-v(x)|^2 dx \int\limits_a^b |g(x)|^2 dx \right)^{1/2} = \|u-v\|_{L^2[a,b]} \|g\|_{L^2[a,b]}.$$

Вибираючи тепер $\delta < \dfrac{\varepsilon}{\|g\|_{L_2[a,b]}}$, одержимо умову неперервності функціоналу $|(l, u-v)| < \varepsilon$.

У просторі лінійних неперервних функціоналів $D'$ вводиться поняття збіжності, яку називають слабкою збіжністю.

*О з н а ч е н н я  4 . Послідовність лінійних неперервних функціоналів $\{l_n\}$, визначених на множині $D$, називається слабко збіжною до функціоналу $l$, якщо послідовність значень функціоналів $\{(l_n, g)\}$ на кожному з елементів $g \in D$ збігається до значення функціоналу $(l, g)$, тобто*

$$\lim_{n \to \infty}(l_n, g) = (l, g), \ g \in D.$$

*З а у в а ж е н н я .* За такого визначення збіжності функціоналів операції додавання і множення на число неперервні (це випливає з властивостей лінійності функціоналів і властивостей числових послідовностей). Тому лінійні неперервні функціонали утворюють лінійний простір зі збіжністю [4].

Поняття функціоналу дозволяє узагальнити класичне означення функції, якщо розглянути лінійний функціонал $l$ вигляду (1.74) на функціональних просторах зі збіжністю.

*О з н а ч е н н я  5 . Лінійний неперервний функціонал $l$, визначений на лінійному функціональному просторі $D$ зі збіжністю, називається узагальненою функцією.*

*П р и к л а д  2 .* Розглянемо лінійний функціонал, що діє за формулою

$$\left(P\frac{1}{x}, g\right) = \text{V.p.} \int\limits_{-\infty}^{\infty} \frac{g(x)}{x} dx = \lim_{\varepsilon \to 0} \left[ \int\limits_{-\infty}^{-\varepsilon} \frac{g(x)}{x} dx + \int\limits_{\varepsilon}^{\infty} \frac{g(x)}{x} dx \right],$$

де $g(x) \in D$, $D$ – простір неперервних фінітних функцій на дійсній осі з неперервними першими похідними.

Покажемо, що функціонал $\left(P\dfrac{1}{x}, g\right)$, $g(x) \in D$, неперервний в $D$ і тому визначає узагальнену функцію.

Лінійність цього функціоналу випливає з лінійності інтеграла. Доведемо його неперервність в $D$. Нехай послідовність $\{g_n(x)\} \subset D$ збігається до нуля і $g_n(x) = 0$, якщо $|x| > A$, тобто $\lim\limits_{n \to \infty} g_n(x) = 0$ і, відповідно, $\lim\limits_{n \to \infty} g'_n(x) = 0$. Тоді одержимо з використанням формули Тейлора і обчислення головного значення інтеграла

$$\left| \left( P\frac{1}{x}, g_n \right) \right| = \left| \text{V.p.} \int_{-\infty}^{+\infty} \frac{g_n(x)}{x} dx \right| = \left| \text{V.p.} \int_{-A}^{A} \frac{g_n(0) + x g'_n(\bar{x})}{x} dx \right| \le$$

$$\le \int_{-A}^{A} \left| g'_n(\bar{x}) \right| dx = 2A \max_{\bar{x} \le A} \left| g'_n(\bar{x}) \right|.$$

Переходячи тут до границі, одержимо $\lim\limits_{n \to \infty} \left( P\frac{1}{x}, g_n \right) = 0$, що підтверджує неперервність розглянутого функціоналу.

Отже, функціонал $\left( P\frac{1}{x}, g \right)$, $g(x) \in D$, визначає узагальнену функцію.

**1.5.2. Слабко збіжні послідовності функцій.** Нехай $D$ – множина функції, визначених на (скінченному або нескінченному) проміжку $(a, b)$ і $M$ – множина функцій, визначених на цьому ж проміжку, такі, що для $g(x) \in D$ і $u(x) \in M$ існують інтеграли (Рімана)

$$\int_{a}^{b} u(x) g(x) dx.$$

Очевидно, що фіксована функція $u(x) \in M$ цим інтегралом визначає функціонал $(l, g)$ на просторі функцій $g(x) \in D$.

Послідовність функціоналів вигляду

$$(l_n, g) = \int_{a}^{b} u_n(x) g(x) dx, \ n = 1, 2, \ldots,$$

визначається відповідною послідовністю функцій $\{u_n(x)\}$.

*З а у в а ж е н н я* . За умови збіжності послідовності функціоналів відповідна послідовність функцій (навіть за умови їх неперервності) не завжди збігається (в розумінні поточкової збіжності чи збіжності за нормою) до функції $u(x) \in M$.

Відповідно до означення 4 формулюється означення слабкої збіжності послідовності функцій**.**

***О з н а ч е н н я   6 .*** *Послідовність функцій $\{u_n(x)\} \subset M$ називається слабко збіжною відносно функціонального простору $D$, якщо для будь-якої функції $g(x) \in D$ існує границя числової послідовності*

$$\left\{ \int_a^b u_n(x)g(x)dx \right\}.$$

*З а у в а ж е н н я   1* . Граничний елемент слабко збіжної послідовності $\{u_n(x)\} \subset M$ не обов'язково належить простору $M$.

Послідовність $\{u_n(x)\} \subset M$ слабко збігається до граничної функції $u(x) \in M$ у двох випадках: а) коли $g(x) \in C[a,b]$, $\{u_n(x)\}$ – послідовність неперервних функцій, яка збігається рівномірно на скінченому проміжку $[a,b]$; б) коли $g(x) \in L_2[a,b]$, елементи послідовності інтегровані з квадратом функції на скінченому проміжку $[a,b]$ і послідовність збігається в середньому до граничної функції.

У першому випадку (за теоремою 4 підрозділу 1.3) послідовність збігається до неперервної функції $u(x)$. Оцінивши відповідний інтеграл з урахуванням обмеженості функції $g(x)$, одержимо

$$\left| \int_a^b [u(x) - u_n(x)]g(x)dx \right| \leq \int_a^b |u(x) - u_n(x)| \, |g(x)| dx \leq$$
$$\leq \sup_{x \in [a,b]} |u(x) - u_n(x)| \sup_{x \in [a,b]} |g(x)| (b-a).$$

Звідси випливає рівність

$$\lim_{n \to \infty} \int_a^b u_n(x)g(x)dx = \int_a^b u(x)g(x)dx .$$

У другому випадку, якщо $g(x) \in L_2[a, b]$, елементи послідовності інтегровні з квадратом функції і послідовність збігається в середньому до граничної функції $u(x)$, то застосувавши нерівність Гельдера до інтеграла $\int_a^b [u_n(x) - u(x)] g(x) dx$, одержимо

$$\left| \int_a^b [u_n(x) - u(x)] g(x) dx \right| \leq \left( \int_a^b |u_n(x) - u(x)|^2 dx \int_a^b |g(x)|^2 dx \right)^{1/2}.$$

Звідси переконуємося, що

$$\lim_{n \to \infty} \int_a^b u_n(x) g(x) dx = \int_a^b u(x) g(x) dx.$$

*З а у в а ж е н н я   2*. Із слабкої збіжності послідовності функцій в $L_2[a, b]$ не випливає її збіжність за нормою $L_2[a, b]$.

*О з н а ч е н н я   7*. *Послідовність $\{u_n(x)\} \subset M$ називається фундаментальною в розумінні слабкої збіжності відносно функціонального простору $D$, якщо для будь-якої функції $g(x) \in D$ і заданого довільного числа $\varepsilon > 0$ знайдеться натуральне число $N$ таке, що якщо тільки $n > N$ і $m > N$, то справджується нерівність*

$$\left| \int_a^b g(x) [u_n(x) - u_m(x)] dx \right| < \varepsilon.$$

*О з н а ч е н н я   8*. *Дві фундаментальні функціональні послідовності $\{u_n(x)\} \subset M$, $\{v_n(x)\} \subset M$ називаються еквівалентними в розумінні слабкої збіжності відносно функціонального простору $D$, якщо для довільної функції $g(x) \in D$ виконується рівність*

$$\lim_{n \to \infty} \int_a^b g(x) [u_n(x) - v_n(x)] dx = 0. \qquad (1.76)$$

*П р и к л а д   3*. Розглянемо дві послідовності $\left\{ u_n(x) = \dfrac{1}{n} \right\}$ і

$\{v_n(x) = \sin nx\}$ неперервних функцій на скінченному проміжку $[a, b]$. Граничними елементами цих послідовностей в розумінні слабкої збіжності відносно множини неперервних функцій на $[a, b]$ буде функція $u(x)$, що тотожньо дорівнює нулю.

Дійсно, $\lim\limits_{n \to \infty} \int\limits_a^b \dfrac{g(x)}{n} dx = 0$, а також відомо (формули (2.56)

розділ 2), що $\lim\limits_{n \to \infty} \int\limits_a^b g(x) \sin nx \, dx = 0$.

Крім того, ці послідовності еквівалентні в розумінні слабкої збіжності відносно множини неперервних функцій на $[a, b]$, оскільки

$$\lim_{n \to \infty} \int\limits_a^b g(x) \left( \dfrac{1}{n} - \sin nx \right) dx = 0.$$

*З а у в а ж е н н я .* Зі слабкої збіжності послідовності не випливає збіжність в середньому цієї послідовності. Послідовність $\{v_n(x) = \sin nx\}$ слабко збігається до нуля на інтервалі $(0, \pi)$ відносно множини неперервних функцій,

$$\lim_{n \to \infty} \int\limits_0^\pi g(x) \sin nx \, dx = 0.$$

Однак вона не збігається в середньому, оскільки $\int\limits_0^\pi (\sin nx - 0)^2 \, dx = \pi$.

**1.5.3. Дельта-функція. Дельтоподібні послідовності функцій.** Відповідно до означення 5 сформулюємо поняття узагальненої функції з використанням граничного елемента слабко збіжної послідовності функцій.

*О з н а ч е н н я  9. Граничний елемент фундаментальної послідовності функцій в сенсі слабкої збіжності відносно функціонального простору $D$, називається узагальненою функцією.*

Введення поняття узагальненої функції дає можливість описати такі ідеалізовані поняття, як густина точкового заряду,

густина матеріальної точки, миттєвий імпульс, інтенсивність сили, прикладеної в точці, тощо.

Розглянемо, наприклад, задачу про опис густини нескінченно довгого стержня з розміщеною в точці $x = x_0$ одиничною масою. В інших точках маси відсутні.

Позначимо через $\delta_r(x, x_0)$ середню густину стержня, яка неперервно розподілена вздовж осі, рис. 1.5.

$$\delta_r(x, x_0) = \begin{cases} \dfrac{1}{r}\left(1 - \dfrac{|x - x_0|}{r}\right), & |x - x_0| \leq r, \\ 0, & |x - x_0| > r. \end{cases} \quad (1.76)$$

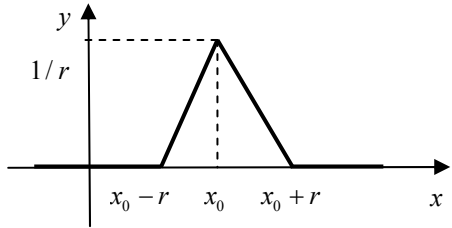

Рис. 1.5.

При цьому маса стержня (сумарна величина густини) залишається незмінною,

$$\int_{-\infty}^{\infty} \delta_r(x, x_0)\,dx = \frac{1}{r}\int_{x_0-r}^{x_0+r}\left(1 - \frac{|x-x_0|}{r}\right)dx = \frac{2}{r}\int_{x_0}^{x_0+r}\left(1 - \frac{x-x_0}{r}\right)dx =$$
$$= 2\int_0^1 (1-t)\,dt = 1.$$

Формально визначимо невідому густину стержня як границю (при $r \to 0$) середньої густини

$$\delta(x, x_0) = \lim_{r \to 0} \delta_r(x, x_0) = \begin{cases} +\infty, & x = x_0, \\ 0, & x \neq x_0. \end{cases} \quad (1.77)$$

Природно вимагати, щоб маса будь-якого відрізка $[a, b]$, що містить точку $x_0$, дорівнювала інтегралу від густини по цьому відрізку,

$$\int\limits_a^b \delta(x, x_0)\,dx = \begin{cases} 1, & x_0 \in [a, b], \\ 0, & x_0 \notin [a, b]. \end{cases}$$

З міркувань класичного означення функції останній вираз суперечить рівності (1.77). Функцію $\delta(x, x_0)$ потрібно вважати або не інтегровною, або інтеграл від неї дорівнює нулю.

Однак у фізиці розглядають «функцію» вигляду (1.77), її називають $\delta$-функцією або функцією Дірака.

Важливий також інший підхід до визначення $\delta$-функції, який ґрунтується на слабкій збіжності послідовностей функцій.

***Л е м а   3*** . *Якщо $g(x)$ – обмежена кусково-неперервна функція на всій осі і $\{\delta_n(x, x_0)\}$ – послідовність функцій* (1.76), *де $r = r_n$ – нескінченно мала числова послідовність, зокрема $r_n = \dfrac{r_0}{n}$, $r_0 = const$, то в кожній точці неперервності функції $g(x)$ справедлива гранична рівність*

$$\lim_{n \to \infty} \int\limits_{-\infty}^{+\infty} \delta_{r_n}(x, x_0) g(x)\,dx = g(x_0). \qquad (1.78)$$

*Д о в е д е н н я* . Дійсно,

$$\left| \int\limits_{-\infty}^{+\infty} \delta_{r_n}(x, x_0) g(x)\,dx - g(x_0) \right| =$$

$$= \left| \frac{1}{r_n} \int\limits_{x_0 - r_n}^{x_0 + r_n} \left(1 - \frac{|x - x_0|}{r_n}\right) g(x)\,dx - g(x_0) \right| =$$

$$= \frac{1}{r_n} \left| \int\limits_{x_0 - r_n}^{x_0 + r_n} \left(1 - \frac{|x - x_0|}{r_n}\right) g(x)\,dx - g(x_0) \int\limits_{x_0 - r_n}^{x_0 + r_n} \left(1 - \frac{|x - x_0|}{r_n}\right) dx \right| =$$

$$= \frac{1}{r_n} \left| \int\limits_{x_0 - r_n}^{x_0 + r_n} \left(1 - \frac{|x - x_0|}{r_n}\right) [g(x) - g(x_0)]\,dx \right|.$$

Врахувавши тут теорему про середнє значення для добутку невід'ємної і обмеженої функцій, одержимо

$$\left| \int_{-\infty}^{+\infty} \delta_{r_n}(x, x_0) g(x)\, dx - g(x_0) \right| =$$

$$= \frac{1}{r_n} \int_{x_0 - r_n}^{x_0 + r_n} \left( 1 - \frac{|x - x_0|}{r_n} \right) dx \, |g(\bar{x}) - g(x_0)| = |g(\bar{x}) - g(x_0)|,$$

де $\bar{x} \in (x_0 - r_n,\, x_0 + r_n)$.

Звідси,

$$\lim_{n \to \infty} \left| \int_{-\infty}^{+\infty} \delta_{r_n}(x, x_0) g(x)\, dx - g(x_0) \right| = 0.$$

Формулу (1.78) доведено.

Властивість сім'ї функцій (1.76), яка виражається рівністю (1.78), покладена в основу визначення $\delta$-функції. Її можна трактувати наступним чином.

1) Послідовність функцій (1.76) слабко збігається до дельта-функції в сенсі означення 6, тобто числова послідовність $\left\{ \int_{-\infty}^{+\infty} \delta_{r_n}(x, x_0) g(x)\, dx \right\}$ для будь-якої кусково-неперервної функції $g(x)$ збігається до значення $g(x_0)$.

2) Якщо вважати, що послідовність функцій (1.76) має своїм граничним елементом дельта-функцію і записати рівність (1.78) у вигляді

$$\int_{-\infty}^{+\infty} \delta(x, x_0) g(x)\, dx = g(x_0),$$

то звідси випливає, що $\delta$-функція ставить у відповідність кожній функції $g(x)$, неперервній на осі, число $g(x_0)$, а отже, $\delta$-функція – функціонал.

Сформулюємо поняття дельта-функції, як граничного елемента слабко збіжної послідовності. Спочатку введемо поняття ядра типу Фейєра.

***О з н а ч е н н я 10***. *Функція* $\Phi(x) = \omega(|x|)$ *називається ядром типу Фейєра, якщо:*

*а) вона неперервна на проміжку* $(-\infty, +\infty)$;

*б)* $(1+x)^{\lambda}|\omega(x)| \leq A < +\infty$, $\lambda > 1$;  (1.79)

*в)* $2\int\limits_{0}^{+\infty} \omega(x)dx = 1$.  (1.80)

*Частинним випадком ядра типу Фейєра є фінітне ядро*

$$\Phi_0(x) = \begin{cases} \omega_0(|x|), & -1 \leq x \leq 1, \\ 0, & |x| > 1, \end{cases} \quad (1.81)$$

*для якого умова* (1.79) *виконується автоматично, а умова* (1.80) *така*

$$2\int\limits_{0}^{1} \omega_0(x)\,dx = 1. \quad (1.82)$$

Функція називається *фінітною*, якщо вона визначена на осі і дорівнює нулю зовні деякого скінченого проміжку.

*З а у в а ж е н н я*. Умова неперервності ядра типу Фейєра і умова (1.79), що забезпечує абсолютну інтегровність цього ядра, є достатніми умовами існування відповідних інтегралів. Вони вибрані з міркувань ефективного використання прикладних теорем, встановлених для неперервних функцій і абсолютно збіжних інтегралів. В окремих застосуваннях, зокрема у фізиці, використовують також кусково-гладкі ядерні функції. Наприклад,

$$\Phi_0(x) = \begin{cases} \dfrac{1}{2}, & -1 \leq x \leq 1, \\ 0, & |x| > 1. \end{cases}$$

*П р и к л а д  4*. Розглянемо ядро Фейєра

$$\Phi(x) = \frac{2}{\pi} \frac{\sin^2 \dfrac{x}{2}}{x^2}.$$

Функція $\Phi(t)$ неперервна і парна. Умова (1.79) виконується оскільки

$$|\omega(t)| = \frac{2}{\pi} \frac{\sin^2 \frac{t}{2}}{t^2} \le \frac{2}{\pi(1+t)^2} \text{ i } \lambda = 2 > 1.$$

Переконаємося у виконанні умови (1.80). Перетворивши підінтегральний вираз і застосовуючи інтегрування за частинами, одержимо

$$\frac{4}{\pi} \int_0^{+\infty} \frac{\sin^2 \frac{t}{2}}{t^2} dt = \frac{2}{\pi} \int_0^{+\infty} (1-\cos t) \frac{d}{dt}\left(-\frac{1}{t}\right) dt = \frac{2}{\pi} \int_0^{+\infty} \frac{\sin t}{t} dt = 1.$$

Тут використано інтеграл Діріхле (прикл. 10, п. 1.2)

$$\int_0^{+\infty} \frac{\sin t}{t} dt = \frac{\pi}{2}.$$

*П р и к л а д  5*. Важливим в аналізі є ядро Пуассона

$$\Phi(x) = \frac{1}{\pi} \frac{1}{1+x^2},$$

для якого виконання умов (1.79) і (1.80) легко перевіряється.

***Т е о р е м а  3.*** *Якщо* $\Phi(x)$ *– ядро типу Фейєра і* $g(x)$ *– обмежена кусково-неперервна функція на дійсній осі, то в кожній точці неперервності функції* $g(x)$ *справедлива формула*

$$\lim_{n \to \infty} \int_{-\infty}^{+\infty} g(x) \frac{1}{r_n} \Phi\left(\frac{x-x_0}{r_n}\right) dx = g(x_0). \qquad (1.83)$$

*Д о в е д е н н я*. Зробимо заміну змінної $x = x_0 + r_n t$ в інтегралі (1.83), одержимо

$$\int_{-\infty}^{+\infty} g(x) \frac{1}{r_n} \Phi\left(\frac{x-x_0}{r_n}\right) dx = \int_{-\infty}^{\infty} g(x_0 + r_n t)\, \omega(|t|)\, dt =$$
$$= 2 \int_0^{+\infty} \frac{g(x_0 + r_n t) + g(x_0 - r_n t)}{2} \omega(t)\, dt.$$

Оскільки функція $g(x)$ обмежена на дійсній осі, $|g(x)| \le M < \infty$, справедлива нерівність

$$\left|\frac{g(x_0 + r_n t) + g(x_0 - r_n t)}{2} \omega(t)\right| \le M |\omega(t)|,$$

тобто існує мажоранта, яка інтегрується (теорема 16 п. 1.2). Тоді за теоремою 17 (п. 1.2) про граничний перехід у невласному інтегралі одержимо, з урахуванням неперервності функції $g(x)$ і умови (1.80),

$$\lim_{n \to \infty} \int_{-\infty}^{+\infty} g(x) \frac{1}{r_n} \Phi\left(\frac{x-x_0}{r_n}\right) dx = \lim_{n \to \infty} \int_{-\infty}^{+\infty} g(x) \frac{1}{r_n} \omega\left(\frac{|x-x_0|}{r_n}\right) dx =$$

$$= 2 \lim_{n \to \infty} \int_0^{+\infty} \frac{g(x_0 + r_n t) + g(x_0 - r_n t)}{2} \omega(t)\, dt =$$

$$= 2 g(x_0) \int_0^{+\infty} \omega(t)\, dt = g(x_0).$$

Теорему доведено.

**О з н а ч е н н я   11 .** *Нехай $D$ – множина функцій, обмежених кусково-неперервних на дійсній осі, $\Phi(x) = \omega(|x|)$ – ядро типу Фейєра і $r_n = r(n)$ – нескінченно мала числова послідовність.*

*Тоді функціональна послідовність*

$$\left\{\delta_n(x - x_0) = \frac{1}{r_n} \Phi\left(\frac{x-x_0}{r_n}\right)\right\} \qquad (1.84)$$

*називається дельтоподібною послідовністю, а її граничний елемент $\delta(x - x_0)$ в сенсі слабкої збіжності відносно множини $D$,*

$$\lim_{n \to \infty} \int_{-\infty}^{+\infty} \delta_n(x - x_0) g(x) dx = g(x_0), \qquad (1.85)$$

*називається $\delta$ – функцією відносно множини $D$ і записуємо*

$$\int_{-\infty}^{+\infty} \delta(x - x_0) g(x)\, dx = g(x_0).$$

Існують інші підходи до визначення дельта-функції [1].

*З а у в а ж е н н я   1 .* Виходячи з властивостей ядра типу Фейєра можна переконатися у справедливості рівності

$$\int_{-\infty}^{+\infty} \delta_n(x - x_0) dx = 1,\, n = 1, 2, \ldots. \qquad (1.86)$$

Дійсно, підстановкою сюди виразу $n$-го члена послідовності (1.84) з урахуванням умови (1.80) прийдемо до рівності

$$\int\limits_{-\infty}^{+\infty}\delta_n(x-x_0)dx = \frac{1}{r_n}\int\limits_{-\infty}^{+\infty}\Phi\left(\frac{x-x_0}{r_n}\right)dx = \frac{1}{r_n}\int\limits_{-\infty}^{+\infty}\omega\left(\frac{|x-x_0|}{r_n}\right)dx =$$

$$= \frac{2}{r_n}\int\limits_{0}^{+\infty}\omega\left(\frac{x-x_0}{r_n}\right)dx = 2\int\limits_{0}^{+\infty}\omega(t)dt = 1.$$

*П р и к л а д  6*. Побудуємо дельтоподібну послідовність вигляду (1.84) з використанням функції

$$\omega(t) = ce^{-t^2},\ 0 \leq t < \infty.$$

Відповідна визначена на всій осі функція $\Phi(t) = ce^{-t^2}$ нескінченно диференційовна. Скориставшись інтегралом (Ейлера – Пуассона) $\int\limits_{0}^{+\infty}e^{-t^2}dt = \frac{\sqrt{\pi}}{2}$, з умови $2\int\limits_{0}^{+\infty}\omega(t)\,dx = 1$, одержимо

$$\omega(t) = \frac{1}{\sqrt{\pi}}e^{-t^2}.$$

За формулою (1.84) запишемо дельтоподібну послідовність

$$\left\{\delta_n(x, x_0) = \frac{1}{\sqrt{\pi}\,r_n}\exp\left[-\frac{(x-x_0)^2}{r_n^2}\right]\right\}.$$

*П р и к л а д  7*. Дельтоподібна послідовність (1.76), що використана для моделювання середньої густини стержня, побудована з використанням неперервної на всій осі фінітної функції (1.81), де $\omega_0(t) = 1 - t$.

Легко переконатися, що $2\int\limits_{0}^{1}\varphi_0(t)\,dx = 1$.

*З а у в а ж е н н я  2*. Не кожна дельтоподібна послідовність обов'язково породжена ядром типу Фейєра. Прикладом такої послідовності є дельтоподібна послідовність

$$\delta_n(x) = \frac{1}{\pi}\frac{\sin\dfrac{x-x_0}{r_n}}{x-x_0},$$

породжена ядром $\Phi(x) = \dfrac{1}{\pi}\dfrac{\sin x}{x}$, для якої умова (1.79) не виконується, оскільки інтеграл Діріхле (приклад 10, п. 1.2) не збігається абсолютно,

$$\frac{2}{\pi}\int\limits_0^{+\infty}\frac{\sin x}{x}dx = 1.$$

***Т е о р е м а   4***. *Нехай $D$ – множина функцій, обмежених кусково-неперервних на дійсній осі, $g(x) \in D$ і $\Phi(x) = \omega(|x|)$ – ядро типу Фейєра.*

*Тоді функціонал*

$$(l_n, g) = \int\limits_{-\infty}^{\infty}\frac{1}{r_n}\Phi\left(\frac{x}{r_n}\right)g(x)dx, \ n = 1, 2, \ldots,$$

*неперервний в $D$.*

*Д о в е д е н н я*. За теоремою 2, якщо лінійний функціонал обмежений в околі нульової точки, то він неперервний в $D$. Нехай $g_k(x) \in D$ і $\lim\limits_{k \to \infty} g_k(x) = 0$. Тоді, очевидно $\lim\limits_{k \to \infty} C_k = 0$, де $C_k = \max\limits_{x \in (-\infty, \infty)}|g_k(x)|$, і з урахуванням оцінки (1.79), знайдемо

$$\left|\int\limits_{-\infty}^{+\infty}\frac{1}{r_n}\Phi\left(\frac{x}{r_n}\right)g_k(x)\,dx\right| \le \int\limits_{-\infty}^{+\infty}\frac{1}{r_n}\left|\Phi\left(\frac{x}{r_n}\right)\right||g_k(x)|\,dx \le$$

$$\le 2AC_k r_n^{\lambda-1}\int\limits_0^{+\infty}\frac{dx}{(r_n + x)^{\lambda}} = \frac{2AC_k}{(\lambda - 1)},$$

де $A = const$, $\lambda > 1$. Звідси

$$\lim\limits_{k \to \infty}\int\limits_{-\infty}^{+\infty}\frac{1}{r_n}\Phi\left(\frac{x}{r_n}\right)g_k(x)dx = 0$$

і, відповідно, функціонал неперервний.

Теорему доведено.

***Т е о р е м а   5***. *В будь-якій області $[\eta, +\infty)$, $\eta > 0$, дельтоподібна послідовність $\{\delta_n(x)\}$ збігається рівномірно до нуля, тобто виконується нерівність*

$$\lim_{n \to \infty} \max_{x \in [\eta, +\infty)} |\delta_n(x)| = 0. \qquad (1.87)$$

*Доведення*. Оскільки функція $\Phi(x)$ неперервна, введемо рівномірну норму члена дельтоподібної послідовності на проміжку $[\eta, +\infty)$, $M_n(\eta) = \max_{x \in [\eta, +\infty)} \delta_n(x)$, $0 < \eta < +\infty$, і оцінимо її з урахуванням нерівності (1.79)

$$M_n(\eta) = \frac{1}{r_n} \max_{x \in [\eta, +\infty)} \left| \omega\left(\frac{x}{r_n}\right) \right| \leq \frac{1}{r_n} \max_{x \in [\eta, +\infty)} \frac{A \varepsilon_n^\lambda}{(r_n + x)^\lambda} =$$
$$= A \max_{x \in [\eta, +\infty)} \frac{r_n^{\lambda-1}}{(r_n + x)^\lambda}.$$

Звідси, оскільки $\lambda > 1$ і $x \geq \eta > 0$, отримаємо $\lim_{n \to \infty} M_n(\eta) = 0$.

Теорему доведено.

***Теорема 6.*** *В будь-якій області* $[\eta, +\infty)$, $\eta > 0$, *дельтоподібна послідовність* $\{\delta_n(x)\}$ *слабко збігається до нуля відносно множини обмежених кусково-неперервних функцій, тобто виконується рівність*

$$\lim_{n \to \infty} \int_\eta^{+\infty} f(x) \delta_n(x) \, dx = 0. \qquad (1.88)$$

*Доведення*. Оцінимо інтеграл у формулі (1.88) з урахуванням властивостей функції $f(x)$, $|f(x)| \leq M < +\infty$, і оцінки (1.79),

$$\left| \int_\eta^{+\infty} f(x) \delta_n(x) \, dx \right| \leq \int_\eta^{+\infty} |f(x)| \frac{1}{r_n} \left| \omega\left(\frac{x}{r_n}\right) \right| dx \leq$$
$$\leq M A \, r_n^{\lambda-1} \int_\eta^{+\infty} \frac{dx}{(r_n + x)^\lambda} = \frac{M A \, r_n^{\lambda-1}}{(\lambda - 1)(r_n + \eta)^{\lambda-1}}.$$

Звідси за скінченності величини $\eta > 0$ і $\lambda > 1$ одержимо рівність (1.88).

Теорему доведено.

***Теорема 7.*** *Дельтоподібні послідовності є фундаментальними і еквівалентними між собою послідовностями*

*відносно функціонального простору D – множини функцій, обмежених кусково-неперервних на дійсній осі.*

Д о в е д е н н я . За означенням 8, якщо $\{\delta_{1n}(x, x_0)\}$, $\{\delta_{2n}(x, x_0)\}$ – дельтоподібні послідовності, що слабко збіжні до $\delta$-функції відносно множини $D$, то для кожної з них справедлива рівність (1.85). Тоді для довільного числа $\varepsilon > 0$ знайдеться натуральне число $N$ таке, що якщо тільки $n > N$, $m > N$ і $g(x)$ – неперервна функція, $g(x) \in D$, то справедливі нерівності

$$\int_{-\infty}^{+\infty} \delta_{in}(x - x_0)\left[g(x) - g(x_0)\right]dx < \frac{\varepsilon}{2},$$

$$\int_{-\infty}^{+\infty} \delta_{im}(x - x_0)\left[g(x) - g(x_0)\right]dx < \frac{\varepsilon}{2}, \quad i = 1, 2.$$

Оцінимо інтеграл

$$\int_{-\infty}^{+\infty} \left[\delta_{in}(x - x_0) - \delta_{im}(x - x_0)\right]g(x)dx$$

з урахуванням цих нерівностей і умови (1.86),

$$\left|\int_{-\infty}^{+\infty}\left[\delta_{in}(x-x_0) - \delta_{im}(x-x_0)\right]g(x)\,dx\right| =$$

$$= \left|\int_{-\infty}^{+\infty}\left[\delta_{in}(x-x_0) - \delta_{im}(x-x_0)\right]\left[g(x) - g(x_0) + g(x_0)\right]dx\right| \leq$$

$$\leq \left|\int_{-\infty}^{+\infty}\delta_{in}(x-x_0)[g(x) - g(x_0)]dx\right| + \left|\int_{-\infty}^{+\infty}\delta_{im}(x-x_0)[g(x) - g(x_0)]dx\right| +$$

$$+ \left|\int_{-\infty}^{+\infty}\left[\delta_{in}(x-x_0) - \delta_{im}(x, x_0)\right]dx\ g(x_0)\right| =$$

$$= \left|\int_{-\infty}^{+\infty}\delta_{in}(x-x_0)\left[g(x) - g(x_0)\right]dx\right| +$$

$$+\left|\int\limits_{-\infty}^{+\infty}\delta_{im}(x-x_0)[g(x)-g(x_0)]dx\right|\leq\frac{\varepsilon}{2}+\frac{\varepsilon}{2}=\varepsilon.$$

Отже, виконується умова фундаментальності послідовностей. Еквівалентність послідовностей підтверджується наступною оцінкою

$$\left|\int\limits_{-\infty}^{+\infty}[\delta_{1n}(x-x_0)-\delta_{2n}(x-x_0)]g(x)dx\right|=$$

$$=\left|\int\limits_{-\infty}^{+\infty}[\delta_{1n}(x-x_0)-\delta_{2n}(x,x_0)][g(x)-g(x_0)+g(x_0)]\,dx\right|\leq$$

$$\leq\left|\int\limits_{-\infty}^{+\infty}\delta_{1n}(x-x_0)[g(x)-g(x_0)]dx\right|+\left|\int\limits_{-\infty}^{+\infty}\delta_{2n}(x-x_0)[g(x)-g(x_0)]dx\right|+$$

$$+\left|\int\limits_{-\infty}^{+\infty}[\delta_{1n}(x-x_0)-\delta_{2n}(x-x_0)]dx\,g(x_0)\right|=$$

$$=\left|\int\limits_{-\infty}^{+\infty}\delta_{1n}(x-x_0)[g(x)-g(x_0)]dx\right|+$$

$$+\left|\int\limits_{-\infty}^{+\infty}\delta_{2n}(x-x_0)[g(x)-g(x_0)]dx\right|\leq\frac{\varepsilon}{2}+\frac{\varepsilon}{2}=\varepsilon.$$

Теорему доведено.

**Т е о р е м а   8** . *Нехай* $\{\delta_{1n}(x-x_0)\}$ *і* $\{\delta_{2n}(x-x_0)\}$ – *дельтоподібні послідовності* (1.84).

*Тоді згортки відповідних елементів цих послідовностей*

$$\delta_n(x-x_0)=\int\limits_{-\infty}^{+\infty}\delta_{1n}(x-t)\delta_{2n}(t-x_0)dt \qquad (1.89)$$

*також утворюють дельтоподібну послідовність, яка слабко збігається до дельта-функції відносно множини D обмежених кусково-неперервних функцій.*

*Д о в е д е н н я* . Покажемо, що для будь-якої обмеженої кусково-неперервної функції $g(x)$, справедлива формула (1.85).

Перетворимо інтеграл у цій формулі з урахуванням виразів (1.84) дельтоподібних функцій наступним чином

$$\int_{-\infty}^{+\infty} \delta_n(x - x_0)g(x)dx = \int_{-\infty}^{+\infty}\int_{-\infty}^{+\infty} \delta_{2n}(x-t)\delta_{1n}(t-x_0)g(x)dt\,dx =$$

$$= \frac{1}{r_n^2} \int_{-\infty}^{+\infty}\int_{-\infty}^{+\infty} \omega_2\!\left(\frac{|x-t|}{r_n}\right)\omega_1\!\left(\frac{|t-x_0|}{r_n}\right)g(x)dt\,dx =$$

$$= \frac{1}{r_n} \int_{-\infty}^{+\infty}\int_{-\infty}^{+\infty} \omega_1\!\left(\frac{|t-x_0|}{r_n}\right)\omega_2(|y|)g(t+r_n y)\,dy\,dt =$$

$$= \int_{-\infty}^{+\infty}\int_{-\infty}^{+\infty} \omega_1(|z|)\omega_2(|y|)g(x_0+r_n z+r_n y)dy\,dz.$$

Оскільки $g(x)$ – обмежена функція, $\max\limits_{x\in(-\infty,\infty)}|g(x)|=M$, для підінтегральної функції справедлива оцінка

$$\big|\omega_1(|z|)\omega_2(|y|)g(x_0+r_n z+r_n y)\big| \le M\,|\omega_1(|z|)|\,|\omega_2(|y|)|,$$

тобто вона має інтегровану мажоранту. Якщо $x_0$ – точка неперервності функції $g(x)$, то підінтегральна функція має границю і за теоремою про граничний перехід у невласному інтегралі одержимо

$$\lim_{n\to\infty}\int_{-\infty}^{+\infty}\delta_n(x-x_0)g(x)dx =$$

$$= \lim_{n\to\infty}\int_{-\infty}^{+\infty}\int_{-\infty}^{+\infty}\omega_1(|z|)\omega_2(|y|)g(x_0+r_n z+r_n y)dy\,dz =$$

$$= g(x_0)\int_{-\infty}^{+\infty}\int_{-\infty}^{+\infty}\omega_1(|z|)\omega_2(|y|)dy\,dz =$$

$$= g(x_0)\int_{-\infty}^{+\infty}\omega_1(|z|)dz\int_{-\infty}^{+\infty}\omega_2(|y|)dy = g(x_0).$$

Отже, послідовність $\{\delta_n(x-x_0)\}$ слабко збігається до дельта-функції відносно множини $D$.

Теорему доведено.

**Т е о р е м а   9**. *Нехай* $\{\delta_n(x-x_0)\}$ – *дельтоподібна послідовність* (1.84), *де* $\Phi(x)=\omega(|x|)$ – *ядро типу Фейєра, що має абсолютно інтегровну похідну* $p$-*го порядку*, $p \geq 1$, *і справедливі оцінки*

$$(1+x)^\lambda \left| \frac{d^m \omega(x)}{dx^m} \right| \leq A < +\infty, \ \lambda > 1, \ m = \overline{0, p}. \qquad (1.90)$$

*Тоді, якщо* $g(x)$ – *функція, що має обмежену кусково-неперервну похідну* $p$-*го порядку, то в кожній точці* $x_0$ *неперервності функції* $g^{(k)}(x)$, $k \leq p$, *справедлива формула*

$$\lim_{n \to \infty} \int_{-\infty}^{+\infty} \delta_n^{(k)}(x-x_0) g(x) dx = (-1)^k g^{(k)}(x_0). \qquad (1.91)$$

*Д о в е д е н н я*. Для будь-якої обмеженої кусково-неперервної функції $g^{(k)}(x)$ і функції $\delta_n(x-x_0)$, внаслідок оцінки (1.90), існують інтеграли $\int_{-\infty}^{+\infty} \delta_n^{(k)}(x-x_0) g(x) dx$, $k \leq p$, в кожній точці неперервності функції $g^{(k)}(x)$. Тому за означенням 11 одержимо рівність

$$\lim_{n \to \infty} \int_{-\infty}^{+\infty} \delta_n(x-x_0) g^{(k)}(x) dx = g^{(k)}(x_0), \ k \leq p. \qquad (1.92)$$

Інтегруємо частинами

$$\int_{-\infty}^{+\infty} \delta_n(x-x_0) g^{(k)}(x) dx = -\int_{-\infty}^{+\infty} \delta_n'(x-x_0) g^{(k-1)}(x) dx = \ldots =$$

$$= (-1)^k \int_{-\infty}^{+\infty} \delta_n^{(k)}(x-x_0) g(x) dx, \ k \leq p.$$

Кожний з цих інтегралів існує внаслідок оцінки (1.90) і обмеженості функції $g^{(k)}(x)$ і абсолютної інтегровності функції $\Phi(x)$. Переходячи в одержаній рівності до границі з урахуванням рівності (1.92), прийдемо до формули (1.91).

Теорему доведено.

*П р и к л а д  8*. Побудуємо дельтоподібну послідовність вигляду (1.81) з використанням функції $\omega_0(t) = c_k(1-t^2)^k$, $0 \leq t \leq 1$. Відповідна фінітна функція має вигляд

$$\Phi(t) = \begin{cases} c_k(1-t^2)^k, & -1 \leq t \leq 1, \\ 0, & |t| > 1. \end{cases}$$

Вона має неперервну похідну $(k-1)$-го порядку і обмежену похідну $k$-го порядку. З умови

$$2c_k \int_0^1 (1-t^2)^k \, dt = 1$$

знайдемо значення сталої

$$2c_k \int_0^1 (1-t^2)^k dt = c_k \int_{-1}^1 (1-t^2)^k dt = c_k \int_{-1}^1 (1-t)^k (1+t)^k dt =$$

$$= c_k \left[ -(1-t)^k \frac{(1+t)^{k+1}}{k+1} \right]_{-1}^1 + c_k \frac{k}{k+1} \int_{-1}^1 (1-t)^{k-1}(1+t)^{k+1} dt = \ldots =$$

$$= c_k \frac{k!}{(k+1)\ldots 2k} \int_{-1}^1 (1+t)^{2k} dt =$$

$$= c_k \frac{k!}{(k+1)\ldots 2k} \frac{1}{2k+1} (1+t)^{2k+1} \Big|_{-1}^1 = c_k \frac{2^{2k+1}(k!)^2}{(2k+1)!} = 1.$$

Звідси $c_k = \dfrac{(2k+1)!}{2^{2k+1}(k!)^2}$.

Враховуючи значення сталої у виразі ядра, запишемо загальний член відповідної дельтоподібної послідовності

$$\delta_n(x-x_0) = \begin{cases} \dfrac{(2k+1)!}{2^{2k+1}(k!)^2 r_n} \left(1 - \dfrac{(x-x_0)^2}{r_n^2}\right)^k, & |x-x_0| \leq r_n, \\ 0, & |x-x_0| > r_n. \end{cases}$$

*П р и к л а д  9*. Знайдемо дельтоподібну послідовність, яка є згорткою двох дельтоподібних послідовностей

$$\delta_n(x) = \int_{-\infty}^{+\infty} \delta_{1n}(x-t)\delta_{2n}(t)dt,$$

де

$$\delta_{1n}(x-x_0) = \delta_{2n}(x-x_0) = \frac{1}{\pi r_n}\frac{1}{1+\frac{(x-x_0)^2}{r_n^2}} = \frac{r_n}{\pi}\frac{1}{r_n^2+(x-x_0)^2}.$$

Обчислимо відповідний інтеграл (ввівши позначення $x = r_n y$ і $t = r_n u$)

$$\delta_n(x) = \frac{1}{\pi^2 r_n^2}\int_{-\infty}^{+\infty}\frac{1}{\left(1+\frac{(t-x)^2}{r_n^2}\right)\left(1+\frac{t^2}{r_n^2}\right)}dt =$$

$$= \frac{1}{\pi^2 r_n}\int_{-\infty}^{+\infty}\frac{1}{[1+(u-y)^2](1+u^2)}du =$$

$$= \frac{1}{\pi^2 r_n}\frac{1}{(y^2+4)}\left[\int_{-\infty}^{\infty}\frac{du}{u^2+1} + \int_{-\infty}^{\infty}\frac{du}{(u-y)^2+1}\right] = \frac{2}{\pi r_n}\frac{1}{(y^2+4)}.$$

Отже,

$$\delta_n(x-x_0) = \frac{2r_n}{\pi}\frac{1}{4r_n^2+(x-x_0)^2}. \qquad (1.93)$$

Оскільки $\{r_n\}$ – довільна нескінченно мала числова послідовність, то, очевидно, згорткою однакових послідовностей

$$\delta_{1n}(x-x_0) = \frac{r_n}{\pi}\frac{1}{r_n^2+(x-x_0)^2},$$

$$\delta_{2n}(x-x_0) = \frac{r_n}{\pi}\frac{1}{r_n^2+(x-x_0)^2}$$

є така ж послідовність (1.93).

### 1.5.4. Завдання до п'ятого параграфа

1. Виходячи з означення неперервності функціоналу на мові околів, довести (теореми 1 і 2): для того щоби лінійний функціонал $l$ був неперервний в $D$, необхідно і достатньо, щоб існував такий окіл нульової точки в $D$, на якому функціонал обмежений.

2. Записати дельтоподібну послідовність функцій, визначених на дійсній осі, і знайти сталу $c$, якщо задана базова функція:

а) $\omega(t) = ce^{-t}$, $t \in [0, +\infty)$; б) $\Phi(t) = \dfrac{c}{1+t^2}$, $t \in (-\infty, +\infty)$;

в) $\omega_0(t) = c\sqrt{1-t^2}$, $t \in [0,1]$; г) $\omega_0(t) = c\cos\dfrac{\pi t}{2}$, $t \in [0,1]$.

3. Довести, що функція $\delta_r(x)$ слабко збігається при $r \to 0$ до дельта-функції відносно множини обмежених кусково-неперервних функцій:

а) $\delta_r(x) = \dfrac{1}{\pi}\dfrac{r}{x^2+r^2}$, $x \in (-\infty, +\infty)$;

б) $\delta_r(x) = \dfrac{r}{\pi x^2}\sin^2\dfrac{x}{r}$, $x \in (-\infty, +\infty)$.

4. Довести, що дельтоподібна послідовність слабко збігається до дельта-функції відносно множини обмежених кусково-неперервних функцій на проміжку $[-\pi, \pi]$:

а) $\delta_n(x) = \dfrac{1}{2\pi n}\dfrac{\sin^2\dfrac{nx}{2}}{\sin^2\dfrac{x}{2}}$; б) $\delta_n(x) = \dfrac{1}{2}\sqrt{\dfrac{n}{\pi}}\cos^{2n}\dfrac{x}{2}$.

5. Знайти дельтоподібну послідовність, яка є згорткою двох дельтоподібних послідовностей:

$$\delta_{1n}(x) = \dfrac{1}{2r_n}e^{-\dfrac{|x|}{r_n}}, \quad \delta_{2n}(x) = \dfrac{1}{2r_n}e^{-\dfrac{|x|}{r_n}}, \qquad x \in (-\infty, +\infty)$$

($\Phi(t) = \dfrac{1}{2}(2|t|+1)e^{-|t|}$).

# Р О З Д І Л  II

# РІВНОМІРНО  ЗБІЖНІ  РЯДИ
________________________________________________

### 2.1. Властивості функціональних рядів

**2.1.1. Збіжність ряду в точці та на проміжку.** Розглянемо послідовність $\{u_n(x)\}$ функцій дійсної змінної, неперервних на деякому проміжку $(a, b)$.

***О з н а ч е н н я  1 .*** *Сума нескінченного числа функцій* $u_n(x)$, $n = 1, 2, \ldots ,$ *називається функціональним рядом*

$$\sum_{n=1}^{\infty} u_n(x) = u_1(x) + u_2(x) + \ldots + u_n(x) + \ldots , \qquad (2.1)$$

*де* $u_1(x)$, $u_2(x)$, … *– члени ряду.*

Область визначення функцій $u_n(x)$, $n = 1, 2, \ldots ,$ є областю визначення функціонального ряду.

***О з н а ч е н н я  2 .*** *Сума перших n членів ряду* (2.1)

$$S_n(x) = \sum_{k=1}^{n} u_k(x) = u_1(x) + u_2(x) + \ldots + u_n(x) \qquad (2.2)$$

*називається n -ою частинною сумою ряду* (2.1), *а вираз*

$$r_n(x) = \sum_{k=n+1}^{\infty} u_k(x) = u_{n+1}(x) + u_{n+2}(x) + \ldots + u_{n+k}(x) + \ldots \qquad (2.3)$$

*називається n -им залишком ряду* (2.1).

Дослідження функціональних рядів еквівалентно дослідженню функціональних послідовностей, оскільки кожному ряду (2.1) можна поставити у відповідність послідовність частинних сум $\{S_n(x)\}$ і кожній послідовності $\{S_n(x)\}$ однозначно відповідає ряд (2.1) з членами

$$u_n(x) = S_n(x) - S_{n-1}(x), \ n = 1, 2, \ldots .$$

***О з н а ч е н н я  3 .*** *Функціональний ряд* (2.1) *називається збіжним в точці* $x$ , *якщо існує скінченна границя послідовності*

*частинних сум* (2.2) *в цій точці*,
$$\lim_{n \to \infty} S_n(x) = S(x). \quad (2.4)$$

*При цьому функція* $S(x)$ *називається сумою ряду* (2.1), *а множина всіх точок його збіжності – деякий проміжок* $(a, b)$ - *називається областю збіжності ряду.*

Можна сформулювати означення збіжності ряду (2.1) (еквівалентне означенню 3), яке ґрунтується на означенні збіжності послідовності.

**О з н а ч е н н я  3'** . *Функціональний ряд* (2.1) *називається збіжним в області* $(a, b)$, *якщо для будь-якого* $\varepsilon > 0$ *і для кожного* $x \in (a, b)$ *існує натуральне число* $N = N(\varepsilon, x)$ *таке, що*

$$|r_n(x)| < \varepsilon$$

*для всіх* $n > N$.

**О з н а ч е н н я  4** . *Ряд* (2.1) *називається абсолютно збіжним в області* $(a, b)$, *якщо для всіх* $x \in (a, b)$ *збігається ряд*

$$\sum_{n=1}^{\infty} |u_n(x)| = |u_1(x)| + |u_2(x)| + \ldots + |u_n(x)| + \ldots .$$

*П р и к л а д  1* . Дослідити на збіжність функціональний ряд

$$\sum_{n=0}^{\infty} \frac{x^n}{n!} .$$

Дослідимо цей ряд на абсолютну збіжність, тобто розглянемо ряд з $n$-м членом $u_n = \frac{|x|^n}{n!}$. За ознакою Даламбера знайдемо

$$\lim_{n \to \infty} \frac{|u_{n+1}|}{|u_n|} = \lim_{n \to \infty} \frac{|x|}{n+1} = 0$$

для будь-якого скінченого $x$.

Отже, ряд збігається абсолютно і, відповідно, збігається в усіх точках дійсної осі.

*П р и к л а д  2* . Дослідити на збіжність функціональний ряд

$$\sum_{n=0}^{\infty} \frac{x^{2n}}{(1+x^2)^n} .$$

Цей ряд збігається для всіх дійсних значень $x$. Члени ряду складають геометричну прогресію зі знаменником $q = \dfrac{x^2}{1+x^2}$. Якщо $x \neq 0$, то $0 < q < 1$. Тоді сума ряду дорівнює

$$S(x) = \frac{1}{1-q} = 1 + x^2.$$

Якщо $x = 0$, то всі члени ряду нульові і його сума $S(0) = 0$.
Отже,

$$S(x) = \begin{cases} 0, & x = 0, \\ 1 + x^2, & x \neq 0. \end{cases}$$

Ряд збігається на всій дійсній осі.

*З а у в а ж е н н я*. Цей приклад ілюструє випадок, коли сума нескінченної кількості неперервних функцій є розривною функцією, тобто, границя при $x \to 0$ суми нескінченної кількості функцій не дорівнює сумі їх границь

$$\lim_{x \to 0} \sum_{n=1}^{\infty} u_n(x) \neq \sum_{n=1}^{\infty} \lim_{x \to 0} u_n(x).$$

**2.1.2. Рівномірно збіжний функціональний ряд.** Нехай функціональний ряд (2.1) збігається до функції $S(x)$ в області $(a, b)$.

*О з н а ч е н н я 5*. *Функціональний ряд* (2.1) *рівномірно збігається в області* $(a, b)$ *до функції* $S(x)$, *якщо послідовність його частинних сум* $\{S_n(x)\}$ *рівномірно збігається до* $S(x)$ *в* $(a, b)$.

Ґрунтуючись на означенні 2 (п. 1.3) рівномірно збіжної послідовності функцій, можна сформулювати означення рівномірної збіжності ряду, еквівалентне означенню 5.

*О з н а ч е н н я 5'*. *Функціональний ряд* (2.1) *рівномірно збігається до функції* $S(x)$ *в області* $(a, b)$, *якщо для як завгодно малого* $\varepsilon > 0$ *знайдеться номер* $N(\varepsilon)$ *такий, що для всіх* $n > N(\varepsilon)$ *і всіх* $x \in (a, b)$ *справджується нерівність*

$$|r_n(x)| < \varepsilon.$$

Сформулюємо основні властивості рівномірно збіжних рядів $[5, 8, 12, 21]$.

***Т е о р е м а  1 (критерій Коші).*** *Для того щоби функціональний ряд* (2.1) *рівномірно збігався в області* $(a,b)$ *до деякої граничної функції, необхідно і достатньо, щоб для як завгодно малого* $\varepsilon > 0$ *знайшовся номер* $N(\varepsilon)$ *такий, що*

$$\left| \sum_{k=m+1}^{n} u_k(x) \right| < \varepsilon \qquad (2.5)$$

*для всіх* $n \geq N(\varepsilon)$, $m \geq N(\varepsilon)$ *і всіх* $x \in (a,b)$.

*Д о в е д е н н я*. Теорема є наслідком теореми 1 (п. 1.3) про рівномірну збіжність послідовностей неперервних функцій, оскільки в лівій частині нерівності (2.5) є різниця $S_n(x) - S_m(x)$ частинних сум ряду (2.1). Послідовність частинних сум $\{S_n(x)\}$ задовольняє умови цієї теореми.

***Т е о р е м а  2.*** *Сума рівномірно збіжного в області* $(a,b)$ *ряду* (2.1) *функцій, неперервних в* $(a,b)$, *неперервна функція в* $(a,b)$.

*Д о в е д е н н я*. Послідовність частинних сум $\{S_n(x)\}$ ряду (2.1) задовольняє в області $(a,b)$ умови теореми 4 (п. 1.3) і тому сума цього ряду є неперервною функцією в $(a,b)$.

***Т е о р е м а  3 (ознака Діні).*** *Якщо всі члени ряду* (2.1) *неперервні і додатні функції на сегменті* $[a,b]$ *і його сума також неперервна функція на* $[a,b]$, *то ряд* (2.1) *збігається рівномірно на* $[a,b]$.

*Д о в е д е н н я*. Легко переконатися, що послідовність частинних сум $\{S_n(x)\}$ ряду (2.1) задовольняє умови теореми 2 (п. 1.3). Дійсно, а) всі члени послідовності $\{S_n(x)\}$ неперервні функції, б) послідовність $\{S_n(x)\}$ не спадає, як сума неперервних і додатних функцій, на $[a,b]$, в) послідовність $\{S_n(x)\}$ збігається до неперервної функції.

Отже, всі умови виконані.

Теорему доведено.

***Т е о р е м а  4 (ознака Вейєрштрасса).*** *Нехай функціональний ряд* (2.1) *визначений на проміжку* $(a,b)$ *і існує*

збіжний числовий ряд $\sum_{k=1}^{\infty} c_k$ такий, що для всіх $x \in (a, b)$ і для кожного номера $k$ справедлива нерівність
$$|u_k(x)| \leq c_k. \qquad (2.6)$$

*Тоді ряд* (2.1) *збігається абсолютно і рівномірно на проміжку* $(a, b)$.

*Д о в е д е н н я .* За критерієм Коші (для числового ряду) яке б не було мале число $\varepsilon > 0$ знайдеться номер $N(\varepsilon)$ такий, що для всіх $n > N(\varepsilon)$ і $m > N(\varepsilon)$ справедлива нерівність

$$\sum_{k=m+1}^{n} c_k < \varepsilon. \qquad (2.7)$$

З нерівностей (2.6) і (2.7) одержимо нерівність
$$\left|\sum_{k=m+1}^{n} u_k(x)\right| < \sum_{k=m+1}^{n} |u_k(x)| < \sum_{k=m+1}^{n} c_k < \varepsilon,$$
яка виконується для всіх $n > N(\varepsilon)$, $m > N(\varepsilon)$ і для всіх $x \in (a, b)$.

Отже, за критерієм Коші функціональний ряд збігається абсолютно і рівномірно.

Теорему доведено.

***Т е о р е м а  5 (ознака Діріхле).*** *Нехай функціональний ряд має вигляд*

$$\sum_{k=1}^{\infty} u_k(x) v_k(x) \qquad (2.8)$$

*і виконуються умови:*

*а) послідовність* $\{v_k(x)\}$ *не зростає на проміжку* $(a, b)$ *і рівномірно збігається до нуля на цій множині,* $\lim_{k \to \infty} v_k(x) = 0$;

*б) послідовність частинних сум* $U_n(x) = \sum_{k=1}^{n} u_k(x)$ *ряду* $\sum_{k=1}^{\infty} u_k(x)$ *рівномірно обмежена.*

*Тоді ряд* (2.8) *збігається рівномірно на проміжку* $(a, b)$.

*Доведення*. Послідовність $\{U_n(x)\}$ рівномірно обмежена на проміжку $(a,b)$, якщо існує дійсне число $M>0$ таке, що для всіх $x \in (a,b)$ і для всіх $n$ виконується нерівність $|U_n(x)| < M$.

Оскільки послідовність частинних сум $\left\{U_n(x) = \sum_{k=1}^{n} u_k(x)\right\}$ рівномірно обмежена, існує $M>0$ таке, що для всіх $x \in (a,b)$ справедлива нерівність $|U_n(x)| \leq M$. Звідси, для будь-яких натуральних $n$ і $p$, ввівши позначення $U_p^n(x) = U_{n+p}(x) - U_n(x) = \sum_{k=n+1}^{n+p} u_k(x)$, маємо оцінку

$$\left|U_p^n(x)\right| \leq \left|U_{n+p}(x) - U_n(x)\right| \leq 2M. \qquad (2.9)$$

Нехай $\varepsilon > 0$. Тоді з умови $\lim_{k \to \infty} v_k(x) = 0$ випливає, що існує номер $N(\varepsilon)$ такий, що для всіх $m > N(\varepsilon)$ і всіх $x \in (a,b)$ справедлива нерівність

$$v_m(x) < \frac{\varepsilon}{2M}. \qquad (2.10)$$

Запишемо перетворення Абеля $[8\,c.\,431, 12\,c.\,511]$ для довільного відрізка ряду (2.8)

$$\sum_{k=m+1}^{m+p} u_k(x) v_k(x) = U_1^m v_{m+1} + \sum_{k=m+1}^{m+p}\left[U_{k-m}^m(x) - U_{k-m-1}^m(x)\right] v_k(x) =$$
$$= U_p^m(x) v_{m+p}(x) + \sum_{k=m+1}^{m+p-1} U_{k-m}^m(x)\left[v_k(x) - v_{k+1}(x)\right], \qquad (2.11)$$

де $U_p^m(x) = \sum_{l=m+1}^{m+p} u_l(x)$.

Оцінивши цей відрізок з урахуванням умови а), нерівностей (2.9), (2.10) і $p = n - m$, одержимо

$$\left|\sum_{k=m+1}^{n} u_k(x) v_k(x)\right| = \left|U_{n-m}^m(x) v_n(x) + \sum_{k=m+1}^{n-1} U_{k-m}^m(x)\left[v_k(x) - v_{k+1}(x)\right]\right| \leq$$

$$\leq 2M\left\{v_n(x)+\sum_{k=m+1}^{n-1}[v_k(x)-v_{k+1}(x)]\right\}\leq 2M\,v_{m+1}<\varepsilon.$$

Отже, за критерієм Коші ряд (2.8) збігається рівномірно.
Теорему доведено.

***Т е о р е м а  6 (ознака Абеля).*** *Нехай функціональний ряд має вигляд* (2.8) *і виконуються умови*:

*а) члени послідовності* $\{v_k(x)\}$ *монотонно спадають (при збільшенні номера $k$) і рівномірно обмежені на проміжку* $(a,b)$;

*б) ряд* $\sum_{k=1}^{\infty}u_k(x)$ *рівномірно збігається на* $(a,b)$.

*Тоді ряд* (2.8) *збігається рівномірно на проміжку* $(a,b)$.

*Д о в е д е н н я* . Оскільки послідовність $\{v_k(x)\}$ рівномірно обмежена, існує $A>0$ таке, що для всіх $x\in(a,b)$ справедлива нерівність $v_k(x)\leq v_{k+1}(x)$, $|v_k(x)|\leq A$, $k=1,2,...$ (функції $v_k(x)$ можуть бути і від'ємними). За рівномірної збіжності ряду $\sum_{k=1}^{\infty}u_k(x)$ для будь-якого $\varepsilon>0$ існує номер $N(\varepsilon)$ такий, що для всіх $m>N(\varepsilon)$, $n>N(\varepsilon)$ і всіх $x\in(a,b)$, справедлива нерівність

$$\left|\sum_{k=m+1}^{n}u_k(x)\right|=\left|U_{n-m}^{m}(x)\right|<\frac{\varepsilon}{3A}. \qquad (2.12)$$

Оцінимо довільний відрізок ряду (2.8) з урахуванням перетворення Абеля (2.11) і нерівності (2.12)

$$\left|\sum_{k=m+1}^{n}u_k(x)v_k(x)\right|=\left|U_{n-m}^{m}(x)v_n(x)+\sum_{k=m+1}^{n-1}U_{k-m}^{m}(x)[v_k(x)-v_{k+1}(x)]\right|\leq$$

$$\leq\frac{\varepsilon}{3A}\left\{|v_n(x)|+\sum_{k=m+1}^{n-1}[v_k(x)-v_{k+1}(x)]\right\}=\frac{\varepsilon}{3A}\left(|v_n(x)|-v_n(x)+v_{m+1}\right)\leq\varepsilon.$$

Ця нерівність виконується для всіх $m>N(\varepsilon)$, $n>N(\varepsilon)$ і всіх $x\in(a,b)$, а отже, за критерієм Коші ряд (2.8) збігається рівномірно.
Теорему доведено.

*П р и к л а д  3* . Дослідити рівномірну збіжність рядів

$$\sum_{k=1}^{\infty}\frac{\sin kx}{k^{\alpha}}, \quad \sum_{k=1}^{\infty}\frac{\cos kx}{k^{\alpha}}, \ \alpha > 0.$$

Для випадку $\alpha > 1$ ряди збігаються (за ознакою Вейєршасса) абсолютно і рівномірно на всій дійсній осі, оскільки справедливі нерівності $\left|\frac{\sin kx}{k^{\alpha}}\right| \leq \frac{1}{k^{\alpha}}$, $\left|\frac{\cos kx}{k^{\alpha}}\right| \leq \frac{1}{k^{\alpha}}$ і ряд $\sum_{k=1}^{\infty}\frac{1}{k^{\alpha}}$ збігається.

Дослідимо збіжність рядів для випадку $0 < \alpha \leq 1$. Послідовність $\left\{v_k(x) = \frac{1}{k^{\alpha}}\right\}$ не зростає для всіх $x \in R$ і рівномірно прямує до нуля. Тому умова а) теореми 6 виконується.

Розглянемо ряд $\sum_{k=1}^{\infty} u_k(x)$, де $u_k(x) = \sin kx$, і послідовність його частинних сум

$$U_n(x) = \sum_{k=1}^{n}\sin kx = \frac{1}{2\sin\frac{x}{2}}\sum_{k=1}^{n} 2\sin\frac{x}{2}\sin kx =$$

$$= \frac{1}{2\sin\frac{x}{2}}\sum_{k=1}^{n}\left[\cos\left(k-\frac{1}{2}\right)x - \cos\left(k+\frac{1}{2}\right)x\right] =$$

$$= \frac{\cos\frac{x}{2} - \cos\left(n+\frac{1}{2}\right)x}{2\sin\frac{x}{2}} = \frac{\sin\frac{n+1}{2}x \sin\frac{n}{2}x}{\sin\frac{x}{2}}.$$

Звідси $|U_n(x)| \leq \frac{1}{\left|\sin\frac{x}{2}\right|}$.

Тоді умова б) теореми 4 виконується на будь-якому сегменті, що не містить точок $x_m = 2\pi m$, $m = 0, \pm 1, \ldots$, оскільки на такому сегменті функція $\left|\sin\frac{x}{2}\right|$ має додатну нижню грань.

Аналогічно знайдемо вираз частинної суми і її оцінку для другого ряду

$$U_n(x) = \sum_{k=1}^{n} \cos kx = \frac{\sin\left(n + \frac{1}{2}\right)x - \sin\frac{x}{2}}{2\sin\frac{x}{2}} = \frac{\sin\frac{nx}{2}\cos\frac{(n+1)x}{2}}{\sin\frac{x}{2}},$$

$$|U_n(x)| \le \frac{1}{\left|\sin\frac{x}{2}\right|}.$$

Тут також виконуються умови теореми 4 на будь-якому сегменті, що не містить точок $x_m = 2\pi m$, $m = 0, \pm 1, \ldots$.

Отже, розглянуті ряди збігається рівномірно на будь-якому сегменті, що не містить точок $x_m = 2\pi m$, $m = 0, \pm 1, \ldots$.

*З а у в а ж е н н я*. Збіжний ряд може втратити рівномірну збіжність або її набути, якщо всі його члени помножити на множник, не залежний від $n$. Наприклад, якщо частинна сума ряду має вигляд

$$S_n(x) = \frac{x}{1 + nx}, \ 0 \le x \le 1, \quad u_n(x) = S_n(x) - S_{n-1}(x),$$

то справедлива оцінка $|S_n(x)| < \frac{1}{n}$ і ряд рівномірно збігається до нуля. Ряд, одержаний з цього ряду множенням на $\frac{1}{x}$, не є рівномірно збіжним. Оскільки його сума – розривна функція, яка приймає нульові значення, якщо $x > 0$, і дорівнює одиниці, якщо $x = 0$.

Однак, якщо помножити рівномірно збіжний ряд на обмежений множник (не залежний від $n$), то одержаний ряд буде також рівномірно збіжним. Це легко показати, виходячи з означення рівномірної збіжності.

**2.1.3. Почленний перехід до границі. Диференціювання та інтегрування рядів.** Розглянемо достатні умови неперервності, диференційовності та інтегровності суми ряду $[8, 12, 21, 24]$.

*Т е о р е м а 7. Якщо члени $u_n(x)$, $n = 1, 2, \ldots$, ряду (2.1) неперервні функції на проміжку $(a, b)$ і цей ряд збігається рівномірно на $(a, b)$, то його сума*

$$S(x) = \sum_{k=1}^{\infty} u_k(x)$$

*також неперервна функція на проміжку* $(a, b)$ *і в будь-якій точці* $x_0 \in (a, b)$ *справедлива формула*

$$\lim_{x \to x_0} \sum_{k=1}^{\infty} u_k(x) = \sum_{k=1}^{\infty} \lim_{x \to x_0} u_k(x). \qquad (2.13)$$

Д о в е д е н н я . Ряд (2.1) збігається в будь-якій точці проміжку $(a, b)$. Розглянемо довільну точку $x_0 \in (a, b)$ і покажемо, що функція $S(x)$ неперервна в цій точці.

Зафіксуємо мале число $\varepsilon > 0$. За умов теореми послідовність частинних сум $\{S_n(x)\}$ рівномірно збігається до функції $S(x)$ в області $(a, b)$. Тому існує номер $N = N(\varepsilon)$ такий, що для всіх $x \in (a, b)$ і всіх $n \geq N$, зокрема для $n = N$, справедлива нерівність

$$|S(x) - S_N(x)| < \frac{\varepsilon}{3}. \qquad (2.14)$$

Функція $S_N(x)$, як сума скінченої кількості неперервних функцій $u_n(x)$, неперервна в точці $x_0$. Тому існує мале число $\delta = \delta(\varepsilon)$ таке, що для всіх $x \in (a, b)$ і $|x - x_0| < \delta$ справедлива нерівність

$$|S_N(x) - S_N(x_0)| < \frac{\varepsilon}{3}. \qquad (2.15)$$

Перетворимо тепер вираз $S(x) - S(x_0) = [S(x) - S_N(x)] + [S_N(x) - S_N(x_0)] + [S_N(x_0) - S(x_0)]$ і оцінимо його з урахуванням нерівностей (2.14), (2.15), якщо тільки $|x - x_0| < \delta$,

$$|S(x) - S(x_0)| \leq |S(x) - S_N(x)| + |S_N(x) - S_N(x_0)| + |S_N(x_0) - S(x_0)| < \varepsilon.$$

Звідси випливає неперервність функції $S(x)$ в точці $x_0$.

Гранична рівність (2.13) випливає з умов неперервності функцій $S(x)$ і $u_n(x)$, $n = 1, 2, \ldots$, а отже,

$$\lim_{x \to x_0} \sum_{k=1}^{\infty} u_k(x) = \lim_{x \to x_0} S(x) = S(x_0) = \sum_{k=1}^{\infty} u_k(x_0) = \sum_{k=1}^{\infty} \lim_{x \to x_0} u_k(x),$$

тобто, за умов теореми можна почленно переходити до границі у ряді (2.1).

Теорему доведено.

**Т е о р е м а   8 .** *Якщо члени $u_n(x)$, $n = 1, 2, \ldots$, ряду* (2.1) *неперервні функції на відрізку $[a, b]$ і він збігається рівномірно на $[a, b]$, то ряд*

$$\sum_{n=1}^{\infty} \int_{a}^{x} u_n(t)\, dt \qquad (2.16)$$

*також збігається рівномірно на $[a, b]$ і, якщо*

$$S(x) = \sum_{k=1}^{\infty} u_k(x),$$

*то*

$$\int_{a}^{x} S(t)\, dt = \sum_{n=1}^{\infty} \int_{a}^{x} u_n(t)\, dt, \quad a \le x \le b. \qquad (2.17)$$

*Д о в е д е н н я .* Покажемо, що ряд (2.16) рівномірно збігається і справедлива формула (2.17).

Розглянемо частинну суму і залишок рівномірно збіжного ряду (2.1)

$$S_n(x) = \sum_{k=1}^{n} u_k(x), \quad r_n(x) = S(x) - S_n(x), \quad a \le x \le b.$$

Фіксуємо довільне число $\varepsilon > 0$. Внаслідок рівномірної збіжності ряду (2.1) знайдеться номер $N = N(\varepsilon)$ такий, що для всіх $n > N$ і для всіх $x \in [a, b]$ виконується нерівність

$$|r_n(x)| < \frac{\varepsilon}{b - a}. \qquad (2.18)$$

Оскільки ряд (2.1) збігається рівномірно, за теоремою 7 сума $S(x)$ цього ряду неперервна функція на відрізку $[a, b]$ і тому інтегровна на будь-якому відрізку $[a, x]$, $a \le x \le b$. Оцінимо залишок ряду у формулі (2.17) з урахуванням нерівності (2.18),

$$\left| \int_{a}^{x} S(t)\, dt - \sum_{k=1}^{n} \int_{a}^{x} u_k(t)\, dt \right| = \left| \int_{a}^{x} S(t)\, dt - \int_{a}^{x} \sum_{k=1}^{n} u_k(t)\, dt \right| =$$

$$=\left|\int_a^x S(t)dt - \int_a^x S_n(t)dt\right| \le \int_a^x |S(t)-S_n(t)|dt = \int_a^x |r_n(t)|dt \le \frac{\varepsilon}{b-a}\int_a^x dt < \varepsilon.$$

Ця нерівність виконується для всіх $n > N$ і всіх $x \in [a,b]$.

Отже, ряд у формулі (2.17) збігається рівномірно до граничної функції.

Теорему доведено.

*З а у в а ж е н н я .* Формулу (2.17) можна записати ще так

$$\int_a^x \sum_{k=1}^{\infty} u_k(t)dt = \sum_{n=1}^{\infty}\int_a^x u_n(t)dt, \quad a \le x \le b,$$

яку розуміємо, що за умов теореми ряд можна почленно інтегрувати.

**Т е о р е м а   9 .** *Нехай члени $u_n(x)$, $n = 1, 2, \ldots$, ряду (2.1) неперервно диференційовні функції на відрізку $[a,b]$ і ряд, складений з їх похідних*

$$\sum_{k=1}^{\infty} u_k'(x), \qquad (2.19)$$

*рівномірно збігається на відрізку $[a,b]$.*

*Тоді, якщо ряд (2.1) збігається хоч би в одній точці $x_0 \in [a,b]$, то він збігається рівномірно на відрізку $[a,b]$, його сума*

$$S(x) = \sum_{k=1}^{\infty} u_k(x)$$

*неперервно диференційована і справедлива формула*

$$S'(x) = \sum_{k=1}^{\infty} u_k'(x). \qquad (2.20)$$

*Д о в е д е н н я .* Оскільки ряд (2.19) збігається рівномірно, його сума $\sigma(x) = \sum_{k=1}^{\infty} u_k'(x)$ неперервна функція. За теоремою 8 цей ряд можна інтегрувати

$$\int_{x_0}^x \sigma(t)dt = \sum_{k=1}^{\infty}\int_{x_0}^x u_k'(t)dt = \sum_{k=1}^{\infty}[u_k(x) - u_k(x_0)], \quad a \le x \le b, \quad (2.21)$$

і одержаний після цього ряд збігається. За умовою збігається також ряд $\sum_{k=1}^{\infty} u_k(x_0)$.

Отже, збігається також сума обох збіжних рядів, тобто збігається ряд $\sum_{k=1}^{\infty} u_k(x)$, $a \leq x \leq b$.

Вираз (2.21) запишемо у вигляді
$$\int_{x_0}^{x} \sigma(t)dt = \sum_{k=1}^{\infty} u_k(x) - \sum_{k=1}^{\infty} u_k(x_0) = S(x) - S(x_0), \quad a \leq x \leq b.$$

Функція, що стоїть у правій частині цієї рівності, має похідну (як інтеграл зі змінною верхньою межею) і тому диференційовною є функція $S(x)$. Після диференціювання, одержимо $\sigma(x) = S'(x)$. Врахувавши тут подання функції $\sigma(x)$ у вигляді ряду, одержимо формулу (2.20).

Покажемо, що ряд (2.1) збігається рівномірно на $[a,b]$. Запишемо цей ряд у вигляді
$$\sum_{k=1}^{\infty} u_k(x) = \sum_{k=1}^{\infty} \int_{x_0}^{x} u_k'(t)\,dt + \sum_{k=1}^{\infty} u_k(x_0).$$

У правій частині цієї рівності маємо рівномірно збіжний на відрізку $[a,b]$ (за теоремою 8) перший ряд і збіжний (числовий) другий ряд. Тому їх сума також рівномірно збігається на цьому відрізку.

Теорему доведено.

*З а у в а ж е н н я*. Формула (2.20), записана у вигляді
$$\left[\sum_{k=1}^{\infty} u_k(x)\right]' = \sum_{k=1}^{\infty} u_k'(x),$$
виражає той факт, що за умов теореми функціональний ряд можна *почленно диференціювати*.

*П р и к л а д   4*. Розглянемо питання про можливість інтегрування ряду
$$\frac{1}{1+x} = \sum_{k=0}^{\infty} (-1)^k x^k, \quad -1 < x < 1.$$

За теоремою Вейєрштрасса ряд рівномірно збігається на будь-якому відрізку $[-q, q]$, $0 < q < 1$, оскільки $\left|(-1)^k x^k\right| \leq q^k$ і $\sum_{k=0}^{\infty} q^k < \infty$. Тому за теоремою 8 його можна інтегрувати на відрізку $[0, x]$, де $x \in [-q, q]$,

$$\ln(1+x) = \int_0^x \frac{dt}{1+t} = \sum_{k=0}^{\infty} \frac{(-1)^k}{k+1} x^{k+1}. \qquad (2.22)$$

Оскільки $q < 1$ довільне число, формула (2.22) справедлива для всіх $x \in (-1, 1)$. При $x = -1$ формула (2.22) не має сенсу. Однак при $x = 1$ справедлива рівність

$$\ln 2 = \sum_{k=0}^{\infty} \frac{(-1)^k}{k+1}.$$

Дійсно, за теоремою Абеля ряд (2.22) можна записати у вигляді $\sum_{k=0}^{\infty} u_k v_k$, де $u_k = \frac{(-1)^k}{k+1}$ – члени збіжного ряду $\sum_{k=0}^{\infty} u_k = \sum_{k=0}^{\infty} \frac{(-1)^k}{k+1}$, який можна розглядати як рівномірно збіжний ряд сталих функцій; $\{v_k = x^{k+1}\}$ – послідовність функцій, члени якої монотонно спадають при збільшенні $k$.

Отже, умови теореми виконані і ряд рівномірно збігається в області $(-1, 1]$.

### 2.2. Степеневі ряди

**2.2.1. Радіус і круг збіжності. Неперервність суми степеневого ряду**

*О з н а ч е н н я  1 . Функціональний ряд вигляду*

$$\sum_{n=0}^{\infty} a_n (x - x_0)^n, \qquad (2.23)$$

*де $a_n$, $x$ і $x_0$ – дійсні числа, називається степеневим рядом.*

*Числа $a_n$, $n = 0, 1, \ldots$, називаються коефіцієнтами степеневого ряду* (2.23).

Провівши заміну $\xi = x - x_0$ в ряді (2.23), одержимо ряд

$$\sum_{n=0}^{\infty} a_n \xi^n.$$

Очевидно, що дослідження збіжності цього ряду еквівалентно дослідженню збіжності ряду (2.23).

Зауважимо, що твердження, які будуть одержані, справедливі також для степеневих рядів від комплексної змінної.

***Т е о р е м а  1 (Абеля).*** *Якщо степеневий ряд*

$$\sum_{n=0}^{\infty} a_n x^n \tag{2.24}$$

*збігається при* $x = x_0$, *то він збігається і при цьому абсолютно в крузі* $|x| < |x_0|$ *і, якщо він розбігається в точці* $x = x_1$, *то він розбігається в області* $|x| > |x_1|$.

*Д о в е д е н н я*. Нехай ряд

$$\sum_{n=0}^{\infty} a_n x_0^n$$

збігається, тоді його $n$-ий член $a_n x_0^n$ прямує до нуля, коли $n \to \infty$ і, відповідно, послідовність $\{a_n x_0^n\}$ обмежена $|a_n x_0^n| \leq M$, $n = 0, 1, \ldots$, $0 < M < \infty$. Тому для $n$-го члена ряду (2.24) одержимо оцінку

$$\left| a_n x^n \right| \leq \left| a_n x_0^n \right| \left| \frac{x}{x_0} \right|^n \leq M \left| \frac{x}{x_0} \right|^n.$$

Якщо $|x| < |x_0|$, то ряд

$$\sum_{n=0}^{\infty} M \left| \frac{x}{x_0} \right|^n = M \sum_{n=0}^{\infty} \left| \frac{x}{x_0} \right|^n$$

збігається, оскільки є геометричною прогресією зі знаменником $q = \left| \dfrac{x}{x_0} \right| < 1$. За ознакою порівняння збігається ряд $\sum_{n=0}^{\infty} |a_n x^n|$, а отже, ряд (2.24) абсолютно збігається в області $|x| < |x_0|$.

Якщо в точці $x = x_1$ ряд (2.24) розбігається, то він розбігається для всіх $|x|>|x_1|$. Якщо не так і в точці $x$, $|x|>|x_1|$, ряд (2.24) збігається, то за першою частиною теореми він збігається в точці $x_1$, що суперечить умові.

Теорему доведено.

*З а у в а ж е н н я* . Будь-який степеневий ряд вигляду (2.24) збігається в точці $x=0$ і $S(0)=a_0$.

**О з н а ч е н н я  2 .** *Величина $R$ (додатне число, нуль або $\infty$) така, що для всіх $x$, для яких $|x|<R$, ряд* (2.24) *збігається, а для всіх $x$, для яких $|x|>R$, ряд* (2.24) *розбігається, називається радіусом збіжності степеневого ряду* (2.24).

*Множина точок $x$, для яких $|x|<R$, називається інтервалом збіжності ряду* (2.24).

**Т е о р е м а  2 .** *Радіус збіжності степеневого ряду* (2.24) *визначається за формулою Коші – Адамара*

$$R = \frac{1}{\overline{\lim_{n \to \infty}} \sqrt[n]{|a_n|}}, \qquad (2.25)$$

*де в знаменнику знаходиться верхня границя послідовності $\{|a_n|\}$.*

*Д о в е д е н н я* . Число $x$ називається верхньою (нижньою) границею числової послідовності $\{x_n\}$, $\overline{\lim_{n \to \infty}} x_n$ ($\underline{\lim_{n \to \infty}} x_n$), якщо існує підпослідовність $\{x_{n_k}\}$ послідовності $\{x_n\}$, збіжна до цього числа, і при цьому інша збіжна підпослідовність послідовності $\{x_n\}$ збігається до числа не більшого (не меншого) від числа $x$.

Приймемо, що $\frac{1}{0}=\infty$ і $\frac{1}{\infty}=0$. Тоді, якщо границя у формулі (2.25) дорівнює нулю, то $R=\infty$, якщо ж вона дорівнює $\infty$, то $R=0$.

За умови $R=0$ степеневий ряд має лише одну точку збіжності $x=0$.

За цих умов радіус збіжності степеневого ряду завжди існує.

Нехай величина $R$ визначається за формулою (2.25). В точці $x=0$ степеневий ряд збігається, тому в цій точці теорема правильна, $|x|<R$.

Розглянемо ряд (2.24), коли $|x| > 0$, і розглянемо ряд, складений з абсолютних величин членів ряду (2.24),

$$\sum_{n=0}^{\infty} \left| a_n x^n \right|. \qquad (2.26)$$

За ознакою Коші ряд (2.26) збігається, якщо $\overline{\lim_{n \to \infty}} \sqrt[n]{\left| a_n x^n \right|} < 1$, і розбігається, якщо $\overline{\lim_{n \to \infty}} \sqrt[n]{\left| a_n x^n \right|} > 1$. Знайдемо цю границю

$$\overline{\lim_{n \to \infty}} \sqrt[n]{\left| a_n x^n \right|} = \overline{\lim_{n \to \infty}} \sqrt[n]{\left| a_n \right| \left| x \right|^n} = |x| \overline{\lim_{n \to \infty}} \sqrt[n]{\left| a_n \right|} = \frac{|x|}{R}.$$

Звідси, якщо $\frac{|x|}{R} < 1$, тобто $|x| < R$, то ряд (2.26) збігається і, відповідно, ряд (2.24) збігається абсолютно, якщо ж $\frac{|x|}{R} > 1$, тобто $|x| > R$, то загальні члени як ряду (2.26), так і ряду (2.24) необмежено зростають і, відповідно, ряд (2.24) розбігається.

Отже, за означенням 1 величина, що визначається формулою (2.25), є радіусом збіжності ряду (2.24).

Теорему доведено.

*З а у в а ж е н н я .* Для визначення радіуса збіжності степеневого ряду можна використати ознаку Даламбера збіжності числового ряду (за умови існування відповідної границі)

$$\lim_{n \to \infty} \left| \frac{a_{n+1}}{a_n} \right| = \frac{1}{R}. \qquad (2.27)$$

***Т е о р е м а  3 .*** *Степеневий ряд* (2.24) *рівномірно збігається на будь-якому сегменті* $[-q, q]$, $0 < q < R$, *і сума цього ряду є неперервною функцією на проміжку* $(-R, R)$.

*Д о в е д е н н я .* Якщо $0 < q < R$, то за теоремою 2 в точці $x = q$ ряд (2.24) збігається, тобто збігається числовий ряд

$$\sum_{n=0}^{\infty} |a_n| q^n.$$

Для будь-якої точки $x$, що лежить на сегменті $-q \le x \le q$,

справедлива оцінка $\left|a_n x^n\right| \le \left|a_n\right| q^n$, тому за ознакою Вейєрштрасса на цьому сегменті ряд (2.24) збігається рівномірно.

Оскільки члени степеневого ряду є неперервними функціями і ряд (2.24) рівномірно збігається на сегменті $-q \le x \le q$, то його сума є неперервною функцією на цьому сегменті. Очевидно, що для будь-якої точки $x$ з інтервалу збіжності $-R < x < R$, можна підібрати таке число $q$, що $|x| < q < R$. Тому сума степеневого ряду неперервна функція в кожній точці інтервалу збіжності цього ряду.

*З а у в а ж е н н я* . Інтервал збіжності ряду (2.23) задається з урахуванням заміни $\xi = x - x_0$ у вигляді $|x - x_0| < R$. При цьому $R$ називається радіусом збіжності ряду (2.23).

**2.2.2. Диференціювання та інтегрування степеневих рядів.** Розглядаємо степеневі ряди (2.23) і (2.24) з дійсними коефіцієнтами і вважаємо, що змінні $x$ і $x_0$ також дійсні.

***Т е о р е м а  4*** . *Степеневий ряд* (2.24*) можна інтегрувати почленно всередині інтервалу збіжності і одержаний при цьому ряд має такий же радіус збіжності.*

*Д о в е д е н н я* . Нехай $R > 0$ – радіус збіжності ряду (2.24) і точка $x$ така, що $|x| < R$. Тоді знайдеться таке число $q > 0$, що $|x| < q < R$. За теоремою 3 ряд рівномірно збігається на сегменті $[-q, q]$ і, відповідно, на сегменті $[0, x]$, а отже, за теоремою 8 (п. 2.1) його можна інтегрувати на сегменті $[0, x]$

В результаті почленного інтегрування ряду (2.24) одержимо степеневий ряд

$$\sum_{n=1}^{\infty} \frac{a_{n-1}}{n} x^n,$$

радіус збіжності якого визначається за формулою (2.25) і є величиною, оберненою до границі

$$\overline{\lim_{n \to \infty}} \sqrt[n]{\frac{|a_{n-1}|}{n}} = \overline{\lim_{n \to \infty}} \sqrt[n]{|a_{n-1}|} = \overline{\lim_{n \to \infty}} \left(\sqrt[n-1]{|a_{n-1}|}\right)^{\frac{n-1}{n}} =$$
$$= \overline{\lim_{n \to \infty}} \sqrt[n-1]{|a_{n-1}|} = \overline{\lim_{n \to \infty}} \sqrt[n]{|a_n|}.$$

Отже, радіус збіжності той самий, що і радіус збіжності ряду

(2.24). Тут враховано рівність $\lim\limits_{n\to\infty}\sqrt[n]{n}=1$.

Теорему доведено.

***Т е о р е м а  5 .*** *Степеневий ряд* (2.24*) можна почленно диференціювати всередині його інтервалу збіжності і одержаний при цьому ряд має такий же радіус збіжності.*

*Д о в е д е н н я .* Нехай $R>0$ – радіус збіжності ряду (2.24). Знайдемо ряд, одержаний почленним диференціюванням ряду (2.24),

$$\sum_{n=1}^{\infty} n\, a_n\, x^{n-1}. \qquad (2.28)$$

Знайдемо радіус збіжності $R'$ цього ряду,

$$\frac{1}{R'}=\overline{\lim_{n\to\infty}}\sqrt[n]{(n+1)|a_{n+1}|}=\lim_{n\to\infty}\sqrt[n]{(n+1)}\ \overline{\lim_{n\to\infty}}\left(\sqrt[n+1]{|a_{n+1}|}\right)^{\frac{n+1}{n}}=$$
$$=\overline{\lim_{n\to\infty}}\sqrt[n+1]{|a_{n+1}|}=\frac{1}{R}.$$

Отже, радіуси збіжності рядів (2.24) і (2.28) співпадають. Тепер за теоремою 3 ряд (2.28) рівномірно збігається на будь-якому сегменті, що лежить всередині інтервалу його збіжності, а отже, за теоремою 9 (п. 2.1) ряд (2.24) можна почленно диференціювати і сума ряду (2.25) дорівнює похідній від суми ряду (2.24).

***Н а с л і д о к .*** *Степеневий ряд всередині його інтервалу збіжності можна диференціювати скільки завгодно разів і одержаний при цьому ряд має той же радіус збіжності, що й вихідний ряд.*

**2.2.3. Розвинення функцій в степеневі ряди.** Розглядаємо функції дійсної змінної, визначені на деякому інтервалі.

***О з н а ч е н н я  3 .*** *Функція $f(x)$ називається аналітичною в точці $x_0$, якщо існує таке число $r>0$, що на проміжку $|x-x_0|<r$ вона зображується степеневим рядом, тобто існують дійсні числа $a_n$, $n=0,1,...$, такі що*

$$f(x)=\sum_{n=0}^{\infty} a_n(x-x_0)^n. \qquad (2.29)$$

***Т е о р е м а  6 .*** *Якщо функція $f(x)$ аналітична в точці $x_0$, тобто подається в околі цієї точки рядом* (2.29) *з радіусом*

*збіжності $R$, то*

$$a_n = \frac{f^{(n)}(x_0)}{n!}, \; n = 0, 1, ..., \qquad (2.30)$$

*тобто*

$$f(x) = \sum_{n=0}^{\infty} \frac{f^{(n)}(x_0)}{n!}(x-x_0)^n. \qquad (2.31)$$

*Д о в е д е н н я .* Диференціюючи обидві частини рівності (2.29), одержимо

$$f^{(n)}(x) = n(n-1)...2 \cdot 1 \cdot a_n + (n+1)n...2a_{n+1}(x-x_0) +$$
$$+ (n+2)(n+1)...3a_{n+2}(x-x_0)^2 + ... \; .$$

Звідси при $x = x_0$ одержимо формулу (2.30), а також однозначність подання коефіцієнтів і, відповідно, єдність розвинення функції в ряд.

Теорему доведено.

*З а у в а ж е н н я .* Для того, щоби функція $f(x)$ розвивалася у степеневий ряд необхідно, щоби вона була нескінченно диференційованою в точці $x_0$. Однак ця умова не є достатньою.

*П р и к л а д   1 .* Розглянемо нескінченно диференційовану в точці $x_0 = 0$ функцію, що не розвивається в степеневий ряд

$$f(x) = \begin{cases} e^{-1/x^2}, & x \neq 0, \\ 0, & x = 0. \end{cases}$$

Дійсно, якщо $x \neq 0$, то

$$f'(x) = \frac{2}{x^3}e^{-1/x^2}, \; f''(x) = \left(\frac{4}{x^6} - \frac{6}{x^4}\right)e^{-1/x^2}, \; f^{(n)}(x) = P_{3n}\left(\frac{1}{x}\right)e^{-1/x^2},$$

де $P_{3n}\left(\frac{1}{x}\right)$ - многочлен за змінною $\frac{1}{x}$.

Знайдемо границі похідних від даної функції при $x \to 0$

$$\lim_{x \to 0} f^{(n)}(x) = \lim_{x \to 0} P_{3n}\left(\frac{1}{x}\right)e^{-1/x^2} = \lim_{t \to +\infty} \frac{P_{3n}(\sqrt{t})}{e^t} = 0, \; n = 1, 2, ... \; .$$

Отже, функція $f(x)$ і її похідні довільного порядку неперервні в точці $x_0 = 0$ і дорівнюють нулю в цій точці. Тому всі коефіцієнти ряду (2.31) в точці $x_0 = 0$ і його сума дорівнює нулю, яка не співпадає з функцією $f(x)$.

Очевидно також, що розглянута функція не є аналітичною в точці $x_0 = 0$.

**О з н а ч е н н я 4 .** *Якщо функція $f(x)$ визначена в деякому околі точки $x_0$ і має в цій точці похідні довільного порядку, то ряд*

$$\sum_{n=0}^{\infty}\frac{f^{(n)}(x_0)}{n!}(x-x_0)^n$$

*називається рядом Тейлора функції $f(x)$ в точці $x_0$.*

Природно виникає запитання за яких умов сума ряду Тейлора даної функції співпадає з самою функцією, для якої цей ряд складений.

**Т е о р е м а 7** . *Для того щоби нескінченно диференційована в деякому околі $(x_0 - \delta, x_0 + \delta)$ функція $f(x)$ могла бути сумою складеного для неї ряду Тейлора необхідно і достатньо, щоби додатковий член у формулі Тейлора для цієї функції прямував до нуля при $n \to \infty$, тобто*

$$\lim_{n\to\infty} R_n(x) = 0, \ x \in (x_0 - \delta, x_0 + \delta). \qquad (2.32)$$

*Н е о б х і д н і с т ь .* Запишемо формулу Тейлора для функції $f(x)$ в околі точки $x_0$

$$f(x) = S_n(x) + R_n(x). \qquad (2.33)$$

Нехай функція $f(x)$ є сумою ряду Тейлора, тобто $\lim_{n\to\infty} S_n(x) = f(x)$. Тоді з формули (2.33) одержимо $\lim_{n\to\infty} R_n(x) = 0$.

*Д о с т а т н і с т ь .* Нехай справджується умова (2.32). Тоді з формули (2.33) випливає рівність $\lim_{n\to\infty}[f(x) - S_n(x)] = 0$, а отже, $\lim_{n\to\infty} S_n(x) = f(x)$.

**Т е о р е м а 8** . *Якщо функція $f(x)$ визначена в деякому околі $(x_0 - \delta, x_0 + \delta)$ точки $x_0$ і має в цій точці похідні довільного порядку, а також існує число $M > 0$ таке, що виконуються нерівності*

$$\left| f^{(k)}(x) \right| \le M, \ k = 1, 2, \ldots, \qquad (2.34)$$

то $f(x)$ у цьому околі розвивається в ряд Тейлора.

Д о в е д е н н я . Запишемо додатковий член формули Тейлора у формі Лагранжа

$$R_n(x) = \frac{f^{(n+1)}(c)}{(n+1)!}(x-x_0)^{n+1}, \ c = x_0 + \theta(x-x_0), \ 0 < \theta < 1 .$$

Тоді, враховуючи нерівність (2.34), одержимо

$$|R_n(x)| \leq \frac{M}{(n+1)!}|x-x_0|^{n+1}, \ n = 1, 2, \ldots .$$

Нехай $x \in (x_0 - \delta, x_0 + \delta)$ – фіксована точка. Оскільки $\lim\limits_{n\to\infty} \dfrac{a^n}{n!} = 0$, яке б не було скінчене число $a$, то $\lim\limits_{n\to\infty} R_n(x) = 0$.

Отже, умова (2.32) виконана.

Теорему доведено.

**2.2.4. Розвинення елементарних функцій в ряд Тейлора.** Розглянемо розвинення деяких елементарних функцій в степеневі ряди Тейлора.

*П р и к л а д  2* . Ряд Тейлора для функції $e^x$.

Функція $f(x) = e^x$ має похідну будь-якого порядку в кожній точці дійсної осі

$$f^{(k)}(x) = e^x, \ k = 1, 2, \ldots .$$

В точці $x = 0$ похідні приймають значення $f^{(k)}(0) = e^0 = 1, \ k = 1, 2, \ldots$ . Підставивши їх у формулу (2.31), одержимо

$$e^x = \sum_{n=0}^{\infty} \frac{x^n}{n!} . \tag{2.35}$$

За теоремою 2 одержаний ряд збігається в інтервалі $(-\infty, \infty)$, $\dfrac{1}{R} = \lim\limits_{n\to\infty} \dfrac{1}{\sqrt[n]{n!}} = 0$, а за теоремою 8 цей ряд збігається до функції $f(x) = e^x$, оскільки яке б не було $x \in (-\infty, \infty)$ існує число $h > 0$ таке, що $|x| \leq h$ і всі похідні від цієї функції обмежені $\left|f^{(k)}(x)\right| \leq e^h = M$ .

*П р и к л а д  3* . Ряд Тейлора для функцій $\operatorname{ch} x$, $\operatorname{sh} x$.

Замінюючи у формулі (2.35) $x$ на $-x$, одержимо

$$e^{-x} = \sum_{n=0}^{\infty} \frac{(-1)^n x^n}{n!}. \qquad (2.36)$$

Комбінуючи рівності (2.35) і (2.36), одержимо

$$\operatorname{ch} x = \frac{e^x + e^{-x}}{2} = \sum_{n=0}^{\infty} \frac{x^{2n}}{(2n)!}, \ \operatorname{sh} x = \frac{e^x - e^{-x}}{2} = \sum_{n=0}^{\infty} \frac{x^{2n+1}}{(2n+1)!}. \qquad (2.37)$$

В силу єдиності розвинень (2.35), (2.36) і визначення функцій $f(x) = \operatorname{ch} x$, $f(x) = \operatorname{sh} x$ ряди (2.35) збігаються до вказаних функцій для кожного $x \in (-\infty, \infty)$.

*П р и к л а д  4* . Ряд Тейлора для функцій $\cos x$, $\sin x$.

Функція $f(x) = \sin x$ визначена на дійсній осі і має похідні всіх порядків

$$f^{(k)}(x) = \sin\left(x + \frac{k\pi}{2}\right), \ k = 1, 2, \ldots \ .$$

Тоді в точці $x = 0$ знайдемо

$$f^{(k)}(0) = \sin\frac{k\pi}{2} = \begin{cases} 0, & k = 2n, \\ (-1)^{n-1}, & k = 2n-1, \ n = 1, 2, \ldots \end{cases}$$

і за формулою (2.31) маємо ряд Тейлора

$$\sin x = \sum_{k=1}^{\infty} \frac{(-1)^{k-1} x^{2n-1}}{(2n-1)!}, \qquad (2.38)$$

який за теоремою 8, оскільки справедлива оцінка $\left|f^{(k)}(x)\right| = \left|\sin\left(x + \frac{k\pi}{2}\right)\right| \leq 1$, $k = 1, 2, \ldots$, збігається до функції $f(x) = \sin x$ для кожного $x \in (-\infty, \infty)$.

Для функції $f(x) = \cos x$ з урахуванням формул

$$f^{(k)}(x) = \cos\left(x + \frac{k\pi}{2}\right), \ k = 1, 2, \ldots,$$

$$f^{(k)}(0) = \cos\frac{k\pi}{2} = \begin{cases} (-1)^n, & k = 2n, \\ 0, & k = 2n-1, \ n = 1, 2, \ldots \end{cases}$$

одержимо ряд Тейлора

$$\cos x = \sum_{k=0}^{\infty} \frac{(-1)^k x^{2k}}{(2k)!}. \qquad (2.39)$$

Цей ряд за теоремою 8 збігається до функції $f(x) = \cos x$ для кожного $x \in (-\infty, \infty)$, оскільки справедлива оцінка

$$\left| f^{(k)}(x) \right| = \left| \cos\left( x + \frac{k\pi}{2} \right) \right| \le 1, \ k = 1, 2, \ldots \ .$$

*П р и к л а д  5* . Ряд Тейлора для функції $(1+x)^m$.

Розглянемо функцію $f(x) = (1+x)^m$, де $m$ – довільне дійсне число, яка називається біномом. Похідна $k$-го порядку від бінома дорівнює

$$f^{(k)}(x) = m(m-1)(m-2)\ldots(m-k+1)(1+x)^{m-k}, \ k = 1, 2, \ldots \ .$$

Підставивши у цю рівність $x = 0$, дістанемо

$$f^{(k)}(0) = m(m-1)(m-2)\ldots(m-k+1), \ k = 1, 2, \ldots \ .$$

Тоді ряд Тейлора для бінома має вигляд

$$f(x) = 1 + \frac{m}{1!}x + \frac{m(m-1)}{2!}x^2 + \ldots +$$
$$+ \frac{m(m-1)(m-2)\ldots(m-k+1)}{k!}x^k + \ldots \ ,$$

який називається біноміальним рядом. Знайдемо за формулою (2.27) радіус збіжності цього ряду

$$\lim_{n \to \infty} \left| \frac{a_{n+1}}{a_n} \right| = \lim_{n \to \infty} \left| \frac{m(m-1)(m-2)\ldots(m-n)n!}{(n+1)!\, m(m-1)(m-2)\ldots(m-n+1)} \right| =$$
$$= \lim_{n \to \infty} \left| \frac{m-n}{n+1} \right| = \lim_{n \to \infty} \left| \frac{m}{n} - 1 \right| \bigg/ \left| 1 + \frac{1}{n} \right| = 1.$$

Отже, радіус збіжності $R = 1$ та інтервал збіжності – $(-1, 1)$.

Можна показати, використавши вираз додаткового члена формули Маклорена у формі Коші, що залишок ряду прямує до нуля, коли $n \to \infty$,

$$\lim_{n \to \infty} r_n(x) = 0, \ x \in (-1, 1).$$

Таким чином, для всіх $x \in (-1, 1)$ справедливе розвинення

$$(1+x)^m = 1 + \frac{m}{1!}x + \frac{m(m-1)}{2!}x^2 + \ldots +$$
$$+ \frac{m(m-1)(m-2)\ldots(m-k+1)}{k!}x^k + \ldots . \qquad (2.40)$$

На кінцях інтервалу збіжність ряду залежить від конкретних значень величини $m$.

*П р и к л а д  6*. Знайти розвинення функції $\dfrac{1}{\sqrt{1-x^2}}$ за степенями $x$.

За формулою (2.40) при $m = -\dfrac{1}{2}$ знайдемо

$$\frac{1}{\sqrt{1+t}} = 1 + \sum_{n=1}^{\infty} \frac{(-1)^n \cdot 1 \cdot 3 \cdot 5 \cdot \ldots \cdot (2n-1)\, t^n}{2^n n!}.$$

Підставивши сюди $t = -x^2$, одержимо потрібний ряд

$$\frac{1}{\sqrt{1-x^2}} = \sum_{n=0}^{\infty} \frac{C_{2n}^n\, x^{2n}}{2^{2n}}, \qquad (2.41)$$

де $C_n^k$ – біноміальні коефіцієнти.

Цей ряд є рядом Тейлора і збігається в інтервалі $(-1, 1)$, оскільки збігається в цій області ряд (2.40).

*П р и к л а д  7*. Записати ряд Тейлора для функції $\ln(1+x)$.

Ряд Тейлора цієї функції одержимо, користуючись геометричною пргресією

$$\frac{1}{1+x} = \sum_{k=0}^{\infty} (-1)^k x^k, \ \ x \in (-1, 1). \qquad (2.42)$$

Оскільки ряд (2.42) рівномірно збігається на будь-якому сегменті $[0, x]$, якщо $x < 1$, його можна почленно інтегрувати

$$\ln(1+x) = \int_0^x \frac{dt}{1+t} = \sum_{k=0}^{\infty} \frac{(-1)^k x^{k+1}}{k+1}. \qquad (2.43)$$

Отже, за теоремою 4 ряд Тейлора функції $f(x) = \ln(1+x)$ збігається до цієї функції в області $(-1, 1)$.

*П р и к л а д  8*. Записати ряди Тейлора для функцій $\arcsin x$ і $\operatorname{arctg} x$.

Ряд Тейлора для першої функції $f(x) = \arcsin x$ одержимо, проінтегрувавши ліву і праву частини ряду Тейлора (2.41), який рівномірно збігається на сегменті $[0, x]$, $|x| < 1$,

$$\arcsin x = \int_0^x \frac{dt}{\sqrt{1-t^2}} = \sum_{n=0}^{\infty} \frac{C_{2n}^n \, x^{2n+1}}{2^{2n}(2n+1)}. \qquad (2.44)$$

Радіус збіжності цього ряду $R = 1$ і його інтервал збіжності – $(-1, 1)$.

Ряд Тейлора функції $f(x) = \operatorname{arctg} x$ одержимо, замінивши $x$ на $x^2$ і інтегруючи рівномірно збіжний на сегменті $[0, x]$, $|x| < 1$, ряд (2.42),

$$\operatorname{arctg} x = \int_0^x \frac{dt}{1+t^2} = \sum_{k=0}^{\infty} \frac{(-1)^k x^{2k+1}}{(2k+1)}, \qquad (2.45)$$

який також має радіус збіжності $R = 1$ і інтервал збіжності $(-1, 1)$.

**2.2.5. Підсумовування рядів. Наближене обчислення значень функцій.** Диференціюванням або інтегруванням відомих розвинень в ряд Тейлора можна одержати розвинення нових функцій у степеневі ряди.

*П р и к л а д  9*. Знайдемо суму степеневого ряду

$$S(x) = x^m \sum_{n=1}^{\infty} n \, x^{n-1}.$$

Радіус збіжності цього ряду дорівнює одиниці, $\frac{1}{R} = \lim_{n \to \infty} \sqrt[n]{n} = 1$, і він абсолютно збігається в області $|x| < 1$. Запишемо його у вигляді

$$\frac{S(x)}{x^m} = \sum_{n=1}^{\infty} n \, x^{n-1}.$$

і знайдемо інтеграл на сегменті $[0, x]$, $|x| < 1$,

$$\int_0^x \frac{S(t)}{t^m} dt = \sum_{n=1}^{\infty} x^n = \frac{x}{1-x}.$$

Диференціюючи цю тотожність, отримаємо

$$\frac{S(x)}{x^m} = \frac{d}{dx}\frac{x}{1-x} = \frac{1}{(1-x)^2}.$$

Звідси знайдемо

$$S(x) = \frac{x^m}{(1-x)^2}, \; |x| < 1.$$

*П р и к л а д  10*. Знайдемо суму ряду

$$S(x) = \sum_{n=1}^{\infty} \frac{x^n}{n^2}.$$

Радіус збіжності цього ряду дорівнює одиниці, $\frac{1}{R} = \lim_{n \to \infty} \sqrt[n]{n^2} = 1$, і він збігається в області $|x| < 1$. Продиференціюємо цей ряд і використаємо розвинення логарифма (2.43), одержимо

$$\frac{dS(x)}{dx} = \sum_{n=1}^{\infty} \frac{x^{n-1}}{n}, \; x\frac{dS(x)}{dx} = \sum_{n=1}^{\infty} \frac{x^n}{n} = -\ln(1-x), \; \frac{dS(x)}{dx} = -\frac{\ln(1-x)}{x}.$$

Звідси з урахуванням рівності $S(0) = 0$ знайдемо

$$S(x) = \int_0^x \frac{\ln(1-t)}{t}\, dt, \; |x| < 1.$$

**Розвинення функцій в степеневі ряди ефективно використовується для наближеного обчислення значень функцій**. Оцінка похибки наближеного обчислення функцій може бути зроблена з використанням як виразу залишку ряду, так і формули для додаткового члена формули Тейлора. В цьому випадку абсолютна похибка обчислення дорівнює модулю залишка ряду або додаткового члена формули Тейлора.

*П р и к л а д  11*. Знайдемо наближені формули для обчислення значень тригонометричних функцій $\sin x$ і $\cos x$. Взявши перші $n$ членів рядів (2.38), (2.39), дістанемо наближені формули

$$\sin x \approx x - \frac{x^3}{3!} + \frac{x^5}{5!} - \ldots + \frac{(-1)^{n-1} x^{2n-1}}{(2n-1)!},$$

$$\cos x \approx 1 - \frac{x^2}{2!} + \frac{x^4}{4!} - \ldots + \frac{(-1)^{n-1} x^{2n-2}}{(2n-2)!}. \qquad (2.46)$$

Оскільки степеневі ряди для цих функцій знакозмінні і збігаються для будь-яких $x \in (-\infty, \infty)$, то похибка не перевищує модуля першого члена $n$-го залишку для цих рядів,

$$|r_n(x)| \leq \frac{|x|^{2n+1}}{(2n+1)!} \text{ для } \sin x;$$

$$|r_n(x)| \leq \frac{|x|^{2n}}{(2n)!} \text{ для } \cos x.$$

Користуючись цими нерівностями, можна підібрати найменше число $n$ таке, щоби з використанням наближених формул (2.46) можна обчислити значення функцій $\sin x$ і $\cos x$ з наперед заданою точністю.

*П р и к л а д 12.* Обчислення логарифмів натуральних чисел. З використанням формули (2.43) одержимо такий ряд

$$\ln\left(\frac{1+x}{1-x}\right) = 2\left(x + \frac{x^3}{3} + \ldots + \frac{x^{2n-1}}{2n-1} + \ldots\right).$$

Підставивши сюди $x = \dfrac{1}{2m+1}$, де $m$ – натуральне число, знайдемо такий числовий ряд:

$$\ln(m+1) = \ln m + 2\left(\frac{1}{2m+1} + \frac{1}{3(2m+1)^3} + \frac{1}{5(2m+1)^5} + \ldots + \right.$$

$$\left. + \frac{1}{(2n-1)(2m+1)^{2n-1}} + \ldots\right).$$

Оцінимо $n$-ий залишок цього ряду,

$$r_n(x) = 2\left(\frac{1}{(2n+1)(2m+1)^{2n+1}} + \frac{1}{(2n+3)(2m+1)^{2n+3}} + \ldots\right) <$$

$$< \frac{2}{(2n+1)(2m+1)^{2n+1}}\left(1 + \frac{1}{(2m+1)^2} + \ldots\right) =$$

$$= \frac{2}{(2n+1)(2m+1)^{2n+1}} \frac{1}{1 - \frac{1}{(2m+1)^2}} =$$

$$= \frac{2}{(2n+1)(2m+1)^{2n-1} 2m(2m+2)} < \frac{1}{m(2n+1)(2m+1)^{2n}}.$$

Отже, маємо наближену формулу для обчислення логарифмів натуральних чисел

$$\ln(m+1) \approx \ln m + 2\left(\frac{1}{2m+1} + \frac{1}{3(2m+1)^3} + \frac{1}{5(2m+1)^5} + \ldots + \right.$$
$$\left. + \frac{1}{(2n-1)(2m+1)^{2n-1}}\right).$$

При цьому похибка буде меншою, ніж

$$r_n < \frac{1}{m(2n+1)(2m+1)^{2n}}.$$

*П р и к л а д  13*. Розглянемо приклад оцінки похибки наближення функції $f(x)$ частинною сумою $S(x)$ ряду Тейлора з використанням додаткового члена формули Тейлора. Абсолютна похибка наближення визначається за формулою

$$|f(x) - S(x)| = |R_n(x)| = \left|\frac{f^{(n+1)}(c)}{(n+1)!}(x - x_0)^{n+1}\right|,$$

де точка $c$ лежить між точками $x_0$ і $x$.

Знайдемо з точністю до 0,001 число $e$. Підставляючи $x = 1$ у формулі (2.35), знайдемо

$$e = 1 + 1 + \frac{1}{2!} + \ldots + \frac{1}{n!} + \ldots.$$

Оцінимо похибку наближення цього числа сумою

$$e \approx 1 + 1 + \frac{1}{2!} + \ldots + \frac{1}{n!}$$

з використанням додаткового члена формули Маклорена. Оскільки $f^{(n+1)}(x) = e^x$, знайдемо $R_n(x) = \frac{e^c}{(n+1)!} x^{n+1}$, де точка $c$ лежить між точками $0$ і $x$. Для $x = 1$ маємо $R_n(1) = \dfrac{e^c}{(n+1)!} < \dfrac{3}{(n+1)!}$, оскільки $0 < c < 1$ і, відповідно, $e^c < 3$. Якщо $n = 6$, то $R_6(1) < \dfrac{3}{7!} < \dfrac{1}{1680} < 0{,}001$. Тоді число $e$ з точністю $0{,}001$ обчислюємо за формулою

$$e \approx 1 + 1 + \frac{1}{2!} + \frac{1}{3!} + \frac{1}{4!} + \frac{1}{5!} + \frac{1}{6!} = 2{,}718.$$

Отже, $e = 2{,}718$ з точністю $0{,}001$.

### 2.3. Тригонометричні ряди. Ряди Фур'є

**2.3.1. Основна тригонометрична система функцій.** Розвинення функції в ряд Тейлора в околі точки вимагає, щоб функція мала похідні будь-якого порядку в цьому околі. Це звужує область застосування степеневих рядів.

Розглянемо дещо інший підхід до розвинення функцій в ряди, що дозволяє розглядати неперервні і навіть кусково-неперервні функції [11, 19, 22].

Нехай на відрізку $[a, b]$ задано дві функції $f(x)$ і $g(x)$.

***О з н а ч е н н я 1 .*** *Функції $f(x)$ і $g(x)$ називаються ортогональними на відрізку $[a, b]$, якщо*

$$\int\limits_a^b f(x)g(x)dx = 0.$$

***О з н а ч е н н я 2 .*** *Скінченна або нескінченна система функцій $\{f_n(x)\}$ називається ортогональною на відрізку $[a, b]$, якщо справджуються співвідношення*

$$\int\limits_a^b f_n(x) f_m(x) dx = \begin{cases} 0, & n \neq m, \\ \alpha_m \neq 0, & n = m. \end{cases} \qquad (2.47)$$

Ортогональною на проміжку $[-\pi, \pi]$ є тригонометрична система функцій

$$1, \cos x, \sin x, \cos 2x, \sin 2x, \ldots, \cos nx, \sin nx, \ldots, \qquad (2.48)$$

яку називають *основною тригонометричною системою функцій.*

Співвідношення ортогональності наступні:

$$\int_{-\pi}^{\pi} \cos nx \cos mx \, dx = \begin{cases} 0, & n \neq m, \\ \pi, & n = m, \end{cases}$$

$$\int_{-\pi}^{\pi} \sin nx \sin mx \, dx = \begin{cases} 0, & n \neq m, \\ \pi, & n = m, \end{cases} \quad (2.49)$$

$$\int_{-\pi}^{\pi} \cos nx \sin mx \, dx = 0,$$

які б не були цілі числа $n$ і $m$.

Для першої функції $f_0(x) = 1$, $n = 0$, маємо такі умови ортогональності

$$\int_{-\pi}^{\pi} \cos mx \, dx = 0, \quad \int_{-\pi}^{\pi} \sin mx \, dx = 0, \quad \int_{-\pi}^{\pi} dx = 2\pi. \quad (2.50)$$

Перевіримо ці рівності. Співвідношення (2.50) є результатом безпосереднього інтегрування.

Використовуючи формули

$$\cos nx \cos mx = \frac{\cos(n+m)x + \cos(n-m)x}{2},$$

$$\cos^2 mx = \frac{1 + \cos 2mx}{2},$$

$$\sin nx \sin mx = \frac{\cos(n-m)x - \cos(n+m)x}{2},$$

$$\sin^2 mx = \frac{1 - \cos 2mx}{2},$$

знайдемо перші два співвідношення (2.49).

Третє співвідношення (2.49) одержимо з використанням формули

$$\cos nx \sin mx = \frac{\sin(n+m)x + \sin(n-m)x}{2}.$$

*О з н а ч е н н я  3 .* *Функціональний ряд*

$$\frac{a_0}{2} + \sum_{k=1}^{\infty}(a_k \cos kx + b_k \sin kx) \qquad (2.51)$$

*називається тригонометричним рядом, а дійсні числа $a_0, a_k, b_k$ ( $k = 1, 2, ...$ ) – коефіцієнтами тригонометричного ряду.*

Нехай тригонометричний ряд (2.51) рівномірно збігається на відрізку $[-\pi, \pi]$. Тоді його сума, яку позначимо через $f(x)$, є неперервна функція і справедлива рівність

$$f(x) = \frac{a_0}{2} + \sum_{k=1}^{\infty}(a_k \cos kx + b_k \sin kx). \qquad (2.52)$$

За теоремою 8 (п. 2.1) його можна почленно інтегрувати на відрізку $[-\pi, \pi]$

$$\int_{-\pi}^{\pi} f(x)dx = \frac{a_0}{2}\int_{-\pi}^{\pi}dx + \sum_{k=1}^{\infty}\left(a_k \int_{-\pi}^{\pi}\cos kx\,dx + b_k \int_{-\pi}^{\pi}\sin kx\,dx\right).$$

Помножимо ліву і праву частини рівності (2.52) на $\cos nx$ і $\sin nx$. Можна показати, що одержані ряди також рівномірно збігаються і їх можна почленно інтегрувати на відрізку $[-\pi, \pi]$. Маємо

$$\int_{-\pi}^{\pi} f(x)\cos nx\,dx = \frac{a_0}{2}\int_{-\pi}^{\pi}\cos nx\,dx +$$
$$+ \sum_{k=1}^{\infty}\left(a_k \int_{-\pi}^{\pi}\cos kx \cos nx\,dx + b_k \int_{-\pi}^{\pi}\sin kx \cos nx\,dx\right),$$
$$\int_{-\pi}^{\pi} f(x)\sin nx\,dx = \frac{a_0}{2}\int_{-\pi}^{\pi}\sin nx\,dx +$$
$$+ \sum_{k=1}^{\infty}\left(a_k \int_{-\pi}^{\pi}\cos kx \sin nx\,dx + b_k \int_{-\pi}^{\pi}\sin kx \sin nx\,dx\right).$$

У правій частині кожної з цих рівностей, з урахуванням співвідношень (2.49), залишиться тільки по одному доданку

$$\int_{-\pi}^{\pi} f(x)dx = \pi a_0, \quad \int_{-\pi}^{\pi} f(x)\cos nx\,dx = \pi a_n, \quad \int_{-\pi}^{\pi} f(x)\sin nx\,dx = \pi b_n.$$

Звідси одержимо формули

$$a_n = \frac{1}{\pi} \int\limits_{-\pi}^{\pi} f(x)\cos nx\, dx, \ n = 0, 1, \ldots,$$

$$b_n = \frac{1}{\pi} \int\limits_{-\pi}^{\pi} f(x)\sin nx\, dx, \ n = 1, 2, \ldots. \qquad (2.53)$$

Нехай тепер задана довільна інтегровна на відрізку $[-\pi, \pi]$ функція $f(x)$. Знайдемо для цієї функції коефіцієнти за формулами (2.53) і складемо тригонометричний ряд (2.51), який запишемо

$$f(x) \sim \frac{a_0}{2} + \sum_{k=1}^{\infty}(a_k \cos kx + b_k \sin kx). \qquad (2.54)$$

*О з н а ч е н н я   4 .* *Коефіцієнти $a_n$ і $b_n$, знайдені за формулами (2.53), називається коефіцієнтами Фур'є для інтегровної функції $f(x)$, а тригонометричний ряд (2.54) називається рядом Фур'є для цієї функції.*

З попереднього викладу випливає наступне твердження.

*Т е о р е м а   1 .* *Якщо функція $f(x)$ розвивається на проміжку $[-\pi, \pi]$ у рівномірно збіжний тригонометричний ряд, то він є рядом Фур'є для функції $f(x)$.*

*Д о в е д е н н я .* За теоремами 7 і 8 (п. 2.1), якщо справедлива рівність (2.52), то функція $f(x)$ неперервна і ряд в цій рівності можна почленно інтегрувати. Звідси отримаємо рівність

$$\int\limits_{-\pi}^{\pi} f(x)\, dx = \pi a_0.$$

Розглянемо рівності

$$f(x)\cos nx = \frac{a_0}{2}\cos nx + \sum_{k=1}^{\infty}(a_k \cos kx \cos nx + b_k \sin kx \cos nx),$$

$$f(x)\sin nx = \frac{a_0}{2}\sin nx + \sum_{k=1}^{\infty}(a_k \cos kx \sin nx + b_k \sin kx \sin nx) \qquad (2.55)$$

і покажемо, що ряди справа збігаються рівномірно. Для цього розглянемо частинну суму

$$S_m(x) = \frac{a_0}{2} + \sum_{k=1}^{m}\left(a_k \cos kx + b_k \sin kx\right)$$

і зафіксуємо довільне число $\varepsilon > 0$. Якщо ряди (2.52) збігається рівномірно, то існує число $N$ таке, що для всіх $m > N$

$$|f(x) - S_m(x)| < \varepsilon.$$

Добутки $S_m(x)\cos nx$ і $S_m(x)\sin nx$, очевидно, є частинними сумами рядів (2.55). Тому зі співвідношень

$$|f(x)\cos nx - S_m(x)\cos nx| = |f(x) - S_m(x)||\cos nx| < \varepsilon,$$
$$|f(x)\sin nx - S_m(x)\sin nx| = |f(x) - S_m(x)||\sin nx| < \varepsilon,$$

справедливих для всіх $m > N$, випливає рівномірна збіжність рядів (2.55). За теоремою 8 ці ряди можна інтегрувати, звідки і випливають формули (2.53).

Отже, ряд (2.52) є рядом Фур'є для функції $f(x)$.

Теорему доведено.

*З а у в а ж е н н я*. Оскільки члени ряду (2.52) періодичні функції з періодом $2\pi$, то сума цього ряду також періодична функція, тобто $S(x + 2\pi) = S(x)$ для всіх $x \in (-\infty, \infty)$.

Якщо функція $f(x)$ неперервна на відрізку $[-\pi, \pi]$ і не є періодичною, однак $f(-\pi) = f(\pi)$, то $S(x)$ – періодична функція, яка співпадає з функцією $f(x)$ на відрізку $[-\pi, \pi]$. Якщо ж $f(-\pi) \neq f(\pi)$, то функції $f(x)$ і $S(x)$ співпадають тільки на проміжку $(-\pi, \pi)$, а точки $x = \pm(2k-1)\pi$, $k = 0, 1, \ldots$, є точками розриву функції $S(x)$.

Отже, сума тригонометричного ряду для неперервної на відрізку $[-\pi, \pi]$ функції $f(x)$ не обов'язково неперервна функція.

***О з н а ч е н н я  5****. Нехай на відрізку $[-\pi, \pi]$ задана кусково-неперервна функція $f(x)$. Введемо $2\pi$-періодичну функцію $f^*(x)$, значення якої співпадають зі значеннями функції $f(x)$ у точках інтервалів неперервності, у точках розриву $x_i$, $i = 1, \ldots, n$, вона приймає значення $f^*(x_i) = [f(x_i + 0) + f(x_i - 0)]/2$ і на кінцях відрізка $[-\pi, \pi]$ приймає такі значення: $f^*(\pi) = f^*(-\pi) = [f(\pi) + f(-\pi)]/2$.*

*Побудова такої функції $f^*(x)$ називається періодичним продовженням функції $f(x)$.*

З а у в а ж е н н я . Коефіцієнти, що визначаються за формулами (2.53), задають у вигляді ряду Фур'є періодичну функцію $f^*(x)$ з періодом $2\pi$. Тому проміжок $[-\pi, \pi]$ можна замінити будь-яким іншим проміжком $[a, a+2\pi]$ довжини $2\pi$, тобто

$$a_n = \frac{1}{\pi}\int\limits_a^{a+2\pi} f^*(x)\cos nx\, dx,\ n = 0, 1, \ldots,$$

$$b_n = \frac{1}{\pi}\int\limits_a^{a+2\pi} f^*(x)\sin nx\, dx,\ n = 1, 2, \ldots.$$

П р и к л а д  1 . Записати ряд Фур'є для функції $f(x)=|x|$ на відрізку $[-\pi, \pi]$. Знайдемо коефіцієнти Фур'є для цієї функції

$$a_0 = \frac{1}{\pi}\int\limits_{-\pi}^{\pi}|x|\,dx = \frac{2}{\pi}\int\limits_0^{\pi} x\,dx = \pi,$$

$$a_n = \frac{1}{\pi}\int\limits_{-\pi}^{\pi}|x|\cos nx\, dx = \frac{2}{\pi}\int\limits_0^{\pi} x\cos nx\, dx =$$

$$= \frac{2}{\pi n}\left[(x\sin nx)\Big|_0^{\pi} - \int\limits_0^{\pi}\sin nx\, dx\right] =$$

$$= \frac{2}{\pi n^2}\cos nx\Big|_0^{\pi} = \frac{2}{\pi n^2}\Big[(-1)^n - 1\Big],\ n = 1, 2, \ldots,$$

$$b_n = \frac{1}{\pi}\int\limits_{-\pi}^{\pi}|x|\sin nx\, dx = 0,\ n = 1, 2, \ldots.$$

Підставивши значення коефіцієнтів у формулу (2.54), одержимо ряд Фур'є

$$|x| \sim \frac{\pi}{2} + \sum_{n=1}^{\infty}\frac{2}{\pi n^2}\Big[(-1)^n - 1\Big]\cos nx.$$

Оскільки коефіцієнти цього ряду по модулю менші, ніж коефіцієнти збіжного числового ряду $\dfrac{2}{\pi}\sum\limits_{n=1}^{\infty}\dfrac{1}{n^2}<\infty$, за теоремою Вейєрштрасса він збігається рівномірно на відрізку $[-\pi,\pi]$. Тому можна записати, перетворивши вираз у квадратних дужках,

$$|x|=\dfrac{\pi}{2}-\dfrac{4}{\pi}\sum\limits_{n=1,3,\ldots}^{\infty}\dfrac{\cos nx}{n^2},\ x\in[-\pi,\pi].$$

Періодичне продовження

$$f^{*}(x)=\dfrac{\pi}{2}-\dfrac{4}{\pi}\sum\limits_{n=1,3,\ldots}^{\infty}\dfrac{\cos nx}{n^2},\ x\in(-\infty,\infty),$$

співпадає з функцією $f(x)=|x|$ на відрізку $[-\pi,\pi]$. На рис. 2.1 показано графік функції $f^{*}(x)$.

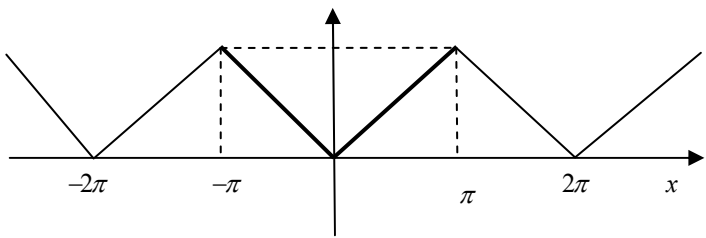

Рис. 2.1.

*П р и к л а д  2*. Розвинути в ряд Фур'є функцію $f(x)=\pi+x$ на відрізку $[-\pi,\pi]$.

Знайдемо коефіцієнти ряду за формулами (2.53)

$$a_0=\dfrac{1}{\pi}\int\limits_{-\pi}^{\pi}(\pi+x)dx=2\pi,\ a_n=\dfrac{1}{\pi}\int\limits_{-\pi}^{\pi}(\pi+x)\cos nx\,dx=0,\ n=1,2,\ldots,$$

$$b_n=\dfrac{1}{\pi}\int\limits_{-\pi}^{\pi}(\pi+x)\sin nx\,dx=\dfrac{2}{\pi}\int\limits_{0}^{\pi}x\sin nx\,dx=$$

$$=\dfrac{2}{n\pi}\left[-(x\cos nx)\Big|_{0}^{\pi}+\int\limits_{0}^{\pi}\cos nx\,dx\right]=$$

$$= \frac{2}{\pi}\left[-\frac{(-1)^n \pi}{n} + \frac{1}{n^2}\sin nx \Big|_0^\pi \right] = \frac{2(-1)^{n+1}}{n}, \ n = 1, 2, \ldots .$$

Підставивши їх у формулу (2.54), одержимо ряд Фур'є функції $f(x) = \pi + x$ на відрізку $[-\pi, \pi]$

$$\pi + x \sim \pi + 2\sum_{n=1}^{\infty} \frac{(-1)^{n+1}}{n}\sin nx.$$

Цей ряд рівномірно збігається на відрізку $[-\pi + r, \pi - r]$, $0 < r < \frac{\pi}{2}$, (прикл. 3, п. 2.1), а отже,

$$\pi + x = \pi + 2\sum_{n=1}^{\infty} \frac{(-1)^{n+1}}{n}\sin nx, \ \ x \in (-\pi, \pi).$$

Графік періодичного продовження функції $f(x) = \pi + x$,

$$f^*(x) = \pi + 2\sum_{n=1}^{\infty} \frac{(-1)^{n+1}}{n}\sin nx, \ \ x \in (-\infty, \infty),$$

показаний на рис. 2.2. У точках $x_n^\pm = \pm(2n-1)\pi$ функція має розриви першого роду.

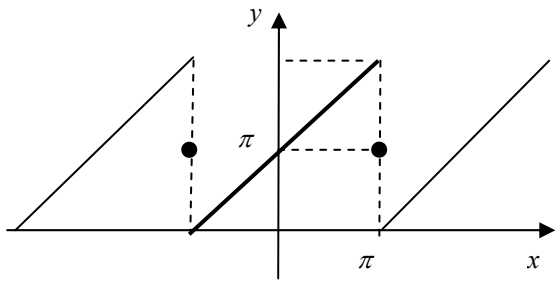

Рис. 2.2.

Поклавши у вираз одержаного ряду $x = \frac{\pi}{2}$, одержимо такий числовий ряд для обчислення числа $\pi$:

$$\frac{\pi}{4} = \sum_{n=1}^{\infty} \frac{(-1)^{n+1}}{n} \sin \frac{n\pi}{2} = 1 - \frac{1}{3} + \frac{1}{5} - \ldots + (-1)^{2n-1} \frac{1}{2n-1} + \ldots \ .$$

**2.3.2. Достатні умови збіжності тригонометричних рядів.** Вивід формул (2.53) ґрунтується на рівномірній збіжності відповідного тригонометричного ряду. Якщо ж припустити тільки існування інтегралів у правих частинах формул (2.53), то одержаний тригонометричний ряд буде рядом Фур'є для заданої функції, однак висновки стосовно його збіжності можна зробити тільки за додаткових умов.

Спочатку розглянемо допоміжні результати.

*Л е м а 1 (Рімана). Нехай* $f(x)$ *– абсолютно інтегровна на відрізку* $[a, b]$ *функція. Тоді справедливі рівності*

$$\lim_{\lambda \to \infty} \int_a^b f(t) \cos \lambda t \, dt = 0, \ \lim_{\lambda \to \infty} \int_a^b f(t) \sin \lambda t \, dt = 0. \qquad (2.56)$$

Доведемо цю лему для дещо жорсткіших умов стосовно функції $f(t)$ (загальний випадок див. [11, 22]). Вважаємо, що $f(t)$ – кусково-гладка функція на відрізку $[a, b]$.

*О з н а ч е н н я 6 . Функцію* $f(x)$ *називають гладкою на відрізку* $[a, b]$, *якщо вона на цьому відрізку має неперервну похідну. При цьому існують праві границі функції і її похідної в точці* $a$ *і ліві границі функції і її похідної в точці* $b$.

Функцію $f(x)$ називають *кусково-гладкою на відрізку* $[a, b]$, якщо вона і її похідна або неперервні на цьому відрізку, або мають на ньому лише скінчену кількість розривів першого роду. При цьому існують ліві і праві границі функції і її похідної в точках розриву, а також праві границі в точці $a$ і ліві границі в точці $b$ функції і її похідної.

Неперервна або розривна функція $f(x)$ називається *кусково-гладкою на всій дійсній осі*, якщо вона кусково-гладка на будь-якому скінченому відрізку.

*Будь-яка кусково-гладка функція на відрізку* $[a, b]$ *має обмежену похідну.* Неперервні функції з обмеженими на відрізку $[a, b]$ похідними належать класу *Ліпшіца*.

Покажемо, що неперервна і кусково-гладка на відрізку $[a, b]$

функція $f(x)$ належить класу Ліпшіца. Для будь-яких точок $x_1, x_2 \in [a, b]$ знайдеться за теоремою Лагранжа точка $\xi$, що лежить між точками $x_1$ і $x_2$, така що $f(x_1) - f(x_2) = f'(\xi)(x_1 - x_2)$. Оскільки $f'(\xi)$ обмежена на цьому відрізку, $|f'(\xi)| \le M$, то звідси випливає нерівність
$$|f(x_1) - f(x_2)| \le M |x_1 - x_2|.$$

Кусково-гладка функція на відрізку $[a, b]$ абсолютно інтегровна на цьому відрізку.

***Л е м а  1'.*** *Нехай $f(x)$ – кусково-гладка на відрізку $[a, b]$ функція.*

*Тоді справедливі рівності* (2.56).

*Д о в е д е н н я*. Відрізок $[a, b]$ розіб'ємо на скінченне число відрізків $[x_{i-1}, x_i]$, $i = 1, 2, ..., n$; $x_0 = a, x_n = b$, на кожному з яких функція неперервна і має неперервні похідні. Тому
$$\int_a^b f(t)\cos\lambda t\, dt = \sum_{i=1}^n \int_{x_{i-1}}^{x_i} f(t)\cos\lambda t\, dt = \frac{1}{\lambda}\sum_{i=1}^n [f(t)\sin\lambda t]\Big|_{x_{i-1}}^{x_i} -$$
$$- \frac{1}{\lambda}\sum_{i=1}^n \int_{x_{i-1}}^{x_i} f'(t)\sin\lambda t\, dt.$$

Останній вираз прямує до нуля, коли $\lambda \to 0$, оскільки множники біля $\frac{1}{\lambda}$ є скінченими величинами.

Аналогічний результат одержимо для інтеграла від функції з множником $\sin\lambda t$.

Лему доведено.

***Н а с л і д о к.*** *Коефіцієнти Фур'є абсолютно інтегрованої на відрізку $[-\pi, \pi]$ функції $f(x)$ прямують до нуля, коли $n \to \infty$,*
$$\lim_{n\to\infty} a_n = 0, \ \lim_{n\to\infty} b_n = 0.$$

*Д о в е д е н н я* випливає безпосередньо з леми 1, якщо прийняти $\lambda = n$, $a = -\pi, b = \pi$.

***О з н а ч е н н я  7.*** *Функція*

$$D_n(t) = \frac{1}{\pi}\left(\frac{1}{2} + \sum_{k=1}^{n} \cos kt\right) = \frac{\sin\left(n + \frac{1}{2}\right)t}{2\pi \sin \frac{t}{2}} \qquad (2.57)$$

*називається ядром Діріхле.*

Ядро Діріхле має такі властивості:
1) $D_n(x)$ – $2\pi$-періодична функція; 2) $D_n(x)$ – парна функція;

3) $\int\limits_{-\pi}^{\pi} D_n(t)\,dt = 1$.

Перші дві властивості випливають безпосередньо з подання ядра Діріхле. Третю властивість ядра одержимо після інтегрування виразу (2.57) з урахуванням зображення у вигляді суми скінченого числа тригонометричних функцій.

***Л е м а  2 .*** *Нехай* $f(x)$ *– абсолютно інтегровна на відрізку* $[\pi, \pi]$ *функція і* $a_n, b_n$ *– коефіцієнти Фур'є цієї функції.*

*Тоді для частинної суми ряду Фур'є цієї функції справедлива формула*

$$S_n(x) = \int\limits_{-\pi}^{\pi} f^*(x-t) D_n(t)\,dt = \int\limits_{0}^{\pi} \left[f^*(x+t) + f^*(x-t)\right] D_n(t)\,dt, \quad (2.58)$$

*де* $D_n(t)$ *– ядро Діріхле;* $f^*(x)$ *–* $2\pi$*-періодичне продовження функції* $f(x)$.

Д о в е д е н н я . Перетворимо вираз частинної суми ряду з урахуванням формул (2.53)

$$S_n(x) = \frac{a_0}{2} + \sum_{k=1}^{n}(a_k \cos kx + b_k \sin kx) =$$

$$= \frac{1}{\pi}\int\limits_{-\pi}^{\pi} f(t)\left[\frac{1}{2} + \sum_{k=1}^{n}(\cos kx \cos kt + \sin kx \sin kt)\right]dt =$$

$$= \frac{1}{\pi}\int\limits_{-\pi}^{\pi} f(t)\left[\frac{1}{2} + \sum_{k=1}^{n}\cos k(t-x)\right]dt.$$

У прикладі 3 (п. 2.1.2) показано, що

$$\frac{1}{2} + \sum_{k=1}^{n} \cos kt = \frac{\sin\left(n+\frac{1}{2}\right)t}{2\sin\frac{t}{2}}.$$

З урахуванням цієї формули одержимо такий вираз для часткової суми ряду

$$S_n(x) = \int_{-\pi}^{\pi} f(t) D_n(x-t) dt = \int_{-\pi}^{\pi} f^*(t) D_n(x-t) dt =$$

$$= \int_{x-\pi}^{x+\pi} f^*(x-u) D_n(u) du = \int_{-\pi}^{\pi} f^*(x-u) D_n(u) du.$$

Тут враховано періодичність функцій $f^*(x)$ і $D_n(x)$ і, відповідно, формули (2.55).

Вираз частинної суми зручно записати з урахуванням парності функції $D_n(x)$ у такому вигляді

$$S_n(x) = \left[\int_0^{\pi} f^*(x-u) D_n(u) du + \int_{-\pi}^{0} f^*(x-u) D_n(u) du\right] =$$

$$= \int_0^{\pi} \left[f^*(x+u) + f^*(x-u)\right] D_n(u) du.$$

Лему доведено.

***Т е о р е м а   2***. *Нехай $f(x)$ – кусково-гладка функція на відрізку $[-\pi, \pi]$.*

*Тоді ряд Фур'є для функції $f(x)$ збігається в усіх точках числової осі. При цьому, в кожній точці неперервності функції $f(x)$ сума ряду дорівнює значенню функції в цій точці*

$$\frac{a_0}{2} + \sum_{k=1}^{\infty} (a_k \cos kx + b_k \sin kx) = f(x),$$

*в кожній точці $x_0$ розриву функції сума ряду дорівнює середньому арифметичному граничних значень функції в цій точці*

$$\frac{a_0}{2} + \sum_{k=1}^{\infty} (a_k \cos kx_0 + b_k \sin kx_0) = \frac{f(x_0 - 0) + f(x_0 + 0)}{2}$$

*і в кінцевих точках $x=-\pi$ і $x=\pi$ справедлива рівність*

$$\frac{a_0}{2}+\sum_{k=1}^{\infty}(a_k\cos kx+b_k\sin kx)=\frac{f(-\pi+0)+f(\pi-0)}{2}.$$

*Д о в е д е н н я .* Запишемо вираз різниці частинної суми ряду і періодичного продовження функції $f(x)$ з урахуванням третьої властивості ядра Діріхле і виразу частинної суми (2.58)

$$S_n(x)-\frac{f^*(x+0)+f^*(x-0)}{2}=\int_0^{\pi}\left[f^*(x+t)+f^*(x+t)\right]D_n(t)dt-$$

$$-\int_0^{\pi}\left[f^*(x+0)+f^*(x-0)\right]D_n(t)dt=$$

$$=\int_0^{\pi}\left[f^*(x+t)-f^*(x+0)\right]D_n(t)dt+$$

$$+\int_0^{\pi}\left[f^*(x-t)-f^*(x-0)\right]D_n(t)dt. \qquad (2.59)$$

Перепишемо перший доданок (2.59) у вигляді

$$\int_0^{\pi}\left[f^*(x+t)-f^*(x+0)\right]D_n(t)dt=I_1^++I_2^+, \qquad (2.60)$$

де $I_1^+=\frac{1}{\pi}\int_0^{\delta}\frac{f^*(x+t)-f^*(x+0)}{2\sin\frac{t}{2}}\sin\left(n+\frac{1}{2}\right)t\,dt$;

$I_2^+=\frac{1}{\pi}\int_{\delta}^{\pi}\frac{f^*(x+t)-f^*(x+0)}{2\sin\frac{t}{2}}\sin\left(n+\frac{1}{2}\right)t\,dt$.

Розглянемо функцію

$$G(t)=\frac{f^*(x+t)-f^*(x+0)}{2\sin\frac{t}{2}}, \quad 0<t\leq\pi,$$

за фіксованого $x$ і покажемо, що вона обмежена на відрізку $[0,\pi]$. На відрізку $[\delta,\pi]$, де $\delta>0$ – досить мале число, функція $G(t)$ обмежена, оскільки обмежений її чисельник і не дорівнює нулю знаменник. Вона має границю при $t \to +0$,

$$\lim_{t \to +0} G(t) = \lim_{t \to +0} \frac{f^*(x+t) - f^*(x+0)}{t} \frac{t}{2\sin\frac{t}{2}} = \frac{df^*(x+0)}{dx},$$

оскільки функція $f(x)$ кусково-гладка. Якщо $G(t)$ довизначити в нулі, $G(0) = \frac{df^*(x+0)}{dx}$, то вона неперервна на відрізку $[0, \delta]$ і, відповідно, обмежена на цьому відрізку.

Отже, існує число $M$ таке, що
$$|G(t)| < M, \quad t \in [0, \pi]$$

Нехай тепер задано довільне число $\varepsilon > 0$. Виберемо $\delta < \frac{\varepsilon \pi}{4M}$ і оцінимо перший доданок $I_1$ у формулі (2.60) з урахуванням оцінки для функції $G(t)$,

$$\left|I_1^+\right| = \frac{1}{\pi}\left|\int_0^\delta \frac{f^*(x+t) - f^*(x+0)}{2\sin\frac{t}{2}} \sin\left(n + \frac{1}{2}\right)t\, dt\right| \leq$$

$$\leq \frac{1}{\pi}\int_0^\delta \left|\frac{f^*(x+t) - f^*(x+0)}{2\sin\frac{t}{2}}\right|\left|\sin\left(n+\frac{1}{2}\right)t\right|dt \leq \frac{M}{\pi}\int_0^\delta dt = \frac{M\delta}{\pi} < \frac{\varepsilon}{4}.$$

Другий доданок $I_2$ у формулі (2.60) при фіксованому $\delta$ за лемою 1 прямує до нуля, коли $n \to \infty$, тобто за вибраного $\varepsilon > 0$ знайдеться номер $N_1 = N_1(\varepsilon)$ такий, що для всіх $n > N_1$ виконується нерівність $|I_2| < \frac{\varepsilon}{4}$. Таким чином, якщо тільки $n > N_1$, для інтеграла в (2.60) одержимо оцінку

$$\left|\int_0^\pi \left[f^*(x+t) - f^*(x+0)\right]D_n(t)dt\right| = \left|I_1^+\right| + \left|I_2^+\right| < \frac{\varepsilon}{2}.$$

Аналогічно оцінимо другий доданок у (2.59), тобто за вибраного $\varepsilon > 0$ знайдеться номер $N_2 = N_2(\varepsilon)$ такий, що для всіх $n > N_2$, виконується нерівність

$$\left| \int_0^\pi \left[ f^*(x-t) - f^*(x-0) \right] D_n(t) dt \right| < \frac{\varepsilon}{2}. \qquad (2.61)$$

Враховуючи ці нерівності при оцінці виразу (2.59), одержимо нерівність

$$\left| S_n(x) - \frac{f^*(x+0) + f^*(x-0)}{2} \right| \leq \left| \int_0^\pi \left[ f^*(x+t) - f^*(x+0) \right] D_n(t) dt \right| +$$

$$+ \left| \int_0^\pi \left[ f^*(x-t) - f^*(x-0) \right] D_n(t) \, dt \right| < \varepsilon,$$

яка виконується, якщо тільки $n > N$, $N = \max\{N_1, N_2\}$.

Остання нерівність підтверджує виконання рівності

$$\frac{a_0}{2} + \sum_{k=1}^\infty (a_k \cos kx + b_k \sin kx) = \frac{f^*(x-0) + f^*(x+0)}{2}.$$

Сформулювавши цю рівність для функції $f(x)$ в точках неперервності, точках розриву і кінцевих точках проміжку, одержимо твердження теореми.

*З а у в а ж е н н я   1*. Вибір чисел $N_1 = N_1(\varepsilon, x)$ і $N_2 = N_2(\varepsilon, x)$ при доведенні теореми 2 залежить не тільки від $\varepsilon$, але і від $x$, оскільки функція $G(t)$ залежить від $x$ як від параметра. Тому оцінка (2.61) справедлива для всіх $n > N(\varepsilon, x)$, $N = \max\{N_1, N_2\}$, і вона не є рівномірною відносно $x$, тобто відповідний ряд не обов'язково рівномірно збіжний. Якщо функція $f^*(x)$ має розриви або $f^*(-\pi) \neq f^*(\pi)$, то збіжність ряду не може бути рівномірною.

*З а у в а ж е н н я   2*. Використана в доведенні теореми 2 умова про кускову-неперервність функції була потрібна лише для того, щоби існував інтеграл від абсолютної величини відношення

$$\frac{f^*(x+t) - f^*(x \pm 0)}{t}$$

як функції від змінної $t$ при фіксованому $x$. Тому, замінивши умову існування лівої і правої похідних функції умовою абсолютної інтегровності цього відношення, можна одержати загальнішу ознаку збіжності ряду.

***Т е о р е м а  3 (Діні).*** *Нехай для абсолютно інтегровної* $2\pi$-*періодичної функції* $f^*(x)$ *у фіксованій точці* $x_0$ *і для деякого* $\delta > 0$ *існує інтеграл*

$$\int_{-\delta}^{\delta}\left|\frac{f^*(x+t)-f^*(x_0)}{t}\right|dx.$$

*Тоді*

$$\lim_{n\to\infty}S_n(x_0)=f^*(x_0),$$

*якщо* $x_0$ – *точка неперервності функції, і*

$$f^*(x_0)=\frac{f^*(x_0+0)+f^*(x_0-0)}{2},$$

*якщо* $x_0$ – *точка розриву першого роду цієї функції.*

*Д о в е д е н н я .* Використовуючи формулу (2.58), знайдемо

$$S_n(x_0)-f^*(x_0)=\int_{-\pi}^{\pi}\left[f^*(x+t)-f^*(x_0)\right]D_n(t)dt = I_3+I_4,$$

де

$$I_3=\int_{-\delta}^{\delta}\frac{f^*(x+t)-f^*(x_0)}{t}\,t\,D_n(t)dt;$$

$$I_4=\frac{1}{\pi}\int_{-\pi}^{-\delta}\left[f^*(x+t)-f^*(x+0)\right]\frac{1}{2\sin\frac{t}{2}}\sin\left(n+\frac{1}{2}\right)dt +$$

$$+\frac{1}{\pi}\int_{\delta}^{\pi}\left[f^*(x+t)-f^*(x+0)\right]\frac{1}{2\sin\frac{t}{2}}\sin\left(n+\frac{1}{2}\right)dt.$$

Перший інтеграл $I_3$ як завгодно малий, внаслідок умови теореми і довільної малості числа $\delta$

$$|I_3|\le \int_{-\delta}^{\delta}\left|\frac{f^*(x+t)-f^*(x_0)}{t}\right||t\,D_n(t)|dt \le \frac{1}{\pi}\int_{-\delta}^{\delta}\left|\frac{f^*(x+t)-f^*(x_0)}{t}\right|dt,$$

а другий інтеграл $I_4$ прямує до нуля при $n\to\infty$ (і фіксованому $\delta$) за лемою 1.

Теорему доведено.

**2.3.3. Розвинення функцій на довільному проміжку.** Нехай функція $f(x)$ задана на довільному проміжку $[-l, l]$, $0 < l < \infty$, і кусково-гладка на ньому. Продовживши її на дійсну вісь, як періодичну функцію з періодом $2l$, одержимо функцію $f^*(x)$ таку, що

$$f^*(x \pm 2l) = f^*(x).$$

Введемо нову змінну $z$ за формулою $\dfrac{z}{\pi} = \dfrac{x}{l}$ і розглянемо функцію $\varphi(z) = f^*\left(\dfrac{l}{\pi}z\right)$, яка періодична з періодом $2\pi$. Дійсно,

$$\varphi(z + 2\pi) = f^*\left[\dfrac{l}{\pi}(z + 2\pi)\right] = f^*\left(\dfrac{l}{\pi}z + 2l\right) = f^*\left(\dfrac{l}{\pi}z\right) = \varphi(z).$$

Запишемо для функції $\varphi(z)$ ряд Фур'є

$$\varphi(z) \sim \dfrac{a_0}{2} + \sum_{n=1}^{\infty}(a_n \cos nz + b_n \sin nz),$$

де

$$a_n = \dfrac{1}{\pi}\int_{-\pi}^{\pi}\varphi(z)\cos nz\,dz, \quad n = 0, 1, \ldots,$$

$$b_n = \dfrac{1}{\pi}\int_{-\pi}^{\pi}\varphi(z)\sin nz\,dz, \quad n = 1, 2, \ldots,$$

і зробимо підстановку $z = \dfrac{\pi}{l}x$. Одержимо для $2l$-періодичної функції $\varphi\left(\dfrac{\pi}{l}x\right) = f^*(x)$ ряд Фур'є

$$f^*(x) \sim \dfrac{a_0}{2} + \sum_{n=1}^{\infty}\left(a_n \cos\dfrac{n\pi}{l}x + b_n \sin\dfrac{n\pi}{l}x\right), \quad x \in (-\infty, +\infty), \quad (2.62)$$

і формули для коефіцієнтів Фур'є

$$a_n = \dfrac{1}{l}\int_{-l}^{l}f(x)\cos\dfrac{n\pi}{l}x\,dx, \quad n = 0, 1, \ldots,$$

$$b_n = \frac{1}{l}\int_{-l}^{l} f(x)\sin\frac{n\pi}{l}x\,dx, \ \ n = 1, 2, \ldots . \qquad (2.63)$$

Для заданої на проміжку $[-l, l]$ функції $f(x)$ формула (2.62) справедлива (в сенсі теореми 2) тільки на цьому проміжку

$$f(x) \sim \frac{a_0}{2} + \sum_{n=1}^{\infty}\left(a_n\cos\frac{n\pi}{l}x + b_n\sin\frac{n\pi}{l}x\right), \quad x \in [-l, l].$$

*П р и к л а д  3*. Розвинути в ряд Фур'є функцію $f(x) = 1 - x^2$ на проміжку $[-1, 1]$.

Коефіцієнти Фур'є шукаємо за формулами (2.63), прийнявши $l = 1$,

$$a_0 = \int_{-1}^{1}(1-x^2)dx = \left(x - \frac{x^3}{3}\right)\bigg|_{-1}^{1} = \frac{4}{3}, \ a_n = \int_{-1}^{1}(1-x^2)\cos n\pi x\,dx =$$

$$= \frac{1}{n\pi}\left[(1-x^2)\sin n\pi x\right]\bigg|_{-1}^{1} + \frac{2}{n\pi}\int_{-1}^{1}x\sin n\pi x\,dx = -\frac{2}{(n\pi)^2}(x\cos n\pi x)\bigg|_{-1}^{1} +$$

$$+ \frac{2}{(n\pi)^2}\int_{-1}^{1}\cos n\pi x\,dx = -\frac{4(-1)^n}{(n\pi)^2} + \frac{2}{(n\pi)^3}\sin n\pi x\bigg|_{-1}^{1} = \frac{4(-1)^{n+1}}{(n\pi)^2},$$

$$n = 1, 2, \ldots ,$$

$$b_n = \int_{-1}^{1}(1-x^2)\sin n\pi x\,dx = 0, \ \ n = 1, 2, \ldots .$$

Оскільки функція $f(x) = 1 - x^2$ неперервна на відрізку $[-1, 1]$ і парна, $f(-1) = f(1)$, то за теоремою 2 справедлива формула

$$1 - x^2 = \frac{2}{3} + \frac{4}{\pi^2}\sum_{n=1}^{\infty}\frac{(-1)^{n+1}}{n^2}\cos n\pi x, \ x \in [-1, 1].$$

Коефіцієнти даного ряду задовольняють нерівність $|a_n| \leq \frac{4}{\pi^2}\frac{1}{n^2}$, при цьому числовий ряд $\sum_{n=1}^{\infty}\frac{1}{n^2}$ збігається. Тому за ознакою Вейєрштрасса ряд Фур'є для функції $f(x) = 1 - x^2$ збігається рівномірно на відрізку $[-1, 1]$.

Графік суми цього ряду наведено на рис. 2.3.

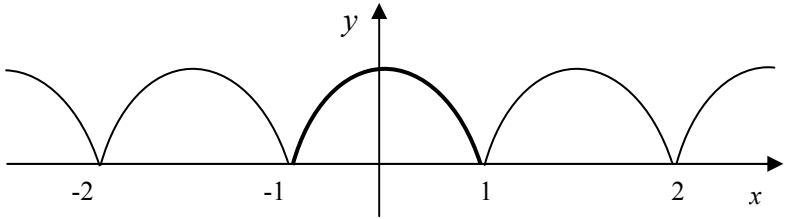

Рис. 2.3.

*П р и к л а д  4* . Розвинути в ряд Фур'є функцію $f(x)=x$ на проміжку $[-2,\,2]$.

Коефіцієнти Фур'є для цієї функції шукаємо за формулами (2.63)

$$a_n = \frac{1}{2}\int_{-2}^{2} x\cos\frac{n\pi}{2}x\,dx = 0,\ n=0,1,\dots,$$

$$b_n = \frac{1}{2}\int_{-2}^{2} x\sin\frac{n\pi}{2}x\,dx = -\frac{1}{n\pi}\left(x\cos\frac{n\pi}{2}x\right)\bigg|_{-2}^{2} + \frac{1}{n\pi}\int_{-2}^{2}\cos\frac{n\pi}{2}x\,dx =$$

$$= -\frac{4(-1)^n}{n\pi} + \frac{2}{(n\pi)^2}\sin\frac{n\pi}{2}x\bigg|_{-2}^{2} = \frac{4(-1)^{n+1}}{n\pi},\ n=1,2,\dots\ .$$

Оскільки $f(-2)\neq f(2)$, розвинення функції $f(x)=x$ в ряд Фур'є справедливе тільки на інтервалі $(-2,\,2)$, в точках $x=\pm 2$ сума ряду дорівнює $\dfrac{f(-2)+f(2)}{2}=0$,

$$x = \frac{4}{\pi}\sum_{n=1}^{\infty}\frac{(-1)^{n+1}}{n}\sin\frac{n\pi}{2}x,\ x\in(-2,\,2).$$

Одержаний ряд рівномірно збігається на будь-якому відрізку $[-2+\delta,\,2-\delta]$, $0<\delta<2$ (прикл. 3, п. 2.1). Графік суми цього ряду наведений на рис. 2.4.

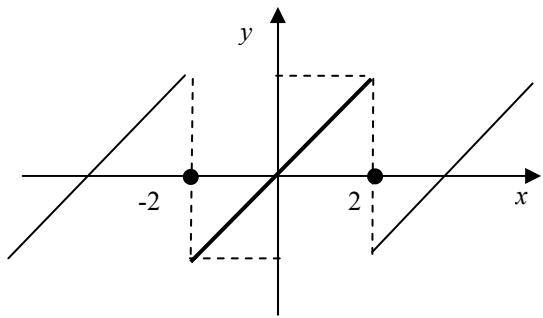

Рис. 2.4.

**2.3.4 Розвинення функцій за косинусами або синусами.** Нехай функція $f(x)$ кусково-гладка на проміжку $[-\pi, \pi]$ і парна. Тоді коефіцієнти Фур'є для функції $f(x)$ знайдемо за формулами (2.53) у вигляді

$$b_n = \frac{1}{\pi}\int_{-\pi}^{\pi} f(x)\sin nx\, dx = 0,$$

$$a_n = \frac{1}{\pi}\int_{-\pi}^{\pi} f(x)\cos nx\, dx = \frac{2}{\pi}\int_{0}^{\pi} f(x)\cos nx\, dx$$

і ряд Фур'є має наступний вигляд

$$f(x) \sim \frac{a_0}{2} + \sum_{n=1}^{\infty} a_n \cos nx, \qquad (2.64)$$

де $a_n = \dfrac{2}{\pi}\int_{0}^{\pi} f(x)\cos nx\, dx$, $n = 0, 1, \ldots$.

Якщо $f(x)$ – кусково-гладка на проміжку $[-\pi, \pi]$ непарна функція, то знайдемо її коефіцієнти Фур'є

$$a_n = \frac{1}{\pi}\int_{-\pi}^{\pi} f(x)\cos nx\, dx = 0,$$

$$b_n = \frac{1}{\pi}\int_{-\pi}^{\pi} f(x)\sin nx\, dx = \frac{2}{\pi}\int_{0}^{\pi} f(x)\sin nx\, dx$$

і ряд Фур'є

$$f(x) \sim \sum_{n=1}^{\infty} b_n \sin nx, \qquad (2.65)$$

де $b_n = \dfrac{2}{\pi} \int\limits_0^{\pi} f(x)\sin nx\, dx$, $n = 1, 2, \ldots$ .

*Нехай тепер кусково-гладка функція $f(x)$ задана на проміжку $[0, \pi]$. Продовживши її на проміжок $[-\pi, 0)$ як парну або непарну функцію, одержимо для $f(x)$ на проміжку $[0, \pi]$ ряди Фур'є (2.64) або (2.65) тільки за косинусами або тільки за синусами.*

Якщо функція $f(x)$ визначена на проміжку $[0, l]$, $0 < l < \infty$, то, провівши відповідну заміну змінної, одержимо для цієї функції наступні ряди Фур'є:

$$f(x) \sim \frac{a_0}{2} + \sum_{n=1}^{\infty} a_n \cos \frac{n\pi}{l} x, \qquad (2.66)$$

де $a_n = \dfrac{2}{l} \int\limits_0^{l} f(x) \cos \dfrac{n\pi}{l} x\, dx$, $n = 0, 1, \ldots$ ;

$$f(x) \sim \sum_{n=1}^{\infty} b_n \sin \frac{n\pi}{l} x, \qquad (2.67)$$

де $b_n = \dfrac{2}{l} \int\limits_0^{l} f(x) \sin \dfrac{n\pi}{l} x\, dx$, $n = 1, 2, \ldots$ .

*З а у в а ж е н н я* . Розглянуті ряди Фур'є для функцій на проміжках $[-l, l]$, $[0, l]$, $[0, \pi]$ є по суті розвиненнями цих функцій в ряди за системами тригонометричних функцій, ортогональних на відповідних проміжках. Поряд з основною ортогональною системою тригонометричних функцій (п. 2.3) маємо ще такі ортогональні системами функцій:

$1, \cos x, \cos 2x, \ldots, \cos nx, \ldots,\quad x \in [0, \pi]$;

$\sin x, \sin 2x, \ldots, \sin nx, \ldots,\quad x \in [0, \pi]$;

$1, \cos \dfrac{\pi}{l} x, \sin \dfrac{\pi}{l} x, \cos \dfrac{2\pi}{l} x, \sin \dfrac{2\pi}{l} x, \ldots, \cos \dfrac{n\pi}{l} x, \sin \dfrac{n\pi}{l} x, \ldots,$

$$x \in [-l, l];$$

$$1, \cos\frac{\pi}{l}x, \cos\frac{2\pi}{l}x, \ldots, \cos\frac{n\pi}{l}x, \ldots, \quad x \in [0, l];$$

$$\sin\frac{\pi}{l}x, \sin\frac{2\pi}{l}x, \ldots, \sin\frac{n\pi}{l}x, \ldots, \quad x \in [0, l].$$

**2.3.5. Принцип локалізації.** З доведення теореми 2 випливає, що збіжність ряду Фур'є функції $f^*(x)$ суттєво залежить від поведінки функції в околі точки $x$ і мало залежить від її поведінки у віддалених точках.

***Т е о р е м а 4 (Рімана).*** *Збіжність чи розбіжність ряду Фур'є функції $f^*(x)$ (яку можна вважати абсолютно інтегровною на відрізку $[-\pi, \pi]$ і $2\pi$-періодичною) в точці $x$ залежить тільки від значень функції в як завгодно малому околі цієї точки.*

*Д о в е д е н н я .* Якщо записати частинну суму ряду у вигляді (2.56) і перетворити її аналогічно з (2.60), то одержимо

$$S_n(x) = \int_{-\pi}^{\pi} f^*(x-t)D_n(t)dt = \int_{-\delta}^{\delta} f^*(x-t)D_n(t)dt + I_2^- + I_2^+.$$

Нехай функція $f^*(x)$ абсолютно інтегровна на відрізку $[-\pi, \pi]$. Тоді функції $f^*(x \pm t)$ абсолютно інтегровні по $t$ на проміжку $[\delta, \pi]$ і інтеграли $I_2^\pm$ прямують до нуля (за лемою 1), коли $n \to \infty$. Тому значення границі при $n \to \infty$ частинної суми визначається інтегралом

$$\int_{-\delta}^{\delta} f^*(x-t)D_n(t)\,dt,$$

який містить значення функції $f^*(x)$ з околу точки $x$. Це і доводить принцип локалізації.

З цієї теореми випливає висновок про те, що два ряди Фур'є можуть вести себе однаково у одному інтервалі і вести себе по різному в іншому інтервалі.

Теорему доведено.

*З а у в а ж е н н я .* Відзначимо, що з рівності сум степеневих рядів в околі деякої точки випливає рівність їх коефіцієнтів і, відповідно, рівність сум в інших точках областей збіжності.

*П р и к л а д   5* . Розглянемо приклад розвинення в ряд Фур'є абсолютно інтегровної (у сенсі невласного інтеграла) функції. Знайти ряд за косинусами парної функції $f(x) = \ln\left(2\cos\dfrac{x}{2}\right)$ на проміжку $(-\pi, \pi)$. На кінцях проміжку функція приймає нескінченно великі значення, однак абсолютно інтегровна на цьому проміжку.

За першою формулою (2.53) при $n = 0$ знайдемо

$$a_0 = \frac{1}{\pi}\int_0^\pi \ln\left(2\cos\frac{x}{2}\right)dx = \ln 2 + \frac{2}{\pi}\int_0^{\pi/2}\ln\cos t\, dt = 0,$$

де  $J = \int\limits_0^{\pi/2}\ln\cos t\, dt = \int\limits_0^{\pi/2}\ln\sin t\, dt = -\dfrac{\pi}{2}\ln 2$  – інтеграл Ейлера [22], який обчислюється методом заміни змінних.

Підставивши в інтеграл $t = 2x$, одержимо

$$J = 2\int_0^{\pi/4}\ln\sin 2t\, dt = \frac{\pi}{2}\ln 2 + 2\int_0^{\pi/4}\ln\sin x\, dx + 2\int_0^{\pi/4}\ln\cos x\, dx.$$

В останньому інтегралі зробимо заміну $x = \dfrac{\pi}{2} - t$, знайдемо

$$J = \frac{\pi}{2}\ln 2 + 2\int_0^{\pi/4}\ln\sin x\, dx - 2\int_{\pi/2}^{\pi/4}\ln\sin x\, dx =$$
$$= \frac{\pi}{2}\ln 2 + 2\int_0^{\pi/2}\ln\sin x\, dx = \frac{\pi}{2}\ln 2 + 2J.$$

Отже, для визначення $J$ одержимо рівняння $J = \dfrac{\pi}{2}\ln 2 + 2J$, звідси $J = -\dfrac{\pi}{2}\ln 2$.

Для інших значень $n > 0$ маємо

$$a_n = \frac{2}{\pi}\int_0^\pi \ln\left(2\cos\frac{x}{2}\right)\cos nx\, dx = \frac{2}{\pi}\left[\ln\left(2\cos\frac{x}{2}\right)\frac{\sin nx}{n}\right]\Bigg|_0^\pi +$$

$$+\frac{1}{n\pi}\int_0^\pi \frac{\sin nx \, \sin\frac{x}{2}}{\cos\frac{x}{2}}\, dx.$$

Проводячи тут заміну $x = \pi - t$ з урахуванням залежності $\sin nx \cos\frac{x}{2} = \frac{1}{2}\left[\sin\left(n+\frac{1}{2}\right)x + \sin\left(n-\frac{1}{2}\right)x\right]$ і формули

$$\frac{\sin\left(n+\frac{1}{2}\right)x}{2\sin\frac{x}{2}} = \frac{1}{2} + \sum_{k=1}^n \cos nx, \quad \frac{\sin\left(n-\frac{1}{2}\right)x}{2\sin\frac{x}{2}} = \frac{1}{2} + \sum_{k=1}^{n-1} \cos nx,$$

одержимо

$$a_n = \frac{(-1)^{n-1}}{n\pi}\int_0^\pi \frac{\sin nx \cos\frac{x}{2}}{\sin\frac{x}{2}}\, dx =$$

$$= \frac{(-1)^{n-1}}{n\pi}\left(\int_0^\pi \frac{\sin\left(n+\frac{1}{2}\right)x}{2\sin\frac{x}{2}}\, dx + \int_0^\pi \frac{\sin\left(n-\frac{1}{2}\right)x}{2\sin\frac{x}{2}}\, dx\right) = \frac{(-1)^{n-1}}{n}.$$

Отже, шукане розвинення має вигляд

$$\ln\left(2\cos\frac{x}{2}\right) = \sum_{n=1}^\infty \frac{(-1)^{n-1}}{n}\cos nx, \quad x \in (-\pi, \pi). \qquad (2.68)$$

Якщо вважати, що в точках $x = (2k+1)\pi$, $k = 0, \pm 1, \ldots,$ функція приймає «значення» $-\infty$ і взяти під логарифмом модуль відповідної функції, то одержимо розвинення, яке справедливе на всій осі

$$\ln\left|2\cos\frac{x}{2}\right| = \sum_{n=1}^\infty \frac{(-1)^{n-1}}{n}\cos nx, \quad x \in (-\infty, \infty).$$

*П р и к л а д   6 .* Якщо в рівності (2.68) зробити заміну $x = \pi - t$, то прийдемо ще до такої рівності

$$\ln\left|2\sin\frac{x}{2}\right| = -\sum_{n=1}^\infty \frac{\cos nx}{n}, \quad x \in (0, 2\pi)$$

і, відповідно,

$$\ln\left|2\sin\frac{x}{2}\right| = -\sum_{n=1}^{\infty}\frac{\cos nx}{n}, \ x \in (-\infty, \infty).$$

При цьому обидві частини цієї рівності перетворюються у нескінченність в точках $x = 2k\pi$, $k = 0, \pm 1, \ldots$.

**2.3.6. Збіжність тригонометричних рядів.** Дослідимо поведінку $2\pi$-періодичної абсолютно інтегровної функції $f(x)$ на окремому проміжку (якщо ж вона задана на відрізку $[-\pi, \pi]$, то досліджуємо її $2\pi$-періодичне продовження) $[11, 19]$.

*Періодична функція абсолютно інтегровна, якщо вона абсолютно інтегровна на будь-якому скінченному проміжку.*

Спочатку сформулюємо доповнення і узагальнення леми 1.

**Л е м а  3 .** *Нехай $f(x)$ – абсолютно інтегровна $2\pi$-періодична функція і $\omega(t)$ – функція з обмеженою похідною на проміжку $[a, b]$.*

*Тоді, яке б не було число $\varepsilon > 0$, для всіх $x$ справедливі нерівності*

$$\int_a^b f(x \pm t)\omega(t)\cos\lambda t\, dt < \varepsilon, \ \int_a^b f(x \pm t)\omega(t)\sin\lambda t\, dt < \varepsilon, \quad (2.69)$$

*якщо тільки $\lambda$ досить велике число, тобто інтеграли рівномірно збігаються при $\lambda \to \infty$ до нуля відносно змінної $x$.*

Це твердження приймемо без доведення $[19, c.108]$. Одержимо відповідне твердження за жорсткіших умов.

**Л е м а  3'.** *Нехай $f(x)$ – неперервна $2\pi$-періодична функція, що має абсолютно інтегровну похідну, і $\omega(t)$ – функція з обмеженою похідною на проміжку $[a, b]$.*

*Тоді, яке б не було число $\varepsilon > 0$, для всіх $x$ справедливі нерівності* (2.69), *якщо тільки число $\lambda$ досить велике.*

*Д о в е д е н н я .* Розглянемо, наприклад, перший інтеграл і використаємо формулу інтегрування за частинами

$$\int_a^b f(x \pm t)\omega(t)\cos\lambda t\, dt = \frac{1}{\lambda}\left[f(x \pm t)\omega(t)\sin\lambda t\right]\Big|_a^b -$$

$$-\frac{1}{\lambda}\int_a^b [f(x\pm t)\omega(t)]' \sin\lambda t\, dt. \qquad (2.70)$$

За умовою вираз у квадратній дужці у правій частині цієї рівності обмежений. Покажемо, що інтеграл у правій частині цього виразу також обмежений. Величини $f(x\pm t)\omega'(t)$ і $\omega(t)$ у квадратних дужках під інтегралом обмежені деякою сталою $M$.

Тоді
$$\left|\int_a^b [f(x\pm t)\omega(t)]' \sin\lambda t\, dt\right| \le \int_a^b |\pm f'(x\pm t)\omega(t) + f(x\pm t)\omega'(t)|\, dt \le$$
$$\le M\int_a^b |f'(x\pm t)|\, dt + M(b-a).$$

Оскільки функція $f(x)$ періодична, то $|f'(x)|$ також періодична функція і справедлива оцінка
$$\left|\int_a^b [f(x\pm t)\omega(t)]' \sin\lambda t\, dt\right| \le M\int_{-\pi}^{\pi} |f'(x\pm t)|\, dt + M(b-a) = const,$$

яке б не було $x$.

Враховуючи обмеженість виразу у квадратних дужках і інтеграла у формулі (2.70), одержимо першу нерівність (2.69). Другу нерівність (2.69) доводимо аналогічно.

Лему доведено.

*Л е м а  4 .* *Інтеграл*
$$I = \int_0^x \frac{\sin mt}{2\sin\frac{t}{2}}\, dt$$

*обмежений для будь-якого* $m$ *і* $x\in[-\pi, \pi]$.

*Д о в е д е н н я*. Запишемо цей інтеграл у вигляді
$$I = \int_0^x \frac{\sin mt}{t}\, dt + \int_0^x \omega(t)\sin mt\, dt,$$
де $\omega(t) = \dfrac{1}{2\sin\dfrac{t}{2}} - \dfrac{1}{t}$.

Застосовуючи правило Лопіталя, переконуємося, що функції $\omega(t)$ і $\omega'(t)$ неперервні (якщо прийняти $\omega(0) = 0$). Тому другий інтеграл у цьому виразі обмежений для будь-якого $m$ і $x \in [-\pi, \pi]$.

Перший інтеграл перетворимо з урахуванням заміни $mt = u$,

$$\int_0^x \frac{\sin mt}{t} dt = \int_0^{mx} \frac{\sin u}{u} du.$$

Цей інтеграл не перевищує площі області, обмеженої першою аркою кривої $y = \frac{\sin x}{x}$ і віссю $Ox$. Крім того, відомо, що

$$\int_0^\infty \frac{\sin x}{x} dx = \frac{\pi}{2}.$$

Отже, обмежений кожний з інтегралів і, відповідно, обмежений інтеграл $I$.

Тепер дослідимо окремі класи функцій.

***Т е о р е м а  5 .*** *Нехай $f(x)$ – неперервна $2\pi$-періодична функція, що має абсолютно інтегровну похідну.*

*Тоді ряд Фур'є цієї функції збігається до $f(x)$ рівномірно для всіх $x$.*

*Д о в е д е н н я .* Запишемо різницю частинної суми ряду і функції з урахуванням формули (2.57) у вигляді

$$S_n(x) - f(x) = \frac{1}{\pi} \int_{-\pi}^{\pi} [f(x+t) - f(x)] \frac{\sin mt}{2\sin\frac{t}{2}} dt, \qquad (2.71)$$

де $m = n + \frac{1}{2}$.

Задамо довільне число $\varepsilon > 0$. Виберемо число $\delta$, $0 < \delta < \pi$, і розіб'ємо інтеграл у (2.71) на три інтеграли $I_1$, $I_2$, $I_3$, відповідно, на відрізках $[-\delta, \delta]$, $[\delta, \pi]$, $[-\pi, -\delta]$. Інтегруючи частинами, одержимо

$$I_1 = \left\{ [f(x+t) - f(x)] \int_0^t g(u) du \right\} \Bigg|_{-\delta}^{\delta} - \int_{-\delta}^{\delta} [f'(x+t) \int_0^t g(u) du] dt,$$

де $g(u) = \dfrac{\sin mu}{2\sin\dfrac{u}{2}}$ – парна функція.

Перший доданок в цьому виразі запишемо у вигляді

$$\{[f(x+\delta)-f(x)]+[f(x-\delta)-f(x)]\}\int\limits_0^\delta g(u)du$$

і оцінимо його абсолютну величину. Внаслідок неперервності функції $f(x)$ і обмеженості інтеграла (за лемою 4), абсолютна величина цього виразу може бути зроблена (за рахунок вибору $\delta$) меншою, ніж $\dfrac{\varepsilon}{2}$. Оцінимо другий доданок з урахуванням твердження леми 4 і достатньо малого $\delta$,

$$\left|\int\limits_{-\delta}^{\delta}[f'(x+t)\int\limits_0^t g(u)du]dt\right| \le \int\limits_{-\delta}^{\delta}|f'(x+t)|\left|\int\limits_0^t g(u)du\right|dt \le$$

$$\le M\int\limits_{-\delta}^{\delta}|f'(x+t)|dt = M\int\limits_{x-\delta}^{x+\delta}|f'(t)|dt < \varepsilon.$$

Тут враховано, що за лемою 4

$$\left|\int\limits_0^t g(u)du\right| < M,\ t\in[-\pi,\pi],\quad M = const,$$

а також враховано, що внаслідок абсолютної інтегровності підінтегральної функції інтеграл $\int\limits_{x-\delta}^{x+\delta}|f'(t)|dt$ можна зробити (за рахунок вибору малого $\delta$) як завгодно малим.

Отже, якщо вибрати $\delta$ достатньо малим, то $|I_1|<\varepsilon$, яке б не було $x$.

Інтеграли $I_2$ і $I_3$ за лемою 1 можна вибрати як завгодно малими для всіх $x$, якщо тільки $n$ достатньо велике.

Наприклад, за лемою 1, прийнявши $\omega(t) = \dfrac{1}{2\sin\dfrac{t}{2}}$, знайдемо

$$|I_2| = \left| \int_\delta^\pi [f(x+t) - f(x)] \frac{\sin mt}{2\sin\frac{t}{2}} dt \right| \leq \left| \int_\delta^\pi f(x+t) \frac{\sin mt}{2\sin\frac{t}{2}} dt \right| +$$

$$+ \left| \int_\delta^\pi f(x) \frac{\sin mt}{2\sin\frac{t}{2}} dt \right| < \varepsilon$$

для всіх $x$ і досить великого $n$.

Аналогічна нерівність справедлива і для третього інтеграла $|I_3| < \varepsilon$.

Тоді для всіх $x$ і досить великого $n$ маємо оцінку для відхилення (2.71)

$$|S_n(x) - f(x)| = \frac{1}{\pi} |I_1 + I_2 + I_3| < \frac{3\varepsilon}{\pi} < \varepsilon,$$

що стверджує рівномірну збіжність ряду Фур'є функції $f(x)$.

Теорему доведено.

**Н а с л і д о к .** *Нехай $f(x)$ – неперервна кусково-гладка $2\pi$-періодична функція.*

*Тоді ряд Фур'є цієї функції збігається до $f(x)$ рівномірно.*

Це твердження випливає з того, що кусково-гладка функція має обмежену похідну і, відповідно, абсолютно інтегровну похідну.

**Т е о р е м а 6 .** *Нехай $f(x)$ – абсолютно інтегровна і $2\pi$-періодична функція, яка неперервна на деякому відрізку $[a, b]$ і має на цьому відрізку абсолютно інтегровну похідну.*

*Тоді ряд Фур'є цієї функції збігається до $f(x)$ рівномірно на будь-якому меншому відрізку $[a + \delta, b - \delta], \ \delta > 0$.*

*Д о в е д е н н я .* Якщо довжина відрізка $[a, b]$ більша або дорівнює $2\pi$, то виконуються умови попередньої теореми і відповідний ряд збігається рівномірно на всій дійсній осі. Тому розглядаємо випадок, коли довжина відрізка $[a, b]$ менша $2\pi$.

Введемо допоміжну $2\pi$-періодичну і неперервну функцію $F(x)$, що дорівнює $f(x)$ на відрізку $[a, b]$ і є лінійною на відрізку $[b, a + 2\pi]$, а також $F(a + 2\pi) = f(a)$. Зовні відрізка $[a, a + 2\pi]$ значення функції $F(x)$ одержуємо періодичним продовженням.

Покладемо $\Phi(x) = f(x) - F(x)$. Ця функція абсолютно інтегровна і $\Phi(x) = 0$ для $x \in [a, b]$ і, очевидно, $f(x) = F(x) + \Phi(x)$.

Запишемо вираз відхилення частинної суми від функції $f(x)$

$$S_n(x) - f(x) = \frac{1}{\pi} \int_{-\pi}^{\pi} [F(x+t) - F(x)] \frac{\sin mt}{2\sin\frac{t}{2}} dt +$$

$$+ \frac{1}{\pi} \int_{-\pi}^{\pi} [\Phi(x+t) - \Phi(x)] \frac{\sin mt}{2\sin\frac{t}{2}} dt = I_1 + I_2, \qquad (2.72)$$

де $m = n + \frac{1}{2}$.

Виберемо довільне число $\varepsilon > 0$. За теоремою 5 ряд Фур'є для неперервної функції $F(x)$ збігається до неї рівномірно. Тоді для всіх $x$ і достатньо великих $n$ справедлива оцінка $|I_1| < \frac{\varepsilon}{2}$.

Нехай тепер $x \in [a+\delta, b-\delta]$. Тоді $\Phi(x) = 0$ і, відповідно,

$$I_2 = \frac{1}{\pi} \int_{-\pi}^{\pi} \Phi(x+t) \frac{\sin mt}{2\sin\frac{t}{2}} dt.$$

Якщо $t \in [-\delta, \delta]$, то для $x \in [a+t, b-t]$ також $\Phi(x+t) = 0$. Тому

$$I_2 = \frac{1}{\pi} \int_{-\pi}^{-\delta} \Phi(x+t) \frac{\sin mt}{2\sin\frac{t}{2}} dt + \frac{1}{\pi} \int_{\delta}^{\pi} \Phi(x+t) \frac{\sin mt}{2\sin\frac{t}{2}} dt.$$

Застосовуючи до кожного з цих інтегралів лему 3, для всіх $x \in [a+\delta, b-\delta]$ і достатньо великих $n$ одержимо оцінку $|I_2| < \frac{\varepsilon}{2}$.

Враховуючи оцінки для інтегралів $I_1$ і $I_2$ в поданні (2.72), маємо нерівність

$$|S_n(x) - f(x)| \le |I_1| + |I_2| < \varepsilon,$$

яка справедлива для всіх $x \in [a+\delta, b-\delta]$, якщо тільки $n$ достатньо велике.

Теорему доведено.

**Н а с л і д о к .** *Нехай $f(x)$ – кусково-гладка $2\pi$-періодична функція, яка неперервна на деякому відрізку $[a, b]$.*

*Тоді ряд Фур'є цієї функції збігається до $f(x)$ рівномірно на будь-якому меншому відрізку $[a+\delta, b-\delta]$, $\delta > 0$.*

Це твердження справедливе, оскільки кусково-гладка функція має абсолютно інтегровну похідну.

*П р и к л а д  7* . Розглянемо кусково-гладку непарну функцію, яка дорівнює $\dfrac{\pi}{4}$, якщо $x \in (0, \pi)$, і дорівнює $-\dfrac{\pi}{4}$, якщо $x \in (-\pi, 0)$.

За формулами (2.53) знайдемо
$$a_k = 0, \quad k = 0, 1, \ldots,$$
$$b_k = \frac{2}{\pi}\int\limits_0^\pi f(x)\sin kx\,dx = \frac{1}{2}\int\limits_0^\pi \sin kx\,dx = -\frac{1}{2k}\cos kx\Big|_0^\pi = \frac{1-(-1)^k}{2k},$$
$$k = 1, 2, \ldots .$$

Тоді ряд Фур'є можна записати у вигляді
$$f(x) \sim \sum_{k=1}^{\infty} \frac{\sin(2k-1)x}{2k-1},$$

який рівномірно збігається до $f(x) = \dfrac{\pi}{4}$, якщо $x \in [\delta, \pi - \delta]$, і

збігається до $f(x) = -\dfrac{\pi}{4}$, якщо $x \in [-\pi + \delta, -\delta]$, $\delta > 0$

$$\frac{\pi}{4} = \sum_{k=1}^{\infty} \frac{\sin(2k-1)x}{2k-1}, \quad x \in (0, \pi),$$

$$-\frac{\pi}{4} = \sum_{k=1}^{\infty} \frac{\sin(2k-1)x}{2k-1}, \quad x \in (-\pi, 0).$$

**Т е о р е м а  7** *. Якщо ряд*
$$\sum_{n=1}^{\infty} \left(|a_n| + |b_n|\right)$$

*збігається, то ряд*

$$f(x) = \frac{a_0}{2} + \sum_{k=1}^{\infty}(a_k \cos kx + b_k \sin kx) \qquad (2.73)$$

*збігається абсолютно, рівномірно і, відповідно, має неперервну суму для якої є рядом Фур'є.*

*Д о в е д е н н я.* Для членів ряду (2.73) справедлива оцінка
$$|a_k \cos kx + b_k \sin kx| \leq |a_k \cos kx| + |b_k \sin kx| \leq |a_k| + |b_k|.$$
Тому за теоремою 4 (п. 2.1) ряд (2.73) збігається абсолютно і рівномірно, а його сума неперервна функція і, відповідно, за теоремою 1 ряд (2.73) є рядом Фур'є для своєї суми.

Теорему доведено.

*Т е о р е м а 8. Нехай $f(x)$ – неперервна кусково-гладка, $2\pi$-періодична функція і*

$$f(x) = \frac{a_0}{2} + \sum_{k=1}^{\infty}(a_k \cos kx + b_k \sin kx)$$

*– її ряд Фур'є.*

*Тоді цей ряд допускає почленне диференціювання в тому розумінні, що ряд Фур'є функції $f'(x)$ можна одержати з вихідного ряду почленним диференціюванням*

$$f'(x) \sim \sum_{k=1}^{\infty}(kb_k \cos kx - ka_k \sin kx).$$

*Д о в е д е н н я.* Функція $f'(x) \in QL[-\pi, \pi]$ розвивається в ряд Фур'є

$$f'(x) \sim \sum_{k=1}^{\infty}(a'_k \cos kx + b'_k \sin kx). \qquad (2.74)$$

Враховуючи рівність $f(-\pi) = f(\pi)$, одержимо

$$a'_0 = \frac{1}{\pi}\int_{-\pi}^{\pi} f'(x)dx = \frac{1}{\pi}[f(\pi) - f(-\pi)] = 0,$$

$$a'_k = \frac{1}{\pi}\int_{-\pi}^{\pi} f'(x)\cos kx\, dx = \frac{1}{\pi}\left\{[f(x)\cos kx]\Big|_{-\pi}^{\pi} + k\int_{-\pi}^{\pi} f(x)\sin kx\, dx\right\} = kb_k,$$

$$b'_k = \frac{1}{\pi}\int_{-\pi}^{\pi} f'(x)\sin kx\, dx = \frac{1}{\pi}\left\{[f(x)\sin kx]\Big|_{-\pi}^{\pi} - k\int_{-\pi}^{\pi} f(x)\cos kx\, dx\right\} = -ka_k.$$

Підстановка цих формул у ряд (2.74) приводить до ряду Фур'є для функції $f'(x)$.

Теорему доведено.

**Т е о р е м а   9**. *Нехай $f(x)$ – абсолютно інтегровна $2\pi$-періодична функція і*

$$f(x) \sim \frac{a_0}{2} + \sum_{k=1}^{\infty} (a_k \cos kx + b_k \sin kx) \qquad (2.75)$$

*– її ряд Фур'є.*

*Тоді цей ряд можна почленно інтегрувати і одержаний при цьому ряд збігається рівномірно*

$$\int_0^x f(t)dt = \frac{a_0}{2}x + \sum_{k=1}^{\infty} \int_0^x (a_k \cos kx + b_k \sin kx)dx$$

*або*

$$\int_0^x \left[ f(t) - \frac{a_0}{2} \right] dt = \sum_{k=1}^{\infty} \frac{b_k}{k} + \sum_{k=1}^{\infty} \left( -\frac{b_k}{k} \cos kx + \frac{k_k}{k} \sin kx \right). \quad (2.76)$$

*Д о в е д е н н я*. Розглянемо функцію

$$\Phi(x) = \int_0^x \left[ f(t) - \frac{a_0}{2} \right] dx, \ -\pi \le x \le \pi .$$

Вона неперервна, має абсолютно інтегровану похідну $\Phi'(x) = f(x) - \frac{a_0}{2}$ і $\Phi(0) = 0$. Її можна періодично продовжити на всю вісь, оскільки $\Phi(\pi) - \Phi(-\pi) = \int_{-\pi}^{\pi} f(t)dt - \pi a_0 = 0$.

Тому за теоремою 5 функція $\Phi(x)$ розвивається в рівномірно збіжний ряд Фур'є

$$\Phi(x) = \frac{A_0}{2} + \sum_{k=1}^{\infty} (A_k \cos kx + B_k \sin kx). \qquad (2.77)$$

Якщо скористатись формулою інтегрування частинами, то матимемо (для $k \ge 1$)

$$A_k = \frac{1}{\pi} \int_{-\pi}^{\pi} \Phi(x) \cos kx\, dx = \frac{1}{\pi} \left[ \Phi(x) \frac{\sin kx}{k} \right] \Bigg|_{-\pi}^{\pi} -$$

$$-\frac{1}{k\pi}\int\limits_{-\pi}^{\pi}\left[f(x)-\frac{a_0}{2}\right]\sin kx\,dx = -\frac{1}{k\pi}\int\limits_{-\pi}^{\pi}f(x)\sin kx\,dx = -\frac{b_k}{k}.$$

Аналогічно одержимо $B_k = \dfrac{a_k}{k}$. Вираз для коефіцієнта $A_0$ одержимо з (2.77) при $x = 0$ у вигляді $0 = \dfrac{A_0}{2} + \sum\limits_{k=1}^{\infty}A_k$ або $\dfrac{A_0}{2} = \sum\limits_{k=1}^{\infty}\dfrac{b_k}{k}$. Підставивши вирази коефіцієнтів в ряд (2.77), одержимо ряд (2.76).

Теорему доведено.

**2.3.7. Оцінка коефіцієнтів ряду Фур'є.** Лема 1 встановлює, що коефіцієнти Фур'є для абсолютно інтегровної $2\pi$-періодичної функції прямують до нуля, коли $n \to \infty$.

Розглянемо поведінку при великому $n$ частинної суми ряду Фур'є для неперервної (абсолютно інтегровної) $2\pi$-періодичної функції.

*Т е о р е м а  10*. *Нехай $f(x)$ – неперервна $2\pi$-періодична функція, $\max\limits_{x\in[-\pi,\pi]}|f(x)| = M$.*

*Тоді справедлива оцінка*
$$S_n(x) \le AM\ln n, \qquad (2.78)$$

де $A = 2 + \dfrac{1+\ln\pi}{\ln 2}$, $n \ge 2$.

*Д о в е д е н н я*. Виходячи з формули (2.58), одержимо
$$|S_n(x)| = \left|\frac{1}{\pi}\int\limits_{-\pi}^{\pi}f^*(x-t)D_n(t)dt\right| \le$$
$$\le \frac{M}{\pi}\int\limits_{-\pi}^{\pi}|D_n(t)|dt = \frac{M}{\pi}\int\limits_{0}^{\pi}\frac{\left|\sin\left(n+\dfrac{1}{2}\right)t\right|}{\sin\dfrac{t}{2}}dt.$$

Проводячи заміну $\left(n+\dfrac{1}{2}\right)t = y$ і використовуючи нерівність

$\sin\dfrac{t}{2} \geq \dfrac{t}{\pi}$, $0 \leq t \leq \pi$, одержимо

$$|S_n(x)| \leq M \int\limits_0^{(n+1/2)\pi} \dfrac{|\sin y|}{y} dy \leq$$

$$\leq M \int\limits_0^1 dy + M \int\limits_1^{(n+1/2)\pi} \dfrac{1}{y} dy = M\left[1 + \ln\left(n + \dfrac{1}{2}\right)\pi\right] \leq$$

$$\leq M(1 + \ln 2\pi n) \leq M \ln n \left(1 + \dfrac{1 + \ln 2\pi}{\ln 2}\right) = MA \ln n, \ n \geq 2.$$

Звідси одержимо нерівність (2.78).

Якщо ряд Фур'є збіжний, то його залишок записується у вигляді $r_n(x) = f(x) - S_n(x)$. Тоді одержимо таку оцінку для залишку:

$$|r_n(x)| = |f(x)| + |S_n(x)| \leq M + AM \ln n = (1+A)M \ln n, \ n \geq 2.$$

Теорему доведено.

*З а у в а ж е н н я .* Теорема 10 справедлива і за слабших умов, а саме за умов абсолютної інтегровності та обмеженості функції.

*Т е о р е м а  11 . Нехай неперервна $2\pi$-періодична функція $f(x)$ має неперервні похідні до $(k-1)$-го порядку $(k \geq 1)$ включно, а $k$-а похідна абсолютно інтегровна.*

*Тоді для коефіцієнтів Фур'є цієї функції справедливі рівності*

$$\lim_{n \to \infty} n^k a_n = \lim_{n \to \infty} n^k b_n = 0. \qquad (2.79)$$

*Д о в е д е н н я .* Розглянемо вирази для коефіцієнтів $a_n$, $b_n$ і застосуємо формулу інтегрування частинами

$$a_n = \dfrac{1}{\pi} \int\limits_{-\pi}^{\pi} f(x) \cos nx \, dx = \left[\dfrac{f(x)\sin nx}{n}\right]\Bigg|_{-\pi}^{\pi} -$$

$$- \dfrac{1}{n\pi} \int\limits_{-\pi}^{\pi} f'(x) \sin nx \, dx = -\dfrac{b_n^{(1)}}{n},$$

$$b_n = \dfrac{1}{\pi} \int\limits_{-\pi}^{\pi} f(x) \sin nx \, dx = -\left[\dfrac{f(x)\cos nx}{n}\right]\Bigg|_{-\pi}^{\pi} +$$

$$+\frac{1}{n\pi}\int\limits_{-\pi}^{\pi}f'(x)\cos nx\,dx = \frac{a_n^{(1)}}{n}.$$

Тут враховано неперервність функції $f(x)$ і позначено через $a_n^{(1)}$ і $b_n^{(1)}$ коефіцієнти Фур'є похідної $f'(x)$. Застосовуючи послідовно $k$ раз ці формули, одержимо

$$a_n = -\frac{b_n^{(1)}}{n} = -\frac{a_n^{(2)}}{n^2} = \frac{b_n^{(3)}}{n^3} = \ldots = \frac{\alpha_n}{n^k},$$

$$b_n = \frac{a_n^{(1)}}{n} = -\frac{b_n^{(2)}}{n^2} = -\frac{a_n^{(3)}}{n^3} = \ldots = \frac{\beta_n}{n^k},$$

де $a_n^{(1)}, a_n^{(2)}, \ldots,$ $b_n^{(1)}, b_n^{(2)}, \ldots$ – коефіцієнти Фур'є функцій $f'(x), f''(x), \ldots$, а через $\alpha_n, \beta_n$ – коефіцієнти Фур'є (з відповідними знаками) функції $f^{(k)}(x)$. Оскільки функція $f^{(k)}(x)$ абсолютно інтегровна, то за лемою 1 маємо $\lim\limits_{n\to\infty}\alpha_n = \lim\limits_{n\to\infty}\beta_n = 0$. Звідси випливають рівності (2.79).

Теорему доведено.

*Н а с л і д о к .* *Нехай неперервна $2\pi$-періодична функція $f(x)$ має неперервні похідні до $(k-1)$-го порядку $(k \geq 1)$ включно, а $k$-а похідна абсолютно інтегровна.*

*Тоді збігається ряд*

$$\sum_{n=1}^{\infty} n^{k-1}\left(|a_n| + |b_n|\right). \tag{2.80}$$

*Д о в е д е н н я .* Запишемо рівності (2.80) з урахуванням нерівностей $\alpha_n \leq \rho_n$, $\beta_n \leq \rho_n$, де $\rho_n = \sqrt{\alpha_n^2 + \beta_n^2}$, у вигляді

$$n^{k-1}\left(|a_n| + |b_n|\right) \leq 2\frac{\rho_n}{n} \leq \rho_n^2 + \frac{1}{n^2}.$$

Оскільки ряди $\sum\limits_{n=1}^{\infty}\rho_n^2$, $\sum\limits_{n=1}^{\infty}n^{-2}$ збігаються, то з останньої нерівності випливає твердження наслідку. Збіжність першого ряду обґрунтовано у наступному розділі.

***Т е о р е м а  12 .*** *Нехай неперервна* $2\pi$ *-періодична функція* $f(x)$ *має неперервні похідні до* $(k-1)$*-го порядку* $(k \geq 1)$ *включно, а* $k$ *-а похідна кусково-гладка.*

*Тоді для коефіцієнтів Фур'є цієї функції справедливі нерівності*

$$|a_n| \leq \frac{A}{n^{k+1}}, \quad |b_n| \leq \frac{A}{n^{k+1}}, \, A = const. \qquad (2.81)$$

*Д о в е д е н н я .* Оскільки $f^{(k)}(x)$ – функція кусково-гладка і, відповідно, $f^{(k+1)}(x)$ – обмежена абсолютно інтегровна функція, то співвідношення (2.81) можна продовжити

$$a_n = -\frac{b_n^{(1)}}{n} = -\frac{a_n^{(2)}}{n^2} = ... = \frac{\alpha_n}{n^{k+1}},$$

$$b_n = \frac{a_n^{(1)}}{n} = -\frac{b_n^{(2)}}{n^2} = ... = \frac{\beta_n}{n^{k+1}}, \qquad (2.82)$$

де абсолютні величини коефіцієнтів $\alpha_n, \beta_n$ задаються (в залежності від парності числа $k$) одним з інтегралів

$$I_1 = \frac{1}{\pi}\left|\int_{-\pi}^{\pi} f^{(k+1)}(x)\cos nx\, dx\right|, \; I_2 = \frac{1}{\pi}\left|\int_{-\pi}^{\pi} f^{(k+1)}(x)\sin nx\, dx\right|.$$

Враховуючи тут обмеженість тригонометричних функцій і інтегровність функції $\left|f^{(k+1)}(x)\right|$, одержимо

$$I_1 = I_2 \leq \frac{1}{\pi}\int_{-\pi}^{\pi}\left|f^{(k+1)}(x)\right|dx = A.$$

Звідси з урахуванням залежностей (2.82) випливають нерівності (2.81).

Теорему доведено.

*П р и к л а д  8 .* Знайдемо оцінку для коефіцієнтів Фур'є $2\pi$-періодичної функції $f^*(x)$, яка на проміжку $[-\pi, \pi]$ задається формулою $f(x) = \left(\pi^2 - x^2\right)^2$.

Легко переконатися, що періодичне продовження функцій $f'(x) = -4x\left(\pi^2 - x^2\right)$ – неперервна функція. Продовження функції $f''(x) = 4\left(3x^2 - \pi^2\right)$ – неперервна кусково-гладка функція, оскільки перша похідна від цього продовження – кусково-неперервна функція. Отже, за теоремою 12 справедлива оцінка $|a_n| = \mathrm{O}\left(\dfrac{1}{n^3}\right)$.

**2.3.8. Застосування функцій комплексної змінної. Комплексна форма ряду Фур'є.** Нехай $F(z)$ – функція комплексної змінної $z = x + iy$, аналітична в крузі $|z| \le 1$, тобто функція не має у крузі і на його межі особливих точок. Тоді функція розвивається у степеневий ряд

$$F(z) = \sum_{n=0}^{\infty} c_n z^n. \qquad (2.83)$$

Вважаємо також, що коефіцієнти цього ряду дійсні числа.

Підставимо $z = e^{ix}$ у формулу (2.83) з урахуванням того, що у точках $|z| = |e^{ix}| = 1$ ряд збігається. Врахувавши при цьому формули Ейлера $e^{ix} = \cos x + i \sin x$ і $e^{inx} = \cos nx + i \sin nx$, одержимо

$$F(e^{ix}) = \sum_{n=0}^{\infty} c_n e^{inx} = c_0 + \sum_{n=0}^{\infty} c_n (\cos nx + i \sin nx) =$$

$$= c_0 + \sum_{n=0}^{\infty} c_n \cos nx + i \sum_{n=0}^{\infty} c_n \sin nx.$$

Виділивши дійсну і уявну частини функції

$$F(e^{ix}) = f(x) + i g(x),$$

матимемо такі їх розвинення у тригонометричні ряди:

$$f(x) = c_0 + \sum_{n=0}^{\infty} c_n \cos nx, \qquad (2.84)$$

$$g(x) = \sum_{n=0}^{\infty} c_n \sin nx.$$

Таким чином, використовуючи відомі розвинення аналітичних функцій, можна одержати розвинення дійсних і уявних частин цих функцій у тригонометричні ряди.

Можна також вирішити і обернену задачу: за заданим тригонометричним рядом побудувати відповідний степеневий ряд,

знайти його суму і, відділивши дійсну і уявну частини суми цього ряду, знайти суму тригонометричного ряду.

*П р и к л а д  9*. Відомо, що

$$e^z = \sum_{n=0}^{\infty} \frac{z^n}{n!}, \quad |z| < \infty.$$

Підставивши $z = e^{ix}$ у ліву і праву частини цієї рівності, одержимо

$$e^{e^{ix}} = e^{\cos x + i\sin x} = e^{\cos x} e^{i\sin x} = e^{\cos x}[\cos(\sin x) + i\sin(\sin x)] =$$
$$= e^{\cos x}\cos(\sin x) + ie^{\cos x}\sin(\sin x),$$

$$\sum_{n=0}^{\infty} \frac{e^{inx}}{n!} = 1 + \sum_{n=1}^{\infty} \frac{\cos nx}{n!} + i\sum_{n=1}^{\infty} \frac{\sin nx}{n!}.$$

Звідси

$$e^{\cos x}\cos(\sin x) = 1 + \sum_{n=1}^{\infty} \frac{\cos nx}{n!}, \quad e^{\cos x}\sin(\sin x) = \sum_{n=1}^{\infty} \frac{\sin nx}{n!}.$$

*П р и к л а д  10*. Знайти суму рядів

$$\sum_{n=0}^{\infty} \rho^n \cos nx, \quad \sum_{n=1}^{\infty} \rho^n \sin nx,$$

де $\rho$ – дійсне число, $|\rho| < 1$.

За ознакою Вейєрштрасса ряди абсолютно і рівномірно збігаються для будь-яких $x$. Побудуємо степеневий ряд і знайдемо його суму

$$\sum_{n=0}^{\infty} \rho^n \cos nx + i\sum_{n=1}^{\infty} \rho^n \sin nx = \sum_{n=0}^{\infty} \rho^n e^{inx} = \sum_{n=0}^{\infty} (\rho z)^n = \frac{1}{1-\rho z},$$
$$|z| \le 1.$$

де $z = e^{ix} = \cos x + i\sin x$.

Виділивши дійсну і уявну частини суми степеневого ряду, одержимо

$$\frac{1}{1-\rho e^{ix}} = \frac{1}{1-\rho\cos x - i\rho\sin x} = \frac{1-\rho\cos x + i\rho\sin x}{(1-\rho\cos x)^2 + \rho^2\sin^2 x}.$$

Отже, маємо такі суми рядів для всіх $x$

$$\frac{1-\rho\cos x}{(1-\rho\cos x)^2 + \rho^2 \sin^2 x} = \sum_{n=0}^{\infty} \rho^n \cos nx,$$

$$\frac{\sin x}{(1-\rho\cos x)^2 + \rho^2 \sin^2 x} = \sum_{n=1}^{\infty} \rho^n \sin nx.$$

*З а у в а ж е н н я .* Проведені в попередньому підрозділі викладки стосуються також аналітичних функцій, що мають на колі $|z|=1$ як особливі, так і неособливі точки. Тоді розглядаємо ряд (2.83), збіжний в точках кола, які не є особливими точками відповідної аналітичної функції. При цьому ряди (2.84) також збіжні для значень аргумента, що відповідають неособливим точкам функції (2.83).

**Знайдемо з урахуванням формул Ейлера комплексну форму ряду Фур'є.** Розглянемо абсолютно інтегровну (на будь-якому скінченному проміжку) $2\pi$-періодичну дійсну функцію $f(x)$ і знайдемо її ряд Фур'є

$$f(x) \sim \frac{a_0}{2} + \sum_{n=1}^{\infty}(a_n \cos nx + b_n \sin nx), \qquad (2.85)$$

де $\quad a_n = \frac{1}{\pi}\int\limits_{a}^{a+2\pi} f(x)\cos nx\, dx,\ n=0, 1, \ldots;\qquad b_n = \frac{1}{\pi}\int\limits_{a}^{a+2\pi} f(x)\sin nx\, dx$,

$n = 1, 2, \ldots$ – коефіцієнти Фур'є для функції $f(x)$.

Виразимо тригонометричні функції через показникові функції за формулами

$$\cos nx = \frac{1}{2}\left(e^{inx} + e^{-inx}\right),\ \sin nx = \frac{i}{2}\left(e^{-inx} - e^{inx}\right)$$

і запишемо ряд (2.85) у вигляді

$$f(x) \sim \frac{a_0}{2} + \sum_{n=1}^{\infty}\left[\frac{1}{2}(a_n + ib_n)e^{-inx} + \frac{1}{2}(a_n - ib_n)e^{inx}\right]$$

або скорочено

$$f(x) \sim \sum_{n=-\infty}^{\infty} c_n e^{inx}, \qquad (2.86)$$

де $c_0 = \frac{1}{2}a_0,\ c_n = \frac{1}{2}(a_n - ib_n),\ c_{-n} = \frac{1}{2}(a_n + ib_n),\ n = 1, 2, \ldots$.

При цьому очевидна залежність
$$c_{-n} = \overline{c}_n,$$
яка справедлива для дійсної функції. Нагадаємо, що якщо $z$ – комплексне число, то $\overline{z}$ – спряжене з ним число.

Якщо у виразах коефіцієнтів ряду (2.86) врахувати формули для коефіцієнтів Фур'є, то одержимо

$$c_n = \frac{1}{2\pi}\int_{-\pi}^{\pi} f(t)e^{-int}dt, \ n = 0, \pm 1, \ldots. \qquad (2.87)$$

За виконання достатніх умов збіжності ряду (2.85) для функції $f(x)$ збігається також ряд (2.86). При цьому під сумою ряду (2.86) розуміємо границю при $n \to \infty$ послідовності частинних сум $S_n(x) = \sum_{k=-n}^{n} c_n e^{inx}$,

$$\lim_{n\to\infty} \sum_{k=-n}^{n} c_n e^{inx} = \lim_{n\to\infty} S_n(x).$$

Таким чином визначена збіжність ряду називається збіжністю *в сенсі головного значення.*

Теорію розвинення функцій в ряди вигляду (2.86) можна також одержати, ґрунтуючись на ортогональності на відрізку $[-\pi, \pi]$ системи функцій $\{e^{inx}\}_{n=0}^{\infty}$. При цьому враховуємо, що за означенням скалярного добутку комплексно значних функцій справедлива формула

$$\left(e^{ikx}, e^{imx}\right) = \int_{-\pi}^{\pi} e^{ikx}\overline{e^{imx}}dx = \int_{-\pi}^{\pi} e^{ikx}e^{-imx}dx = \begin{cases} 0, \ k \neq m, \\ 2\pi, \ k = m. \end{cases}$$

### 2.4. Подвійні тригонометричні ряди

Нехай задана функція $f(x, y)$ двох дійсних змінних $x$ та $y$. Вважаємо, що вона має період $2\pi$ за обома змінними і абсолютно інтегровна у квадраті $Q = \{(x, y): |x| \leq \pi; |y| \leq \pi\}$.

Система функції
$$\{1, \cos mx, \sin mx, \cos ny, \sin ny,$$
$$\cos mx \cos ny, \sin mx \cos ny, \qquad (2.87)$$

$\cos mx \sin ny, \sin mx \sin ny\}$ $(m = 1, 2, ..., n = 1, 2, ...)$

є основною тригонометричною системою для випадку двох змінних.

Функції системи (2.87) ортогональні у квадраті $Q$, як і в будь-якому квадраті $\{(x, y): a \le x \le a + 2\pi; b \le x \le b + 2\pi\}$.

Маємо такі норми для функцій системи (2.87)

$\|1\| = 2\pi$, $\|\cos mx\| = \sqrt{2}\,\pi$, $\|\sin mx\| = \sqrt{2}\,\pi$,

$\|\cos mx \cos ny\| = \pi$, $\|\sin mx \cos ny\| = \pi$,

$\|\cos mx \sin ny\| = \pi$, $\|\sin mx \sin ny\| = \pi$.

Для коефіцієнтів Фур'є функції $f(x, y)$ за системою (2.87) одержимо вирази

$$A_{00} = \frac{1}{\|1\|^2} \iint_Q f(x, y) dxdy = \frac{1}{4\pi^2} \iint_Q f(x, y) dxdy,$$

$$A_{m0} = \frac{1}{\|\cos mx\|^2} \iint_Q f(x, y) \cos mx\, dxdy = \frac{1}{2\pi^2} \iint_Q f(x, y) \cos mx\, dxdy,$$

$$A_{0n} = \frac{1}{\|\cos ny\|^2} \iint_Q f(x, y) \cos ny\, dxdy = \frac{1}{2\pi^2} \iint_Q f(x, y) \cos ny\, dxdy,$$

$$B_{m0} = \frac{1}{\|\sin mx\|^2} \iint_Q f(x, y) \sin mx\, dxdy = \frac{1}{2\pi^2} \iint_Q f(x, y) \sin nx\, dxdy,$$

$$B_{0n} = \frac{1}{\|\sin ny\|^2} \iint_Q f(x, y) \sin ny\, dxdy =$$

$$= \frac{1}{2\pi^2} \iint_Q f(x, y) \sin ny\, dxdy; \qquad (2.88)$$

$$a_{mn} = \frac{1}{\pi^2} \iint_Q f(x, y) \cos mx\, \cos ny\, dxdy,$$

$$b_{mn} = \frac{1}{\pi^2} \iint_Q f(x, y) \sin mx\, \cos ny\, dxdy,$$

$$c_{mn} = \frac{1}{\pi^2} \iint\limits_{Q} f(x, y) \cos mx \sin ny \, dxdy,$$

$$d_{mn} = \frac{1}{\pi^2} \iint\limits_{Q} f(x, y) \sin mx \sin ny \, dxdy \qquad (2.89)$$

$$(m = 1, 2, \ldots, n = 1, 2, \ldots).$$

Прийнявши позначення

$$A_{00} = \frac{1}{4} a_{00}, \quad A_{m0} = \frac{1}{2} a_{m0}, \quad A_{0n} = \frac{1}{2} a_{0n}, \quad B_{m0} = \frac{1}{2} b_{m0}, \quad A_{0n} = \frac{1}{2} c_{0n},$$

коефіцієнти (2.88) можна обчислювати за формулами (2.89). Тоді ряд Фур'є функції $f(x, y)$ за системою (2.87) запишемо у вигляді

$$f(x, y) \sim \sum_{m=0}^{\infty} \sum_{n=0}^{\infty} \lambda_{mn} (a_{mn} \cos mx \cos ny + b_{mn} \sin mx \cos ny + $$
$$+ c_{mn} \cos mx \sin ny + d_{mn} \sin mx \sin ny), \qquad (2.90)$$

де

$$\lambda_{mn} = \begin{cases} 1/4, & m = n = 0, \\ 1/2, & m > 0, n = 0 \cup m = 0, n > 0, \\ 1, & m > 0, n > 0. \end{cases}$$

Частинну суму ряду (2.90) запишемо за аналогією з рядом функції однієї змінної у вигляді

$$S_{MN}(x, y) = \sum_{m=0}^{M} \sum_{n=0}^{N} \lambda_{mn} (a_{mn} \cos mx \cos ny + b_{mn} \sin mx \cos ny +$$
$$+ c_{mn} \cos mx \sin ny + d_{mn} \sin mx \sin ny). \qquad (2.91)$$

Використавши тут формули (2.89), одержимо

$$S_{MN}(x, y) = \frac{1}{\pi^2} \sum_{m=0}^{M} \sum_{n=0}^{N} \lambda_{mn} \iint\limits_{Q} f(s, t) \cos m(s-x) \cos n(t-y) dsdt =$$

$$= \frac{1}{\pi^2} \iint\limits_{Q} f(s, t) \left[ \frac{1}{2} + \sum_{m=1}^{M} \cos m(s-x) \right] \left[ \frac{1}{2} + \sum_{m=1}^{M} \cos n(t-y) \right] dsdt$$

або з врахуванням формули (2.57)

$$S_{MN}(x, y) =$$

$$= \frac{1}{\pi^2} \iint_Q f(s,t) \frac{\sin\left[\left(M+\frac{1}{2}\right)(s-x)\right]\sin\left[\left(N+\frac{1}{2}\right)(t-y)\right]}{4\sin\frac{s-x}{2}\sin\frac{t-y}{2}} dsdt.$$

Якщо врахувати періодичність підінтегральних функцій, то вираз частинної суми запишемо у аналогічному з одновимірним випадком вигляді:

$$S_{MN}(x,y) = \frac{1}{\pi^2} \iint_Q f(x+u, y+v) D_M(u) D_N(v) dudv, \qquad (2.92)$$

де $D_M(u) = \dfrac{\sin\left(M+\frac{1}{2}\right)u}{2\sin\frac{u}{2}}; \quad D_N(v) \dfrac{\sin\left(N+\frac{1}{2}\right)v}{2\sin\frac{v}{2}}$.

Ряд (2.90) збігається до функції $f(x,y)$ в точці $(x,y)$, якщо
$$\lim_{\substack{M \to \infty \\ N \to \infty}} S_{MN}(x,y) = f(x,y). \qquad (2.93)$$

Розглянемо достатні умови розвинення функції в ряд за системою (2.87).

*Т е о р е м а  1 . Нехай $f(x,y)$ – $2\pi$-періодична за обома змінними і неперервна в квадраті $Q$ функція з неперервними частинними похідними $\dfrac{\partial f}{\partial x}$, $\dfrac{\partial f}{\partial y}$.*

*Тоді, якщо в деякому околі точки $(x,y)$ неперервною є похідна $\dfrac{\partial^2 f}{\partial x \partial y}$, то ряд Фур'є функції $f(x,y)$ збігається до $f(x,y)$ в цій точці.*

Д о в е д е н н я . Потрібно довести рівність (2.93), яку запишемо у вигляді

$$\lim_{\substack{M \to \infty \\ N \to \infty}} \frac{1}{\pi^2} \iint_Q f(x+u, y+v) D_M(u) D_N(v) dudv = f(x,y). \quad (2.94)$$

Розглядаючи функцію $f(x,y)$, як функцію одного змінного, за теоремою 6 (п. 2.3) справедливі рівності

$$\lim_{M \to \infty} \frac{1}{\pi^2} \iint_Q f(x+u, y) D_M(u) D_N(v) \, dudv = f(x, y),$$

$$\lim_{N \to \infty} \frac{1}{\pi^2} \iint_Q f(x, y+v) D_M(u) D_N(v) \, dudv = f(x, y).$$

Враховуючи, що

$$\frac{1}{\pi^2} \iint_Q f(x, y) D_M(u) D_N(v) dudv = f(x, y),$$

рівність (2.94) еквівалентна такій рівності

$$\lim_{\substack{M \to \infty \\ N \to \infty}} \frac{1}{\pi^2} \iint_Q \bigl[ f(x+u, y+v) - f(x+u, y) -$$
$$- f(x, y+v) + f(x, y) \bigr] D_M(u) D_N(v) dudv = 0. \qquad (2.95)$$

Отже, задача полягає у встановленні рівності (2.95). Спочатку встановимо абсолютну інтегровність наступної функції змінних $u$ і $v$

$$\varphi(u, v) = \frac{f(x+u, y+v) - f(x+u, y) - f(x, y+v) + f(x, y)}{4 \sin \frac{u}{2} \sin \frac{v}{2}} =$$

$$= \frac{f(x+u, y+v) - f(x+u, y) - f(x, y+v) + f(x, y)}{uv} \cdot \frac{uv}{4 \sin \frac{u}{2} \sin \frac{v}{2}}.$$

Оскільки існує змішана друга похідна від функції $f(x, y)$ у точці $(x, y)$, то відношення

$$\frac{f(x+u, y+v) - f(x+u, y) - f(x, y+v) + f(x, y)}{uv} =$$
$$= \frac{\Delta_y f(x+u, y) - \Delta_y f(x, y)}{uv} = \frac{\Delta_x f(x, y+v) - \Delta_x f(x, y)}{uv} \qquad (2.96)$$

є обмеженим при $u \to 0$ і $v \to 0$, тобто існує $\delta > 0$ таке, що

$$\frac{f(x+u, y+v) - f(x+u, y) - f(x, y+v) + f(x, y)}{uv} \le M = const$$

для всіх $-\delta \le u \le \delta$ і $-\delta \le v \le \delta$. Оскільки функція $f(x, y)$ неперервна, це відношення абсолютно інтегровна функція в області $-\delta \le u \le \delta$, $-\delta \le v \le \delta$. Зовні цієї області відношення є абсолютно інтегрованою функцією, оскільки функція $f(x,y)$ неперервна і $|u| > 0$ і $|v| > 0$.

Отже, відношення (2.96) абсолютно інтегровна функція в області $Q$.

Функція
$$\frac{uv}{4\sin\frac{u}{2}\sin\frac{v}{2}} \qquad (2.97)$$

неперервна при $u \neq 0$, $v \neq 0$ і прямує до одиниці при $u \to 0$, $v \to 0$, оскільки виконується рівність $\lim\limits_{\alpha \to 0}\frac{\sin\alpha}{\alpha} = 1$.

Отже, функція $\varphi(u,v)$ дорівнює добутку абсолютно інтегровної функції (2.96) і обмеженої функції (2.97), а тому сама є абсолютно інтегровною.

Тоді інтеграл у рівності (2.95) запишемо у вигляді
$$\frac{1}{\pi^2}\iint_Q [f(x+u, y+v) - f(x+u, y) -$$
$$- f(x, y+v) + f(x,y)] D_M(u) D_N(v) du dv =$$
$$= \frac{1}{\pi^2}\iint_Q \varphi(u,v)\sin\left(M+\frac{1}{2}\right)u \sin\left(N+\frac{1}{2}\right)v \, du dv.$$

Звідси, спираючись на двовимірне узагальнення леми 1 (п. 2.3), одержимо рівність (2.95) і, відповідно, (2.94).

Теорему доведено.

За аналогією з теоремою 5 (п. 2.3) з використанням результатів теореми 1 може бути доведена теорема про рівномірну збіжність подвійного ряду (2.90).

*Т е о р е м а 2*. *Нехай $f(x,y)$ – $2\pi$-періодична за обома змінними неперервна функція з неперервними частинними похідними $\frac{\partial f}{\partial x}$, $\frac{\partial f}{\partial y}$ і $\frac{\partial^2 f}{\partial x \partial y}$.*

*Тоді ряд Фур'є функції $f(x,y)$ збігається до $f(x,y)$ абсолютно і рівномірно.*

*П р и к л а д .* Нехай функція $f(x, y) = xy$ задана у квадраті $Q$. Якщо її продовжити, як періодичну функцію за обома змінними з періодом $2\pi$, то справджуються умови теореми 1. Коефіцієнти Фур'є знайдемо за формулами

$$a_{mn} = 0, \ b_{mn} = 0, \ c_{mn} = 0, \ (m = 0, 1, \ldots, n = 0, 1, \ldots);$$

$$d_{mn} = \frac{1}{\pi^2} \iint\limits_Q f(x, y) \sin mx \, \sin ny \, dxdy = \frac{4(-1)^{m+n}}{mn}$$

$$(m = 1, 2, \ldots, n = 1, 2, \ldots).$$

За формулою (2.90) знайдемо ряд

$$xy = \frac{4}{\pi^2} \sum_{m=1}^{\infty} \sum_{n=1}^{\infty} \frac{\sin mx \, \sin ny}{mn}, \ x \in (-\pi, \pi), \ y \in (-\pi, \pi).$$

Оскільки функція $f(x, y)$ зображена у вигляді добутку двох функцій, залежних окремо від змінних $x$ та $y$, її ряд зображується також у вигляді добутку двох рядів.

### 2.4. Завдання до другого розділу

1. Дослідити на рівномірну збіжність ряд $1 + \sum\limits_{k=1}^{\infty} \left( x^k - x^{k-1} \right), \ x \in [0, 1]$.

(Знайдемо часткову суму ряду $S_n(x) = x^n$ і її границю

$$S(x) = \lim_{n \to \infty} S_n(x) = \begin{cases} 1, & x = 1, \\ 0, & 0 \leq x < 1. \end{cases}$$

Отже, ряд збігається на відрізку $[0, 1]$, однак нерівномірно, оскільки $\sup\limits_{x \in [0, 1]} |S(x) - S_n(x)| = \sup\limits_{x \in [0, 1]} |x^n| = 1$ не прямує до нуля, коли $n \to \infty$).

Ряд рівномірно збігається на будь-якому відрізку $[0, q]$, де $0 < q < 1$, оскільки $\sup\limits_{x \in [0, q]} |S(x) - S_n(x)| = \sup\limits_{x \in [0, q]} |x^n| = q^n \underset{n \to \infty}{\to} 0$).

2. Дослідити на рівномірну збіжність ряди:

a) $\sum\limits_{n=1}^{\infty} \frac{n}{x^n}$ (ряд рівномірно збіжний на множині $|x| > 1$);

б) $\sum_{n=1}^{\infty} \dfrac{n^p \sin nx}{1+n^2}$ (при $p < 1$ ряд абсолютно і рівномірно збіжний в області $(-\infty, \infty)$; при $1 \leq p < 2$ ряд рівномірно збіжний на будь-якому сегменті, що не містить точок $x_m = 2\pi m$, $m = 0, \pm 1, \ldots$; при $p \geq 2$ ряд розбіжний);

в) $\sum_{n=1}^{\infty} \dfrac{(-1)^{n+1}}{n+x^2}$ (ряд рівномірно збіжний в області $(-\infty, \infty)$);

г) $\sum_{n=1}^{\infty} e^{-n^2 x^2} \sin nx$ (ряд рівномірно збіжний в області $(-\infty, \infty)$);

д) $\sum_{n=1}^{\infty} \dfrac{x \sin nx}{n(1+nx^2)}$ (ряд рівномірно збіжний в області $(-\infty, \infty)$).

3. Дослідити на рівномірну збіжність ряди, одержані від почленного диференціювання рядів:

а) $\sum_{k=1}^{\infty} \dfrac{\sin kx}{k^{\alpha}}$; б) $\sum_{k=1}^{\infty} \dfrac{\cos kx}{k^{\alpha}}$.

4. Дослідити на рівномірну збіжність ряди:

а) $\sum_{k=2}^{\infty} \dfrac{\sin kx}{\ln k}$; б) $\sum_{k=2}^{\infty} \dfrac{\cos kx}{\ln k}$.

5. Показати, виходячи з означення рівномірної збіжності, що якщо рівномірно збіжний ряд помножити на обмежений множник, не залежний від $n$, то одержаний при цьому ряд рівномірно збігається.

6. Знайти радіус збіжності рядів:

а) $\sum_{n=1}^{\infty} n^2 x^n$; б) $\sum_{n=1}^{\infty} \dfrac{x^n}{n^2}$; в) $\sum_{n=1}^{\infty} \dfrac{x^n}{n^n}$; г) $\sum_{n=1}^{\infty} \dfrac{x^n}{(n+1)(n+2)}$.

7. Дослідити ряди на рівномірну збіжність в області $[-1, q]$, $0 < q < 1$:

а) $\sum_{n=1}^{\infty} x^n$; б) $\sum_{n=1}^{\infty} nx^n$.

8. Показати, що на кінцях інтервалу $(-1, 1)$ ряд (2.40): а) розбігається при $m \leq 1$ в точці $x = 1$; б) збігається не абсолютно при $-1 < m < 0$ в точці $x = 1$; в) збігається абсолютно при $m > 0$ в точці $x = 1$; г) розбігається при $m < 0$ в точці $x = -1$; д) збігається абсолютно в точці $x = -1$.

9. Знайти суму ряду $S(x) = \sum_{n=1}^{\infty} n^2 x^n$ ($S(x) = \dfrac{x(1+x)}{(1-x)^3}$).

10. Використовуючи біномний ряд, обчислити $\sqrt[3]{30} = 3\sqrt[3]{1+\frac{1}{9}}$ з точністю 0,0001 ( $\sqrt[3]{30} = 3{,}1155$ ).

11. Обчислити наближено інтеграл $I = \int\limits_0^{1/3} e^{-x^2}\,dx$ з точністю до 0,001 ( $I = 0{,}321$ ).

12. Довести, виходячи з означення, якщо ряд (2.52) рівномірно збігається на відрізку $[-\pi, \pi]$ і $\varphi(x)$ – неперервна на цьому відрізку функція, то рівномірно збіжним на цьому відрізку є також ряд

$$f(x)\varphi(x) = \frac{a_0}{2}\varphi(x) + \sum_{k=1}^{\infty}[a_k \cos kx\,\varphi(x) + b_k \sin kx\,\varphi(x)].$$

13. Показати, що справедливі розвинення:

а) $e^{\alpha x} = \dfrac{2}{\pi}\operatorname{sh}\alpha\pi\left[\dfrac{1}{2\alpha} + \sum_{n=1}^{\infty}\dfrac{(-1)^n}{\alpha^2 + n^2}(\alpha\cos nx - n\sin nx)\right], \quad x \in (-\pi, \pi)$;

б) $\cos\alpha x = \dfrac{2}{\pi}\sin\alpha\pi\left[\dfrac{1}{2\alpha} + \sum_{n=1}^{\infty}(-1)^n\dfrac{\alpha\cos nx}{\alpha^2 - n^2}\right], \; x \in [-\pi, \pi]$;

в) $\sin\alpha x = \dfrac{2}{\pi}\sin\alpha\pi\sum_{n=1}^{\infty}(-1)^n\dfrac{\alpha\sin nx}{\alpha^2 - n^2}, \; x \in (-\pi, \pi)$;

г) $f(x) = \dfrac{2h}{\pi}\left[\dfrac{1}{2} + \sum_{n=1}^{\infty}\dfrac{\sin nh}{nh}\cos nx\right], \; x \in [0, \pi]$, де $f(x) = \begin{cases} 1, & 0 \le x \le h, \\ 0, & h < x \le \pi; \end{cases}$

д) $1 = \dfrac{4}{\pi}\sum_{n=1}^{\infty}\dfrac{\sin(2n-1)\pi x}{2n-1}, \; x \in (0, 1)$;

е) $x = 2\sum_{n=1}^{\infty}(-1)^{n-1}\dfrac{\sin nx}{n}, \; x \in (-\pi, \pi)$;

є) $f(x) = \dfrac{2h}{\pi}\left[\dfrac{1}{2} + \sum_{n=1}^{\infty}\left(\dfrac{\sin nh}{nh}\right)^2\cos nx\right], \; x \in [0, \pi]$,

де $f(x) = \begin{cases} 1 - \dfrac{x}{2h}, & 0 \le x \le h, \\ 0, & h < x \le \pi. \end{cases}$

14. Показати справедливість розвинень:

а) $x = \pi - 2\sum_{n=1}^{\infty}\frac{1}{n}\sin nx$, $x \in (0, 2\pi)$;

б) $x^2 = \frac{4\pi^2}{3} - 4\sum_{n=1}^{\infty}\left(\frac{1}{n^2}\cos nx - \frac{\pi}{n}\sin nx\right)$, $x \in (0, 2\pi)$.

15. Показати, що для коефіцієнтів Фур'є функції $f(x)$ справедлива оцінка:

а) $|a_n| = \mathrm{O}\left(\frac{1}{n^2}\right)$ на проміжку $[-1, 1]$, якщо $f(x) = x^6$;

б) $|a_n| = \mathrm{O}\left(\frac{1}{n}\right)$ на проміжку $[-\pi, \pi]$, якщо $f(x) = \pi x + x^2$;

в) $|b_n| = \mathrm{O}\left(\frac{1}{n^3}\right)$ на проміжку $[-\pi, \pi]$, якщо $f(x) = \pi^2 x - x^3$.

16. Використовуючи розвинення функцій у прикладі 9 (п. 2.3), знайти суми рядів:

а) $1 + \sum_{n=2,4,\ldots}^{\infty}\frac{\cos nx}{n!}$; б) $\sum_{n=1,3,\ldots}^{\infty}\frac{\sin nx}{n!}$; в) $\sum_{n=1,3,\ldots}^{\infty}\frac{\cos nx}{n!}$,

г) $\sum_{n=2,4,\ldots}^{\infty}\frac{\sin nx}{n!}$; д) $1 + \sum_{n=1}^{\infty}(-1)^n\frac{\cos 2nx}{(2n)!}$;

е) $\sum_{n=1}^{\infty}(-1)^{n-1}\frac{\sin(2n-1)x}{(2n-1)!}$; є) $\sum_{n=1}^{\infty}(-1)^{n-1}\frac{\cos(2n-1)x}{(2n-1)!}$.

17. Інтегруючи ряд $\sum_{n=1}^{\infty}\frac{\sin nx}{n} = \begin{cases}(-\pi-x)/2, & -2\pi < x < 0, \\ (\pi-x)/2, & 0 < x < 2\pi, \\ 0, & x = 0, x = \pm 2\pi\end{cases}$

з врахуванням рівності $\sum_{n=1,\ldots}^{\infty}\frac{1}{n^2} = \frac{\pi^2}{6}$, знайти розвинення:

а) $\sum_{n=1}^{\infty}\frac{\cos nx}{n^2} = \begin{cases}\dfrac{2\pi^2 + 6\pi x + 3x^2}{12}, & -2\pi \leq x \leq 0, \\ \dfrac{2\pi^2 - 6\pi x + 3x^2}{12}, & 0 \leq x \leq 2\pi;\end{cases}$

б) $\sum_{n=1}^{\infty}\frac{\sin nx}{n^3} = \begin{cases}\dfrac{2\pi^2 x + 3\pi x^2 + x^3}{12}, & -2\pi \leq x \leq 0, \\ \dfrac{2\pi^2 x - 3\pi x^2 + x^3}{12}, & 0 \leq x \leq 2\pi.\end{cases}$

# Р О З Д І Л  III

# ЗБІЖНІ В СЕРЕДНЬОМУ РЯДИ
________________________________________

### 3.1. Узагальнений ряд Фур'є

**3.1.1. Ортогональні системи функцій. Евклідів простір.** У попередньому розділі розглянуто ортогональні системи тригонометричних функцій і ряди за цими системами. Дослідимо ряди за довільними ортогональними системами функцій $[5, 9, 11]$.

Розглянемо лінійний простір $E$ функцій, інтегровних на проміжку $(a, b)$, і нехай

$$\varphi_0(x), \varphi_1(x), \varphi_2(x), ..., \varphi_n(x), ...$$

– система функцій цього простору.

Система $\{\varphi_n(x)\}$ ортогональна на проміжку $(a, b)$, якщо виконуються співвідношення

$$\int_a^b \varphi_k(x)\, \varphi_n(x)\, dx = \begin{cases} 0, & k \neq n, \\ \alpha_n \neq 0, & k = n. \end{cases}$$

Поняття ортогональності безпосередньо пов'язане з поняттям скалярного добутку.

*Скалярним добутком двох функцій* $f(x) \in E$ *і* $\varphi(x) \in E$ *називається інтеграл*

$$(f, \varphi) = \int_a^b f(x)\varphi(x)\, dx. \qquad (3.1)$$

*Легко переконатися, що виконуються умови:*
1) $(f, \varphi) = (\varphi, f)$;
2) $(f, f) \geq 0$ *і з рівності* $(f, f) = 0$ *випливає, що* $f(x) = 0$ *на* $[a, b]$, *хіба що за виключенням скінченного числа точок;*
3) $(\alpha f_1 + \beta f_2, \varphi) = \alpha(f_1, \varphi) + \beta(f_2, \varphi)$.

Простір зі скалярним добутком є нормованим простором, тобто, якщо у функціональному просторі $E$ визначений скалярний

добуток, то величина

$$\| f \| = \sqrt{(f, f)} = \left( \int_a^b f^2(x) dx \right)^{1/2} \qquad (3.2)$$

є нормою цього простору.

Дійсно, у підрозділі 1.4 розглянуто функціональні простори з нормою $\| f \|_{L_p} = \left( \int_a^b | f(x) |^p dx \right)^{1/p}$, $p = 1, 2, ...$ . Тому вираз (3.2) є нормою простору $L_2(a, b)$.

Отже, простір $E$ є нормованим простором з нормою
$$\| f \| = \| f \|_{L_2}$$

і це – простір $L_2(a, b)$, або його підпростори – $CL_2[a, b]$, $QL_2[a, b]$, $L'_2(a, b)$.

**О з н а ч е н н я   1 .** *Простір з нормою, породженою скалярним добутком, називається евклідовим простором.*

Функцію $f(x) \in E$, що справджує норму (3.2), називають інтегровною в середньому або інтегровною з квадратом на проміжку $(a, b)$.

Класичним *прикладом* евклідового простору є простір кусково-неперервних на проміжку $(a, b)$ функцій зі значеннями в точках розриву, що є середніми арифметичними її граничних значень зліва і справа у точках розриву.

*Прикладом* евклідового простору є простір $R^n$ (дійсний простір векторів $x = (x_1, x_2, ... x_n)$) зі скалярним добутком $(x, y) = \sum_{k=1}^{n} x_k y_k$ .

**Т е о р е м а   1 .** *Ортогональна система функцій $\{\varphi_n(x)\} \subset E$ лінійно незалежна в $E$.*

*Д о в е д е н н я .* Система елементів $\{f_n\}$ з деякого простору є лінійно незалежною за нормою цього простору, якщо будь-яке скінченне число елементів цієї системи є лінійно незалежною системою за нормою, тобто рівність

$$c_1 f_1 + c_2 f_2 + \ldots c_k f_k \overset{E}{=} o$$

можлива лише за значень $c_1 = c_2 = \ldots = c_k = 0$, де $o$ – нульовий елемент простору.

Розглянемо лінійну комбінацію скінченного числа елементів ортогональної системи $\{\varphi_n(x)\} \subset E$ і нехай

$$\|c_1\varphi_1(x) + c_2\varphi_2(x) + \ldots c_k\varphi_k(x) - o\| =$$

$$= \left[\int_a^b [c_1\varphi_1(x) + c_2\varphi_2(x) + \ldots c_k\varphi_k(x) - o]^2 dx\right]^{1/2} = 0.$$

Врахувавши тут ортогональність системи, отримаємо $c_1^2\|\varphi_1\|^2 + c_2^2\|\varphi_2\|^2 + \ldots c_k^2\|\varphi_k\|^2 = 0$. Звідки, оскільки $\|\varphi_m\| \neq 0$, випливає $c_1 = c_2 = \ldots = c_k = 0$.

Теорему доведено.

*Н а с л і д о к.* *Якщо для довільної системи функцій $\{f_n(x)\}_1^N \subset E$ визначник*

$$G(f_1, \ldots, f_N) = \begin{vmatrix} (f_1, f_1) & (f_1, f_2) & \ldots & (f_1, f_N) \\ (f_2, f_1) & (f_2, f_2) & \ldots & (f_2, f_N) \\ \ldots & \ldots & \ldots & \ldots \\ (f_N, f_1) & (f_N, f_2) & \ldots & (f_N, f_N) \end{vmatrix}$$

*дорівнює нулю, то система лінійно залежна.*

*Визначник $G(f_1, \ldots, f_N)$ називається визначником Грамма даної системи.*

*Д о в е д е н н я.* Розглянемо лінійну системи $N$ рівнянь з $N$ невідомими $c_i$, $i = 1, 2, \ldots, N$,

$(c_1 f_1 + \ldots + c_N f_N, f_i) = 0$, $i = 1, 2, \ldots, N$, або
$c_1(f_1, f_i) + \ldots + c_N(f_N, f_i) = 0$, $i = 1, 2, \ldots, N$.

Визначник цієї системи є транспонованим визначником Грамма, який дорівнює нулеві. Отже, система має нетривіальний розв'язок $c_i$, $i = 1, 2, \ldots, N$. Помножимо рівняння цієї системи, відповідно, на $c_i$ і просумуємо по $i$ від 1 до $N$,

$$(c_1 f_1 + \ldots + c_N f_N, c_1 f_1 + \ldots + c_N f_N) = 0.$$

Звідси $c_1 f_1 + \ldots + c_N f_N \overset{E}{=} o$, що доводить лінійну залежність системи.

Твердження доведено.

Одночасно з ортогональною системою функцій можна ввести ортонормовану систему функцій.

*О з н а ч е н н я  2 .* *Система функцій $\{\widetilde{\varphi}_n(x)\} \subset E$ називається ортонормованою (ОНС), якщо вона ортогональна і норма кожного її елемента дорівнює одиниці,*

$$\int_a^b \widetilde{\varphi}_n(x)\widetilde{\varphi}_m(x)dx = \begin{cases} 0, & n \neq m, \\ 1, & n = m. \end{cases} \quad (3.3)$$

Очевидно кожну ортогональну систему функцій можна нормувати. Якщо система $\{\varphi_n(x)\}$ ортогональна в $E$, то система $\left\{\widetilde{\varphi}_n(x) = \dfrac{\varphi_n(x)}{\|\varphi_n\|}\right\}$ ортонормована в $E$.

*П р и к л а д  1 .* Ортогональну на відрізку $[-\pi, \pi]$ тригонометричну систему функцій

$$1, \cos x, \sin x, \ldots, \cos nx, \sin x, \ldots$$

звести до ортонормованої системи.

Якщо врахувати норми елементів $\|1\| = \sqrt{2\pi}$, $\|\cos nx\| = \|\sin nx\| = \sqrt{\pi}$, то одержимо відповідну ортонормовану систему функцій

$$\frac{1}{\sqrt{2\pi}}, \frac{\cos x}{\sqrt{\pi}}, \frac{\sin x}{\sqrt{\pi}}, \ldots, \frac{\cos nx}{\sqrt{\pi}}, \frac{\sin x}{\sqrt{\pi}}, \ldots .$$

**3.1.2. Многочлени Лежандра.** Розглянемо систему многочленів Лежандра $[11]$, ортогональну на відрізку $[-1, 1]$,

$$P_0(x) = 1, \ P_n(x) = \frac{1}{2^n n!} \frac{d^n (x^2 - 1)^n}{dx^n}, \ n = 1, 2, \ldots .$$

1. Покажемо, що многочлен $P_n(x)$ ортогональний на відрізку $[-1, 1]$ до будь-якого многочлена $\sigma_m(x)$ степеня $m < n$. Інтегруючи частинами, одержимо

$$2^n n!(P_n, \sigma_m) = 2^n n! \int_{-1}^{1} P_n(x)\sigma_m(x)dx = \frac{d^{n-1}(x^2-1)^n}{dx^{n-1}}\sigma_m(x)\Big|_{-1}^{1} -$$

$$- \int_{-1}^{1} \frac{d^{n-1}(x^2-1)^n}{dx^{n-1}}\sigma'_m(x)dx = \ldots = (-1)^m \int_{-1}^{1} \frac{d^{n-m}(x^2-1)^n}{dx^{n-m}}\sigma_m^{(m)}(x)dx.$$

Звідси, оскільки $\sigma_m^{(m)}(x) = const$, маємо $(P_n, \sigma_m) = 0$ і, відповідно, $(P_n, P_m) = 0$ при $m \neq n$.

Знайдемо норму многочлена $P_n(x)$. Інтегруючи частинами інтеграл у виразі скалярного добутку, одержимо

$$(P_n, P_n) = \frac{1}{2^{2n}(n!)^2} \int_{-1}^{1} \left[(x^2-1)^n\right]^{(n)}\left[(x^2-1)^n\right]^{(n)} dx = \ldots =$$

$$= \frac{(-1)^n}{2^{2n}(n!)^2} \int_{-1}^{1} \left[(x^2-1)^n\right]^{(2n)} (x^2-1)^n dx = \frac{(-1)^n (2n)!}{2^{2n}(n!)^2} \int_{-1}^{1} (x^2-1)^n dx.$$

Аналогічно, інтегруючи частинами інтеграл

$$\int_{-1}^{1} (x^2-1)^n dx = \int_{-1}^{1} (x-1)^n (x+1)^n dx = \ldots =$$

$$= \frac{(-1)^n (n!)^2}{(2n)!} \int_{-1}^{1} (x+1)^{2n} dx = \frac{(-1)^n (n!)^2 2^{2n+1}}{(2n+1)!},$$

одержимо $(P_n, P_n) = \dfrac{2}{2n+1}$, тобто

$$\|P_n\| = \sqrt{\frac{2}{2n+1}}.$$

Отже, ортонормована на відрізку $[-1, 1]$ система многочленів Лежандра має вигляд

$$\widetilde{P}_n(x) = \sqrt{\frac{2n+1}{2}} \frac{1}{2^n n!} \frac{d^n (x^2-1)^n}{dx^n}, \ n = 0, 1, \ldots.$$

Розглянемо ще інші властивості многочленів Лежандра.

2. Твірна функція системи многочленів Лежандра

$$\Psi(x,\alpha) = \sum_{n=0}^{\infty} P_n(x)\alpha^n$$

має вигляд

$$\Psi(x,\alpha) = \frac{1}{\sqrt{1+\alpha^2 - 2\alpha x}}.$$

Розвиваючи функцію $\Psi(x,\alpha)$ за степенями змінної $\alpha \in (-1, 1)$ і $x \in [-1, 1]$ з використанням формули $\frac{1}{\sqrt{1+t}} = \sum_{k=0}^{\infty} \frac{(-1)^k C_{2k}^k t^k}{2^{2k}}$, $t \in (-1, 1)$, знайдемо

$$\Psi(x,\alpha) = \sum_{k=0}^{\infty} \frac{(-1)^k C_{2k}^k}{2^{2k}} \left(\alpha^2 - 2\alpha x\right)^k =$$

$$= \sum_{k=0}^{\infty} \frac{(-1)^k C_{2k}^k}{2^{2k}} \sum_{m=0}^{k} C_k^m (-2x)^{k-m} \alpha^{k+m} =$$

$$= \sum_{k=0}^{\infty} \frac{(-1)^k C_{2k}^k}{2^{2k}} \sum_{n=k}^{2k} C_k^{n-k} (-2x)^{2k-n} \alpha^n =$$

$$= \sum_{n=0}^{\infty} \frac{\alpha^n}{2^n} \sum_{k=[(n+1)/2]}^{n} (-1)^{k-n} C_{2k}^k C_k^{n-k} x^{2k-n} =$$

$$= \sum_{n=0}^{\infty} \frac{\alpha^n}{2^n} \sum_{k=0}^{[n/2]} (-1)^k C_{2(n-k)}^{n-k} C_{n-k}^k x^{n-2k}$$

або, врахувавши співвідношення $C_{2(n-k)}^{n-k} C_{n-k}^k = C_n^k C_{2(n-k)}^n$,

$$\Psi(x,\alpha) = \sum_{n=0}^{\infty} \frac{\alpha^n}{2^n} \sum_{k=0}^{[n/2]} (-1)^k C_n^k C_{2(n-k)}^n x^{n-2k}.$$

Звідси одержимо загальну формулу для многочленів Лежандра

$$P_n(x) = \frac{1}{2^n} \sum_{k=0}^{[n/2]} (-1)^k C_n^k C_{2(n-k)}^n x^{n-2k}, \ n = 0, 1, \ldots. \qquad (3.4)$$

Для парних та непарних значень індексів маємо такі формули:

$$P_{2n}(x) = \frac{1}{2^{2n}} \sum_{k=0}^{n} (-1)^{n-k} C_{2n}^{n-k} C_{2(n+k)}^{2n} x^{2k},$$

$$P_{2n+1}(x) = \frac{1}{2^{2n+1}} \sum_{k=0}^{n} (-1)^{n-k} C_{2n+1}^{n-k} C_{2(n+1+k)}^{2n+1} x^{2k+1}, \ n = 0, 1, \ldots.$$

3. Справедливі формули
$$P_n(-x) = (-1)^n P_n(x), \ P_n(1) = 1, \ P_n(-1) = (-1)^n,$$
які випливають з формули (3.4) з урахуванням комбінаторної тотожності
$$\sum_{k=0}^{[n/2]} (-1)^k C_n^k C_{2(n-k)}^n = 2^n.$$

4. Многочлени Лежандра задовольняють рекурентне співвідношення
$$(n+1)P_{n+1}(x) = (2n+1)x P_n(x) - n P_{n-1}(x), \ n > 0. \qquad (3.5)$$

З цього співвідношення за першими двома многочленами $P_0 = 1$ і $P_1 = x$ можна визначити систему многочленів Лежандра.

Доведемо виконання рівності (3.5). Оскільки $xP_n(x)$ – многочлен $(n+1)$-го степеня, знайдемо коефіцієнти його розвинення за многочленами Лежандра
$$xP_n(x) = a_0 P_0(x) + \ldots + a_n P_n(x) + \frac{\mu_n}{\mu_{n+1}} P_{n+1}(x), \qquad (3.6)$$

де $\mu_n = \frac{1}{2^n} C_{2n}^n$ – старший член многочлена $P_n(x)$. Тут враховано рівність коефіцієнтів (зліва і справа) при $x^{n+1}$.

Помножимо рівність (3.6) на $P_k(x)$, $k < n-1$, і проінтегруємо на відрізку $[-1, 1]$. Оскільки $xP_k(x)$ – многочлен степеня $< n$, то
$$\int_{-1}^{1} x P_k(x) P_n(x) dx = 0 \text{ і, відповідно, } a_k \int_{-1}^{1} P_k^2(x) dx = 0.$$

Отже,
$$a_k = 0, \quad k = 0, 1, \ldots, n-2,$$
а з властивості 1 (рівність містить або тільки парні або тільки непарні степені змінної) маємо $a_n = 0$. Тоді рівність (3.6) прийме вигляд
$$x P_n(x) = a_{n-1} P_{n-1}(x) + \frac{\mu_n}{\mu_{n+1}} P_{n+1}(x).$$

Для визначення коефіцієнта $a_{n-1}$ помножимо рівність на $P_{n-1}(x)$ і проінтегруємо на відрізку $[-1, 1]$ (з повторним її використанням)

$$a_{n-1}\|P_{n-1}\|^2 = \int_{-1}^{1} x\, P_{n-1}(x) P_n(x) dx =$$

$$= \int_{-1}^{1}\left(a_{n-2} P_{n-2}(x) + \frac{\mu_{n-1}}{\mu_n} P_n(x)\right) P_n(x) dx = \frac{\mu_{n-1}}{\mu_n}\|P_n\|^2.$$

Отже,

$$x\, P_n(x) = \frac{\mu_{n-1}}{\mu_n} \frac{\|P_n\|^2}{\|P_{n-1}\|^2} P_{n-1}(x) + \frac{\mu_n}{\mu_{n+1}} P_{n+1}(x).$$

Врахувавши тут формули $\|P_n\|^2 = \dfrac{2}{2n+1}$ і $\mu_n = \dfrac{1}{2^n}\dfrac{(2n)!}{(n!)^2}$, одержимо формулу (3.5).

5. Можна переконатися, що твірна $\Psi(x, \alpha)$ задовольняє диференціальне рівняння

$$\left(\frac{1}{\alpha} - \alpha\right)\frac{\partial \Psi}{\partial x} - 2\alpha \frac{\partial \Psi}{\partial \alpha} = \Psi\,.$$

Підставивши в це рівняння розвинення твірної за поліномами Лежандра і прирівнявши коефіцієнти біля степенів змінної $\alpha$, одержимо рекурентне співвідношення

$$(2n+1) P_n(x) = \frac{dP_{n+1}(x)}{dx} - \frac{dP_{n-1}(x)}{dx}. \qquad (3.7)$$

6. Кожний многочлен задовольняє диференціальне рівняння

$$\frac{d}{dx}\left[(1 - x^2)\frac{dP_n(x)}{dx}\right] + n(n+1) P_n(x) = 0\,. \qquad (3.8)$$

Для виведення формули (3.8) використаємо тотожність

$$\frac{dz}{dx}(x^2 - 1) = 2nxz\,,$$

де $z = (x^2 - 1)^n$.

Продиференціюємо цю тотожність $n+1$ раз. Знайдемо з використанням формули Лейбніца вирази окремо для лівої та правої частин цієї тотожності

$$\frac{d^{n+1}}{dx^{n+1}}\left[\frac{dz}{dx}(x^2-1)\right]=(x^2-1)\frac{d^{n+2}z}{dx^{n+2}}+2(n+1)x\frac{d^{n+1}z}{dx^{n+1}}+n(n+1)\frac{d^n z}{dx^n},$$

$$\frac{d^{n+1}}{dx^{n+1}}[z(-2nx)]=-2nx\frac{d^{n+1}z}{dx^{n+1}}-2n(n+1)\frac{d^n z}{dx^n}.$$

Прирівнявши ці вирази, одержимо

$$(x^2-1)\frac{d^{n+2}z}{dx^{n+2}}+2x\frac{d^{n+1}z}{dx^{n+1}}-n(n+1)\frac{d^n z}{dx^n}=0$$

і, відповідно, одержимо рівняння (3.8).

7. Справедливе зображення многочленів у вигляді контурного інтеграла

$$P_n(x)=\frac{1}{2\pi i}\int_L \left(x+it\sqrt{1-x^2}\right)^n \frac{dt}{\sqrt{t^2-1}}, \qquad (3.9)$$

де $\dfrac{1}{\sqrt{t^2-1}}=\sum\limits_{k=0}^{\infty}\dfrac{C_{2k}^k}{2^{2k}}\dfrac{1}{t^{2k+1}}$; $L$ – додатно орієнтований замкнутий контур, що охоплює початок координат і точки $t_{1,2}=\pm 1$.

Безпосередньо обчислюючи інтеграл в (3.9), одержимо

$$\frac{1}{2\pi i}\int_L \left(x+it\sqrt{1-x^2}\right)^n \frac{dt}{\sqrt{t^2-1}}=$$

$$=\frac{1}{2\pi i}\int_L \sum_{k=0}^{\infty}\sum_{l=0}^{n}\frac{i^l C_{2k}^k C_n^l}{2^{2k}}x^{n-l}(1-x^2)^{l/2}\frac{dt}{t^{2k-l+1}}=$$

$$=\sum_{k=0}^{[n/2]}\frac{(-1)^k C_{2k}^k C_n^{2k}}{2^{2k}}x^{n-2k}(1-x^2)^k=\sum_{k=0}^{[n/2]}\sum_{m=0}^{k}\frac{(-1)^m C_{2k}^k C_n^{2k} C_k^m}{2^{2k}}x^{n-2m}=$$

$$=\sum_{m=0}^{[n/2]}(-1)^m x^{n-2m}\sum_{k=m}^{[n/2]}\frac{C_{2k}^k C_n^{2k} C_k^m}{2^{2k}}.$$

Враховуючи в останньому виразі комбінаторну рівність

$$\sum_{k=m}^{[n/2]}\frac{C_{2k}^k C_n^{2k} C_k^m}{2^{2k}}=\frac{C_n^m C_{2(n-m)}^n}{2^n},$$

одержимо праву частину формули (3.4).

Якщо у формулі (3.9) контур $L$ – еліпс $t = \dfrac{1}{2}\left(\rho\, e^{i\varphi} + \dfrac{1}{\rho} e^{-i\varphi}\right)$, $1 < \rho < \infty$, $0 \leq \varphi < 2\pi$, з півосями $a = \dfrac{1}{2}\left(\rho + \dfrac{1}{\rho}\right)$, $b = \dfrac{1}{2}\left(\rho - \dfrac{1}{\rho}\right)$, то знайдемо

$$P_n(x) = \frac{1}{2\pi}\int_0^{2\pi}\left[x + \frac{i}{2}\left(\rho\, e^{i\varphi} + \frac{1}{\rho} e^{-i\varphi}\right)\sqrt{1-x^2}\right]^n d\varphi.$$

Одержаний інтеграл не залежить від $\rho$, тому при $\rho \to 1$ матимемо

$$P_n(x) = \frac{1}{2\pi}\int_0^{2\pi}\left(x + i\cos\varphi\sqrt{1-x^2}\right)^n d\varphi$$

або

$$P_n(x) = \frac{1}{\pi}\int_0^{\pi}\left(x + i\cos\varphi\sqrt{1-x^2}\right)^n d\varphi. \qquad (3.10)$$

Інтегральне зображення (3.10) полінома Лежандра називається *формулою Лапласа.*

8. Справедлива оцінка для поліномів Лежандра
$$|P_n(x)| < 1,\; x \in (-1, 1).$$

Дійсно, з формули (3.10) отримаємо при $x \in (-1, 1)$

$$|P_n(x)| \leq \frac{1}{\pi}\int_0^{\pi}\left|x + i\cos\varphi\sqrt{1-x^2}\right|^n d\varphi = \frac{1}{\pi}\int_0^{\pi}\left(\cos^2\varphi + x^2\sin^2\varphi\right)^{n/2} d\varphi <$$

$$< \frac{1}{\pi}\int_0^{\pi}\left(\cos^2\varphi + \sin^2\varphi\right)^{n/2} d\varphi = 1.$$

Отже, з урахуванням властивості 3 можна стверджувати, що
$$|P_n(x)| \leq 1,\; x \in [-1, 1].$$

9. З ортогональності многочленів випливає наступна властивість: *нулі многочлена $P_n(x)$, $n \geq 1$, дійсні, різні і лежать в інтервалі $(-1, 1)$.*

Дійсно, внаслідок ортогональності многочленів $P_n(x)$ і $P_0(x)$, маємо

$$\int\limits_{-1}^{1} P_n(x)dx = 0.$$

Отже, многочлен $P_n(x)$ міняє знак на проміжку $(-1, 1)$. Він може мати, як многочлен степеня $n$, не більше ніж $n$ різних дійсних коренів.

Припустимо, що на проміжку $(-1, 1)$ многочлен $P_n(x)$ міняє знак $m$ $(m < n)$ разів у точках $x_1, x_2, \ldots, x_m$. Тоді і многочлен $\sigma(x) = (x - x_1)(x - x_2)\ldots(x - x_m)$ міняє знак в цих точках і, очевидно,

$$\int\limits_{-1}^{1} P_n(x)\sigma(x)dx \neq 0.$$

З іншого боку многочлен $P_n(x)$ ортогональний до будь-якого многочлена степеня $< n$. Ми прийшли до протиріччя, яке заперечує прийняте припущення $m < n$. Отже, $m = n$, тобто многочлен $P_n(x)$ має $n$ дійсних різних коренів.

**3.1.3. Ряд Фур'є за ортогональною системою функцій.** Нехай $E$ – евклідів простір, елементами якого є функції, визначені на проміжку $(a, b)$, і $\{\varphi_n(x)\} \subset E$ – ортогональна система функцій.

Нехай функція $f(x)$ є сумою ряду

$$f(x) = \sum_{k=0}^{\infty} c_k \varphi_k(x), \qquad (3.11)$$

де $c_k$ – коефіцієнти ряду (дійсні числа), які потрібно знайти.

Припустимо, що результат множення функції $f(x)$ і членів ряду (3.11) на функції $\varphi_n(x)$ можна почленно інтегрувати на проміжку $(a, b)$. Тоді, внаслідок ортогональності системи $\{\varphi_n(x)\}$, одержимо

$$\int\limits_a^b f(x)\varphi_n(x)dx = c_n \int\limits_a^b \varphi_n^2(x)dx = c_n \|\varphi_n\|, \ n = 0, 1, \ldots.$$

Звідси знайдемо

$$c_n = \frac{1}{\|\varphi_n\|^2}(f, \varphi_n) = \frac{1}{\|\varphi_n\|^2} \int\limits_a^b f(x)\varphi_n(x)dx, \ n = 0, 1, \ldots. \quad (3.12)$$

Очевидно, що не для кожної інтегровної на проміжку $(a, b)$ функції $f(x)$ ряд (3.11) збігається.

Нехай $f(x)$ інтегровна на проміжку $(a, b)$ функція. Тоді за формулами (3.12) шукаємо коефіцієнти, які називаються *коефіцієнтами Фур'є* для функції $f(x)$, і будуємо *ряд Фур'є* для цієї функції. Поки не встановлено, що ряд збігається до функції $f(x)$, записуємо

$$f(x) \sim \sum_{k=0}^{\infty} c_k \varphi_k(x). \qquad (3.13)$$

Якщо функції системи $\{\varphi_n(x)\}$ неперервні на відрізку $[a, b]$ і ряд (3.13) збігається рівномірно, то його сума неперервна функція і ряд можна почленно інтегрувати (теорема 2 п. 2.1). Цей результат можна сформулювати у вигляді твердження, яке узагальнює теорему 1 (п. 2.3).

***Т е о р е м а  2 .*** *Якщо неперервна на проміжку $(a, b)$ функція $f(x)$ розвивається на $(a, b)$ у рівномірно збіжний ряд (3.13) за ортогональною системою $\{\varphi_n(x)\}$ неперервних функцій, то цей ряд є рядом Фур'є для функції $f(x)$.*

*Д о в е д е н н я .* Розглянемо ряд (3.13), коефіцієнти якого визначаються за формулами (3.12). Оскільки його члени неперервні функції і ряд збігається рівномірно, то за теоремою 2 (п. 2.1) його сума $S(x)$ неперервна функція і рівність

$$S(x) = \sum_{k=0}^{\infty} c_k \varphi_k(x).$$

можна інтегрувати попередньо помноживши на неперервні функції $\varphi_n(x)$. Скориставшись властивістю ортогональності системи $\{\varphi_n(x)\}$, одержимо формули (3.12), в яких всі функції неперервні, тому $f(x) \equiv S(x),\ x \in (a, b)$.

Теорему доведено.

**3.1.4. Апроксимація функції в середньому. Нерівність Бесселя.** Розглядаємо (для простоти викладу) евклідів простір $E \equiv L_2'(a, b)$, хоч викладені тут результати справедливі також для простору $L_2(a, b)$. Нехай $\{\widetilde{\varphi}_n(x)\} \subset E$ – ортонормована система функцій. Розглянемо поліном $n$-го порядку за цією системою

$$\sigma_n(x) = \sum_{k=0}^{n} \gamma_k \widetilde{\varphi}_k(x), \qquad (3.14)$$

де $\gamma_k$ ($k = 0, 1, ..., n$) – сталі.

За умовою, оскільки існує норма $\|\widetilde{\varphi}_k(x)\| = 1$, функції системи $\{\widetilde{\varphi}_k(x)\}$ інтегровні з квадратом, а також інтегровною з квадратом є $f(x) - \sigma_n(x)$, $f(x) \in E$. Це випливає з властивостей норми.

Розглянемо питання про найкраще наближення функції $f(x) \in E$ в розумінні *середнього квадратичного* (за нормою простору $E$) скінченною сумою (3.14).

За яких значень коефіцієнтів $\gamma_k$ інтеграл

$$\left( \int_a^b \left| f(x) - \sum_{k=0}^{n} \gamma_k \widetilde{\varphi}_k(x) \right|^2 dx \right)^{1/2} = \left\| f(x) - \sum_{k=0}^{n} \gamma_k \widetilde{\varphi}_k(x) \right\|$$

приймає найменше значення?

Розглянемо також $n$-у частинну суму Фур'є для функції $f(x)$ за системою $\{\widetilde{\varphi}_k(x)\}$

$$S_n(x) = \sum_{k=0}^{n} \widetilde{c}_k \widetilde{\varphi}_k(x), \qquad (3.15)$$

де

$$\widetilde{c}_k = (f, \widetilde{\varphi}_k) = \int_a^b f(x) \widetilde{\varphi}_k(x) dx. \qquad (3.16)$$

***Т е о р е м а 3***. *Серед всіх многочленів вигляду* (3.14) *найкраще наближення для функції* $f(x) \in E$ *в розумінні середнього квадратичного дає сума Фур'є цієї функції, тобто*

$$\min_{\gamma_k} \left\| f(x) - \sum_{k=0}^{n} \gamma_k \widetilde{\varphi}_k(x) \right\| = \left\| f(x) - \sum_{k=0}^{n} \widetilde{c}_k \widetilde{\varphi}_k(x) \right\|.$$

*Д о в е д е н н я*. Знайдемо оцінку в розумінні середнього квадратичного для відхилення полінома $\sigma_n(x)$ від функції $f(x)$ на проміжку $(a, b)$

$$\left\| f(x) - \sum_{k=0}^{n} \gamma_k \widetilde{\varphi}_k(x) \right\|^2 = \left( f - \sum_{k=0}^{n} \gamma_k \widetilde{\varphi}_k, f - \sum_{m=0}^{n} \gamma_m \widetilde{\varphi}_m \right) =$$

$$= (f, f) - 2\sum_{k=1}^{n} \gamma_k (f, \widetilde{\varphi}_k) + \sum_{k=1}^{n} \gamma_k^2 =$$

$$= \sum_{k=1}^{n} (f, \widetilde{\varphi}_k)^2 - 2\sum_{k=1}^{n} \gamma_k (f, \widetilde{\varphi}_k) + \sum_{k=1}^{n} \gamma_k^2 + (f, f) - \sum_{k=1}^{n} (f, \widetilde{\varphi}_k)^2 =$$

$$= \sum_{k=1}^{n} \left[ (f, \widetilde{\varphi}_k)^2 - 2\gamma_k (f, \widetilde{\varphi}_k) + \gamma_k^2 \right] + (f, f) - \sum_{k=1}^{n} (f, \widetilde{\varphi}_k)^2 =$$

$$= \sum_{k=1}^{n} \left[ (f, \widetilde{\varphi}_k) - \gamma_k \right]^2 + (f, f) - \sum_{k=1}^{n} (f, \widetilde{\varphi}_k)^2 .$$

Оскільки останні два доданки в цьому співвідношенні не залежать від коефіцієнтів $\gamma_k$ ($k = 0, 1, ..., n$), найменше значення відхилення досягається, коли перший доданок дорівнює нулю, тобто $\gamma_k = (f, \widetilde{\varphi}_k)$.

Отже, квадратичне відхилення буде мінімальним, якщо коефіцієнти многочлена (3.14) є коефіцієнтами Фур'є функції $f(x)$.

Теорему доведено.

З останнього співвідношення одержимо рівність

$$\left\| f(x) - \sum_{k=0}^{n} \widetilde{c}_k \widetilde{\varphi}_k(x) \right\|^2 = (f, f) - \sum_{k=1}^{n} (f, \widetilde{\varphi}_k)^2 .$$

З цієї рівності, оскільки її ліва частина невід'ємне число, випливає нерівність

$$\sum_{k=1}^{n} (f, \widetilde{\varphi}_k)^2 \leq (f, f),$$

яка справедлива для довільного $n$. Ліва частина цієї нерівності – частинна сума, яка зростає зі збільшенням $n$ і обмежена, а отже, має скінченну границю.

Тому справедлива нерівність

$$\sum_{k=1}^{\infty} (f, \widetilde{\varphi}_k)^2 \leq (f, f), \qquad (3.17)$$

яка називається *нерівністю Бесселя*.

Запишемо з урахуванням формули (3.16) нерівність (3.17) у розгорнутому вигляді

$$\sum_{k=1}^{\infty} \widetilde{c}_k^2 \le \int_a^b f^2(x)dx. \qquad (3.18)$$

Якщо система $\{\varphi_k(x)\}$ лише ортогональна на проміжку $(a,b)$, то формула (3.18) має вигляд

$$\sum_{k=1}^{\infty} c_k^2 \|\varphi_k\|^2 \le \int_a^b f^2(x)dx. \qquad (3.19)$$

*З а у в а ж е н н я .* З нерівностей (3.18) і (3.19) випливає, що для інтегровної з квадратом функції $f(x)$ збігаються числові ряди $\sum_{k=1}^{\infty} \widetilde{c}_k^2$ (у випадку розвинення за ортонормованою системою) і $\sum_{k=1}^{\infty} c_k^2 \|\varphi_k\|^2$ (у випадку розвинення за ортогональною системою).

Звідси за необхідною умовою збіжності рядів маємо
$$\lim_{n \to \infty} \widetilde{c}_n = 0$$
(у випадку розвинення за ортонормованою системою) і
$$\lim_{n \to \infty} c_n \|\varphi_n\| = 0$$
(у випадку розвинення за ортогональною системою).

**3.1.5. Збіжність в середньому ряду Фур'є.** Нехай в евклідовому просторі $E \equiv L_2'(a,b)$ задана ортонормована система $\{\widetilde{\varphi}_k(x)\}$. Розглянемо функціональний ряд

$$\sum_{k=1}^{\infty} \widetilde{c}_k \widetilde{\varphi}_k(x) \qquad (3.20)$$

і його частинну суму

$$S_n(x) = \sum_{k=1}^{n} \widetilde{c}_k \widetilde{\varphi}_k(x) \qquad (3.21)$$

з метою зображення функції $f(x) \in E$.

*О з н а ч е н н я 3 .Функціональний ряд* (3.20) *називається збіжним до функції* $f(x) \in E$ *в середньому (за нормою простору $E$), якщо*

$$\lim_{n\to\infty}\int_a^b \left[f(x)-S_n(x)\right]^2 dx = 0 \qquad (3.22)$$

( $\lim\limits_{n\to\infty}\|f(x)-S_n(x)\|_E = 0$, або $f(x)\stackrel{E}{=}\lim\limits_{n\to\infty}S_n(x)$ )

*і записуємо*

$$f(x) \stackrel{E}{=} \sum_{k=1}^{\infty} \widetilde{c}_k \widetilde{\varphi}_k(x). \qquad (3.23)$$

***Теорема 4***. *Якщо ряд* (3.20) *збігається в середньому до* $f(x)\in E$, *то коефіцієнти* $\widetilde{c}_k$ *однозначно визначаються за формулою*

$$\widetilde{c}_k = (f,\widetilde{\varphi}_k),\ k=1,2,\ldots\,.$$

*Доведення*. Розглянемо скалярний добуток $\left(f - \sum\limits_{k=0}^{n}\gamma_k \widetilde{\varphi}_k,\widetilde{\varphi}_m\right)$ при $m < n$ і перетворимо його з урахуванням ортонормованості системи $\{\widetilde{\varphi}_n(x)\}$

$$\left|\left(f - \sum_{k=0}^{n}\widetilde{c}_k\widetilde{\varphi}_k,\widetilde{\varphi}_m\right)\right| = \left|(f,\widetilde{\varphi}_m) - \sum_{k=0}^{n}\widetilde{c}_k(\widetilde{\varphi}_k,\widetilde{\varphi}_m)\right| = \left|(f,\widetilde{\varphi}_m) - \widetilde{c}_m\right|.$$

Застосуємо до цього скалярного добутку нерівність Гельдера

$$\left|\left(f - \sum_{k=0}^{n}\widetilde{c}_k\widetilde{\varphi}_k,\widetilde{\varphi}_m\right)\right| \leq \left\|f - \sum_{k=0}^{n}\widetilde{c}_k\widetilde{\varphi}_k\right\|\|\widetilde{\varphi}_m\| = \left\|f - \sum_{k=0}^{n}\widetilde{c}_k\widetilde{\varphi}_k\right\|.$$

За умовою для достатньо малого $\varepsilon > 0$ існує номер $N = N(\varepsilon)$ такий, що

$$\left\|f(x) - \sum_{k=0}^{n}\widetilde{c}_k\widetilde{\varphi}_k(x)\right\| < \varepsilon.$$

Тоді з урахуванням одержаних оцінок знайдемо нерівність

$$\left|(f,\widetilde{\varphi}_m) - \widetilde{c}_m\right| = \left|\left(f - \sum_{k=0}^{n}\widetilde{c}_k\widetilde{\varphi}_k,\widetilde{\varphi}_m\right)\right| \leq \left\|f - \sum_{k=0}^{n}\widetilde{c}_k\widetilde{\varphi}_k\right\| \leq \varepsilon,$$

яка справедлива для всіх $n > N$. Оскільки число $\varepsilon$ довільне, матимемо

$$\widetilde{c}_m = (f,\widetilde{\varphi}_m).$$

Теорему доведено.

***Теорема 5***. *Якщо функціональний ряд* (3.20) *на проміжку* $(a,b)$ *збігається до функції* $f(x)$ *рівномірно, то він збігається і в середньому на цьому відрізку до* $f(x)$.

*Доведення*. Нехай ряд (3.20) на проміжку $(a,b)$ збігається рівномірно. Тоді для $\varepsilon > 0$ існує натуральне $N = N(\varepsilon)$ таке, що при $n > N$ і всіх $x \in (a,b)$ виконується нерівність

$$|r_n(x)| = |f(x) - S_n(x)| < \sqrt{\frac{\varepsilon}{b-a}}.$$

У такому разі при $n > N$ справджується нерівність

$$\int_a^b [f(x) - S_n(x)]^2 dx < \frac{\varepsilon}{b-a} \int_a^b dx = \varepsilon.$$

Це показує, що виконується рівність (3.22).

Теорему доведено.

**3.1.6. Повні системи функцій.** Розглянемо ортогональну на проміжку $(a,b)$ систему $\{\varphi_n(x)\}$ функцій в евклідовому просторі $E = L_2'(a,b)$.

***Означення 4***. *Система* $\{\varphi_n(x)\}$ *називається повною в E (за нормою простору E), якщо для будь-якої функції* $f(x) \in E$ *справедлива рівність* (3.22),

$$\lim_{n \to \infty} \int_a^b \left[ f(x) - \sum_{k=1}^n c_k \varphi_k(x) \right]^2 dx = 0, \qquad (3.24)$$

*де* $c_k$ – *коефіцієнти Фур'є для функції* $f(x)$ *за системою* $\{\varphi_n(x)\}$.

***Теорема 6***. *Якщо ортогональна система функцій* $\{\varphi_n(x)\} \subset E$ *повна, то будь-яка функція* $f(x) \in E$ *визначається своїм рядом Фур'є*.

*Доведення*. Нехай поряд з функцією $f(x) \in L_2'(a,b)$ існує функція $F(x) \in L_2'(a,b)$ така, що справджується рівність (3.24)

$$\lim_{n \to \infty} \int_a^b [F(x) - S_n(x)]^2 dx = 0. \qquad (3.25)$$

Оцінимо величину $\int\limits_a^b [F(x)-f(x)]^2 dx \geq 0$ з урахуванням нерівності $(a+b)^2 \leq 2(a^2+b^2)$,

$$\int\limits_a^b [F(x)-f(x)]^2 dx = \int\limits_a^b \{[F(x)-S_n(x)]+[S_n(x)-f(x)]\}^2 dx \leq$$

$$\leq 2\int\limits_a^b [F(x)-S_n(x)]^2 dx + 2\int\limits_a^b [S_n(x)-f(x)]^2 dx.$$

Звідси з урахуванням (3.24) і (3.25) одержимо

$$\lim_{n\to\infty}\int\limits_a^b [F(x)-f(x)]^2 dx = 0.$$

З додатної визначеності підінтегральної функції випливає, що в точках її неперервності справедлива рівність $F(x)=f(x)$. Функції $F(x)$ і $f(x)$ можуть мати лише скінченне число точок розриву. Тому вони співпадають всюди, хіба-що за виключенням скінченого числа точок.

Теорему доведено.

***Т е о р е м а  7***. *Для того, щоби ортогональна система функцій $\{\varphi_n(x)\}\subset E$ була повною (за нормою простору $E$), необхідно і достатньо, щоби для будь-якої функції $f(x)\in E$ виконувалася рівність*

$$\int\limits_a^b f^2(x)dx = \sum_{k=1}^\infty c_k^2\|\varphi_k\|^2, \qquad (3.26)$$

*де $c_k$ – коефіцієнти Фур'є для функції $f(x)$ за системою $\{\varphi_n(x)\}$.*

Рівність (3.23) називається *рівністю Парсеваля* (рівнянням повноти)

*Н е о б х і д н і с т ь*. Нехай система $\{\varphi_n(x)\}$ повна в $E$. Тоді для будь-якої функції $f(x)\in E$ справедлива рівність

$$\lim_{n\to\infty}\int\limits_a^b \left[f(x)-\sum_{k=1}^n c_k\varphi_k(x)\right]^2 dx = 0.$$

Враховуючи тут формули (3.12) і ортогональність системи $\{\varphi_n(x)\}$, одержимо

$$\lim_{n\to\infty}\int_a^b\left[f(x)-\sum_{k=1}^n c_k\varphi_k(x)\right]^2 dx = \lim_{n\to\infty}\left[\int_a^b f^2(x)dx - \right.$$

$$\left. -2\sum_{k=1}^n c_k\int_a^b f(x)\varphi_k(x)dx + \sum_{k=1}^n c_k^2\int_a^b \varphi_k^2(x)dx\right] =$$

$$= \lim_{n\to\infty}\left[\int_a^b f^2(x)dx - \sum_{k=0}^n c_k^2\|\varphi_k\|^2\right]. \qquad (3.27)$$

Звідси з урахуванням рівності нулю останнього виразу маємо рівність (3.26).

*Д о с т а т н і с т ь* . Нехай для довільної функції $f(x)\in E$ справджується рівність Парсеваля (3.26). Тоді, з останнього виразу (3.27) одержимо умову (3.24) повноти системи $\{\varphi_n(x)\}$.

Теорему доведено.

*З а у в а ж е н н я* . Повні ортогональні (зокрема, ортонормовані) системи в $E$ ще називають ортогональними (ортонормованими) базами в $E$ (ОНБ).

**Узагальнена умова повноти.** Властивість ортогональності системи $\{\varphi_n(x)\}$ дозволяє одержати узагальнення умови (3.26).

***Т е о р е м а   8*** *. Якщо система* $\{\varphi_n(x)\}\subset E$ *повна і функції* $f(x)\in E$, $g(x)\in E$, *тобто*

$$f(x)\sim\sum_{k=0}^\infty c_k\varphi_k(x),\quad g(x)\sim\sum_{k=0}^\infty d_k\varphi_k(x),$$

*то справедлива рівність*

$$\int_a^b f(x)g(x)dx = \sum_{k=0}^\infty c_k d_k\|\varphi_k\|^2. \qquad (3.28)$$

*Д о в е д е н н я* . Розглянемо функції $f(x)+g(x)$ і $f(x)-g(x)$, які також інтегровані з квадратом і, відповідно, мають коефіцієнти Фур'є $c_k+d_k$, $c_k-d_k$. Умови повноти для цих функцій наступні

$$\int\limits_a^b [f(x) \pm g(x)]^2 dx = \sum_{k=1}^{\infty}(c_k \pm d_k)^2 \|\varphi_k\|^2.$$

Врахувавши тут рівність $(a+b)^2 - (a-b)^2 = 4ab$, одержимо

$$4\int\limits_a^b f(x)g(x)dx = \sum_{k=0}^{\infty} 4c_k d_k \|\varphi_k\|^2.$$

Звідси маємо рівність (3.28), яку також називають *рівністю Парсеваля*.

Теорему доведено.

**Н а с л і д о к .** *З теореми безпосередньо випливає, що ряд Фур'є функції $f(x)$, інтегровної з квадратом, після множення всіх його членів на функцію $g(x)$, інтегровну з квадратом, можна інтегрувати почленно на проміжку $(a, b)$.*

*При цьому одержимо формулу для інтегралу від добутку двох функцій.*

**Т е о р е м а 9.** *Якщо система $\{\varphi_n(x)\} \subset E$ повна, то ряд Фур'є для кожної функції $f(x) \in E$ можна інтегрувати почленно, незалежно від того, збігається він чи ні і справедлива формула*

$$\int\limits_{x_1}^{x_2} f(x)dx = \sum_{k=0}^{\infty} c_k \int\limits_{x_1}^{x_2} \varphi_k(x)dx, \qquad (3.29)$$

*де $x_1, x_2 \in (a, b)$*

*Д о в е д е н н я .* Прийнявши для визначеності, що $x_1 < x_2$, і врахувавши нерівність Коші – Буняковського

$$\left(\int\limits_a^b f(x)g(x)dx\right)^2 \le \int\limits_a^b f^2(x)dx \int\limits_a^b g^2(x)dx,$$

знайдемо оцінку різниці лівої частини формули (3.29) і відповідної частинної суми ряду у правій частині цієї формули

$$\left|\int\limits_{x_1}^{x_2} f(x)dx - \sum_{k=0}^{n} c_k \int\limits_{x_1}^{x_2} \varphi_k(x)dx\right| \le \int\limits_{x_1}^{x_2} \left|f(x) - \sum_{k=0}^{n} c_k \varphi_k(x)\right|dx \le$$

$$\leq \int_a^b \left| f(x) - \sum_{k=0}^n c_k \varphi_k(x) \right| dx \leq \left\{ \int_a^b \left[ f(x) - \sum_{k=0}^n c_k \varphi_k(x) \right]^2 dx \int_a^b dx \right\}^{1/2} =$$

$$= \left\{ (b-a) \int_a^b \left[ f(x) - \sum_{k=0}^n c_k \varphi_k(x) \right]^2 dx \right\}^{1/2}.$$

Оскільки система $\{\varphi_n(x)\}$ повна і виконується рівність (3.24), маємо

$$\lim_{n \to \infty} \left| \int_{x_1}^{x_2} f(x) dx - \sum_{k=0}^n c_k \int_{x_1}^{x_2} \varphi_k(x) dx \right| = 0,$$

що рівносильно (3.29).

Теорему доведено.

### 3.2. Повнота тригонометричної системи функцій

**3.2.1. Збіжність рядів за нормою.** У другому розділі викладено достатні умови рівномірної збіжності тригонометричних рядів для різних класів функцій. Показано, що умови неперервності функції не достатньо для рівномірної збіжності її ряду Фур'є. Існують неперервні функції з розбіжними рядами Фур'є. Однак це не стосується функцій і відповідних їм тригонометричних рядів, коефіцієнти яких не є коефіцієнтами Фур'є або функцій зі збіжними в середньому їх рядами Фур'є $[9, 10, 18, 21, 23]$.

Розглянемо допоміжні твердження. Наступне твердження називається теоремою Вейєрштрасса про рівномірну апроксимацію неперервної функції тригонометричними поліномами.

*Л е м а  1 (Вейєрштрасс)*. Нехай функція $f(x)$ неперервна на відрізку $[-\pi, \pi]$ і $f(-\pi) = f(\pi)$.

*Тоді, яке б не було $\varepsilon > 0$, існує тригонометричний поліном $\sigma_n(x)$, що рівномірно для всіх $x \in [-\pi, \pi]$ виконується нерівність*

$$|f(x) - \sigma_n(x)| \leq \varepsilon. \qquad (3.30)$$

*Д о в е д е н н я*. Розіб'ємо відрізок $[-\pi, \pi]$ на $m$ частин точками

$$x_0 = -\pi < x_1 < x_2 < \ldots < x_{m-1} < x_m = \pi$$

і побудуємо неперервну функцію $g(x)$, для якої $g(x_k) = f(x_k)$, $k = 0, 1, \ldots m$, і яка лінійна на кожному з відрізків $[x_{k-1}, x_k]$. Графік функції $y = g(x)$ є ламана крива з вершинами на кривій $y = f(x)$. Оскільки $y = f(x)$ – неперервна функція, відрізки виберемо настільки малими, щоби для довільного $x \in [-\pi, \pi]$ виконувалася нерівність

$$|f(x) - g(x)| \leq \frac{\varepsilon}{2}. \tag{3.31}$$

Очевидно періодичні продовження функцій $f(x)$ і $g(x)$ є неперервними на всій дійсній осі, крім того функція $g(x)$ кусково-гладка. За теоремою 5 (п. 2.3) функція $g(x)$ розвивається в рівномірно збіжний ряд, а отже, для досить великих $n$ і всіх $x$ справедлива нерівність

$$|g(x) - \sigma_n(x)| \leq \frac{\varepsilon}{2}, \tag{3.32}$$

де $\sigma_n(x) = \frac{\alpha_0}{2} + \sum_{k=1}^{n}(\alpha_k \cos kx + \beta_k \sin kx)$ – $n$-а частинна сума ряду Фур'є для функції $g(x)$.

Тоді для величини $f(x) - \sigma_n(x)$ з урахуванням нерівностей (3.31) і (3.32) одержимо оцінку

$$|f(x) - \sigma_n(x)| = |[f(x) - g(x)] + [g(x) - \sigma_n(x)]| \leq$$
$$\leq |f(x) - g(x)| + |g(x) - \sigma_n(x)| \leq \varepsilon,$$

яка справедлива для досить великих $n$ і всіх $x$.

Лему доведено.

*З а у в а ж е н н я 1*. Розглянемо нескінченно малу числову послідовність $\{\varepsilon_k\}$ і для кожного $\varepsilon = \varepsilon_k$ побудуємо за лемою 1 поліном $\sigma_n = \sigma_n^k$, де $n = n(\varepsilon_k)$. Тоді одержимо послідовність $\{\sigma_n^k\}$ тригонометричних поліномів, яка збігається до функції $f(x)$ рівномірно на відрізку $[-\pi, \pi]$. Ґрунтуючись на цьому сформулюємо лему 1 ще так: *за виконання умов леми 1 функція $f(x)$ розвивається в рівномірно збіжний ряд на проміжку $[-\pi, \pi]$,* *членами якого є тригонометричні поліноми,*

$$f(x) = \lim_{k \to \infty} \sigma_n^k(x), \ \textit{або} \ f(x) = \sigma_n^1(x) + \sum_{k=1}^{\infty} \left[\sigma_n^{k+1}(x) - \sigma_n^k(x)\right].$$

*З а у в а ж е н н я   2 .* Систему функцій у просторі $E$ називають *лінійно щільною* в $E$, якщо будь-який елемент з $E$ можна наблизити з довільною точністю скінченними лінійними комбінаціями елементів системи. В цьому розумінні система тригонометричних поліномів є *лінійно щільною* в просторі $2\pi$-періодичних неперервних функцій.

**Н а с л і д о к .** *Якщо функція $f(x)$ неперервна на відрізку $[a, b]$, де $-\pi < a$ і $b = \pi$, або $-\pi = a$ і $b < \pi$, то, яке б не було $\varepsilon > 0$, існує тригонометричний поліном $\sigma_n(x)$, що для всіх $x \in [a, b]$ справедлива нерівність* (3.30).

*Д о в е д е н н я .* Нехай $-\pi = a$ і $b < \pi$. Продовжимо функцію $f(x)$ на відрізок $[-\pi, \pi]$ так, щоби збереглася неперервність і для продовженої функції $f^*(x)$ виконувалася рівність $f^*(\pi) = f(-\pi)$. Тоді функція $f^*(x)$ задовольняє умови теореми, а отже справедлива нерівність. Аналогічно розглядаємо випадок, коли $-\pi < a$ і $b = \pi$.

**Л е м а   2 .** *Якщо функція $f(x) \subset L_2'(-\pi, \pi)$ (інтегрована з квадратом на інтервалі $(-\pi, \pi)$), то, яке б не було мале число $\varepsilon > 0$, існує неперервна на відрізку $[-\pi, \pi]$ функція $F(x)$, що задовольняє умову $F(-\pi) = F(\pi)$ і така, що*

$$\int_{-\pi}^{\pi} [f(x) - F(x)]^2 dx \leq \varepsilon. \qquad (3.33)$$

*Д о в е д е н н я .* Функція $f(x)$ може мати лише скінченне число точок розриву (в яких функція може приймати і необмежені односторонні значення). Кожну таку точку і точку $x = \pi$ помістимо у проміжок такої довжини, щоби сума інтегралів від функції $f^2(x)$ на цих проміжках була меншою $\dfrac{\varepsilon}{4}$.

Введемо спочатку допоміжну функцію $\Phi(x)$, що дорівнює $f(x)$ зовні цих проміжків і дорівнює нулеві всередині них. Функція

Ф$(x)$ обмежена і має скінченне число точок розриву. Тому

$$\int_a^b [f(x) - \Phi(x)]^2 dx \leq \frac{\varepsilon}{4}. \qquad (3.34)$$

Сумарну довжину $l$ цих проміжків виберемо, щоб виконувалась умова

$$4M^2 l \leq \frac{\varepsilon}{4}, \text{ де } M = \max_{x \in [a,b]} |\Phi(x)|.$$

Тепер введемо неперервну функцію $F(x)$, що дорівнює $\Phi(x)$ зовні згаданих проміжків, лінійна всередині кожного з них і $F(-\pi) = F(\pi)$. Тоді

$$\int_a^b [F(x) - \Phi(x)]^2 dx \leq 4M^2 l \leq \frac{\varepsilon}{4}. \qquad (3.35)$$

Для інтегралу в (3.33) з урахуванням нерівностей (3.34), (3.35) і нерівності $(a+b)^2 \leq 2(a^2 + b^2)$ одержимо

$$\int_a^b [f(x) - F(x)]^2 dx = \int_a^b \{[f(x) - \Phi(x)] + [\Phi(x) - F(x)]\}^2 dx \leq$$

$$\leq 2\int_a^b [f(x) - \Phi(x)]^2 dx + 2\int_a^b [\Phi(x) - F(x)]^2 dx \leq \varepsilon,$$

яка показує справедливість нерівності (3.33).

Лему доведено.

***Т е о р е м а  1 .*** *Якщо функція $f(x) \subset L_2'(-\pi, \pi)$, то її ряд Фур'є збігається в середньому до цієї функції, тобто справедлива рівність*

$$\lim_{n \to \infty} \int_{-\pi}^{\pi} [f(x) - S_n(x)]^2 dx = 0, \qquad (3.36)$$

*де $S_n(x) = \dfrac{a_0}{2} + \sum_{k=1}^{n} (a_k \cos kx + b_k \sin kx)$ – $n$-а частинна сума ряду Фур'є для функції $f(x)$.*

*Д о в е д е н н я .* За лемою 2 для функції $f(x)$ і довільного $\varepsilon > 0$ існує неперервна на відрізку $[-\pi, \pi]$ функція $F(x)$, що задовольняє умову $F(-\pi) = F(\pi)$ і така, що

$$\int\limits_{-\pi}^{\pi} [f(x) - F(x)]^2 dx \le \frac{\varepsilon}{2}. \qquad (3.37)$$

За лемою 1 для функції $F(x)$ знайдеться тригонометричний многочлен $\sigma_n(x)$ такий, що для всіх $x \in [-\pi, \pi]$ справедлива нерівність

$$|F(x) - \sigma_n(x)| \le \frac{\varepsilon}{4\pi}.$$

З цієї нерівності знайдемо

$$\int\limits_{-\pi}^{\pi} [F(x) - \sigma_n(x)]^2 dx \le \frac{\varepsilon}{2}. \qquad (3.38)$$

Оцінимо інтеграл в (3.36) з урахуванням нерівностей (3.37), (3.38) і мінімальної властивості многочлена Фур'є (теорема 3 п. 3.1)

$$\int\limits_{-\pi}^{\pi} [f(x) - S_n(x)]^2 dx = \int\limits_{-\pi}^{\pi} \{[f(x) - F(x)] + [F(x) - S_n(x)]\}^2 dx \le$$

$$\le 2 \int\limits_{-\pi}^{\pi} [f(x) - F(x)]^2 dx + 2 \int\limits_{-\pi}^{\pi} [F(x) - S_n(x)]^2 dx \le$$

$$\le 2 \int\limits_{-\pi}^{\pi} [f(x) - F(x)]^2 dx + 2 \int\limits_{-\pi}^{\pi} [F(x) - \sigma_n(x)]^2 dx \le 2\varepsilon.$$

Тут також використано елементарну нерівність $(a+b)^2 \le 2(a^2 + b^2)$.

Оскільки $\varepsilon > 0$ довільне мале число, з одержаної нерівності випливає формула (3.36).

Теорему доведено.

***Т е о р е м а  2 .*** *Тригонометрична система*

$$\frac{1}{2}, \ \cos x, \ \sin x, \ ..., \cos kx, \ \sin kx, \ ... \qquad (3.39)$$

*повна в просторі $L'_2(-\pi, \pi)$ і справедлива рівність Парсеваля*

$$\frac{a_0^2}{2} + \sum_{k=1}^{\infty} (a_k^2 + b_k^2) = \frac{1}{\pi} \int\limits_{-\pi}^{\pi} f^2(x) dx, \qquad (3.40)$$

*де $a_k$, $b_k$ – коефіцієнти Фур'є функції $f(x) \in L'_2(-\pi, \pi)$.*

*Д о в е д е н н я .* Повнота системи функцій (3.39) випливає з теореми 1, оскільки за означенням 4 (п. 3.1) рівність (3.36) визначає повну систему функцій.

Встановимо рівність (3.40). Враховуючи у виразі інтегралу в (3.36) ортогональність системи (3.39), одержимо

$$\frac{1}{\pi}\int_{-\pi}^{\pi}[f(x)-S_n(x)]^2 dx = \frac{1}{\pi}\int_{-\pi}^{\pi}f^2(x)dx - \frac{a_0^2}{2} - \sum_{k=1}^{n}\left(a_k^2 + b_k^2\right).$$

Тому

$$\lim_{n\to\infty}\left[\frac{1}{\pi}\int_{-\pi}^{\pi}f^2(x)dx - \frac{a_0^2}{2} - \sum_{k=1}^{n}\left(a_k^2 + b_k^2\right)\right]=0$$

або

$$\lim_{n\to\infty}\left[\frac{a_0^2}{2} + \sum_{k=1}^{n}\left(a_k^2 + b_k^2\right)\right] = \frac{1}{\pi}\int_{-\pi}^{\pi}f^2(x)dx.$$

Оскільки під знаком границі стоїть *n*-а частинна сума збіжного ряду, маємо рівність (3.40).

Теорему доведено.

***Т е о р е м а   3 .*** *Нехай $f(x)$ – неперервна, кусково-гладка на відрізку $[-\pi, \pi]$ функція, така, що $f(-\pi)= f(\pi)$ і*

$$f(x)= \frac{a_0}{2}+\sum_{k=1}^{\infty}\left(a_k \cos kx + b_k \sin kx\right)$$

*– її ряд Фур'є.*

*Тоді цей ряд допускає почленне диференціювання*

$$f'(x) \overset{L_2}{=} \sum_{k=1}^{\infty}\left(kb_k \cos kx - ka_k \sin kx\right) \qquad (3.41)$$

*і одержаний ряд збігається у середньому.*

*Д о в е д е н н я .* Функція $f'(x)$ належить простору кусково-неперервних функцій і її ряд збігається в середньому (теорема 1)

$$f'(x) \overset{L_2}{=} \sum_{k=1}^{\infty}\left(a'_k \cos kx + b'_k \sin kx\right). \qquad (3.42)$$

Тут $a'_0 = \dfrac{1}{\pi}\int_{-\pi}^{\pi}f'(x)dx = \dfrac{1}{\pi}\left[f(\pi)- f(\pi)\right]= 0$,

$$a'_k = \frac{1}{\pi}\int_{-\pi}^{\pi} f'(x)\cos kx\, dx = \frac{1}{\pi}\left\{\left[f(x)\cos kx\right]\Big|_{-\pi}^{\pi} + k\int_{-\pi}^{\pi} f(x)\sin kx\, dx\right\} = kb_k,$$

$$b'_k = \frac{1}{\pi}\int_{-\pi}^{\pi} f'(x)\sin kx\, dx = \frac{1}{\pi}\left\{\left[f(x)\sin kx\right]\Big|_{-\pi}^{\pi} - k\int_{-\pi}^{\pi} f(x)\cos kx\, dx\right\} = -ka_k.$$

Підставивши ці формули в ряд (3.42), одержимо ряд (3.41).

Теорему доведено.

### 3.2.2. Укорочені розвинення

**Т е о р е м а  5 .** *Система косинусів* $\{\cos nx\}_{n=0}^{\infty}$ *і система синусів* $\{\sin nx\}_{n=1}^{\infty}$ *повні в просторі* $L'_2(0, \pi)$.

*Д о в е д е н н я .* Якщо $f(x) \in L'_2(0, \pi)$, то, очевидно парне продовження $f^+(x)$ цієї функції належить $L'_2(-\pi, \pi)$. Тоді справедливе розвинення

$$f^+(x) \stackrel{L'_2}{=} \frac{a_0}{2} + \sum_{k=1}^{\infty} a_k \cos kx$$

і одержаний ряд збігається в середньому на проміжку $(-\pi, \pi)$ і на меншому проміжку $(0, \pi)$. Зокрема

$$\lim_{n\to\infty}\int_0^{\pi}\left|f(x) - \left(\frac{a_0}{2} + \sum_{k=1}^{n} a_k \cos kx\right)\right|^2 dx = 0$$

або

$$f(x) \stackrel{L_2(0,\pi)}{=} \frac{a_0}{2} + \sum_{k=1}^{\infty} a_k \cos kx. \qquad (3.43)$$

Система функцій $\{\cos nx\}_{n=0}^{\infty}$ ортогональна на $[0, \pi]$, тому рівність (3.43) гарантує єдиність коефіцієнтів $a_k$. Отже, система $\{\cos nx\}_{n=0}^{\infty}$ повна в просторі $L'_2(0, \pi)$.

Аналогічно проводиться доведення повноти системи $\{\sin nx\}_{n=1}^{\infty}$.

Теорему доведено.

### 3.3. Наближення неперервних функцій алгебраїчними многочленами

**3.3.1. Повнота системи невід'ємних цілих степенів змінної.** Попередньо розглянуто задачу про рівномірну апроксимацію неперервної на проміжку функції тригонометричними поліномами (коефіцієнти яких не є коефіцієнтами Фур'є цієї функції).

Розглянемо допоміжне твердження, яке має назву теореми Вейєрштрасса про рівномірне наближення неперервної функції алгебраїчними многочленами $[9, 11, 22]$.

*Л е м а  1 (Вейєрштрасс).* Якщо функція $f(x)$ неперервна на відрізку $[a, b]$, то, яке б не було число $\varepsilon > 0$, знайдеться алгебраїчний многочлен
$$P_n(x) = c_0 + c_1 x + c_2 x^2 + \ldots + c_n x^n,$$
що для всіх $x \in [a, b]$ виконується нерівність
$$|f(x) - P_n(x)| < \varepsilon. \qquad (3.44)$$

*Д о в е д е н н я.* Спочатку, використавши підстановку
$$x = a + \frac{t}{\pi}(b - a),$$
одержимо функцію $f^*(t)$, неперервну на відрізку $[0, \pi]$. Потім, продовживши її на відрізок $[-\pi, 0]$ як парну функцію
$$f^*(-t) = f^*(t),\ 0 < x \leq \pi,$$
матимемо функцію, що задовольняє умови леми 1 (п. 3.2). Тоді знайдеться такий тригонометричний поліном $\sigma_n(t)$, що для будь-якого числа $\varepsilon > 0$ і всіх $t \in [-\pi, \pi]$ справджується нерівність
$$\left|f^*(t) - \sigma_n(t)\right| < \frac{\varepsilon}{2}. \qquad (3.45)$$

Зобразивши кожну тригонометричну функцію полінома $\sigma_n(t)$ у вигляді степеневого ряду, збіжного на всій дійсній осі, одержимо степеневий ряд
$$\sigma_n(t) = \sum_{m=0}^{\infty} c_m^{(n)} t^m.$$

На проміжку $[-\pi, \pi]$ цей ряд збігається рівномірно. Тому для вибраного $\varepsilon > 0$ і всіх $t \in [-\pi, \pi]$ $n$-а частинна сума $P_n^*(t)$ цього ряду при досить великому $n$ справджує нерівність

$$\left|\sigma_n(t) - P_n^*(t)\right| < \frac{\varepsilon}{2}. \qquad (3.46)$$

Використовуючи нерівності (3.45) і (3.46), одержимо

$$\left|f^*(t) - P_n^*(t)\right| \leq \left|f^*(t) - \sigma_n(t)\right| + \left|\sigma_n(t) - P_n^*(t)\right| \leq \varepsilon.$$

Повертаючись тут знову до змінної $x$, одержимо нерівність (3.44).

Лему доведено.

*З а у в а ж е н н я   1*. Доведеній лемі (як і для випадку рівномірного наближення неперервної функції тригонометричним поліномом) можна дати інше формулювання: *функція $f(x)$, неперервна на відрізку $[a, b]$, розвивається на цьому відрізку в рівномірно збіжний ряд, членами якого є алгебраїчні многочлени.*

*З а у в а ж е н н я   2*. Ґрунтуючись на лемі 1, можна стверджувати, що система невід'ємних цілих степенів змінної $x$ є *лінійно щільною* в просторі неперервних функцій (див. зауваження 2 до леми 1, п. 3.2).

*Т е о р е м а   1*. *Система невід'ємних цілих степенів змінної $x$ повна в розумінні середнього квадратичного на множині функцій, неперервних на будь-якому відрізку.*

*Д о в е д е н н я*. Нехай функція $f(x)$ неперервна на відрізку $[a, b]$. Тоді за лемою 1 для кожного $\varepsilon > 0$ існує многочлен $P_n(x)$, що для всіх $x \in [a, b]$ виконується нерівність

$$|f(x) - P_n(x)| < \frac{\varepsilon}{\sqrt{b-a}}.$$

Ґрунтуючись на цій нерівності, для квадратичного відхилення маємо оцінку

$$\left(\int_a^b |f(x) - P_n(x)|^2 dx\right)^{1/2} < \left(\frac{\varepsilon^2}{b-a}\int_a^b dx\right)^{1/2} = \varepsilon.$$

Одержана нерівність показує, що системи степенів $x$ повна в розумінні середньо квадратичного на відрізку $[a, b]$.

Теорему доведено.

З будь-якою лінійно незалежною системою функцій в евклідовому просторі можна пов'язати ортогональну систему функцій. Процес побудови ортогональної системи функцій із заданої системи лінійно незалежних функцій називається *ортогоналізацією за Шмідтом*.

*Т е о р е м а   2 .* Нехай в евклідовому просторі $E$ задана система $\{f_k\}$ лінійно незалежних елементів.

Тоді в $E$ існує єдина ортонормована система (ОНС) $\varphi_k$ елементів, яка задовольняє умови:

а) кожний елемент $\varphi_n$ є лінійною комбінацією перших $n$ елементів системи $\{f_k\}$,

$$\varphi_n = \sum_{k=1}^{n} a_n^k f_k, \text{ при цьому } a_n^n > 0;$$

б) кожний елемент $f_n$ однозначно виражається через перші $n$ елементів системи $\{\varphi_k\}$

$$f_n = \sum_{k=1}^{n} b_n^k \varphi_k, \text{ при цьому } b_n^n > 0;$$

в) якщо система $\{f_k\}$ – повна в $E$, тобто довільний елемент простору $E$ може бути наближений (за нормою $E$) з будь-якою точністю скінченною лінійною комбінацією системи $\{f_k\}$, то і система $\{\varphi_k\}$ – повна в $E$.

*Д о в е д е н н я .* Оскільки система $\{f_k\}$ лінійно незалежна, ні один з елементів $f_k$ не є нуль-елементом. Приймемо

$$\varphi_1 = \frac{f_1}{\|f_1\|}.$$

Вибираючи елемент $\varphi_1$, запишемо вираз елемента $f_2$ у вигляді

$$f_2 = c_2^1 \varphi_1 + h_2$$

і підберемо сталу $c_2^1$ так, щоб елемент $h_2$ був ортогональний до $\varphi_1$, тобто

$$(f_2, \varphi_1) = c_2^1.$$

Зазначимо, що $h_2 \neq o$, оскільки у протилежному випадку елементи $f_1$ і $f_2$ – лінійно залежні. Приймемо

$$\varphi_2 = \frac{h_2}{\|h_2\|} = \frac{f_2 - (f_2, \varphi_1)\varphi_1}{\|f_2 - (f_2, \varphi_1)\|}.$$

Отже, при $n = 2$ побудована система $\{\varphi_1, \varphi_2\}$, що стверджується теоремою.

За методом індукції, нехай вже побудована ортонормована система $\{\varphi_1, \varphi_2, \ldots \varphi_{n-1}\}$, кожний елемент якої виражається через відповідні елементи системи $\{f_1, f_2, \ldots f_{n-1}\}$. Запишемо вираз елемента $f_n$ у вигляді

$$f_n = c_n^1 \varphi_1 + c_n^2 \varphi_2 + \ldots c_n^{n-1} \varphi_{n-1} + h_n \qquad (3.47)$$

і підберемо коефіцієнти $c_n^k$ так, щоби елемент $h_n$ був ортогональний до всіх $\varphi_k$, тобто з умов

$$(f_n, \varphi_k) = \left(c_n^1 \varphi_1 + c_n^2 \varphi_2 + \ldots c_n^{n-1} \varphi_{n-1}, \varphi_k\right), \; k = 1, 2, \ldots, n-1.$$

Одержимо $(f_n, \varphi_k) = c_n^k$. Оскільки $h_n \neq o$ (у протилежному випадку система $\{f_1, f_2, \ldots f_n\}$ – лінійно залежна), виберемо

$$\varphi_n = \frac{h_n}{\|h_n\|} = \frac{f_n - \sum_{k=1}^{n-1}(f_n, \varphi_k)\varphi_k}{\left\|f_n - \sum_{k=1}^{n-1}(f_n, \varphi_k)\varphi_k\right\|}. \qquad (3.48)$$

Отже, побудовано ортогональну систему $\{\varphi_1, \varphi_2, \ldots \varphi_n\}$ з вихідної системи $\{f_1, f_2, \ldots f_n\}$ так, як того вимагає твердження теореми. Зокрема, формулами (3.47), (3.48) встановлюють лінійні залежності між елементами цих систем. При цьому коефіцієнти біля старших членів елементів в цих поданнях додатні.

Якщо елемент $f$ простору $E$ наближений з довільною точністю скінченною лінійною комбінацією елементів системи $\{f_k\}$, то з відповідної апроксимаційної нерівності з використанням залежностей (3.47) одержимо апроксимаційну нерівність з цією ж точністю для елемента $f$ і скінченної лінійної комбінації елементів системи $\{\varphi_k\}$.

Теорему доведено.

### 3.3.2. Повнота системи многочленів Лежандра.

***Т е о р е м а 3 .*** *Система нормованих многочленів Лежандра*

$$p_n(x) = \sqrt{\frac{2}{2n+1}} P_n(x) = \sqrt{\frac{2}{2n+1}} \frac{1}{2^n n!} \frac{d^n}{dx^n}(x^2-1)^n$$

*повна в просторі* $CL_2[-1,1]$.

*Д о в е д е н н я .* Система степенів $\{x^n\}_{n=0}^{\infty}$ за теоремою 1 лінійно незалежна і повна в $CL_2[-1,1]$. Можна переконатися, що многочлени Лежандра побудовано за схемою ортогоналізації системи степенів і відповідні залежності задаються формулами

$$p_n(x) = \sqrt{\frac{2}{2n+1}} \frac{1}{2^n} \sum_{k=0}^{[n/2]} (-1)^k C_n^k C_{2(n-k)}^n x^{n-2k},$$

$$x^n = \sqrt{\frac{2n+1}{2}} \sum_{k=0}^{[n/2]} \frac{2^{n-2k}(2n-4k+1)}{(2n-2k+1)} \frac{C_n^k}{C_{2(n-k)}^{n-k}} p_{n-2k}(x). \quad (3.49)$$

Коефіцієнт біля найвищого степеня $x^n$ і коефіцієнт біля старшого многочленна $p_n(x)$ у виразах (3.49) додатні.

Отже, система многочленів Лежандра повна в просторі $CL_2[-1,1]$.

Теорему доведено.

*Н а с л і д о к .* *Оскільки система многочленів Лежандра повна в просторі* $CL_2[-1,1]$, *для кожної неперервної на відрізку* $[-1,1]$ *функції* $f(x)$ *справедливе розвинення в ряд Фур'є, яке збігається в середньому*,

$$f(x) \stackrel{L_2}{=} \sum_{k=0}^{\infty} c_k P_k(x),$$

*де* $c_k = \dfrac{(f, P_k)}{\|P_k\|} = \dfrac{2k+1}{2} \int_{-1}^{1} f(x) P_k(x) dx$.

Важливими є умови, за яких ряд Фур'є за многочленами Лежандра збігається рівномірно. Відзначимо, що з рівномірної збіжності ряду Фур'є функції $f(x)$ не випливає збіжності ряду до функції $f(x)$. Однак, якщо розвинення одержано за повною ортогональною системою, то рівномірна збіжність гарантує збіжність до $f(x)$. Розглянемо цей факт.

**Т е о р е м а  4 .** *Нехай система функцій $\{f_n(x)\}$ повна в просторі $CL_2[a,b]$.*

*Тоді, якщо ряд Фур'є функції $f(x) \in CL_2[a,b]$,*

$$\sum_{k=1}^{n} c_k f_k(x), \quad c_k = (f, f_k), \qquad (3.50)$$

*збігається рівномірно, то $f(x)$ є сумою цього ряду не тільки в розумінні середньоквадратичної збіжності, але і в розумінні рівномірної збіжності.*

*Д о в е д е н н я .* Нехай ряд (3.50) збігається до неперервної на $[a, b]$ функції $\sigma(x)$, оскільки він збігається рівномірно. За нерівністю трикутника одержимо

$$\left(\int_a^b |f(x) - \sigma(x)|^2 dx\right)^{1/2} \le \left(\int_a^b \left|f(x) - \sum_{k=1}^n c_k f_k(x)\right|^2 dx\right)^{1/2} +$$

$$+ \left(\int_a^b \left|\sum_{k=1}^n c_k f_k(x) - \sigma(x)\right|^2 dx\right)^{1/2}.$$

Перший доданок при $n \to \infty$ прямує до нуля, оскільки ряд (3.50) збігається в середньому до функції $f(x)$. Другий доданок також прямує до нуля, оскільки ряд (3.50) збігається рівномірно до функції $\sigma(x)$. Звідси випливає рівність

$$\int_a^b |f(x) - \sigma(x)|^2 dx = 0.$$

Оскільки функції $f(x)$ і $\sigma(x)$ неперервні, то $f(x) \equiv \sigma(x)$.

Теорему доведено.

*Н а с л і д о к .* *Повертаючись до системи многочленів Лежандра, можна стверджувати, що рівномірно збіжний ряд Фур'є для функції $f(x) \in CL_2[a, b]$ за системою многочленів Лежандра збігається до функції $f(x)$,*

$$f(x) = \sum_{k=0}^{\infty} c_k P_k(x),$$

де $c_k = \dfrac{2k+1}{2} \int\limits_{-1}^{1} f(x) P_k(x) dx$.

Для одержання відповіді на питання умов рівномірної збіжності ряду Фур'є функції $f(x)$ розглянемо наступне твердження.

**Т е о р е м а  5 .** *Якщо функція $f(x)$ має неперервні другі похідні на відрізку $[-1,1]$, то її ряд Фур'є за многочленами Лежандра збігається рівномірно до функції $f(x)$ на відрізку $[-1,1]$.*

*Д о в е д е н н я .* Запишемо ряд Фур'є функції за поліномами Лежандра і оцінимо його коефіцієнти,

$$\sum_{k=0}^{\infty} c_k P_k(x), \qquad (3.51)$$

де $c_k = \dfrac{2k+1}{2} \int\limits_{-1}^{1} f(t) P_k(t) dt$.

Використовуючи рекурентні співвідношення (3.7), формулу інтегрування за частинами і формули $P_n(1) = 1$, $P_n(-1) = (-1)^n P_n(1)$, знайдемо при $k \geq 2$

$$c_k = \frac{2k+1}{2} \int\limits_{-1}^{1} f(t) P_k(t) dt = \frac{1}{2} \int\limits_{-1}^{1} f(t) \left[ \frac{dP_{k+1}(t)}{dt} - \frac{dP_{k-1}(t)}{dt} \right] dt =$$

$$= \frac{1}{2} \left[ f(t) P_{k+1}(t) - f(t) P_{k-1}(t) \right]\Big|_{-1}^{1} - \frac{1}{2} \int\limits_{-1}^{1} \frac{df(t)}{dt} \left[ P_{k+1}(t) - P_{k-1}(t) \right] dt =$$

$$= -\frac{1}{2} \int\limits_{-1}^{1} \frac{df(t)}{dt} \left[ P_{k+1}(t) - P_{k-1}(t) \right] dt =$$

$$= -\frac{1}{2} \int\limits_{-1}^{1} \frac{df(t)}{dt} \left[ \frac{\dfrac{dP_{k+2}(t)}{dt} - \dfrac{dP_k(t)}{dt}}{2k+3} - \frac{\dfrac{dP_k(t)}{dt} - \dfrac{dP_{k-2}(t)}{dt}}{2k-1} \right] dt =$$

$$= \frac{1}{2}\int_{-1}^{1} \frac{d^2 f(t)}{dt^2} \left[ \frac{P_{k+2}(t) - P_k(t)}{2k+3} - \frac{P_k(t) - P_{k-2}(t)}{2k-1} \right] dt.$$

Звідси отримаємо оцінку

$$|c_k| \le \frac{1}{2(2k-1)} \left[ \int_{-1}^{1} \left| \frac{d^2 f(t)}{dt^2} \right| |P_{k+2}(x)| dt + 2\int_{-1}^{1} \left| \frac{d^2 f(t)}{dt^2} \right| |P_k(x)| dt + \right.$$
$$\left. + \int_{-1}^{1} \left| \frac{d^2 f(t)}{dt^2} \right| |P_{k-2}(x)| dt \right].$$

Застосовуючи тут нерівність Коші – Буняковського, отримаємо

$$|c_k| \le \frac{1}{2(2k-1)} \left( \int_{-1}^{1} \left| \frac{d^2 f(t)}{dt^2} \right|^2 dt \right)^{1/2} \left[ \left( \int_{-1}^{1} |P_{k+2}(x)|^2 dt \right)^{1/2} + \right.$$
$$\left. + 2\left( \int_{-1}^{1} |P_k(x)|^2 dt \right)^{1/2} + \left( \int_{-1}^{1} |P_{k-2}(x)|^2 dt \right)^{1/2} \right]$$

або з врахуванням формули $\|P_n\|^2 = \frac{2}{2n+1}$

$$|c_k| \le \frac{\|f''\|}{2(2k-1)} \left( \sqrt{\frac{2}{2k+5}} + 2\sqrt{\frac{2}{2k+1}} + \sqrt{\frac{2}{2k-3}} \right) \le$$
$$\le \frac{2\|f''\|}{(2k-1)} \sqrt{\frac{2}{2k-3}} \le \frac{2\sqrt{2}}{(2k-3)^{3/2}} < \frac{1}{(k-1)^{3/2}}.$$

Використовуючи цю оцінку і враховуючи, що $|P_k(x)| \le 1$ при $|x| \le 1$, одержимо збіжний числовий ряд

$$\sum_{k=2}^{\infty} \frac{1}{(k-1)^{3/2}},$$

мажорантний для ряду (3.51). Отже, ряд (3.51) збігається рівномірно і, оскільки він збігається також в середньому до $f(x)$, то за теоремою 4 він збігається і в розумінні рівномірної збіжності до $f(x)$.

Теорему доведено.

**3.3.3. Ортогональність з вагою.** В аналізі поряд з евклідовим простором $CL[a,b]$ використовують евклідові простори, норма в яких визначається скалярним добутком

$$(f,g) = \int_a^b f(x)g(x)h(x)\,dx, \qquad (3.52)$$

де $h(x)$ – неперервна на проміжку $(a,b)$ вагова функція (вага).

Якщо вважати, що елементами відповідного лінійного простору є неперервними на $(a,b)$ функціями $f(x)$, які після множення на $\sqrt{h(x)}$ залишаються неперервними на $(a,b)$, то вираз (3.52) виражає скалярний добуток в розумінні (3.1). Тоді за теоремою 1 вираз (3.52) визначає норму

$$\|f\| = \left(\int_a^b f^2(x)h(x)\,dx\right)^{1/2}. \qquad (3.53)$$

При цьому введений простір стає евклідовим простором, який позначають через $CL_{2,h}(a,b)$, тобто у ньому справедливі поняття і твердження евклідових просторів. Простір $CL_{2,h}(a,b)$ містить систему степенів $\{x^n\}_{n=0}^{\infty}$, яка лінійно незалежна і повна в ньому. Застосовуючи процес ортогоналізації за Шмідтом, одержимо систему $\{P_{n,h}(x)\}_{n=0}^{\infty}$ многочленів, ортогональних в $CL_{2,h}(a,b)$. За цією системою можна будувати ряди Фур'є, збіжність яких за нормою простору випливає з загальної теорії.

Розглянемо приклади систем многочленів, ортогональних з вагою.

**Система похідних від многочленів Лежандра.** Розглянемо систему похідних $k$-го порядку від многочленів Лежандра

$$\left\{G_n^{(k,k)}(x) = \frac{d^k}{dx^k}P_{n+k}(x)\right\}_{n=0}^{\infty} \text{ або } \left\{G_{n-k}^{(k,k)}(x) = \frac{d^k}{dx^k}P_n(x)\right\}_{n=0}^{\infty}. \qquad (3.54)$$

Система (3.54) ортогональна на відрізку $[-1,1]$ з вагою $h(x) = (1-x^2)^k$, тобто

$$\int\limits_{-1}^{1} G_{m-k}^{(k,\,k)}(x) G_{n-k}^{(k,\,k)}(x)\left(1-x^2\right)^k dx = 0, \ m \neq n. \qquad (3.55)$$

Покажемо виконання рівності (3.55) для випадку $k = 1$. Інтегруючи частинами інтеграл в (3.55) з використанням рівняння (3.8), одержимо

$$\int\limits_{-1}^{1} G_{m-1}^{(1,\,1)} G_{n-1}^{(1,\,1)}\left(1-x^2\right) dx = \int\limits_{-1}^{1} \frac{dP_m}{dx}\frac{dP_n}{dx}\left(1-x^2\right) dx =$$

$$= \left[P_m(x)\frac{dP_n}{dx}\left(1-x^2\right)\right]\Bigg|_{-1}^{1} + n(n+1)\int\limits_{-1}^{1} P_m(x) P_n(x)\, dx = 0.$$

Диференціюванням $k$ раз рівняння (3.8) прийдемо до такого рівняння відносно похідних від многочленів Лежандра

$$\left(1-x^2\right)\frac{d^2 G_{n-k}^{(k,\,k)}}{dx^2} - 2(k+1)x\frac{dG_{n-k}^{(k,\,k)}}{dx} +$$

$$+ (n-k)(n-k+2k+1) G_{n-k}^{(k,\,k)} = 0. \qquad (3.56)$$

Знайдемо квадрат норми многочлена $G_{n-k}^{(k,\,k)}$. Інтегруючи частинами, одержимо

$$\left\| G_{n-k}^{(k,\,k)} \right\|^2 = \int\limits_{-1}^{1}\left[G_{n-k}^{(k,\,k)}\right]^2\left(1-x^2\right)^k dx = \int\limits_{-1}^{1}\left[\frac{d^k}{dx^k}P_n(x)\right]^2\left(1-x^2\right)^k dx =$$

$$= -\int\limits_{-1}^{1}\frac{d^{k-1}P_n}{dx^{k-1}}\frac{d}{dx}\left[\left(1-x^2\right)^k \frac{d^k P_n}{dx^k}\right] dx.$$

Диференціальне рівняння (3.56) після множення на $\left(1-x^2\right)^k$ можна записати у вигляді

$$\frac{d}{dx}\left[\left(1-x^2\right)^{k+1}\frac{d^{k+1}P_n}{dx^{k+1}}\right] = -(n-k)(n+k+1)\left(1-x^2\right)^k \frac{d^k P_n}{dx^k}.$$

Враховуючи цю залежність, одержимо рекурентне співвідношення

$$\left\| G_n^{(k,\,k)} \right\|^2 = (n-k+1)(n+k)\int\limits_{-1}^{1}\left[\frac{d^{k-1}P_n}{dx^{k-1}}\right]^2\left(1-x^2\right)^{k-1} dx$$

або

$$\left\|G_n^{(k,k)}\right\|^2 = (n-k+1)(n+k)\left\|G_n^{(k-1,k-1)}\right\|^2.$$

Звідси

$$\left\|G_n^{(k,k)}\right\|^2 = \frac{n!}{(n-k)!}\frac{(n+k)!}{n!}\left\|G_n^{(0,0)}\right\|^2 = \frac{(n+k)!}{(n-k)!}\frac{2}{2n+1}.$$

Враховуючи тепер рівності (3.55), запишемо співвідношення ортогональності

$$\int_{-1}^{1} G_{m-k}^{(k,k)}(x)\, G_{n-k}^{(k,k)}(x)(1-x^2)^k\, dx = \begin{cases} 0, & m \neq n, \\ \dfrac{(n+k)!}{(n-k)!}\dfrac{2}{2n+1}, & m = n. \end{cases} \quad (3.57)$$

Система похідних від многочленів Лежандра повна в просторі $CL_2[-1,1]$, оскільки виконуються умови теореми 2.

**Система приєднаних функцій Лежандра.** В аналізі також використовують приєднані функції Лежандра

$$P_n^k(x) = \sqrt{(1-x^2)^k}\,\frac{d^k}{dx^k}P_n(x) = \sqrt{(1-x^2)^k}\; G_{n-k}^{(k,k)}.$$

Після заміни функції в рівнянні (3.56) одержимо таке рівняння відносно приєднаних функцій Лежандра

$$\left(1-x^2\right)\frac{d^2 P_n^k}{dx^2} - 2x\frac{dP_n^k}{dx} + \left[n(n+1) - \frac{k^2}{1-x^2}\right]P_n^k = 0.$$

Із співвідношення (3.55) випливає ортогональність системи приєднаних функцій

$$\int_{-1}^{1} P_m^k(x)P_n^k(x)\,dx = \begin{cases} 0, & m \neq n, \\ \dfrac{(n+k)!}{(n-k)!}\dfrac{2}{2n+1}, & m = n. \end{cases}$$

Можна показати, що система $\{P_n^k(x)\}$ приєднаних функцій $k$-го порядку також повна в просторі $CL_2[-1,1]$.

Дійсно, нехай $f(x)$ – неперервна на відрізку $[-1,1]$ з інтегрованим квадратом. Для довільного малого $\varepsilon > 0$ підберемо неперервну на відрізку $[-1,1]$ функцію $\varphi(x)$, тотожньо рівною нулеві на досить малих відрізках $[-1, -1+\eta]$ і $[1-\eta, 1]$, $\eta > 0$, таку що

$$\int\limits_{-1}^{1}[f(x)-\varphi(x)]^2 dx < \varepsilon. \qquad (3.58)$$

Оскільки будь-який поліном за степенями $x$ можна записати у вигляді лінійної комбінації поліномів $\dfrac{d^k}{dx^k}P_n(x)$, за теоремою Вейєрштрасса для довільної неперервної на відрізку $[-1,1]$ функції $\varphi(x)$ і заданого $\varepsilon > 0$ справедлива нерівність

$$\left| \frac{\varphi(x)}{\sqrt{(1-x^2)^k}} - \sum_{n=k}^{N} c_n \frac{d^k}{dx^k} P_n(x) \right| < \varepsilon,$$

де $N$ – досить велике натуральне число, $c_n$ – сталі.

Помноживши цю нерівність на $\sqrt{(1-x^2)^k}$, одержимо

$$\left| \varphi(x) - \sum_{n=k}^{N} c_n \frac{d^k}{dx^k} P_n^k(x) \right| < \varepsilon.$$

Звідси випливає така нерівність:

$$\int\limits_{-1}^{1}\left[ \varphi(x) - \sum_{n=k}^{N} c_n P_n^k(x) \right]^2 dx < \varepsilon. \qquad (3.59)$$

Врахувавши нерівності (3.58) і (3.59), знайдемо оцінку

$$\int\limits_{-1}^{1}\left[ f(x) - \sum_{n=k}^{N} c_n P_n^k(x) \right]^2 dx =$$

$$= \int\limits_{-1}^{1}\left\{ [f(x)-\varphi(x)] + \left[ \varphi(x) - \sum_{n=k}^{N} c_n P_n^k(x) \right] \right\}^2 dx \le$$

$$\le 2\int\limits_{-1}^{1}[f(x)-\varphi(x)]^2 dx + 2\int\limits_{-1}^{1}\left[ \varphi(x) - \sum_{n=k}^{N} c_n P_n^k(x) \right]^2 dx < 4\varepsilon.$$

Ця нерівність показує, що система приєднаних функцій $k$-го порядку повна в просторі $CL_2[-1,1]$.

**Система многочленнів Чебишова.** В теорії наближень важливу роль відіграють многочлени Чебишова

$$T_n(x) = \cos(n \arccos x), \ n = 0, 1, \ldots . \qquad (3.60)$$

Система многочленів $\{T_n(x)\}_{n=0}^{\infty}$, ортогональна на відрізку $[-1, 1]$ з вагою $h(x) = \dfrac{1}{\sqrt{1-x^2}}$.

Многочлени Чебишова можна також визначити за рекурентними формулами
$$T_0(x) = 1, \ T_1(x) = x,$$
$$T_n(x) = 2x T_{n-1}(x) - T_{n-2}(x), \ n = 2, 3, \ldots,$$

або за формулою
$$T_n(x) = \frac{(-1)^n \sqrt{1-x^2}}{(2n-1)!!} \frac{d^n}{dx^n} \sqrt{(1-x^2)^{2n-1}}, \ n = 0, 1, \ldots,$$

де $(2n-1)!! = \begin{cases} 1, & n = 0, \\ 1 \cdot 3 \cdot 5 \cdot \ldots \cdot (2n-1), & n = 1, 2, \ldots . \end{cases}$

Звідси знайдемо вирази многочленів у розгорнутому вигляді
$$T_0(x) = 1, \ T_n(x) = \frac{n}{2} \sum_{k=0}^{[n/2]} \frac{(-1)^k 2^{n-2k} C_{n-k}^k}{n-k} x^{n-2k}, \ n = 1, 2, \ldots .$$

Оскільки функції $T_n(\cos t) = \cos nt$ задовольняють рівняння $\dfrac{d^2 y}{dt^2} + n^2 y = 0$, то, зробивши заміну змінної $t = \arccos x$, прийдемо до такого рівняння
$$(1 - x^2) \frac{d^2 y}{dx^2} - x \frac{dy}{dx} + n^2 y = 0,$$

якому задовольняє многочлен $T_n(x)$. Останнє рівняння можна записати ще у такому вигляді
$$\sqrt{1-x^2} \frac{d}{dx}\left(\sqrt{1-x^2} \frac{dy}{dx}\right) + n^2 y = 0 . \qquad (3.61)$$

Легко перевіряється ортогональність системи (3.60). Проводячи заміну $x = \cos\varphi$, $0 \le \varphi \le \pi$, одержимо
$$\int_{-1}^{1} T_n(x) T_m(x) \frac{dx}{\sqrt{1-x^2}} = \int_{\pi}^{0} T_n(\cos\varphi) T_m(\cos\varphi) \frac{(-\sin\varphi)}{|\sin\varphi|} d\varphi =$$

$$= \frac{1}{2}\int\limits_{-\pi}^{\pi} \cos n\varphi \cos m\varphi \, d\varphi = \begin{cases} 0, & m \neq n, \\ \pi/2, & m = n, \\ \pi, & m = n = 0. \end{cases}$$

Система (3.60) справджує умови теореми 2 і повна в просторі $CL_2[-1,1]$.

*З а у в а ж е н н я*. Многочлени Чебишова і многочлени Лежандра можна одержати $[11]$, як частинні випадки, з многочленів Гегенбауера $G_n^{(\sigma,\sigma)}$. При $\sigma = -\frac{1}{2}$ і $\sigma = 0$ многочлени $G_n^{(\sigma,\sigma)}$ перетворюються в многочлени $T_n(x)$ і многочлени $P_n(x)$ відповідно. Якщо $\sigma = k$, де $k$ – ціле невід'ємне число, то многочлени Гегенбауера $G_{n-k}^{(k,k)}$, $n \geq k$, стають похідними від многочленів Лежандра $G_{n-k}^{(k,k)}(x) = \dfrac{d^k}{dx^k} P_n(x)$.

Многочлен $G_n^{(\sigma,\sigma)}$ задовольняє рівняння

$$\frac{1}{(1-x^2)^\sigma} \frac{d}{dx}\left[(1-x^2)^{\sigma+1} \frac{dy}{dx}\right] + \lambda y = 0, \text{ або}$$

$$(1-x^2)\frac{d^2 y}{dx^2} - 2x(\sigma+1)\frac{dy}{dx} + \lambda y = 0,$$

де $h(x) = (1-x^2)^\sigma$; $\lambda = n(n+2\sigma+1)$.

Система $\{G_n^{(\sigma,\sigma)}\}_{n=0}^\infty$ є результатом ортогоналізації системи степенів $\{x^n\}_{n=0}^\infty$ і тому вона повна у просторі $CL_2[-1,1]$.

**Системи многочленів ортогональні з вагою на нескінченному проміжку.** У прикладних задачах мають застосування також функції, ортогональні з вагою на нескінченному або пів-нескінченному проміжку. Їх вивчення, як і вивчення раніше розглянутих систем функцій, диктується крайовими задачами математичної фізики, розв'язання яких ґрунтується на розв'язанні відповідних задач Штурма – Ліувілля $[11]$.

Наведемо для прикладу найважливіші системи функцій і відповідні їм диференціальні рівняння.

1. Система $\{H_n(x)\}_{n=0}^{\infty}$ многочленів Ерміта повна в просторі $CL_{2,h}(-\infty, +\infty)$ і задовольняє диференціальне рівняння

$$\frac{1}{h(x)}\frac{d}{dx}\left(h(x)\frac{dy}{dx}\right)+\lambda y \equiv \frac{d^2 y}{dx^2}-2x\frac{dy}{dx}+\lambda y = 0,$$

де $h(x)=e^{-x^2}$; $\lambda = 2n$.

2. Система $\{L_n(x)\}_{n=0}^{\infty}$ многочленів Лагерра повна в просторі $CL_{2,h}(0, +\infty)$ і задовольняє диференціальне рівняння

$$\frac{1}{h(x)}\frac{d}{dx}\left(h(x)\frac{dy}{dx}\right)+\lambda y \equiv x\frac{d^2 y}{dx^2}+(1-x)\frac{dy}{dx}+\lambda y = 0,$$

де $h(x)=e^{-x}$; $\lambda = n$.

За аналогією з приєднаними функціями Лежандра побудовано системи функцій Ерміта і функцій Лагерра

$$\Psi_n(x)= e^{-\frac{x^2}{2}}H_n(x), \qquad W_n(x)= e^{-\frac{x}{2}}L_n(x).$$

### 3.4. Ряди за системами функцій двох змінних

**3.4.1. Повнота системи функцій двох змінних.** Нехай функція двох змінних $f(x,y)$ інтегровна з квадратом у прямокутнику $D = \{a \leq x \leq b, c \leq y \leq d\}$

$$\iint\limits_D f^2(x,y)\,dxdy < +\infty.$$

Задана також система $\{u_n(x,y)\}_{n=1}^{\infty}$ неперервних в області $D$ функцій і функція $h(x,y)$ неперервна і додатна в $D$.

Така система $\{u_n(x,y)\}_{n=1}^{\infty}$ називається ортогональною з вагою $h(x,y)$, якщо виконується рівність

$$\iint\limits_D u_n(x,y)u_m(x,y)h(x,y)\,dxdy = \begin{cases} 0, & m \neq n, \\ C_n \neq 0, & m = n \end{cases}$$

і називається ортонормованою з вагою $h(x,y)$, якщо

$$\iint\limits_D u_n(x,y)u_m(x,y)h(x,y)dxdy = \begin{cases} 0, & m \neq n, \\ 1, & m = n. \end{cases}$$

Ортонормована система $\{u_n(x,y)\}_{n=1}^{\infty}$ – повна у просторі функцій інтегрованих з квадратом у прямокутнику $D$, якщо справедлива рівність Парсеваля

$$\iint\limits_D f^2(x,y)h(x,y)dxdy = \sum_{n=1}^{\infty} c_n^2,$$

де $c_n$ – коефіцієнти Фур'є функції $f(x,y)$,

$$c_n = \iint\limits_D f(x,y)u_n(x,y)h(x,y)dxdy.$$

Ряд Фур'є для функції $f(x,y)$ за системою $\{u_n(x,y)\}_{n=1}^{\infty}$ має вигляд

$$f(x,y) \sim \sum_{n=1}^{\infty} c_n u_n(x,y).$$

В задачах математичної фізики використовують ортогональні системи функцій від двох змінних, які є добутком функції тільки від $x$ та функції тільки від $y$. Характерним для таких систем є те, що властивість їх повноти випливає з властивості повноти систем функцій однієї змінної.

*Т е о р е м а   1 .  Нехай $\{\varphi_m(x)\}_{m=1}^{\infty}$ – повна система неперервних на відрізку $[a,b]$ функцій, ортонормована з вагою $h_1(x)$, і нехай система неперервних функцій $\{\psi_n(y)\}_{n=1}^{\infty}$ повна і ортонормована з вагою $h_2(y)$ на відрізку $[c,d]$.*

*Тоді система функцій двох змінних*

$$\{u_{mn}(x,y) = \varphi_m(x)\psi_n(y)\}_{m=1,n=1}^{\infty}$$

*повна і ортонормована з вагою $h(x,y) = h_1(x)h_2(y)$ в замкнутому прямокутнику $a \le x \le b$, $c \le y \le d$.*

*Д о в е д е н н я .* Спочатку перевіримо ортонормованість системи $\{u_{mn}(x,y)\}_{m=1,n=1}^{\infty}$ з вагою $h(x,y)$

$$\iint\limits_D u_{mn}(x,y)u_{m'n'}(x,y)h(x,y)dxdy = \int_a^b \varphi_m(x)\varphi_{m'}(x)h_1(x)dx \cdot$$

$$\cdot \int\limits_c^d \psi_n(y)\psi_{n'}(y)h_2(y)dy = \begin{cases} 0, & |m'-m|+|n'-n| \neq 0, \\ 1, & m = m', n = n'. \end{cases}$$

Тепер покажемо, що для функції $f(x, y)$, неперервної в області $D$ виконується рівність Парсеваля. Внаслідок повноти системи $\{\varphi_m(x)\}_{m=1}^{\infty}$ маємо

$$\int\limits_a^b f^2(x, y)h_1(x)dx = \sum_{m=1}^{\infty} g_m^2(y), \qquad (3.62)$$

де

$$g_m(y) = \int\limits_a^b f(x, y)\varphi_m(x)h_1(x)dx.$$

Ліва частина рівності (3.62) і $g_n(y)$ неперервні функції. Тому за ознакою Діні (теорема 3, п. 2.1) ряд (3.62) рівномірно збігається.

Якщо помножити рівність (3.62) на $h_2(y)$ і проінтегрувати по $y$ у межах від $c$ до $d$, то одержимо

$$\int\limits_a^b\int\limits_c^d f^2(x, y)h_1(x)h_2(y)dxdy = \sum_{m=1}^{\infty}\int\limits_c^d g_m^2(y)h_2(y)dy.$$

Внаслідок повноти системи функцій $\{\psi_n(y)\}_{n=1}^{\infty}$, маємо

$$\int\limits_c^d g_m^2(y)dy = \sum_{n=1}^{\infty} c_{mn}^2,$$

де

$$c_{mn} = \int\limits_a^b\int\limits_c^d f(x, y)\varphi_m(x)\psi_n(y)h_1(x)h_2(y)dxdy. \qquad (3.63)$$

Тому справедлива рівність Парсеваля

$$\int\limits_a^b\int\limits_c^d f^2(x, y)h_1(x)h_2(y)dxdy = \sum_{n=1}^{\infty}\sum_{m=1}^{\infty} c_{mn}^2. \qquad (3.64)$$

Теорему доведено.

Ряд Фур'є для заданої у прямокутнику $a \leq x \leq b$, $c \leq y \leq d$ й інтегровної з квадратом функції $f(x, y)$ за ортонормованою системою $\{u_{mn}(x, y)\}_{m=1, n=1}^{\infty}$ наступний:

$$f(x, y) \sim \sum_{m=1}^{\infty}\sum_{n=1}^{\infty} c_{mn} u_{mn}(x, y), \qquad (3.65)$$

де $c_{mn}$ – коефіцієнти Фур'є (3.63) функції $f(x, y)$.

**3.4.2. Повнота подвійної системи тригонометричних функцій.** Розглянемо тригонометричну систему типу (2.87), ортонормовану у прямокутнику $Q = \{|x| \le p\,; |y| \le q\}$,

$$\begin{cases} u_{00}^a(x, y) = \dfrac{1}{4pq},\ u_{m0}^a(x, y) = \dfrac{1}{2pq}\cos\dfrac{m\pi x}{p}, \\[4pt] u_{m0}^b(x, y) = \dfrac{1}{2pq}\sin\dfrac{m\pi x}{p},\ u_{0n}^c(x, y) = \dfrac{1}{2pq}\cos\dfrac{n\pi y}{q}, \\[4pt] u_{0n}^b(x, y) = \dfrac{1}{2pq}\sin\dfrac{n\pi y}{q},\ u_{mn}^a(x, y) = \dfrac{1}{pq}\cos\dfrac{m\pi x}{p}\cos\dfrac{n\pi y}{q}, \\[4pt] u_{mn}^b(x, y) = \dfrac{1}{pq}\sin\dfrac{m\pi x}{p}\cos\dfrac{n\pi y}{q}, u_{mn}^c(x, y) = \dfrac{1}{pq}\cos\dfrac{m\pi x}{p}\sin\dfrac{n\pi y}{q}, \\[4pt] u_{mn}^d(x, y) = \dfrac{1}{pq}\sin\dfrac{m\pi x}{p}\sin\dfrac{n\pi y}{q} \end{cases} (m = 1, 2, \ldots, n = 1, 2, \ldots). \qquad (3.66)$$

За теоремою 1 система (3.66) повна в просторі функцій, заданих у прямокутнику $Q$ і інтегровних з квадратом. Для функції $f(x, y)$ з цього простору маємо подвійний ряд Фур'є

$$f(x, y) \sim \sum_{m=0}^{\infty}\sum_{n=0}^{\infty}\bigl[a_{mn} u_{mn}^a(x, y) + b_{mn} u_{mn}^b(x, y) + \\ + c_{mn} u_{mn}^c(x, y) + d_{mn} u_{mn}^d(x, y)\bigr] \qquad (3.67)$$

і справедлива рівність (3.64), яка для тригонометричної системи функцій подібно до рівності (3.40) запишеться у вигляді

$$\sum_{m=0}^{\infty}\sum_{n=0}^{\infty}\left(a_{mn}^2 + b_{mn}^2 + c_{mn}^2 + d_{mn}^2\right) = \iint_Q f^2(x, y)\,dxdy, \qquad (3.68)$$

де

$$a_{mn} = \iint_Q f(x, y) u_{mn}^a(x, y)\,dxdy,\quad b_{mn} = \iint_Q f(x, y) u_{mn}^b(x, y)\,dxdy,$$

$$c_{mn} = \iint\limits_Q f(x,y) u^c_{mn}(x,y)\, dxdy, \quad d_{mn} = \iint\limits_Q f(x,y) u^d_{mn}(x,y)\, dxdy$$

$$(m = 0, 1, \ldots, n = 0, 1, \ldots).$$

Рівність (3.68) є рівністю Парсеваля для тригонометричної системи функцій у прямокутнику $Q$.

Отже, якщо функція $f(x, y)$ інтегровна з квадратом у прямокутнику $Q$, зокрема неперервна, то подвійний ряд Фур'є (3.67) збігається в середньому до $f(x, y)$

$$\lim_{\substack{M \to \infty \\ N \to \infty}} \iint\limits_Q \left\{ f(x,y) - \sum_{m=0}^{M} \sum_{n=0}^{N} \left[ a_{mn} u^a_{mn}(x,y) + b_{mn} u^b_{mn}(x,y) + \right. \right.$$

$$\left. \left. + c_{mn} u^c_{mn}(x,y) + d_{mn} u^d_{mn}(x,y) \right] \right\}^2 dxdy = 0.$$

### 3.5. Завдання до третього розділу

1. Показати, що зі збіжності ряду

$$\varphi(t) \stackrel{L_2(-\pi, \pi)}{=} \frac{a_0}{2} + \sum_{k=1}^{\infty} (a_k \cos kt + b_k \sin kt)$$

випливає збіжність такого ряду

$$f(x) \stackrel{L_2(-l, l)}{=} \frac{a_0}{2} + \sum_{k=1}^{\infty} \left( a_k \cos \frac{k\pi}{l} t + b_k \sin \frac{k\pi}{l} t \right),$$

де $\varphi(t) = f\left(\frac{lt}{\pi}\right)$; $a_k$, $b_k$ — коефіцієнти Фур'є функції $\varphi(t)$, і навпаки.

2. Довести, використовуючи рівність Парсеваля, що для неперервної на проміжку $[0, \pi]$ функції $f(x)$ з похідною, інтегровною з квадратом, і за виконання однієї з умов

(а) $\int\limits_0^{\pi} f(x)dx = 0$ або (б) $f(0) = f(\pi) = 0$

справедлива нерівність [22, c. 596] $\int\limits_0^{\pi} [f'(x)]^2 dx \geq \int\limits_0^{\pi} [f(x)]^2 dx$.

3. Довести, що збіжний тригонометричний ряд $\sum_{n=2}^{\infty} \frac{\sin nx}{\ln n}$ не є рядом Фур'є $[22, c.\,624]$.

4. Довести, що якщо $2\pi$-періодичні функції $f(x)$ і $g(x)$ розвиваються у тригонометричні ряди

$$f(x) \stackrel{L_2}{=} \frac{a_0}{2} + \sum_{k=1}^{\infty}(a_k \cos kx + b_k \sin kx), \quad g(x) = \frac{c_0}{2} + \sum_{k=1}^{\infty} c_k \cos kx,$$

то справедлива рівність

$$\int_{-\pi}^{\pi} f(t)g(x-t)dt = \frac{a_0 c_0}{2} + \sum_{k=1}^{\infty} c_k(a_k \cos kx + b_k \sin kx).$$

5. Показати, що система функцій

$$\psi_0(x) = \frac{1}{\sqrt{\pi}\sqrt[4]{1-x^2}}, \quad \psi_n(x) = \frac{2^n T_n(x)}{\sqrt{2\pi}\sqrt[4]{1-x^2}}, \quad n = 1, 2, \ldots,$$

ортонормована на сегменті $[-1, 1]$.

6. Показати, що використовувана в теорії ймовірностей система Радемахера

$$\psi_n(x) = \varphi(2^n x), \ n = 0, 1, \ldots,$$

де $\varphi(t) = \operatorname{sgn}(\sin 2\pi t)$, ортонормована на сегменті $[0, 1]$.

7. Показати, що для многочленів Чебишова другого роду $U_n(x) = \frac{1}{n+1}\frac{d}{dx}T_{n+1}(x)$, $-1 \le x \le 1$, справедлива формула

$$U_n(x) = \sum_{k=0}^{[n/2]}(-1)^k 2^{n-2k} C_{n-k}^k x^{n-2k}, \ n = 0, 1, \ldots.$$

8. Показати, що многочлени Чебишова другого роду $U_n(x)$ задовольняють диференціальне рівняння

$$(1-x^2)\frac{d^2 y}{dx^2} - 3x\frac{dy}{dx} + n(n+2)y = 0.$$

9. Скориставшись рівнянням (3.61) і формулою інтегрування за частинами, вивести формулу

$$\int_{-1}^{1} U_n(x)U_m(x)\sqrt{1-x^2}\,dx = \frac{n+1}{m+1}\int_{-1}^{1}\frac{T_{m+1}T_{n+1}}{\sqrt{1-x^2}}dx = \begin{cases} 0, & m \ne n, \\ \frac{\pi}{2}, & m = n. \end{cases}$$

10. Записати розвинення функції $f(x, y) = xy$ у прямокутнику $Q = \{|x| \le p\,;\,|y| \le q\}$ за системою (3.66).

# Р О З Д І Л  IV

# СЛАБКО  ЗБІЖНІ  РЯДИ
______________________________________________________________

### 4.1. Згладжування функцій

**4.1.1. Оператори згладжування.** Важливе місце у математичному аналізі займають лінійні інтегральні перетворення, які встановлюють відповідність між двома множинами функцій і операцій між ними. Такими, наприклад, є інтегральні перетворення Фур'є, Лапласа, Бесселя та інші, з використанням яких лінійні диференціальні рівняння зводяться до лінійних алгебраїчних рівнянь. Інтегральні перетворення ефективно використовуються також для побудови методів підсумовування тригонометричних рядів $[2, 11, 18]$.

Якщо кожному елементу з множини $D$ функцій ставиться у відповідність певний елемент множини $M$ функцій, то задано оператор $K$. Оператор вигляду

$$Kf = \int_a^b K(x,t) f(t) dt, \quad x \in (a,b),$$

є лінійним і називається *інтегральним оператором*, а функція $K(x,t)$ – *ядром* оператора.

Очевидно, якщо ядро оператора не залежить від $x$, то одержимо функціонал, тобто лінійний функціонал є частинним випадком лінійного оператора.

Якщо $\{\delta_n(x-t)\}$ – дельтоподібна послідовність вигляду (1.84),

$$\delta_n(x-t) = \frac{1}{r_n} \Phi\left(\frac{x-t}{r_n}\right), \qquad (4.1)$$

де $\Phi(t)$ – ядро типу Фейєра (означення 10, п. 1.5), $r_n > 0$ – нескінченно мала числова послідовність, і $f(t)$ – обмежена кусково-неперервна функція, то в кожній точці неперервності функції $f(t)$ справедлива формула (1.85),

$$\lim_{n \to \infty} \int_{-\infty}^{+\infty} \delta_n(x-t) f(t) \, dt = f(x). \qquad (4.2)$$

Інтеграл $\int_{-\infty}^{+\infty} \delta_n(x-t) f(t) \, dt$ згідно з теоремою 4 (п. 1.5) є неперервним інтегральним оператором. Перетворимо цей інтеграл наступним чином

$$\int_{-\infty}^{+\infty} \delta_n(x-t) f(t) \, dt = \int_{-\infty}^{+\infty} \frac{1}{r_n} \Phi\left(\frac{x-t}{r_n}\right) f(t) \, dt =$$

$$= \int_{-\infty}^{+\infty} \frac{1}{r_n} \Phi\left(\frac{t}{r_n}\right) f(x+t) \, dt = \int_0^{+\infty} [f(x+t) + f(x-t)] \frac{1}{r_n} \omega\left(\frac{t}{r_n}\right) dt.$$

Відповідно до цієї формули введемо позначення

$$S_r(f; x) = \int_{-\infty}^{+\infty} \frac{1}{r} \Phi\left(\frac{x-t}{r}\right) f(t) \, dt = \int_{-\infty}^{+\infty} f(x+t) \frac{1}{r} \Phi\left(\frac{t}{r}\right) dt =$$

$$= \int_{-\infty}^{+\infty} f(x+rt) \Phi(t) \, dt = \int_0^{+\infty} [f(x+rt) + f(x-rt)] \omega(t) \, dt, \qquad (4.3)$$

де $r > 0$ – мала величина.

У частинному випадку, якщо $\Phi(t) = \Phi_0(t)$ – фінітне ядро (1.81), то

$$S_r^0(f; x) = \int_{-r}^{r} \frac{1}{r} \Phi_0\left(\frac{x-t}{r}\right) f(t) \, dt = \int_{-r}^{r} f(x+t) \frac{1}{r} \Phi_0\left(\frac{t}{r}\right) dt =$$

$$= \int_{-1}^{1} f(x+rt) \Phi_0(t) \, dt = \int_0^1 [f(x+rt) + f(x-rt)] \omega_0(t) \, dt. \qquad (4.4)$$

*О з н а ч е н н я 1.* *Інтегральний оператор*

$$S_r(f; x) = \int_{-\infty}^{+\infty} f(x+rt) \Phi(t) \, dt, \qquad (4.5)$$

*в якому $\Phi(t)$ – ядро типу Фейєра, що справджує умови теореми 9 (п. 1.5), тобто $\Phi(t)$ має абсолютно інтегровну похідну $p$-го порядку, $p \geq 1$, і виконуються оцінки*

$$(1+t)^{\lambda}\left|\frac{d^m\omega(t)}{dt^m}\right| \le A < \infty, \ \lambda > 1, \ m = \overline{0, p}, \qquad (4.6)$$

*називається оператором згладжування (усереднювання).*

*Функція $S_r(f; x)$ називається усередненням функції $f(x)$.*

Функцію $\delta_r(x-t) = \frac{1}{r}\Phi\left(\frac{x-t}{r}\right)$ при $r \to +0$ будемо називати *дельтоподібною функцією*. Вона за умови (1.86) справджує рівність

$$\int_{-\infty}^{+\infty} \frac{1}{r}\Phi\left(\frac{x-t}{r}\right)dt = 1.$$

Відзначимо основні властивості операторів згладжування.

***Т е о р е м а  1 .*** *Якщо $f(x)$ – обмежена кусково-неперервна функція і $\Phi(x) = \omega(|x|)$ – ядро оператора згладжування, то справедливі рівності:*

$$\lim_{r \to +0} S_r(f; x_0) = f(x_0)$$

*в точках неперервності функції $f(x)$;*

$$\lim_{r \to +0} S_r(f; x_0) = \frac{f(x_0 + 0) + f(x_0 - 0)}{2}$$

*в точках розриву першого роду цієї функції.*

*Д о в е д е н н я .* Перетворимо з урахуванням формули (4.3) вираз

$$S_r(f; x_0) - \frac{f(x_0 + 0) + f(x_0 - 0)}{2} =$$

$$= \int_0^{+\infty} [f(x_0 + t) + f(x_0 - t)]\frac{1}{r}\omega\left(\frac{t}{r}\right)dt -$$

$$- [f(x_0 + 0) + f(x_0 - 0)]\int_0^{\infty} \frac{1}{r}\omega\left(\frac{t}{r}\right)dt =$$

$$= \int_0^{+\infty} [f(x_0 + t) - f(x_0 + 0)]\frac{1}{r}\omega\left(\frac{t}{r}\right)dt +$$

$$+ \int\limits_0^{+\infty} [f(x_0 - t) - f(x_0 - 0)] \frac{1}{r} \omega\left(\frac{t}{r}\right) dt.$$

Виберемо довільне число $\varepsilon > 0$. Оскільки функція $f(x)$ неперервна зліва і справа в точці $x_0$, для будь-якого $\varepsilon > 0$ можна вибрати $\eta > 0$ таке, що якщо тільки $t < \eta$, то виконуються нерівності

$$|f(x_0 \pm t) - f(x_0 \pm 0)| < \frac{\varepsilon}{4M}, \qquad (4.7)$$

де $M = \dfrac{1}{r} \int\limits_0^{+\infty} \left|\omega\left(\dfrac{t}{r}\right)\right| dt$.

Розіб'ємо кожний з інтегралів в одержаному виразі оператора усереднювання на два інтеграли

$$I_1^{\pm} = \int\limits_0^{\eta} [f(x_0 \pm t) - f(x_0 \pm 0)] \frac{1}{r} \omega\left(\frac{t}{r}\right) dt,$$

$$I_2^{\pm} = \int\limits_{\eta}^{+\infty} [f(x_0 \pm t) - f(x_0 \pm 0)] \frac{1}{r} \omega\left(\frac{t}{r}\right) dt.$$

Для першого інтегралу з урахуванням нерівності (4.7) одержимо оцінку

$$\left|I_1^{\pm}\right| < \frac{\varepsilon}{M} \int\limits_0^{\eta} \frac{1}{r} \left|\omega\left(\frac{t}{r}\right)\right| dt < \frac{\varepsilon}{4}.$$

Для другого інтегралу справджуються умови теореми 6 (п. 1.5) і тому існує число $r_1 > 0$ таке, що $\left|I_2^{\pm}\right| < \dfrac{\varepsilon}{4}$. Отже, справедлива оцінка

$$\left|S_r(f; x) - \frac{f(x_0 + 0) + f(x_0 - 0)}{2}\right| <$$
$$< \left|I_1^+\right| + \left|I_2^+\right| + \left|I_1^-\right| + \left|I_2^-\right| < \varepsilon \qquad (4.8)$$

для всіх $r$ таких, що $0 < r < r_1$. З цієї оцінки випливає друге твердження теореми.

Якщо функція неперервна в точці $x_0$ і, відповідно, $f(x_0 + 0) = f(x_0 - 0) = f(x_0)$, то з нерівності (4.8) одержимо перше твердження теореми.

Теорему доведено.

***Т е о р е м а 2****. Якщо $f(x)$ – обмежена кусково-неперервна функція і $\Phi(x) = \omega(|x|)$ – ядро оператора згладжування, то функція $S_r(f;x)$ має обмежену похідну $p$-го порядку при $r > 0$ і справедлива формула.*

$$\frac{d^p}{dx^p} S_r(f;x) = \frac{1}{r} \int_{-\infty}^{+\infty} \frac{d^p}{dx^p} \Phi\left(\frac{x-t}{r}\right) f(t) dt.$$

*Д о в е д е н н я*. За умов теореми функція $\Phi(t)$ має абсолютно інтегровану похідну $p$-го порядку, $p \geq 1$, яка справджує умову (4.6). Знайдемо похідну $p$-го порядку від функції $S_r(f;x)$ і покажемо, що диференціювання по параметру може проводитись під знаком невласного інтеграла, тобто справедлива формула

$$\frac{d^p}{dx^p} S_r(f;x) = \frac{d^p}{dx^p} \int_{-\infty}^{+\infty} \frac{1}{r} \Phi\left(\frac{x-t}{r}\right) f(t) dt =$$

$$= \frac{1}{r} \int_{-\infty}^{+\infty} \frac{d^p}{dx^p} \Phi\left(\frac{x-t}{r}\right) f(t) dt.$$

Внаслідок обмеженості функції $f(x)$, $\max_{x \in (-\infty, \infty)} |f(x)| = M$, і оцінки (4.8) похідної $p$-го порядку від функції $\Phi(x)$, справедлива нерівність

$$\left| \frac{d^p}{dx^p} \Phi\left(\frac{x-t}{r}\right) f(t) \right| \leq \frac{MA\,r^{\lambda-1}}{(r+|x-t|)^\lambda}, \quad \lambda > 1.$$

Оскільки підінтегральна функція має інтегровну мажоранту, відповідний інтеграл збігається.

Теорему доведено.

*П р и к л а д 1*. Ядро оператора усереднювання за Соболєвим [17, 20]

$$\Phi(t) = \begin{cases} c_0 \exp\left(-\dfrac{1}{1-t^2}\right), & |t| \le 1, \\ 0, & |t| > 1 \end{cases}$$

– нескінченно диференційовна функція, де $c_0$ – стала, що визначається з умови $\int\limits_{-1}^{1} \Phi(t)dt = 1$.

Покажемо, наприклад, що $\Phi(t)$ має похідну будь-якого порядку у точці $t=1$. Запишемо вираз цієї функції в інтервалі $(-1, 1)$ у вигляді

$$e^{-\frac{1}{1-t^2}} = e^{-\frac{1}{2(1-t)}} e^{-\frac{1}{2(1+t)}}$$

і знайдемо похідну $n$-го порядку від цієї функції за формулою Лейбніца

$$\Phi^{(n)}(t) = \sum_{k=0}^{n} C_n^k \left[e^{-\frac{1}{2(1-t)}}\right]^{(k)} \left[e^{-\frac{1}{2(1+t)}}\right]^{(n-k)}.$$

Очевидно, перший множник в цьому виразі записується у вигляді лінійної комбінації доданків вигляду

$$P_m\left(\frac{1}{1-t}\right) e^{-\frac{1}{2(1-t)}},$$

де $P_m\left(\dfrac{1}{1-t}\right)$ – поліном деякого степеня відносно $\dfrac{1}{1-t}$.

Застосувавши правило Льопіталя до кожного з цих доданків при $t \to 1-0$, одержимо

$$\lim_{t \to 1-0} P_m\left(\frac{1}{1-t}\right) e^{-\frac{1}{2(1-t)}} = \lim_{x \to \infty} P_m(x) e^{-\frac{x}{2}} = 0.$$

Тоді

$$\lim_{t \to 1-0} \left[e^{-\frac{1}{1-t^2}}\right]^{(n)} = 0$$

і, відповідно, $\lim\limits_{t \to 1-0} \Phi^{(n)}(t) = 0$. Похідна будь-якого порядку від функції $\Phi(t)$ справа в точці $t=1$ також дорівнює нулеві.

Аналогічно можна показати рівність нулеві похідних будь-якого порядку від функції $\Phi(t)$ у точці $t = -1$.

***Т е о р е м а 3***. *Усереднення тригонометричних функцій справджують рівності*

$$S_r(1; x_0) = 1,$$
$$S_r(\cos kx; x_0) = \varphi(kr)\cos kx_0, \quad (4.9)$$
$$S_r(\sin kx; x_0) = \varphi(kr)\sin kx_0,$$

*де*

$$\varphi(z) = 2\int_0^\infty \omega(t)\cos zt\, dt \quad (4.10)$$

*– функція, що справджує граничну рівність*

$$\lim_{z \to 0} \varphi(z) = 1. \quad (4.11)$$

Д о в е д е н н я . У справедливості формул (4.9) переконуємося підстановкою тригонометричних функцій у формулу (4.3)

$$S_r(1; x_0) = 2\int_0^{+\infty} \frac{1}{r}\omega\left(\frac{t}{r}\right)dt = 1,$$

$$S_r(\cos kt; x_0) = \int_0^{+\infty} [\cos k(x_0 + t) + \cos k(x_0 - t)]\frac{1}{r}\omega\left(\frac{t}{r}\right)dt =$$

$$= 2\cos kx_0 \int_0^{+\infty} \frac{1}{r}\omega\left(\frac{t}{r}\right)\cos kt\, dt = \varphi(kr)\cos kx_0,$$

$$S_r(\sin kt; x_0) = \int_0^{+\infty} [\sin k(x_0 + t) + \sin k(x_0 - t)]\frac{1}{r}\omega\left(\frac{t}{r}\right)dt =$$

$$= 2\sin kx_0 \int_0^{+\infty} \frac{1}{r}\omega\left(\frac{t}{r}\right)\cos kt\, dt = \varphi(k\varepsilon)\sin kx_0.$$

Гранична рівність (4.11) випливає з рівномірної збіжності інтегралу (4.10), внаслідок властивостей (1.79) і (1.80) ядра типу Фейєра,

$$\lim_{z \to 0}\varphi(z) = 2\int_0^{+\infty}\lim_{z \to 0}\cos zt\,\omega(t)dt = 2\int_0^{+\infty}\omega(t)dt = 1.$$

Теорему доведено.

***Н а с л і д о к  1 .*** *Якщо ядро оператора згладжування – фінітна функція, то формула* (4.10) *набуде вигляду*

$$\varphi_0(z) = 2\int_0^1 \omega_0(t)\cos zt\, dt. \qquad (4.12)$$

Формула (4.12) випливає з (4.10), якщо врахувати подання (1.81).

***Т е о р е м а  4 .*** *Якщо ядро оператора згладжування є згорткою двох ядерних функцій операторів згладжування*

$$\frac{1}{r_1 r_2}\int_{-\infty}^{+\infty}\Phi_1\!\left(\frac{x-t}{r_1}\right)\Phi_2\!\left(\frac{t}{r_2}\right)dt,$$

*то усереднення тригонометричних функцій справджують рівності*

$$S_{r_1,r_2}(1; x_0) = 1,$$
$$S_{r_1,r_2}(\cos kx; x_0) = \varphi_1(kr_1)\varphi_2(kr_2)\cos kx_0, \qquad (4.13)$$
$$S_{r_1,r_2}(\sin kx; x_0) = \varphi_1(kr_1)\varphi_2(kr_2)\sin kx_0,$$

*де* $\varphi_i(z_i) = \int_{-\infty}^{+\infty}\Phi_i(t)\cos z_i t\, dt = 2\int_0^{+\infty}\omega_i(t)\cos z_i t\, dt$.

***Д о в е д е н н я .*** Покажемо правильність другої формули (4.13)

$$S_{r_1,r_2}(\cos kx; x_0) = \int_{-\infty}^{+\infty}\cos k(x_0+x)\frac{1}{r_1 r_2}\int_{-\infty}^{+\infty}\Phi_1\!\left(\frac{x-t}{r_1}\right)\Phi_2\!\left(\frac{t}{r_2}\right)dt\,dx.$$

Змінивши порядок інтегрування, що можливо внаслідок абсолютної збіжності інтегралів, і провівши елементарні перетворення з урахуванням інтегрування парних і непарних функцій, одержимо

$$S_{r_1,r_2}(\cos kx; x_0) = \int_{-\infty}^{+\infty}\frac{1}{r_2}\Phi_2\!\left(\frac{t}{r_2}\right)dt\int_{-\infty}^{+\infty}\cos k(x_0+x)\frac{1}{r_1}\Phi_1\!\left(\frac{x-t}{r_1}\right)dx =$$

$$= \cos kx_0\int_{-\infty}^{+\infty}\frac{1}{r_2}\Phi_2\!\left(\frac{t}{r_2}\right)dt\int_{-\infty}^{+\infty}\cos kx\,\frac{1}{r_1}\Phi_1\!\left(\frac{x-t}{r_1}\right)dx +$$

$$+\sin kx_0 \int\limits_{-\infty}^{+\infty} \frac{1}{r_2}\Phi_2\left(\frac{t}{r_2}\right)dt \int\limits_{-\infty}^{+\infty} \sin kx \frac{1}{r_1}\Phi_1\left(\frac{x-t}{r_1}\right)dx =$$

$$= \cos kx_0\, \varphi_1(kr_1) \int\limits_{-\infty}^{+\infty} \frac{1}{r_2}\Phi_2\left(\frac{t}{r_2}\right)\cos kt\, dt +$$

$$+ \sin kx_0\, \varphi_1(kr_1) \int\limits_{-\infty}^{+\infty} \frac{1}{r_2}\Phi_2\left(\frac{t}{r_2}\right)\sin kt\, dt = \varphi_1(kr_1)\,\varphi_2(kr_2)\cos kx_0.$$

Аналогічно доводимо першу і третю формули (4.13).

Теорему доведено.

*Л е м а  1 .* Якщо $\psi(x)$ – *абсолютно інтегровна на проміжку* $(\eta, \infty)$ *функція, то*

$$\lim_{z\to\infty}\int\limits_{\eta}^{+\infty}\psi(t)\cos zt\, dt = 0, \quad \lim_{z\to\infty}\int\limits_{\eta}^{+\infty}\psi(t)\sin zt\, dt = 0, \qquad (4.14)$$

*де* $\eta$ – *довільне число.*

*Д о в е д е н н я .* Сформульоване твердження є поширенням леми Рімана (п. 2.3) на випадок нескінченного проміжку. Нехай $\varepsilon > 0$ – мале число і $B$ – достатньо велике число. Розіб'ємо інтеграл у формулі (4.13) на два інтеграли

$$\int\limits_{\eta}^{+\infty}\psi(t)\cos zt\, dt = \int\limits_{\eta}^{B}\psi(t)\cos zt\, dt + \int\limits_{B}^{+\infty}\psi(t)\cos zt\, dt.$$

Виберемо число $B$ таке, що

$$\left|\int\limits_{B}^{+\infty}\psi(t)\cos zt\, dt\right| \le \frac{\varepsilon}{2}.$$

Перший інтеграл задовольняє умови леми 1 (п. 2.3). Тому для достатньо великих значень параметра $z$ справедлива нерівність

$$\left|\int\limits_{\eta}^{B}\psi(t)\cos zt\, dt\right| < \frac{\varepsilon}{2}.$$

Внаслідок цих нерівностей, одержимо оцінку

$$\left| \int_{\eta}^{\infty} \psi(t)\cos zt\,dt \right| \leq \left| \int_{\eta}^{B} \psi(t)\cos zt\,dt \right| + \left| \int_{B}^{\infty} \psi(t)\cos zt\,dt \right| \leq \varepsilon.$$

Звідси випливає перша рівність (4.14). Аналогічно доводимо другу рівність (4.14).

Лему доведено.

**Т е о р е м а  5 .** *Нехай* $\Phi(x) = \omega(|x|)$ – *ядро оператора згладжування.*

*Тоді для функції* $\varphi(z)$ *при* $z \to +\infty$ *справедлива оцінка*

$$|\varphi(z)| = o\left(\frac{1}{z^p}\right), \qquad (4.15)$$

*тобто функція* $\varphi(z)$ *є нескінченно малою вищого порядку малості при* $z \to +\infty$ *ніж* $\frac{1}{z^p}$.

*Д о в е д е н н я .* Оцінку (4.15) одержимо з формули (4.10) з використанням формули інтегрування частинами

$$|\varphi(z)| = 2\left| \int_{0}^{+\infty} \omega(t)\cos zt\,dt \right| = \left| \frac{2}{z} \int_{0}^{+\infty} \omega'(t)\sin zt\,dt \right| = \ldots$$

$$= \frac{2}{z^{p-1}}\left| \int_{0}^{+\infty} \omega^{(p-1)}(t)\sin zt\,dt \right| = \frac{2}{z^p}\left| \int_{0}^{+\infty} \omega^{(p)}(t)\cos zt\,dt \right|,$$

якщо $p$ – парне число, і

$$|\varphi(z)| = \frac{2}{z^p}\left| \int_{0}^{+\infty} \omega^{(p)}(t)\sin zt\,dt \right|,$$

якщо $p$ – непарне число.

Звідси з урахуванням твердження леми 1 і, відповідно, граничної рівності (4.14) для $p$-ої похідної від функції $\omega(t)$ одержимо оцінку (4.15).

Теорему доведено.

Аналогічне твердження для випадку оператора згладжування (4.4) з фінітним ядром $\Phi_0(x)$ справджується за дещо послаблених умов.

**Т е о р е м а   6**. *Нехай* $\Phi_0(x)$ – *фінітне ядро типу Фейєра, що має кусково-гладку похідну $p$-го порядку, $p \geq 1$.*

*Тоді для функції $\varphi_0(z)$ справедлива при $z \to +\infty$ оцінка*

$$|\varphi(z)| = O\left(\frac{1}{z^{p+1}}\right). \qquad (4.16)$$

*Д о в е д е н н я*. До інтеграла (4.12) за аналогією з теоремою 12 (п. 2.3) застосуємо інтегрування частинами. Оскільки $\Phi^{(p)}(x)$ – функція кусково-гладка і, відповідно, $\Phi_0^{(p+1)}(x)$ – обмежена абсолютно інтегровна функція, то

$$|\varphi(z)| = \frac{\alpha_z}{z^{p+1}},$$

де абсолютна величина коефіцієнта $\alpha_z$ задається одним з інтегралів

$$I_1 = 2\left|\int_0^1 \omega_0^{(p+1)}(x)\cos zx\, dx\right|, \ I_2 = 2\left|\int_0^1 \omega_0^{(p+1)}(x)\sin zx\, dx\right|.$$

Враховуючи тут обмеженість тригонометричних функцій і інтегровність функції $\left|\omega_0^{(p+1)}(x)\right|$, одержимо

$$I_1 \leq 2\int_0^1 \left|\omega_0^{(p+1)}(x)\right|dx = A, \ I_2 \leq A$$

і, відповідно, оцінку (4.16).

Теорему доведено.

*З а у в а ж е н н я*. З умов теореми 6 про існування кусково-гладкої похідної від фінітної ядерної функції $\Phi_0(x)$ випливають рівності $\Phi_0^{(m)}(\pm 1) = 0$, $m = \overline{0, p-1}$, і, відповідно, $\omega_0^{(m)}(\pm 1) = 0$, $m = \overline{0, p-1}$.

**4.1.2. Оператори згладжування з нульовими моментними характеристиками.** Розглянемо оператор згладжування з фінітним ядром (4.4). Однак вважаємо, що ядро $\omega_0$, крім зазначених в означенні 1 умов, задовольняє ще такі умови:

$$\int\limits_0^1 t^k \omega_0(t)dt = 0, \quad \int\limits_0^1 t^{k+1}|\omega_0(t)|dt = A < +\infty, \quad k=1,...,q. \quad (4.17)$$

***Т е о р е м а*** *7 . Нехай* $f(x)$ *– кусково-неперервна функція з кусково-неперервними похідними до* $(q+1)$*-го порядку включно в деякому околі точки* $x_0 \in (a,b)$ *і* $\Phi(x)$ *– фінітне ядро* (4.4) *оператора згладжування, що справджує додаткові умови* (4.17).

*Тоді в точці* $x_0$ *і для* $r > 0$ *такого, що в інтервалах* $(x_0-r, x_0]$, $[x_0, x_0+r)$ *функція і її відповідні похідні неперервні, справедлива нерівність*

$$\left| \int\limits_0^r [f(x_0+t) + f(x_0-t)]\frac{1}{r}\omega_0\left(\frac{t}{r}\right)dt - \frac{f(x_0+0)+f(x_0-0)}{2} \right| \le$$

$$\le \frac{2Ar^{q+1}}{(q+1)!} \max_{x\in(a,b)} \left| f^{(q+1)}(x) \right|. \quad (4.18)$$

*Д о в е д е н н я .* Запишемо формулу Тейлора для функції $f(x)$ зліва і справа в точці $x_0$ у вигляді

$$f(x_0 \pm t) = \sum_{i=0}^{q} f^{(i)}(x_0 \pm 0)\frac{(\pm t)^i}{i!} +$$
$$+ f^{(q+1)}(x_0 \pm \theta^{\pm} t)\frac{(\pm t)^{n+1}}{(n+1)!}, \quad 0 < \theta^{\pm} < 1.$$

Застосуємо оператор усереднення до функції $f(x)$ в точці $x_0$ з урахуванням одержаної формули і умов (4.17),

$$\int\limits_0^r [f(x_0+t)+f(x_0-t)]\frac{1}{r}\omega_0\left(\frac{t}{r}\right)dt = \frac{f(x_0+0)+f(x_0-0)}{2} +$$

$$+ \int\limits_0^r \left[ f^{(q+1)}(x_0+\theta^+ t) + (-1)^{q+1} f^{(q+1)}(x_0 - \theta^- t) \right]\frac{1}{r}\omega_0\left(\frac{t}{r}\right)t^{q+1}dt =$$

$$= \frac{f(x_0+0)+f(x_0-0)}{2} +$$

$$+ r^{q+1}\int\limits_0^1 \left[ f^{(q+1)}(x_0+\theta^+ r\xi) + (-1)^{q+1} f^{(q+1)}(x_0-\theta^- r\xi) \right]\xi^{q+1}\omega_0(\xi)d\xi.$$

Оцінюючи цей вираз з урахуванням теореми про середнє, одержимо нерівність (4.18),

$$\left| \int\limits_0^r [f(x_0+t) + f(x_0-t)] \frac{1}{r} \omega_0\left(\frac{t}{r}\right) dt - \frac{f(x_0+0) + f(x_0-0)}{2} \right| \leq$$

$$\leq \frac{r^{q+1}}{(q+1)!} \int\limits_0^1 \left| f^{(q+1)}(x_0 + \theta^+ r t) + (-1)^{q+1} f^{(q+1)}(x_0 - \theta^- r t) \right| t^{q+1} |\omega_0(t)| dt \leq$$

$$\leq \frac{2r^{q+1}}{(q+1)!} \max_{x \in (a,b)} \left| f^{(q+1)}(x) \right| \int\limits_0^1 t^{q+1} |\omega_0(t)| dt = \frac{2Ar^{q+1}}{(q+1)!} \max_{x \in (a,b)} \left| f^{(q+1)}(x) \right|.$$

Теорему доведено.

***Н а с л і д о к .*** Якщо

$$f(x) = \begin{cases} P_q(x), & x < x_0, \\ Q_q(x), & x > x_0, \end{cases}$$

*де $P_q(x)$, $Q_q(x)$ – поліноми степеня $\leq q$, то для будь-якого $r > 0$ за виконання умов теореми виконується рівність*

$$\int\limits_0^r [P_q(x_0+t) + Q_q(x_0-t)] \frac{1}{r} \omega_0\left(\frac{t}{r}\right) dt = \frac{P_q(x_0) + Q_q(x_0)}{2}.$$

*Д о в е д е н н я .* Цей результат випливає з нерівності (4.18), якщо врахувати рівності $P_n^{(q+1)}(x) \equiv 0$ і $Q_n^{(q+1)}(x) \equiv 0$.

*П р и к л а д  2 .* Функція $\omega_0(t) = \frac{\pi^2}{4}(1-t)\cos \pi t$, $0 \leq t \leq 1$, – фінітне ядро оператора згладжування з нульовим моментом першого порядку $\int\limits_0^1 t\,\omega_0(t)\,dt = 0$. Перевіримо безпосередньо справедливість формули (4.18) для $q = 1$.

Дійсно,

$$\int\limits_0^r [P_1(x_0+t) + Q_1(x_0-t)] \frac{1}{r} \omega_0\left(\frac{t}{r}\right) dt =$$

$$= \int\limits_0^1 [P_1(x_0+rt) + Q_1(x_0-rt)] \omega_0(t) dt =$$

$$= \frac{\pi^2}{4}\int_0^1 [P_1(x_0+rt)+Q_1(x_0-rt)](1-t)\cos\pi t\,dt = \frac{\pi^2}{4}\int_0^1 R_2(x)\cos\pi t\,dt,$$

де $R_2(x)=[P_1(x_0+rt)+Q_1(x_0-rt)](1-t)$.

Інтегруючи два рази частинами, одержимо

$$\frac{\pi^2}{4}\int_0^1 R_2(x)\cos\pi t\,dt = -\frac{\pi}{4}\int_0^1 R_2'(x)\sin\pi t\,dt = -\frac{1}{4}[R_2'(1)+R_2'(0)]=$$

$$=\frac{1}{4}[P_1(x_0+r)+Q_1(x_0-r)-rP_1'(x_0)+rQ_1'(x_0)+P_1(x_0)+Q_1(x_0)]=$$

$$=\frac{P_1(x_0)+Q_1(x_0)}{2}.$$

Зауважимо, що формула (4.18) для цього випадку справедлива для будь-якого $r>0$.

### 4.2. Періодичні ядерні функції

Поряд з дельтоподібними послідовностями функцій з однією точкою локалізації, визначених на нескінченному проміжку, можна будувати періодичні дельтоподібні послідовності функцій (з періодично розміщеними точками локалізації). Вирішення цієї задачі ґрунтується на усередненні тригонометричних функцій $[2, 7, 18]$.

**Т е о р е м а 1.** *Нехай* $\Phi(x)=\omega(|x|)$ – *ядро оператора згладжування.*

*Тоді ряд*

$$\frac{1}{\pi}\left[\frac{1}{2}+\sum_{k=1}^{\infty}\varphi(kr)\cos kx\right]=\delta_r^*(x) \qquad (4.19)$$

*при* $r>0$ *рівномірно збігається і його сума є* $2\pi$-*періодичною неперервною функцією за змінною* $x$ *з абсолютно інтеґровною похідною* $p$-*го порядку,* $p\geq 1$, *і справедлива формула*

$$\delta_r^*(x)=\frac{1}{r}\sum_{k=-\infty}^{\infty}\omega\!\left(\frac{|x+2k\pi|}{r}\right). \qquad (4.20)$$

*Для випадку фінітного ядра сума ряду* (4.20) *на основному проміжку періодичності наступна*

$$\delta_r^{0*}(x) = \begin{cases} \dfrac{1}{r}\omega_0\!\left(\dfrac{|x|}{r}\right), & -r \le x \le r, \\ 0, & r < |x| \le \pi. \end{cases} \qquad (4.21)$$

*Доведення.* Запишемо частинну суму ряду (4.19) і перетворимо її з урахуванням формули (4.10), виразу ядра Діріхле

$$D_n(t) = \frac{1}{\pi}\left(\frac{1}{2} + \sum_{k=1}^{n}\cos kt\right) = \frac{\sin\!\left(n+\dfrac{1}{2}\right)t}{2\pi\sin\dfrac{t}{2}}$$

і заміною змінних,

$$\delta_{rn}^{*}(x) = \frac{1}{\pi}\left[\frac{1}{2} + \sum_{k=1}^{n}\varphi(kr)\cos kx\right] = \frac{2}{\pi}\int_0^\infty \omega(t)\left(\frac{1}{2} + \sum_{k=1}^{n}\cos krt\,\cos kx\right)dt =$$

$$= \frac{1}{\pi r}\int_{-\infty}^{\infty}\omega\!\left(\frac{|t|}{r}\right)\!\left[\frac{1}{2} + \sum_{k=1}^{n}\cos k(x-t)\right]dt = \frac{1}{r}\int_{-\infty}^{\infty}\omega\!\left(\frac{|t|}{r}\right)D_n(x-t)\,dt =$$

$$= \frac{1}{r}\left[\int_{-\pi}^{\pi}\omega\!\left(\frac{|t|}{r}\right)D_n(x-t)\,dt + \sum_{k=1}^{\infty}\int_{(2k-1)\pi}^{(2k+1)\pi}\omega\!\left(\frac{|t|}{r}\right)D_n(x-t)\,dt + \right.$$

$$\left.+ \sum_{k=1}^{\infty}\int_{-(2k+1)\pi}^{-(2k-1)\pi}\omega\!\left(\frac{|t|}{r}\right)D_n(x-t)\,dt\right] = \frac{1}{r}\left[\int_{-\pi}^{\pi}\omega\!\left(\frac{|t|}{r}\right)D_n(x-t)\,dt + \right.$$

$$\left.+ \sum_{k=1}^{\infty}\int_{-\pi}^{\pi}\omega\!\left(\frac{|2k\pi+t|}{r}\right)D_n(x-t)\,dt + \sum_{k=1}^{\infty}\int_{-\pi}^{\pi}\omega\!\left(\frac{|2k\pi-t|}{r}\right)D_n(x-t)\,dt\right] =$$

$$= \frac{1}{r}\int_{-\pi}^{\pi}\left\{\omega\!\left(\frac{|t|}{r}\right) + \sum_{k=1}^{\infty}\left[\omega\!\left(\frac{|2k\pi+t|}{r}\right) + \omega\!\left(\frac{|2k\pi-t|}{r}\right)\right]\right\}D_n(x-t)\,dt =$$

$$= \frac{1}{r}\int_{-\pi}^{\pi}\sum_{k=-\infty}^{+\infty}\omega\!\left(\frac{|2k\pi+t|}{r}\right)D_n(x-t)\,dt = \frac{1}{r}\int_{-\pi}^{\pi}\sum_{k=-\infty}^{+\infty}\omega\!\left(\frac{|2k\pi+t|}{r}\right)D_n(x-t)\,dt.$$

Звідси, оскільки за оцінкою ядра типу Фейєра ряд в одержаній формулі збігається рівномірно, існує границя при $n \to \infty$ послідовності

$$\delta^*_{r\,n}(x) = \int_{-\pi}^{\pi} \delta^*_r(t) D_n(x-t)\,dt,$$

а отже, справедлива формула (4.20).

Ряд (4.20) і ряди, складені з похідних до $p$-го порядку від функції $\delta^*_r(x)$ і, відповідно, від функції $\omega(|t|)$, збігаються на відрізку $[-\pi, \pi]$, оскільки за виконання умови (4.6) збігаються відповідні невласні інтеграли від функції $\omega(|t|)$ і її похідних. При цьому ряди, складені з похідних до $(p-1)$-го порядку від функції $\omega(|t|)$, рівномірно збігаються на відрізку $[-\pi, \pi]$.

Можливими точками розриву функції $\delta^*_r(t)$ і її похідних є точки $x = \pm\pi$. Легко переконатися, що $\delta^{*(m)}_r(-\pi) = \delta^{*(m)}_r(\pi)$, $m = \overline{0, p-1}$, і легко перевіряється також періодичність цих функцій.

За теоремою 2 (п. 2.3) границі $S_n^{*(m)}$, $m = \overline{1, p}$, при $n \to \infty$ є функції $\delta^{*(m)}_r(x)$.

Формула (4.21) випливає з (4.20), оскільки $\omega_0\!\left(\dfrac{|t|}{r}\right) = 0$, якщо $|t| > r$, і, відповідно, $\omega_0\!\left(\dfrac{2k\pi \pm t}{r}\right) = \omega_0\!\left(\dfrac{2k\pi}{r} \pm \dfrac{t}{r}\right) = 0$.

Теорему доведено.

**Т е о р е м а  2**. *Функція $\delta^*_r(x)$ є дельтоподібною функцією відносно множини $D$ $2\pi$-періодичних кусково-неперервних функцій і*

$$S^*_r(f; x) = \int_{-\pi}^{\pi} \delta^*_r(x-t) f(t)\,dt \qquad (4.22)$$

*– оператор згладжування.*

*Д о в е д е н н я*. Нехай $f(x) \in D$. Розглянемо інтегральний оператор (4.22) з використанням періодичної ядерної функції $\delta^*_r(x)$. Зробивши заміну з урахуванням періодичності підінтегральних функцій, одержимо

$$\int_{-\pi}^{\pi}\delta_r^*(x-t)f(t)dt = \int_{-\pi+x}^{\pi+x}\delta_r^*(u)f(x-u)du = \int_{-\pi}^{\pi}\delta_r^*(u)f(x-t)dt =$$

$$= \frac{1}{r}\int_{-\pi}^{\pi}\left\{\omega\left(\frac{|t|}{r}\right) + \sum_{k=1}^{\infty}\left[\omega\left(\frac{2k\pi+t}{r}\right) + \omega\left(\frac{2k\pi-t}{r}\right)\right]\right\}f(x-t)dt.$$

Провівши тут перетворення, зворотно до перетворення часткової суми при доведенні теореми 1, прийдемо до оператора згладжування (4.3)

$$S_r^*(f;x) = \int_{-\infty}^{\infty} f(x-t)\frac{1}{r}\omega\left(\frac{|t|}{r}\right)dt.$$

За теоремою 1, в кожній точці неперервності функції $f(x)$ справедлива рівність $\lim_{r\to 0} S_r(f;x) = f(x)$, а отже, $\delta_r^*(x)$ є дельтоподібною функцією відносно множини $D$.

Теорему доведено.

***Теорема 3.*** *Нехай $f(x)$ – $2\pi$-періодична кусково-гладка функція, а*

$$f(x) \sim \frac{a_0}{2} + \sum_{k=1}^{\infty} a_k \cos kx + b_k \sin kx$$

*– її ряд Фур'є.*

*Тоді, якщо функція $\Phi(x)$ ядро оператора згладжування, то для усереднення функції $f(x)$ і її похідних порядку $m$, $1 \leq m \leq p-1$, справедливі при $r > 0$ розвинення*

$$S_r^*(f;x) = \frac{a_0}{2} + \sum_{k=1}^{\infty}\varphi(kr)(a_k\cos kx + b_k\sin kx),$$

$$\frac{d^m}{dx^m}S_r^*(f;x) = \sum_{k=1}^{\infty}\varphi(kr)\frac{d^m}{dx^m}(a_k\cos kx + b_k\sin kx), \quad (4.23)$$

*які рівномірно збігаються.*

***Доведення.*** Члени другого ряду (4.23) при $r > 0$ задовольняють внаслідок оцінок (4.16) і (2.81) таку нерівність:

$$\left|\varphi(kr)\frac{d^m}{dx^m}(a_k\cos kx + b_k\sin kx)\right| \leq \frac{A}{k^p}k^m(|a_k|+|b_k|) \leq$$

$$\leq \frac{A}{k^{p-m}} \frac{2M}{k} = \frac{M_0}{k^{p-m+1}},$$

де $A$, $M$, $M_0$ – сталі величини.

Оскільки за умовою $p - m + 1 \geq 2$, ряд збігається рівномірно і його сума неперервна функція. Тому можлива перестановка операцій диференціювання і підсумовування

$$\sum_{k=1}^{\infty} \varphi(kr) \frac{d^m}{dx^m}(a_k \cos kx + b_k \sin kx) =$$
$$= \frac{d^m}{dx^m}\left[\frac{a_0}{2} + \sum_{k=1}^{\infty} \varphi(kr)(a_k \cos kx + b_k \sin kx)\right].$$

Врахувавши тут подання (4.20), одержимо

$$\frac{d^m}{dx^m}\left[\frac{a_0}{2} + \sum_{k=1}^{\infty} \varphi(kr)(a_k \cos kx + b_k \sin kx)\right] =$$
$$= \frac{1}{\pi} \frac{d^m}{dx^m} \int_{-\pi}^{\pi} f(t)\left[\frac{1}{2} + \sum_{k=1}^{\infty} \varphi(kr)\cos k(x-t)\right]dt =$$
$$= \frac{d^m}{dx^m} \int_{-\pi}^{\pi} f(t)\delta_r^*(x-t)dt.$$

Останній вираз є похідною $m$-го порядку від усереднення функції $f(x)$.

Теорему доведено.

*П р и к л а д 1*. Нехай $\Phi(x) = \frac{2}{\pi} \frac{\sin^2 \frac{x}{2}}{x^2}$ – ядро Фейєра (прикл. 4, п. 1.5). Знайдемо періодичний аналог цього ядра.

За формулою (4.20), прийнявши $r = \frac{1}{n}$, одержимо

$$\delta_{1/n}^*(x) = \frac{1}{2\pi n}\left\{\frac{4\sin^2 \frac{nx}{2}}{x^2} + \sum_{k=1}^{\infty}\left[\frac{\sin^2 n\left(k\pi + \frac{x}{2}\right)}{\left(k\pi + \frac{x}{2}\right)^2} + \frac{\sin^2 n\left(k\pi - \frac{x}{2}\right)}{\left(k\pi - \frac{x}{2}\right)^2}\right]\right\} =$$

$$= \frac{\sin^2 \frac{nx}{2}}{2\pi n} \left\{ \frac{4}{x^2} + \sum_{k=1}^{\infty} \left[ \frac{1}{\left(k\pi + \frac{x}{2}\right)^2} + \frac{1}{\left(k\pi - \frac{x}{2}\right)^2} \right] \right\}.$$

Враховуючи тут формулу $[21, с.\,473]$

$$\frac{1}{\sin^2 x} = \sum_{k=-\infty}^{\infty} \frac{1}{(x+k\pi)^2},$$

знайдемо

$$\delta_{1/n}^*(x) = \frac{1}{2\pi n} \frac{\sin^2 \frac{nx}{2}}{\sin^2 \frac{x}{2}}. \qquad (4.24)$$

Функція (4.24) називається також ядром Фейєра.

Знайдемо вираз ряду (4.19) для ядра Фейєра. Спочатку знайдемо коефіцієнти цього ряду за формулою (4.10)

$$\varphi(z) = \frac{4}{\pi} \int_0^{+\infty} \frac{\sin^2 \frac{t}{2}}{t^2} \cos zt\, dt = \frac{2}{\pi} \int_0^{+\infty} \frac{1-\cos t}{t^2} \cos zt\, dt =$$

$$= \frac{1}{\pi} \left[ 2\int_0^{+\infty} \frac{\cos zt - 1}{t^2}dt - \int_0^{+\infty} \frac{\cos(z-1)t - 1}{t^2}dt - \int_0^{+\infty} \frac{\cos(z+1)t - 1}{t^2}dt \right].$$

Зробимо заміни: $|z|t = u$ у першому інтегралі; $|z-1|t = u$ у другому інтегралі; $|z+1|t = u$ у третьому інтегралі. Знайдемо

$$\varphi(z) = \frac{4}{\pi}\int_0^{+\infty} \frac{1-\cos u}{t^2}du = \frac{2}{\pi}\int_0^{+\infty}\frac{1-\cos u}{u^2}du\left[\frac{1}{2}|z+1| + \frac{1}{2}|z-1| - |z|\right] =$$

$$= \begin{cases} 0, & z \leq -1, \\ 1-|z|, & -1 \leq z \leq 1, \\ 0, & z \geq 1. \end{cases}$$

Тепер, якщо у виразі функції $\varphi(z)$ приймемо $z = kr$, $r = \dfrac{1}{n}$, то

$$\varphi(k\varepsilon) = \varphi\left(\frac{k}{n}\right) = \begin{cases} 1 - \dfrac{k}{n}, & k \leq n, \\ 0, & k \geq n \end{cases}$$

і, відповідно,

$$\delta^*_{1/n}(x) = \frac{1}{\pi}\left[\frac{1}{2} + \sum_{k=1}^{\infty}\varphi(k\varepsilon)\cos kx\right] = \frac{1}{\pi}\left[\frac{1}{2} + \sum_{k=1}^{n}\left(1 - \frac{k}{n}\right)\cos kx\right].$$

Якщо $f(x)$ – $2\pi$-періодична кусково-гладка функція, то її усереднення з використанням ядра Фейєра набуде вигляду

$$S^*_{1/n}(f;x) = \frac{1}{2\pi n}\int_{-\pi}^{\pi} f(x-t)\frac{\sin^2\dfrac{nt}{2}}{\sin^2\dfrac{t}{2}}\,dt =$$

$$= \frac{a_0}{2} + \sum_{k=1}^{n}\left(1 - \frac{k}{n}\right)(a_k\cos kx + b_k\sin kx).$$

Тоді за теоремою 1, оскільки $\delta^*_{1/n}(x)$ – ядро типу Фейєра, в точці неперервності функції $f(x)$ маємо

$$f(x) = \lim_{n\to\infty}\left[\frac{a_0}{2} + \sum_{k=1}^{n}\left(1 - \frac{k}{n}\right)(a_k\cos kx + b_k\sin kx)\right].$$

*П р и к л а д 2*. Розглянемо ядро

$$\Phi(x) = \frac{1}{\pi}\frac{\sin x}{x},$$

яке породжує дельтоподібну функцію, однак не є ядром типу Фейєра (зауваж. 3, п. 1.5.3).

За формулою (4.20), прийнявши $r = \dfrac{1}{n}$, одержимо

$$D^*_{1/n}(x) = \frac{1}{\pi n}\left\{\frac{\sin nx}{x} + \sum_{k=1}^{\infty}\left[\frac{\sin n(2k\pi + x)}{2k\pi + x} + \frac{\sin n(2k\pi - x)}{2k\pi - x}\right]\right\} =$$

$$= \frac{\sin nx}{\pi}\left[\frac{1}{x} + \sum_{k=1}^{\infty}\left(\frac{1}{2k\pi + x} - \frac{1}{2k\pi - x}\right)\right] =$$

$$= \frac{\sin nx}{\pi} \frac{\cos\frac{x}{2}}{2\sin\frac{x}{2}} = \frac{1}{2}\left[\frac{1}{2\pi}\frac{\cos\left(n+\frac{1}{2}\right)x}{\sin\frac{x}{2}} + \frac{1}{2\pi}\frac{\cos\left(n-\frac{1}{2}\right)x}{\sin\frac{x}{2}}\right].$$

Тут враховано формулу $[21, c. 472]$

$$\frac{\cos x}{\sin x} = \sum_{k=-\infty}^{\infty} \frac{1}{x - k\pi}.$$

Отже,

$$\delta_{1/n}^*(x) = \frac{1}{2}[D_n(x) + D_{n-1}(x)],$$

де $D_n(x) = \frac{1}{2\pi}\dfrac{\cos\left(n+\frac{1}{2}\right)x}{\sin\frac{x}{2}}$ – ядро Діріхле.

*П р и к л а д   3 .* Знайдемо періодичний аналог ядра Пуассона (прикл. 5, п. 1.5)

$$\Phi(x) = \frac{1}{\pi}\frac{1}{1+x^2}.$$

Коефіцієнти ряду (4.19) цієї функції визначаються за формулою (інтеграли Лапласа $[21, c. 721, 729]$)

$$\varphi(nr) = 2\int_0^\infty \Phi(x)\cos nrx\,dx = \rho^n,$$

де $\rho = e^{-r}$, при цьому виконується рівність $\lim\limits_{r\to+0}\rho = 1$.

За формулами (4.20) з урахуванням формули прикладу 10 (п. 2.3) і (4.19) знайдемо

$$\delta_r^*(x) = \frac{r}{\pi}\sum_{k=-\infty}^{\infty}\frac{1}{r^2 + (x+2k\pi)^2},$$

$$\delta_r^*(x) = \frac{1}{\pi}\left(\frac{1}{2} + \sum_{k=1}^{\infty}\rho^k\cos kx\right) = \frac{1}{2\pi}\frac{1-\rho^2}{1+\rho^2 - 2\rho\cos x}$$

і одержимо формулу

$$\frac{e^{2r}-1}{2r\left(1+e^{2r}-2e^r\cos x\right)} = \sum_{k=-\infty}^{\infty}\frac{1}{r^2+(x+2k\pi)^2}. \qquad (4.25)$$

Якщо $2\pi$-періодична функція $f(x)$ розвивається в ряд Фур'є, то усереднення цієї функції має вигляд

$$S_r^*(f;x) = \int_{-\pi}^{\pi}\delta_r^*(x-t)f(t)dt = \frac{1}{2\pi}\int_{-\pi}^{\pi}f(t)\frac{1-\rho^2}{1+\rho^2-2\rho\cos(x-t)}\,dt =$$
$$= \frac{a_0}{2} + \sum_{k=1}^{n}\rho^k\left(a_k\cos kx + b_k\sin kx\right).$$

### 4.3. Узагальнені методи підсумовування рядів

**4.3.1. Підсумовування тригонометричних рядів.** Розглянемо інтегральні оператори згладжування вигляду

$$S_r(f;x) = \int_{-\infty}^{+\infty} f(x+rt)\Phi(t)\,dt, \qquad (4.26)$$

ядерні функції яких справджують додаткові умови гладкості (4.6).

Інтегральне перетворення кусково-неперервної функції з використанням оператора згладжування за теоремою 2 (п. 4.1) є достатньо гладкою функцією і в кожній точці неперервності функції $f(x)$ за теоремою 1 (п. 4.1) справджується рівність

$$\lim_{r\to+0}\int_{-\infty}^{+\infty}f(x+rt)\Phi(t)dt = f(x).$$

Важливим також є твердження теореми 6 (п. 1.5) про те, що основний вклад в інтеграл (4.26) при малих значеннях $r$ дає лише як завгодно малий окіл точки $x$, тобто для будь-якого числа $\eta > 0$ справедлива рівність

$$\lim_{r\to+0}\int_{\eta}^{+\infty}f(x+rt)\omega(|t|)dt = 0.$$

Крім того, застосування оператора згладжування до тригонометричного ряду

$$\frac{a_0}{2}+\sum_{k=1}^{\infty}(a_k\cos kx+b_k\sin kx), \qquad (4.27)$$

ґрунтуючись на властивості про перетворення тригонометричних функцій (теорема 3, п. 4.1), полягає у множенні коефіцієнтів ряду на множник $\varphi(kr)$,

$$\frac{a_0}{2}+\sum_{k=1}^{\infty}\varphi(kr)(a_k\cos kx+b_k\sin kx).$$

Зроблені зауваження обґрунтовують можливість застосування математичного апарату інтегральних операторів згладжування для аналізу функцій простору, на якому визначені ці оператори. Це дозволяє розглядати не тільки питання збіжності ряду Фур'є для функції з заданого простору, а також питання підсумовування тригонометричного ряду незалежно від того, збігається він чи розбігається $[2, 7, 18, 19, 22]$.

Розглянемо тригонометричний ряд (4.27).

Сумою (або класичною сумою) функціонального ряду (4.27) в області $E$ (проміжку $(a,b)$) за означенням 3 (п. 2.1) називається скінченна границя при $n\to\infty$ послідовності частинних сум
$$\lim_{n\to\infty}S_n(x)=S(x),$$
де $S_n(x)=\dfrac{a_0}{2}+\sum_{k=1}^{n}(a_k\cos kx+b_k\sin kx)$.

Нехай задана також послідовність $\{\varphi_k(r)\}$, $\varphi_0(r)=1$, функцій, визначених на множині $\{r\}$ з точкою згущення $r_0$, таких що ряд

$$S_r(f;x)=\frac{a_0}{2}+\sum_{k=1}^{\infty}\varphi_k(r)(a_k\cos kx+b_k\sin kx) \qquad (4.28)$$

рівномірно збігається відносно $x$ при $r\neq r_0$ $[7]$.

***О з н а ч е н н я   1 .*** *Скінченна границя при $r\to r_0$ суми ряду* (4.28),
$$\lim_{r\to r_0}S_r(x)=S^*(x), \quad x\in E^*, \qquad (4.29)$$
*називається узагальненою сумою ряду* (4.27), *якщо $S^*(x)=S(x)$ для всіх $x\in E$ і, крім того, $E\subset E^*$.*

*При цьому ряд* (4.27) *називається слабко збіжним в області* $E^*$ *(або збіжним у розумінні узагальненого підсумовування), а метод підсумовування* $\{\varphi_k(r)\}$ *називається регулярним.*

Широкий клас методів підсумовування рядів складають методи $\{\varphi(kr)\}$ і їх частинні випадки $\{\varphi_0(kr)\}$ при $r \to +0$, в яких функції $\varphi(kr)$ і $\varphi_0(kr)$ визначаються за формулами (4.10) і (4.12),

$$\varphi(kr) = 2\int_0^{+\infty} \omega(t)\cos krt\, dt,$$

$$\varphi_0(kr) = 2\int_0^1 \omega_0(t)\cos krt\, dt, \qquad (4.30)$$

де $\Phi(x) = \omega(|x|)$ – ядро оператора згладжування і $\Phi_0(x) = \omega_0(|x|)$ – фінітне ядро оператора згладжування, що справджують умови (4.6).

**Т е о р е м а  1 .** *Ряд Фур'є кусково-неперервної функції* $f(x)$ *періоду* $2\pi$ *підсумовується методом* $\{\varphi(kr)\}$ *до цієї функції в кожній точці її неперервності і до значення* $\dfrac{f(x+0)+f(x-0)}{2}$ *в точках розриву першого роду.*

*Д о в е д е н н я .* Оскільки в точках неперервності

$$\frac{f(x+0)+f(x-0)}{2} = f(x),$$

достатньо довести рівність

$$\lim_{r\to 0} S_r(f;x) = \frac{f(x+0)+f(x-0)}{2}.$$

За теоремою 1 (п. 4.2) ряд (4.19) збігається рівномірно, а за лемою 1 (п. 2.3) коефіцієнти Фур'є функції $f(x)$ прямують при $n \to \infty$ до нуля. Тому ряд (4.28) для функції $f(x)$ збігається рівномірно при $r > 0$.

Дійсно, ряд

$$\frac{1}{2} + \sum_{k=1}^{\infty} \varphi(kr)\cos k(t-x)$$

збігається рівномірно при $r > 0$ і фіксованому $x$, оскільки він мажорується збіжним рядом

$$\frac{1}{2} + \sum_{k=1}^{\infty} |\varphi(kr)|.$$

Тому його можна почленно інтегрувати, помноживши на функцію $f(x)$,

$$\frac{1}{\pi} \int_{-\pi}^{\pi} f(t) \left[ \frac{1}{2} + \sum_{k=1}^{\infty} \varphi(kr) \cos k(t-x) \right] dt =$$

$$= \frac{a_0}{2} + \sum_{k=1}^{\infty} \varphi(kr)(a_k \cos kx + b_k \sin kx).$$

Отже, одержаний ряд рівномірно збігається і його сума дорівнює $S_r(f; x)$.

Перетворимо цей ряд з урахуванням виразів для коефіцієнтів Фур'є функції $f(x)$, формул (4.20) і (4.30), а також періодичності розглядуваних функцій

$$S_r(f; x) = \frac{a_0}{2} + \sum_{k=1}^{\infty} \varphi_k(r)(a_k \cos kx + b_k \sin kx) =$$

$$= \frac{1}{\pi} \int_{-\pi}^{\pi} f(t) \left[ \frac{1}{2} + \sum_{k=1}^{\infty} \varphi_k(r) \cos k(x-t) \right] dt =$$

$$= \frac{1}{r} \int_{-\pi}^{\pi} \left\{ \omega\left( \frac{|t|}{r} \right) + \sum_{k=1}^{\infty} \left[ \omega\left( \frac{2k\pi + t}{r} \right) + \omega\left( \frac{2k\pi - t}{r} \right) \right] \right\} f(x+t) dt =$$

$$= \int_{-\infty}^{+\infty} f(x+t) \frac{1}{r} \omega\left( \frac{|t|}{r} \right) dt = \int_{0}^{+\infty} [f(x+rt) + f(x-rt)] \omega(t) dt.$$

Отже,

$$S_r(f; x) = \int_{0}^{+\infty} [f(x+rt) + f(x-rt)] \omega(t) dt. \qquad (4.31)$$

За теоремою 1 (п. 4.1) в точці розриву першого роду функції $f(x)$ існує границя

$$\lim_{r \to 0} \int_{0}^{+\infty} [f(x+rt) + f(x-rt)] \omega(t) dt = \frac{f(x+0) + f(x-0)}{2}.$$

Теорему доведено.

***Т е о р е м а  2 .*** *Ряд Фур'є кусково-неперервної функції* $f(x)$ *періоду* $2\pi$ *рівномірно підсумовується методом* $\{\varphi(kr)\}$ *до* $f(x)$ *на кожному відрізку, строго внутрішньому до відрізка неперервності цієї функції.*

*Д о в е д е н н я .* Нехай відрізок $[a, b]$ строго внутрішній до відрізка неперервності функції $f(x)$. Ряд (4.27) рівномірно підсумовується на відрізку $[a, b]$, якщо для будь-якого $\varepsilon > 0$ існує $r_0 > 0$ таке, що для всіх $r$, $0 < r < r_0$, і всіх $x \in [a, b]$ виконується нерівність

$$|f(x) - S_r(f; x)| < \varepsilon.$$

Використовуючи інтегральне зображення (4.31) ряду (4.28), одержимо

$$S_r(f; x) - f(x) = \frac{2}{r} \int_0^{+\infty} \left[\frac{f(x+t) + f(x-t)}{2} - f(x)\right] \omega\left(\frac{t}{r}\right) dt. \quad (4.32)$$

Виберемо $\eta > 0$ таке, що справджується нерівність

$$\left|\frac{f(x+t) + f(x-t)}{2} - f(x)\right| < \frac{\varepsilon}{2M} \quad (4.33)$$

для всіх $x \in [a, b]$ і всіх $t$, $0 < t < \eta$. Тут (за визначенням ядра типу Фейєра)

$$M = \frac{2}{r} \int_0^{+\infty} \left|\omega\left(\frac{t}{r}\right)\right| dt < +\infty. \quad (4.34)$$

Розіб'ємо інтеграл у формулі (4.32) на два інтеграли

$$I_1 = \frac{2}{r} \int_0^{\eta} \left[\frac{f(x+t) + f(x-t)}{2} - f(x)\right] \omega\left(\frac{t}{r}\right) dt,$$

$$I_2 = \frac{2}{r} \int_{\eta}^{+\infty} \left[\frac{f(x+t) + f(x-t)}{2} - f(x)\right] \omega\left(\frac{t}{r}\right) dt.$$

Для першого інтегралу за виконання нерівності (4.33) маємо для всіх $x \in [a, b]$ оцінку

$$|I_1| = \frac{2}{r}\int_0^\eta \left|\frac{f(x+t)+f(x-t)}{2} - f(x)\right| \left|\omega\left(\frac{t}{r}\right)\right| dt \leq \frac{\varepsilon}{2M}\frac{2}{r}\int_0^\eta \left|\omega\left(\frac{t}{r}\right)\right| dt \leq \frac{\varepsilon}{2}.$$

Для другого інтегралу, внаслідок теореми 6 (п. 1.5), існує число $r_0 > 0$ таке, що $|I_2| < \frac{\varepsilon}{2}$, яке б не було $x \in [a,b]$.

Отже, оцінюючи праву частину формули (4.32), одержимо нерівність
$$|S_r(f;x) - f(x)| \leq |I_1| + |I_2| < \varepsilon,$$
яка справедлива для всіх $x \in [a,b]$ і $r \in (0, r_0)$.

Теорему доведено.

**Н а с л і д о к .** *Ряд Фур'є неперервної функції $f(x)$ періоду $2\pi$ рівномірно підсумовується методом $\{\varphi(kr)\}$ до $f(x)$.*

Відомо (теорема 10, п. 2.3), що ряд Фур'є навіть неперервної функції $f(x)$ може бути розбіжним. Водночас частинні суми усереднення (4.28) функції $f(x)$ є для неї рівномірними наближеннями.

**Т е о р е м а   3 .** *Якщо $2\pi$-періодична кусково-неперервна функція $f(x)$ має на проміжку $(a-h, b+h)$, $a-h < b+h$, $h > 0$, неперервну похідну $m$-го порядку, $1 \leq m \leq p-1$, то її ряд Фур'є*

$$f(x) \sim \frac{a_0}{2} + \sum_{k=1}^\infty (a_k \cos kx + b_k \sin kx) \qquad (4.35)$$

*почленно продиференційований $m$ раз, рівномірно підсумовується методом $\{\varphi(kr)\}$ до $f^{(m)}(x)$ на відрізку $[a,b]$.*

*Д о в е д е н н я .* Рівномірне підсумовування $m$ раз почленно продиференційованого ряду Фур'є функції $f(x)$ на відрізку $[a,b]$ можливе, якщо для будь-якого $\varepsilon > 0$ існує $r_0 > 0$ таке, що для всіх $r$, $0 < r < r_0$, і всіх $x \in [a,b]$ виконується нерівність

$$\left| f^{(m)}(x) - \frac{d^m}{dx^m} S_r(f;x) \right| < \varepsilon, \qquad (4.36)$$

де

$$\frac{d^m}{dx^m} S_r(f;x) =$$
$$= \sum_{k=1}^{\infty} \varphi_k(r) k^m \left[ a_k \cos\left(kx + \frac{m\pi}{2}\right) + b_k \sin\left(kx + \frac{m\pi}{2}\right) \right]. \quad (4.37)$$

Ряд (4.37) за теоремою 2 (п. 4.1) рівномірно збігається відносно $x$ при $r > 0$ і $1 \le m \le p-1$. Справедлива також формула

$$\frac{d^m}{dx^m} S_r(f;x) = \frac{1}{r} \int_{-\infty}^{+\infty} \frac{d^m}{dx^m} \Phi\left(\frac{x-t}{r}\right) f(t) dt,$$

тому відповідний інтеграл збігається рівномірно при $1 \le m \le p-1$. Отже, справедлива формула

$$\sum_{k=1}^{\infty} \varphi_k(r) k^m \left[ a_k \cos\left(kx + \frac{m\pi}{2}\right) + b_k \sin\left(kx + \frac{m\pi}{2}\right) \right] = \quad (4.38)$$
$$= \frac{1}{r} \int_{-\infty}^{+\infty} \frac{d^m}{dx^m} \Phi\left(\frac{x-t}{r}\right) f(t) dt, \quad 1 \le m \le p-1.$$

За заданим $\varepsilon > 0$ виберемо $\eta > 0$ таке, що кожна точка $x \in [a,b]$ належить проміжку $(a-h, b+h)$ разом зі своїм околом $[x-\eta, x+\eta]$ і справджується нерівність

$$\left| f^{(m)}(x+u) - f^{(m)}(x) \right| < \frac{\varepsilon}{2M} \quad (4.39)$$

для всіх $x \in [a,b]$, якщо тільки $|u| < \eta$, де $M$ – стала, що визначається формулою (4.34).

Розіб'ємо інтеграл у формулі (4.38) на два інтеграли і перетворимо їх з використанням методів інтегрування частинами та заміни змінної

$$I_1 = \frac{1}{r} \left[ \int_{-\infty}^{x-\eta} \frac{d^m}{dx^m} \Phi\left(\frac{x-t}{r}\right) f(t) dt + \int_{x+\eta}^{+\infty} \frac{d^m}{dx^m} \Phi\left(\frac{x-t}{r}\right) f(t) dt \right] =$$
$$= \frac{1}{r} \int_{\eta}^{+\infty} \left[ f(x+u) + (-1)^m f(x-u) \right] \frac{d^m}{du^m} \Phi\left(\frac{u}{r}\right) du,$$

$$I_2 = \frac{1}{r} \int\limits_{x-\eta}^{x+\eta} \frac{d^m}{dx^m} \Phi\left(\frac{x-t}{r}\right) f(t)\, dt = \frac{(-1)^m}{r} \int\limits_{-\eta}^{\eta} f(x+u) \frac{d^m}{du^m} \Phi\left(\frac{u}{r}\right) du =$$

$$= A + \frac{1}{r} \int\limits_{-\eta}^{\eta} f^{(m)}(x+u) \Phi\left(\frac{u}{r}\right) du,$$

де

$$A = \frac{(-1)^m}{r} \left[ \sum_{k=0}^{m-1} (-1)^k f^{(k)}(x+u) \frac{d^{m-k-1}}{du^{m-k-1}} \Phi\left(\frac{u}{r}\right) \right]\bigg|_{-\eta}^{\eta}.$$

Оцінимо відхилення у формулі (4.36) з урахуванням нерівності (4.39) і формул для інтегралів $I_1$ і $I_2$

$$\left| \frac{d^m}{dx^m} S_r(f; x) - f^{(m)}(x) \right| = \left| I_1 + A + \frac{1}{r} \int\limits_{-\eta}^{\eta} f^{(m)}(x+u) \Phi\left(\frac{u}{r}\right) du - \right.$$

$$\left. - f^{(m)}(x) \left[ \frac{1}{r} \int\limits_{-\eta}^{\eta} \Phi\left(\frac{u}{r}\right) du + \frac{2}{r} \int\limits_{\eta}^{+\infty} \Phi\left(\frac{u}{r}\right) du \right] \right| = \left| I_1 + A - \right.$$

$$\left. - f^{(m)}(x) \frac{2}{r} \int\limits_{\eta}^{+\infty} \Phi\left(\frac{u}{r}\right) du + \frac{1}{r} \int\limits_{-\eta}^{\eta} \left[ f^{(m)}(x+u) - f^{(m)}(x) \right] \Phi\left(\frac{u}{r}\right) du \right| \le$$

$$\le |I_1| + |A| + \left| f^{(m)}(x) \right| \frac{2}{r} \int\limits_{\eta}^{+\infty} \left| \Phi\left(\frac{u}{r}\right) \right| du + \frac{\varepsilon}{2rM} \int\limits_{-\eta}^{\eta} \left| \Phi\left(\frac{u}{r}\right) \right| du \le$$

$$\le |I_1| + |A| + \left| f^{(m)}(x) \right| \frac{2}{r} \int\limits_{\eta}^{+\infty} \left| \Phi\left(\frac{u}{r}\right) \right| du + \frac{\varepsilon}{2}.$$

За теоремами 6 і 9 (п. 1.5) та обмеженості функції $f^{(m)}(x)$, $x \in (a, b)$, перший, другий та третій доданки у цій нерівності можна зробити шляхом вибору числа $r = r_0 > 0$ як завгодно малими, а саме меншими, ніж $\dfrac{\varepsilon}{2}$.

Тоді для відхилення функції $f^{(m)}(x)$ від похідної $m$-го порядку від суми ряду (4.38) одержимо нерівність

$$\left|\frac{d^m}{dx^m}S_r(f;x)-f^{(m)}(x)\right|\leq\frac{\varepsilon}{2}+\frac{\varepsilon}{2}=\varepsilon,$$

яка виконується для всіх $x\in(a,b)$ і $r\in(0,r_0)$.

Теорему доведено.

*Т е о р е м а  4 .  Нехай $2\pi$-періодична кусково-неперервна функція $f(x)$ задається многочленом степеня меншого або рівного $q$ на проміжку $(a-h,b+h)$, $a-h<b+h$, $h>0$, ряд (4.35) – її ряд Фур'є і $\Phi_0(x)$ – фінітне ядро оператора згладжування, що задовольняє додаткові умови (4.17)*

$$\int_0^1 t^k\omega_0(t)dt=0,\quad \int_0^1 t^{k+1}|\omega_0(t)|dt=A<\infty,\quad k=1,...,q. \quad (4.40)$$

*Тоді справедлива формула*

$$f(x)=\frac{a_0}{2}+\sum_{k=1}^{\infty}\varphi_k(r)(a_k\cos kx+b_k\sin kx) \quad (4.41)$$

*для всіх $x$ і $r$ таких, що $x\in[a,b]$ і $[x-r,x+r]\subset(a-h,b+h)$.*

*Д о в е д е н н я .* Розглянемо усереднення функції $f(x)$ в точці $x\in(a,b)$ з фінітною ядерною функцією

$$S_r^0(f;x)=\int_{-1}^1 f(x+rt)\Phi_0(t)dt=\int_0^1[f(x+rt)+f(x-rt)]\omega_0(t)dt=$$

$$=\frac{a_0}{2}+\sum_{k=1}^{\infty}\varphi_k(r)(a_k\cos kx+b_k\sin kx). \quad (4.42)$$

Зобразивши функцію $f(x+rt)$ в околі точки $x$ у вигляді

$$f(x+rt)=\sum_{i=0}^{q}f^{(i)}(x)\frac{(rt)^i}{i!},$$

одержимо

$$S_r^0(f;x)=\sum_{i=0}^{q}\frac{f^{(i)}(x)r^i}{i!}\int_0^1[1+(-1)^i]t^i\omega_0(t)dt.$$

Врахувавши тут умови (4.40), матимемо

$$S_r^0(f;x) = f(x).$$

Звідси з урахуванням останнього виразу (4.42) одержимо формулу (4.41).

Теорему доведено.

Наведемо деякі найважливіші методи підсумовування рядів $[2, 14, 17, 18, 21]$, що використовуються у математичному аналізі.

*П р и к л а д   1* . **Метод підсумовування середніми арифметичними (метод Фейєра).** Розглянемо ряд Фур'є кусково-неперервної функції

$$f(x) \sim \frac{a_0}{2} + \sum_{k=1}^{\infty}(a_k \cos kx + b_k \sin kx) \qquad (4.43)$$

і його частинну суму

$$S_n(x) = \frac{a_0}{2} + \sum_{k=1}^{\infty}(a_k \cos kx + b_k \sin kx). \qquad (4.44)$$

Для середнього арифметичного частинних сум одержимо вираз

$$\sigma_n(x) = \frac{S_0(x) + S_1(x) + ... + S_{n-1}(x)}{n} =$$
$$= \frac{a_0}{2} + \sum_{k=1}^{n}\left(1 - \frac{k}{n}\right)(a_k \cos kx + b_k \sin kx). \qquad (4.45)$$

Звідси маємо вираз загального члена послідовності $\{\varphi(kr)\}$,

$$\varphi(kr) = \varphi\left(\frac{k}{n}\right) = \begin{cases} 1 - \dfrac{k}{n}, & k \leq n, \\ 0, & k \geq n. \end{cases}$$

Врахувавши у формулі (4.45) вирази коефіцієнтів Фур'є для функції $f(x)$, прийдемо до інтегральної форми для середнього арифметичного часткових сум (прикл. 1, п. 4.2)

$$\sigma_n(x) = \frac{1}{2\pi n}\int_{-\pi}^{\pi} f(x-t)\frac{\sin^2 \dfrac{nt}{2}}{\sin^2 \dfrac{t}{2}}dt \qquad (4.46)$$

або з використанням формули (4.3), прийнявши $r = \dfrac{1}{n}$,

$$\sigma_n(x) = \frac{2}{\pi} \int\limits_{-\infty}^{+\infty} f\left(x - \frac{t}{n}\right) \frac{\sin^2 \frac{t}{2}}{t^2} dt. \qquad (4.47)$$

Інтеграл у формулі (4.47) є оператором усереднювання (4.3) з ядром Фейєра $\Phi(t) = \frac{2}{\pi} \frac{\sin^2 \frac{t}{2}}{t^2}$, а функція $\Phi_{1/n}^*(t) = \frac{1}{2\pi n} \frac{\sin^2 \frac{nt}{2}}{\sin^2 \frac{t}{2}}$ - періодичне ядро Фейєра.

*П р и к л а д  2*. **Метод Пуассона (метод степеневих множників).** Узагальнена сума тригонометричного ряду (4.27), підсумовуваного методом Пуассона, визначається за формулою
$$\lim_{\rho \to 1-0} \sigma_\rho(x) = \sigma(x),$$
де
$$\sigma_\rho(x) = \frac{a_0}{2} + \sum_{k=1}^{n} \rho^k (a_k \cos kx + b_k \sin kx). \qquad (4.48)$$

Ядро оператора усереднювання (4.3) наступне (прикл. 3, п. 4.2)
$$\Phi(t) = \frac{1}{\pi} \frac{1}{1+t^2}.$$

Загальний член послідовності $\{\varphi(kr)\}$ записується у вигляді
$$\varphi(kr) = 2 \int\limits_0^{+\infty} \Phi(t) \cos krt \, dt = \rho^k,$$
де $\rho = e^{-r}$.

Періодичний аналог ядра оператора такий
$$\Phi_r^*(x) = \frac{1}{\pi}\left(\frac{1}{2} + \sum_{k=1}^{\infty} \rho^k \cos kx\right) = \frac{1}{2\pi} \frac{1-\rho^2}{1+\rho^2 - 2\rho \cos x}.$$

Якщо $2\pi$-періодична функція $f(x)$ розвивається в ряд Фур'є, то усереднення цієї функції має вигляд
$$S_r^*(f;x) = \frac{1}{2\pi} \int\limits_{-\pi}^{\pi} f(t) \frac{1-\rho^2}{1+\rho^2 - 2\rho \cos(x-t)} dt =$$

$$= \frac{a_0}{2} + \sum_{k=1}^{n} \rho^k \left(a_k \cos kx + b_k \sin kx\right).$$

Почленно продиференційований $m$ раз ряд Фур'є $2\pi$-періодичної функції $f(x)$, що має в точці $x$ похідну $m$-го порядку $f^{(m)}(x)$, підсумовується за теоремою 3 методом Пуассона до значення $f^{(m)}(x)$, де $m$ – довільне натуральне число.

*З а у в а ж е н н я 1*. При почленному диференціюванні ряду Фур'є не завжди одержується ряд Фур'є. Так, розглянемо функцію $f(x) = \frac{x}{2}$ і її ряд Фур'є

$$\frac{x}{2} = \sum_{n=1}^{\infty} (-1)^{n-1} \frac{\sin nx}{n}, \quad x \in (-\pi, \pi),$$

а також одержані з нього (при одноразовому і дворазовому почленному диференціюванні) ряди (які не є рядами Фур'є):

$$\sum_{n=1}^{\infty} (-1)^{n-1} \cos nx, \qquad \sum_{n=1}^{\infty} (-1)^{n} n \sin nx.$$

За теоремою 3 узагальнені суми цих рядів, відповідно, дорівнюють $\sigma = \frac{1}{2}$ і $\sigma = 0$.

Дійсно, за методом Пуассона маємо (прикл. 3, п. 4.2)

$$S_r^*(f'; x) = \sum_{n=1}^{\infty} (-1)^{n-1} \rho^n \cos nx = \frac{\rho(\rho + \cos x)}{1 + 2\rho \cos x + \rho^2},$$

$$S_r^*(f''; x) = \sum_{n=1}^{\infty} (-1)^{n-1} n \rho^n \cos nx = \frac{\rho \sin x (\rho^2 - 1)}{\left(1 + 2\rho \cos x + \rho^2\right)^2}.$$

Звідси, оскільки $\rho = e^{-r}$,

$$\lim_{\substack{r \to +0 \\ (\rho \to 1-0)}} S_r^*(f'; x) = \frac{1}{2}, \quad \lim_{\substack{r \to +0 \\ (\rho \to 1-0)}} S_r^*(f''; x) = 0, \quad x \in (-\pi, \pi).$$

*П р и к л а д 3*. **Метод Рімана** визначається послідовністю функцій $\left\{ \left( \frac{\sin kr_0}{kr_0} \right)^2 \right\}$ при $r_0 \to +0$. Ядром оператора згладжування

$S_r^0(f;x)$, де $r = 2r_0$, є фінітна функція

$$\omega_0(|t|) = \begin{cases} 1 - |t|, & |t| \leq 1, \\ 0, & |t| > 1. \end{cases}$$

Якщо $2\pi$-періодична кусково-неперервна функція $f(x)$ неперервна в точці $x$, то

$$\lim_{r \to +0} S_r^0(f;x) = \lim_{r_0 \to +0}\left[\frac{a_0}{2} + \sum_{k=1}^{\infty}\left(\frac{\sin kr_0}{kr_0}\right)^2 (a_k \cos kx + b_k \sin kx)\right] = f(x)$$

і за теоремою 3

$$\lim_{r \to +0} S_r^0(f';x) = \lim_{r_0 \to +0}\left[\sum_{k=1}^{\infty}\left(\frac{\sin kr_0}{kr_0}\right)^2 (-ka_k \sin kx + kb_k \cos kx)\right] = f'(x).$$

*П р и к л а д  4*. **Метод Рісса** характеризується послідовністю функцій $\{\varphi(kr)\}$ з загальним членом

$$\varphi(kr) = \varphi_p\left(\frac{k}{n}\right) = \begin{cases} \left[1 - \left(\frac{k}{n}\right)^2\right]^p, & k \leq n, \\ 0, & k > n, \end{cases}$$

де $p$ – натуральне число.

Ядерними функціями відповідних операторів усереднювання є функції Бесселя першого роду $J_m(x)$ порядку $m = p + \frac{1}{2}$ [19, 22]. Їх можна одержати з рекурентного співвідношення

$$J_{m-1}(x) + J_{m+1}(x) = \frac{2m}{x} J_m(x)$$

за виразами функцій

$$J_{1/2}(x) = \sqrt{\frac{2}{\pi x}} \sin x, \quad J_{-1/2}(x) = \sqrt{\frac{2}{\pi x}} \cos x.$$

Наприклад, з попередньої формули при $m = \frac{1}{2}$ і $m = \frac{3}{2}$ знайдемо

$$J_{-1/2}(x) + J_{3/2}(x) = \frac{1}{x} J_{1/2}(x), \quad J_{1/2}(x) + J_{5/2}(x) = \frac{3}{x} J_{3/2}(x).$$

Звідси

$$J_{3/2}(x) = \sqrt{\frac{2}{\pi x^3}}(\sin x - x\cos x),$$

$$J_{5/2}(x) = \sqrt{\frac{2}{\pi x^5}}\left[(3-x^2)\sin x - 3x\cos x\right].$$

За формулою (4.10) з використанням відповідної формули [22] знайдемо

$$\varphi_1\left(\frac{k}{n}\right) = 2\int\limits_0^{+\infty} J_{3/2}(x)\cos\frac{k}{n}x\,dx = \begin{cases} 1-\left(\frac{k}{n}\right)^2, & k \leq n, \\ 0, & k > n, \end{cases}$$

$$\varphi_2\left(\frac{k}{n}\right) = 2\int\limits_0^{+\infty} J_{5/2}(x)\cos\frac{k}{n}x\,dx = \begin{cases} \left[1-\left(\frac{k}{n}\right)^2\right]^2, & k \leq n, \\ 0, & k > n. \end{cases} \quad (4.49)$$

Зауважимо, що хоч інтеграли у формулах (4.49) умовно збігаються, відповідні послідовності визначають також узагальнені методи підсумовування рядів (зауваж. 1, п. 1.5).

*П р и к л а д  5 .* **Метод** $\left\{\rho^{k^2}\right\}$. Загальний член підсумовуючої послідовності визначається за формулою [22]

$$\varphi(kr) = \rho^{k^2} = \frac{1}{\sqrt{\pi}}\int\limits_0^{+\infty}\exp\left(-\frac{t^2}{4}\right)\cos rt\,dt\,,$$

де   $\Phi(x) = \dfrac{1}{2\sqrt{\pi}}\exp\left(-\dfrac{t^2}{4}\right)$   –   ядерна   функція   оператора згладжування; $\rho = e^{-r^2}$, $\lim\limits_{r\to +0}\rho = 1$.

Якщо в точці $x$ існує узагальнена сума ряду (4.27), то вона визначається за формулою

$$\sigma(x) = \lim_{\rho \to 1-0}\left[\frac{a_0}{2} + \sum_{k=1}^n \rho^{k^2}\left(a_k\cos kx + b_k\sin kx\right)\right].$$

Почленно продиференційований $m$ раз ряд Фур'є $2\pi$-періодичної функції $f(x)$, що має в точці $x$ похідну $m$-го порядку

$f^{(m)}(x)$, за теоремою 3 підсумовується сформульованим методом до значення $f^{(m)}(x)$, де $m$ – довільне натуральне число.

*З а у в а ж е н н я 2*. Для функції $\varphi(z)$ справджується оцінка (4.16). Тому вона може бути використана для конструювання нового ядра типу Фейєра і, відповідно, нового методу підсумовування рядів. Так, функція $\varphi(z) = \dfrac{\sin^2 z}{z^2}$, що визначає метод Рімана, є базовою для методу Фейєра.

У таблиці 1 і таблиці 2 наведено ще декілька методів підсумовування рядів. Члени $\varphi(kr)$ і $\varphi_0(kr)$ послідовностей підсумовування одержано за формулами (4.30), в яких відповідні ядерні функції $\omega(t)$ і $\omega_0(t)$ також наведені в таблицях.

Таблиця 1.

| № | Ядерна функція, $t \in [0, 1]$ | Метод підсумовування |
|---|---|---|
| 2. | $\omega_0(t) = \dfrac{3}{2}(1-t)^2$ | $\varphi_0(kr) = \dfrac{6}{(kr)^2}\left(1 - \dfrac{\sin kr}{kr}\right)$ |
| 3. | $\omega_0(t) = \dfrac{3}{4}(1-t^2)$ | $\varphi_0(kr) = \dfrac{3}{(kr)^2}\left(\dfrac{\sin kr}{kr} - \cos kr\right)$ |
| 4. | $\omega_0(t) = \dfrac{\pi^2}{4}(1-t)\cos \pi t$ | $\varphi_0(kr) = \dfrac{1+(kr/\pi)^2}{(1+kr/\pi)^2}\left(\dfrac{\cos(kr/2)}{1 - kr/\pi}\right)^2$ |
| 5. | $\omega_0(t) = \dfrac{2^{m-1}(m!)^2}{(2m)!}(1+\cos \pi t)^m$ | $\varphi_0(kr) = \dfrac{\sin kr}{kr}\prod\limits_{i=1}^{m}\left[1 - \left(\dfrac{kr}{i\pi}\right)^2\right]^{-1}$ |
| 6. | $\omega_0(t) = \dfrac{2^{n-1}(n!)^2}{(2n)!}(1+\cos \pi t)^n$ | $\varphi_{nk} = \begin{cases} \dfrac{(n!)^2}{(n+k)!(n-k)!}, & k \le n, \\ 0, & k > n, \end{cases} \quad r = \pi$ |

Таблиця 2.

| № | Ядерна функція, $t \in [0, \infty)$ | Метод підсумовування |
|---|---|---|
| 1. | $\omega(t) = \dfrac{1}{2\operatorname{ch}(\pi t/2)}$ | $\varphi(kr) = \dfrac{1}{\operatorname{ch} kr}$ |
| 2. | $\omega(t) = \dfrac{\pi}{4\operatorname{ch}^2(\pi t/2)}$ | $\varphi(kr) = \dfrac{kr}{\operatorname{sh} kr}$ |
| 3. | $\omega(t) = \dfrac{1}{2} e^{-t}$ | $\varphi(kr) = \dfrac{1}{1 + (kr)^2}$ |
| 4. | $\omega(t) = e^{-t} \cos t$ | $\varphi(kr) = \dfrac{1 + (kr/\sqrt{2})^2}{1 + (kr/\sqrt{2})^4}$ |

*П р и к л а д  6*. Функція $f(x) = \pi - |x|$ неперервна і кусково-гладка на проміжку $[-\pi, \pi]$. Її ряд Фур'є має наступний вигляд

$$f(x) = \frac{4}{\pi} \sum_{k=1,3,\ldots}^{\infty} \frac{\cos kx}{k^2}.$$

Однак вже два рази почленно продиференційований ряд цієї функції є розбіжним (у класичному розумінні суми) рядом, він навіть не є рядом Фур'є.

Функція $f(x)$ неперервна і її можна диференціювати скільки завгодно разів на проміжках $(-\pi, 0)$ і $(0, \pi)$. Знайдемо узагальнену суму $q$ разів почленно продиференційованого ряду функції $f(x)$ на цих проміжках

$$\frac{d^q f(x)}{dx^q} \sim \frac{4}{\pi} \sum_{k=1,3,\ldots}^{\infty} k^{q-2} \cos\left(kx + \frac{q\pi}{2}\right).$$

За теоремою 3 цей ряд підсумовується методом $\{\varphi(kr)\}$, якщо $q \leq p - 1$. Зокрема, методами $\{\rho^k\}$, $\{\rho^{k^2}\}$ при $\rho \to 1 - 0$ і $\left\{\dfrac{1}{\operatorname{ch} kr}\right\}$, $\left\{\dfrac{kr}{\operatorname{sh} kr}\right\}$ при $r \to +0$ підсумовується диференційований довільне число разів ряд в точках проміжків $(-\pi, 0)$ і $(0, \pi)$.

Для $q \geq 2$ на проміжках $(-\pi, 0)$, $(0, \pi)$ маємо

$$\frac{4}{\pi} \lim_{\rho \to 1-0} \sum_{k=1,3,\ldots}^{\infty} k^{q-2} \rho^k \cos\left(kx + \frac{q\pi}{2}\right) = \frac{d^q f(x)}{dx^q} = 0,$$

$$\frac{4}{\pi} \lim_{\rho \to 1-0} \sum_{k=1,3,\ldots}^{\infty} k^{q-2} \rho^{k^2} \cos\left(kx + \frac{q\pi}{2}\right) = \frac{d^q f(x)}{dx^q} = 0,$$

$$\frac{4}{\pi} \lim_{r \to 0} \sum_{k=1,3,\ldots}^{\infty} \frac{k^{q-2}}{\operatorname{ch} kr} \cos\left(kx + \frac{q\pi}{2}\right) = \frac{d^q f(x)}{dx^q} = 0,$$

$$\frac{4}{\pi} \lim_{r \to 0} \sum_{k=1,3,\ldots}^{\infty} \frac{k^{q-2} kr}{\operatorname{sh} kr} \cos\left(kx + \frac{q\pi}{2}\right) = \frac{d^q f(x)}{dx^q} = 0.$$

Одержані граничні співвідношення за теоремою 3 рівномірно збігаються до нуля на будь-якому відрізку $[a, b]$, строго внутрішньому до проміжків $(-\pi, 0)$, $(0, \pi)$. З цього твердження випливає, що яке б не було мале число $\varepsilon > 0$, існує число $r_0$ таке, що

$$\frac{4}{\pi} \left| \sum_{k=1,3,\ldots}^{\infty} k^{q-2} \varphi(kr) \cos\left(kx + \frac{q\pi}{2}\right) \right| < \varepsilon$$

для всіх $x \in [a, b]$ і $0 < r < r_0$.

**4.3.2. Підсумовування рядів за системою тригонометричних функцій, ортогональних на довільному проміжку.** Нехай функція $f(x)$ абсолютно інтегровна на проміжку $[-l, l]$. Зробимо заміну $z = \frac{\pi}{l} x$ і розглянемо функцію $g(z) = f\left(\frac{l}{\pi} z\right)$, $z \in [-\pi, \pi]$, яку розвинемо в ряд Фур'є

$$g(z) \sim \sum_{k=1}^{\infty} a_k \cos kz + b_k \sin kz.$$

Застосуємо оператор згладжування (4.26)

$$S_r(g; z) = \int_{-\infty}^{\infty} g(z+rt)\Phi(t)\,dt$$

до функції $g(x)$ та її ряду і перейдемо до змінної $x$ і, відповідно, до функції $f(x)$

$$\int_{-\infty}^{+\infty} f\left(x + \frac{rl}{\pi}t\right)\Phi(t)\,dt =$$

$$= \sum_{k=1}^{\infty} \int_{-\infty}^{+\infty} \cos krt\; \Phi(t)dt \left(a_k \cos\frac{k\pi}{l}x + b_k \sin\frac{k\pi}{l}x\right).$$

Ввівши позначення $r_0 = \dfrac{rl}{\pi}$, одержимо $\int\limits_{-\infty}^{+\infty} f(x+r_0 t)\Phi(t)\,dt =$

$$= \sum_{k=1}^{\infty} \int_{-\infty}^{+\infty} \cos\frac{k\pi r_0}{l}t\; \Phi(t)dt \left(a_k \cos\frac{k\pi}{l}x + b_k \sin\frac{k\pi}{l}x\right)$$

або

$$\int_{-\infty}^{+\infty} f(x+r_0 t)\Phi(t)dt = \sum_{k=1}^{\infty} \varphi(\lambda_k r_0)(a_k \cos\lambda_k x + b_k \sin\lambda_k x),$$

де $\lambda_k = \dfrac{k\pi}{l}$.

Отже, узагальнене підсумовування ряду за системою тригонометричних функцій, ортогональних на проміжку $[-l, l]$, полягає у множенні коефіцієнтів ряду на множники $\varphi(\lambda_k r_0)$.

*З а у в а ж е н н я 3.* Оскільки $r_0 \to +0$ у виразі $\varphi(\lambda_k r_0)$, то прямує до нуля також ця величина з деяким множником. Тому для підсумовування тригонометричного ряду на проміжку $[-l, l]$ можна використовувати послідовність $\{\varphi(kr)\}$ при $r \to +0$.

### 4.4. Підсумовування подвійних рядів

Нехай $f(x, y)$ – обмежена $2\pi$-періодична функція з рядом Фур'є

$$f(x, y) \sim \sum_{m=0}^{\infty}\sum_{n=0}^{\infty} \lambda_{mn} \left( a_{mn} \cos mx \cos ny + b_{mn} \sin mx \cos ny + \right.$$
$$\left. + c_{mn} \cos mx \sin ny + d_{mn} \sin mx \sin ny \right), \qquad (4.50)$$

де

$$\lambda_{mn} = \begin{cases} 1/4, & m = n = 0, \\ 1/2, & m > 0, n = 0 \cup m = 0, n > 0, \\ 1, & m > 0, n > 0; \end{cases}$$

$$a_{mn} = \frac{1}{\pi^2} \iint_Q f(x, y) \cos mx \cos ny \, dx dy, \qquad (4.51)$$

$$b_{mn} = \frac{1}{\pi^2} \iint_Q f(x, y) \sin mx \cos ny \, dx dy,$$

$$c_{mn} = \frac{1}{\pi^2} \iint_Q f(x, y) \cos mx \sin ny \, dx dy,$$

$$d_{mn} = \frac{1}{\pi^2} \iint_Q f(x, y) \sin mx \sin ny \, dx dy;$$

$$Q = \{(x, y): -\pi \le x \le \pi, -\pi \le y \le \pi\}.$$

Застосовуючи оператори згладжування (4.26) за кожною зі змінних до ряду (4.50), одержимо рівномірно збіжний ряд

$$S_r(f; x, y) = \sum_{m=0}^{\infty}\sum_{n=0}^{\infty} \lambda_{mn} \varphi_1(mr) \varphi_2(nr) \left( a_{mn} \cos mx \cos ny + \right.$$
$$\left. + b_{mn} \sin mx \cos ny + c_{mn} \cos mx \sin ny + d_{mn} \sin mx \sin ny \right), \quad (4.52)$$

де $\{\varphi_1(mr)\}$, $\{\varphi_2(nr)\}$ – послідовності, що визначаються за формулою (4.10) або (4.12).

Покажемо, що ряд (4.52) для обмеженої функції $f(x, y)$ рівномірно збігається. Оскільки ряди

$$\frac{1}{2} + \sum_{m=1}^{\infty} |\varphi_1(mr)|, \quad \frac{1}{2} + \sum_{n=1}^{\infty} |\varphi_2(nr)|,$$

збігаються при $r > 0$, коефіцієнти Фур'є $a_{mn}$, $b_{mn}$, $c_{mn}$, $d_{mn}$ і тригонометричні функції обмежені величини, тому члени ряду (4.52) не більші, ніж члени збіжного ряду

$$\sum_{m=0}^{\infty}\sum_{n=0}^{\infty}\lambda_{mn}|\varphi_1(mr)||\varphi_2(nr)|.$$

Отже, ряд (4.52) при $r > 0$ збігається рівномірно.

**Л е м а  1 .** *Якщо $f(x, y)$ – $2\pi$-періодична обмежена функція, $\Phi_1(x)$ і $\Phi_1(x)$ – ядра операторів згладжування, то для суми ряду* (4.39) *справедлива формула*

$$S_r(f;x,y) = \int\limits_{-\infty}^{+\infty}\int\limits_{-\infty}^{+\infty} f(u,v)\frac{1}{r^2}\Phi_1\left(\frac{u-x}{r}\right)\Phi_2\left(\frac{v-y}{r}\right)dudv. \qquad (4.53)$$

*Д о в е д е н н я .* Оскільки ряд (4.52) збігається рівномірно, перетворимо його з урахуванням виразів коефіцієнтів Фур'є функції $f(x, y)$

$$S_r(f;x,y) =$$
$$= \frac{1}{\pi^2}\sum_{m=0}^{\infty}\sum_{n=0}^{\infty}\lambda_{mn}\varphi_1(mr)\varphi_2(nr)\iint\limits_{Q}\cos m(u-x)\cos n(v-y)f(u,v)dudv =$$
$$= \frac{1}{\pi^2}\iint\limits_{Q}\left[\sum_{m=0}^{\infty}\sum_{n=0}^{\infty}\lambda_{mn}\varphi_1(mr)\varphi_2(nr)\cos m(u-x)\cos n(v-y)\right]\cdot$$
$$\cdot f(u,v)dudv = \frac{1}{\pi^2}\iint\limits_{Q} f(u,v)\left[\frac{1}{2} + \sum_{m=0}^{\infty}\varphi_1(mr)\cos m(u-x)\right]\cdot$$
$$\cdot\left[\frac{1}{2} + \sum_{n=1}^{\infty}\varphi_2(nr)\cos n(v-y)\right]dudv =$$
$$= \iint\limits_{Q}\frac{1}{r}\Phi_1^*\left(\frac{u-x}{r}\right)\frac{1}{r}\Phi_2^*\left(\frac{v-y}{r}\right)f(u,v)dudv.$$

Провівши в одержаному інтегралі заміну змінних з урахуванням періодичності підінтегральних функцій і формули (4.20), одержимо формулу (4.53).

Лему доведено.

Розглянемо достатні умови існування узагальненої суми тригонометричного ряду Фур'є (4.50).

**Т е о р е м а  1 .** *Якщо $f(x, y)$ – неперервна $2\pi$-періодична функція за обома змінними, то її ряд Фур'є* (4.50) *рівномірно*

*підсумовується методом* $\{\varphi_1(mr)\varphi_2(nr)\}$ *до* $f(x,y)$ *в* $R^2$.

*Д о в е д е н н я .* Ряд (4.50) рівномірно підсумовується в області $R^2$, якщо для як завгодно малого $\varepsilon > 0$ існує $r_0 > 0$ таке, що для всіх $r$, $0 < r < r_0$, і всіх $(x,y) \in R^2$ виконується нерівність
$$|f(x,y) - S_r(f; x, y)| < \varepsilon.$$

Використовуючи формулу (4.53), запишемо рівність
$$S_r(f; x, y) - f(x, y) =$$
$$= \int_{-\infty}^{+\infty}\int_{-\infty}^{+\infty} [f(x+u, y+v) - f(x,y)] \frac{1}{r^2} \Phi_1\left(\frac{u}{r}\right) \Phi_2\left(\frac{v}{r}\right) dudv. \qquad (4.54)$$

Оскільки функція $f(x,y)$ неперервна і періодична, виберемо $\eta > 0$ таке, що справджується нерівність
$$|f(x+u, y+v) - f(x,y)| < \frac{\varepsilon}{2M} \qquad (4.55)$$

для всіх $(x,y) \in R^2$ і всіх $(u,v)$ з квадрата $D_\eta = \{(u,v) : |u| < \eta; |v| < \eta\}$. Тут (за визначенням ядра типу Фейєра)
$$M = \frac{1}{r^2} \int_{-\infty}^{+\infty} \left|\omega_1\left(\frac{u}{r}\right)\right| du \int_{-\infty}^{+\infty} \left|\omega_1\left(\frac{v}{r}\right)\right| dv < +\infty. \qquad (4.56)$$

Розіб'ємо інтеграл у формулі (4.54) на два інтеграли
$$I_1 = \frac{1}{r^2} \int_{-\eta}^{\eta}\int_{-\eta}^{\eta} [f(x+u, y+v) - f(x,y)] \frac{1}{r^2} \Phi_1\left(\frac{u}{r}\right) \Phi_2\left(\frac{v}{r}\right) dudv,$$
$$I_2 = \frac{1}{r^2} \iint_{D_\infty} [f(x+u, y+v) - f(x,y)] \frac{1}{r^2} \Phi_1\left(\frac{u}{r}\right) \Phi_2\left(\frac{v}{r}\right) dudv,$$

де $D_\infty = R^2 \setminus D_\eta$.

Для першого інтегралу за виконання нерівності (4.55) маємо для всіх $(x,y) \in R^2$ оцінку
$$|I_1| = \frac{1}{r^2} \int_{-\eta}^{\eta}\int_{-\eta}^{\eta} |f(x+u, y+v) - f(x,y)| \frac{1}{r^2} \left|\Phi_1\left(\frac{u}{r}\right)\right| \left|\Phi_2\left(\frac{v}{r}\right)\right| dudv \leq$$

$$\leq \frac{\varepsilon}{2r^2 M} \int\limits_{-\eta}^{\eta}\int\limits_{-\eta}^{\eta} \left|\Phi_1\left(\frac{u}{r}\right)\right|\left|\Phi_2\left(\frac{v}{r}\right)\right| du\,dv \leq \frac{\varepsilon}{2}.$$

Для другого інтегралу, внаслідок обмеженості функції $f(x, y)$ і узагальнення теореми 6 (п. 1.5), існує число $r_0 > 0$ таке, що $|I_2| < \frac{\varepsilon}{2}$, яка б не була точка $(x, y) \in R^2$.

Отже, оцінюючи праву частину формули (4.54), одержимо нерівність
$$|S_r(f; x, y) - f(x, y)| \leq |I_1| + |I_2| < \varepsilon,$$
яка справедлива для всіх $(x, y) \in R^2$ і $r \in (0, r_0)$.

Теорему доведено.

***Т е о р е м а  2 .*** *Нехай неперервна $2\pi$-періодична за обома змінними функція $f(x, y)$ має в точці $(x_0, y_0)$ неперервні частинні похідні $q$-го порядку $\dfrac{\partial^q f}{\partial x^s \partial^{q-s} y}$, $0 \leq s \leq q$, $1 \leq q \leq p-1$.*

*Тоді ряд Фур'є функції $f(x, y)$, почленно продиференційований $s$ раз за змінною $x$ і $q-s$ раз за змінною $y$, підсумовується методом $\{\varphi_1(mr)\varphi_2(nr)\}$ до $\dfrac{\partial^q f}{\partial x^s \partial^{q-s} y}$ у точці $(x_0, y_0)$.*

*Д о в е д е н н я .* Почленно диференціюємо $q$ раз ліву частину (ряд (4.52)) і праву частину формули (4.53)

$$\frac{\partial^q S_r(f; x, y)}{\partial x^s \partial^{q-s} y} = \sum_{m=0}^{\infty}\sum_{n=0}^{\infty} \lambda_{mn} m^s n^{q-s} \varphi_1(mr)\varphi_2(nr) \cdot$$
$$\cdot \left[ a_{mn} \cos\left(mx + \frac{s\pi}{2}\right)\cos\left(ny + \frac{(q-s)\pi}{2}\right) + \right.$$
$$+ b_{mn} \sin\left(mx + \frac{s\pi}{2}\right)\cos\left(ny + \frac{(q-s)\pi}{2}\right) +$$
$$+ c_{mn} \cos\left(mx + \frac{s\pi}{2}\right)\sin\left(ny + \frac{(q-s)\pi}{2}\right) +$$

$$+ d_{mn} \sin\left(mx + \frac{s\pi}{2}\right) \sin\left(ny + \frac{(q-s)\pi}{2}\right)\Bigg] = \quad (4.57)$$

$$= \frac{1}{r^2} \int\limits_{-\infty}^{+\infty} \int\limits_{-\infty}^{+\infty} f(u,v) \frac{d^s}{dx^s} \Phi_1\left(\frac{u-x}{r}\right) \frac{d^{q-s}}{dy^{q-s}} \Phi_2\left(\frac{v-y}{r}\right) du\,dv =$$

$$= \frac{(-1)^q}{r^2} \int\limits_{-\infty}^{+\infty} \int\limits_{-\infty}^{+\infty} f(u,v) \frac{d^s}{du^s} \Phi_1\left(\frac{u-x}{r}\right) \frac{d^{q-s}}{dv^{q-s}} \Phi_2\left(\frac{v-y}{r}\right) du\,dv.$$

Рівність у цій формулі обґрунтовується тим, що за умови $r > 0$ одержані після диференціювання ряд і інтеграл рівномірно збігаються відносно змінної $x$.

Дійсно, внаслідок неперервності функції $f(x,y)$ і, відповідно, інтегровності з квадратом у прямокутнику $Q$, а також умови (3.68) повноти тригонометричної системи функцій, справедливі рівності

$$\lim_{\substack{m \to \infty \\ n \to \infty}} a_{mn} = 0, \ \lim_{\substack{m \to \infty \\ n \to \infty}} b_{mn} = 0, \ \lim_{\substack{m \to \infty \\ n \to \infty}} c_{mn} = 0, \ \lim_{\substack{m \to \infty \\ n \to \infty}} d_{mn} = 0.$$

Ряди, членами яких є величини $m^{p-1}|\varphi_1(mr)|$, $n^{p-1}|\varphi_2(nr)|$, за наслідком теореми 11 (п. 2.3) збігаються, оскільки функції $\Phi_i^*(x)$ є ядрами операторів згладжування. Тому збігається ряд

$$\sum_{m=0}^{\infty} \sum_{n=0}^{\infty} \lambda_{mn} m^{p-1} n^{p-1} |\varphi_1(mr)| \, |\varphi_2(nr)|$$

і, оскільки за умовою $q \le p-1$, $s \le p-1$ і $q-s \ge p-1$, збігається також ряд

$$\sum_{m=0}^{\infty} \sum_{n=0}^{\infty} \lambda_{mn} m^s n^{q-s} |\varphi_1(mr)| \, |\varphi_2(nr)| \left(|a_{mn}| + |b_{mn}| + |c_{mn}| + |d_{mn}|\right),$$

члени якого не менші від членів ряду (4.57). Отже, ряд (4.57) збігається рівномірно.

За узагальненням (на випадок двовимірного інтегралу) теореми 16 (Вейєрштрасс, п. 1.2) інтеграл у формулі (4.57) рівномірно збігається відносно $x$, оскільки функція $f(x,y)$ обмежена (періодична і неперервна) і справедлива оцінка (1.90) для похідних від ядра оператора згладжування.

Покажемо тепер, що границя при $r \to 0$ інтегралу (4.57) у точці $(x_0, y_0)$ дорівнює відповідній похідній від функції $f(x, y)$ у цій точці, тобто для довільного $\varepsilon > 0$ існує $r_0 > 0$ таке, що для всіх $r$, $0 < r < r_0$ виконується нерівність

$$\left| \frac{\partial^q S_r(f; x_0, y_0)}{\partial x^s \partial^{q-s} y} - \frac{\partial^q f(x_0, y_0)}{\partial x^s \partial^{q-s} y} \right| < \varepsilon.$$

Оскільки частинні похідні неперервні в точці $(x_0, y_0)$, існує число $\eta > 0$ таке, що для як завгодно малого числа $\varepsilon > 0$ і для всіх $(u, v)$, $-\eta < u < \eta, -\eta < v < \eta$, справджується нерівність

$$\left| \frac{\partial^q f(x_0 + u, y_0 + v)}{\partial x^s \partial^{q-s} y} - \frac{\partial^q f(x_0, y_0)}{\partial x^s \partial^{q-s} y} \right| < \frac{\varepsilon}{2M}, \qquad (4.58)$$

де $M = \dfrac{1}{r^2} \int\limits_{-\infty}^{+\infty} \int\limits_{-\infty}^{+\infty} \left|\Phi_1\left(\dfrac{u}{r}\right)\right| \left|\Phi_2\left(\dfrac{v}{r}\right)\right| du dv$.

Розіб'ємо інтеграл (4.57) в точці $(x_0, y_0)$ на два інтеграли

$$I_1 = \frac{(-1)^q}{r^2} \int\limits_{x_0-\eta}^{x_0+\eta} \int\limits_{y_0-\eta}^{y_0+\eta} f(u, v) \frac{d^s}{du^s}\Phi_1\left(\frac{u-x_0}{r}\right) \frac{d^{q-s}}{dv^{q-s}}\Phi_2\left(\frac{v-y_0}{r}\right) du dv,$$

$$I_2 = \frac{(-1)^q}{r^2} \iint\limits_{D_\infty^0} f(u, v) \frac{d^s}{du^s}\Phi_1\left(\frac{u-x_0}{r}\right) \frac{d^{q-s}}{dv^{q-s}}\Phi_2\left(\frac{v-y_0}{r}\right) du dv,$$

де $D_\infty^0 = R^2 \setminus D^0$, $D^0 = \{(u, v): |u - x_0| < \eta, |v - y_0| < \eta\}$.

Перетворимо перший інтеграл з використанням формул інтегрування частинами і заміни змінної

$$I_1 = \frac{1}{r^2} \int\limits_{x_0-\eta}^{x_0+\eta} \int\limits_{y_0-\eta}^{y_0+\eta} \frac{\partial^q f(u, v)}{\partial u^s \partial v^{q-s}} \Phi_1\left(\frac{u-x_0}{r}\right) \Phi_2\left(\frac{v-y_0}{r}\right) du dv + A =$$

$$= \frac{1}{r^2} \int\limits_{-\eta}^{\eta} \int\limits_{-\eta}^{\eta} \frac{\partial^q f(x_0+u, y_0+v)}{\partial u^s \partial v^{q-s}} \Phi_1\left(\frac{u}{r}\right) \Phi_2\left(\frac{v}{r}\right) du dv + A,$$

де

$$A = \frac{(-1)^q}{r^2} \left[ \sum_{l=0}^{s-1} \sum_{g=0}^{q-s-1} (-1)^{l+g} \frac{\partial^{l+g} f}{\partial u^l \partial v^g} \frac{\partial^{s-1-l} \Phi_1}{\partial u^{s-1-l}} \frac{\partial^{q-s-1-g} \Phi_2}{\partial v^{q-s-1-g}} \right]\bigg|_{x_0-\eta}^{x_0+\eta} \bigg|_{y_0-\eta}^{y_0+\eta}.$$

Знайдемо відхилення похідних від суми ряду (4.57) і похідних від функції $f(x, y)$ в точці $(x_0, y_0)$

$$\frac{\partial^q S_r(f; x_0, y_0)}{\partial x^s \partial^{q-s} y} - \frac{\partial^q f(x_0, y_0)}{\partial x^s \partial^{q-s} y} =$$

$$= \frac{1}{r^2} \int\limits_{-\eta}^{\eta} \int\limits_{-\eta}^{\eta} \frac{\partial^q f(x_0+u, y_0+v)}{\partial u^s \partial v^{q-s}} \Phi_1\left(\frac{u}{r}\right) \Phi_2\left(\frac{v}{r}\right) du dv + A + I_2 -$$

$$- \frac{\partial^q f(x_0, y_0)}{\partial x^s \partial^{q-s} y} \frac{1}{r^2} \int\limits_{-\infty}^{+\infty}\int\limits_{-\infty}^{+\infty} \Phi_1\left(\frac{u}{r}\right)\Phi_2\left(\frac{v}{r}\right) du dv =$$

$$= \frac{1}{r^2} \int\limits_{-\eta}^{\eta}\int\limits_{-\eta}^{\eta} \frac{\partial^q f(x_0+u, y_0+v)}{\partial u^s \partial v^{q-s}} \Phi_1\left(\frac{u}{r}\right)\Phi_2\left(\frac{v}{r}\right) du dv + A + I_2 -$$

$$- \frac{\partial^q f(x_0, y_0)}{\partial x^s \partial^{q-s} y} \frac{1}{r^2} \left[ \int\limits_{-\eta}^{\eta}\int\limits_{-\eta}^{\eta} \Phi_1\left(\frac{u}{r}\right)\Phi_2\left(\frac{v}{r}\right) du dv + \iint\limits_{D_\infty^0} \Phi_1\left(\frac{u}{r}\right)\Phi_2\left(\frac{v}{r}\right) du dv \right] =$$

$$= \frac{1}{r^2} \int\limits_{-\eta}^{\eta}\int\limits_{-\eta}^{\eta} \left[ \frac{\partial^q f(x_0+u, y_0+v)}{\partial u^s \partial v^{q-s}} - \frac{\partial^q f(x_0, y_0)}{\partial u^s \partial v^{q-s}} \right] \Phi_1\left(\frac{u}{r}\right)\Phi_2\left(\frac{v}{r}\right) du dv +$$

$$+ A + I_2 - \frac{\partial^q f(x_0, y_0)}{\partial x^s \partial^{q-s} y} \frac{1}{r^2} \iint\limits_{D_\infty^0} \Phi_1\left(\frac{u}{r}\right)\Phi_2\left(\frac{v}{r}\right) du dv.$$

Оцінимо це відхилення з урахуванням нерівності (4.58)

$$\left| \frac{\partial^q S_r(f; x_0, y_0)}{\partial x^s \partial^{q-s} y} - \frac{\partial^q f(x_0, y_0)}{\partial x^s \partial^{q-s} y} \right| \leq$$

$$\leq \frac{1}{r^2} \int\limits_{-\eta}^{\eta}\int\limits_{-\eta}^{\eta} \left| \frac{\partial^q f(x_0+u, y_0+v)}{\partial u^s \partial v^{q-s}} - \frac{\partial^q f(x_0, y_0)}{\partial u^s \partial v^{q-s}} \right| \left|\Phi_1\left(\frac{u}{r}\right)\right|\left|\Phi_2\left(\frac{v}{r}\right)\right| du dv +$$

$$+ |A| + |I_2| + \left|\frac{\partial^q f(x_0, y_0)}{\partial x^s \partial^{q-s} y}\right| \frac{1}{r^2} \iint\limits_{D_\infty^0} \left|\Phi_1\left(\frac{u}{r}\right)\right| \left|\Phi_2\left(\frac{v}{r}\right)\right| du dv \le$$

$$\le \frac{\varepsilon}{2} + |A| + |I_2| + \left|\frac{\partial^q f(x_0, y_0)}{\partial x^s \partial^{q-s} y}\right| \frac{1}{r^2} \iint\limits_{D_\infty^0} \left|\Phi_1\left(\frac{u}{r}\right)\right| \left|\Phi_2\left(\frac{v}{r}\right)\right| du dv .$$

Другий, третій і четвертий доданки за теоремами 6 і 9 (п. 1.5), а також внаслідок обмеженості величини $\dfrac{\partial^q f(x_0, y_0)}{\partial x^s \partial^{q-s} y}$ можна зробити за рахунок вибору $r_0 > 0$ як завгодно малими. Виберемо їх меншими, ніж $\dfrac{\varepsilon}{2}$. При цьому встановлена нерівність буде виконуватися для всіх $r \in (0, r_0)$.

Тоді для відхилення похідних від суми ряду (4.57) і похідних від функції $f(x, y)$ в точці $(x_0, y_0)$ знайдемо оцінку

$$\left|\frac{\partial^q S_r(f; x_0, y_0)}{\partial x^s \partial^{q-s} y} - \frac{\partial^q f(x_0, y_0)}{\partial x^s \partial^{q-s} y}\right| \le \frac{\varepsilon}{2} + \frac{\varepsilon}{2} = \varepsilon ,$$

яка виконується для всіх $r \in (0, r_0)$.

Теорему доведено.

### 4.7. Завдання до четвертого розділу

1. Показати, що якщо $f(x)$ – кусково-неперервна абсолютно інтегровна на дійсній осі функція, то $F(x) = \sum\limits_{k=-\infty}^{\infty} f(x + kl)$ – періодична функція з періодом $l$.

2. Виходячи з теореми 1 (п. 4.2), вивести формулу (Пуассона)
$$\sum_{k=-\infty}^{\infty} e^{-k^2 x^2} = \frac{\sqrt{\pi}}{x} \sum_{k=-\infty}^{\infty} e^{-\frac{k^2 \pi^2}{x^2}} , \ x \ne 0 .$$

3. Виходячи з теореми 1 (п. 4.2) з використанням формули (4.25) вивести формули (інший підхід [21, *с.* 473; 22, *с.* 452]):
$$\frac{1}{1-e^x} = \frac{1}{2} - x \sum_{k=-\infty}^{\infty} \frac{1}{x^2 + 4k^2 \pi^2} , \ \frac{1}{1+e^x} = \frac{1}{2} - x \sum_{k=-\infty}^{\infty} \frac{1}{x^2 + (2k+1)^2 \pi^2} .$$

4. Показати, що якщо $2\pi$-періодична функція неперервна і задовольняє умову Ліпшіца порядку $\lambda$, $0 < \lambda < 1$ (для будь-якого $h > 0$ справедлива нерівність $|f(x+h) - f(x)| \leq Mh^\lambda$), то для часткової суми (4.45) виконується нерівність

$$|\sigma_n(x) - f(x)| \leq \frac{C_\lambda M}{n^\lambda}, \quad \text{де } C_\lambda \leq \frac{4}{\pi} \int_0^{+\infty} \frac{\sin^2 \frac{t}{2}}{t^{2-\lambda}} dt < \frac{3}{1-\lambda}.$$

5. Показати, що якщо $2\pi$-періодична функція неперервна і задовольняє умову Ліпшіца порядку $\lambda = 1$, то для часткової суми (4.45) виконується нерівність $|\sigma_n(x) - f(x)| \leq CM \frac{\ln n}{n}$.

6. Показати справедливість наступних зображень періодичних розвинень дельтоподібних функцій:

а) $\delta_r(x) = \frac{1}{\pi}\left[\frac{1}{2} + \sum_{k=1}^{\infty} \frac{\cos kx}{1 + (kr)^2}\right] = \frac{1}{2r} \frac{\operatorname{ch} \frac{\pi - |x|}{r}}{\operatorname{sh} \frac{\pi}{r}}$, $x \in [-\pi, \pi]$;

б) $\delta_r(x) = \frac{1}{\pi}\left(\frac{1}{2} + \sum_{k=1}^{\infty} \frac{\cos kx}{e^{kr}}\right) = \frac{1}{2\pi} \frac{\operatorname{sh} r}{\operatorname{ch} r - \cos x}$, $x \in [-\pi, \pi]$.

7. Показати, що

$$\sum_{k=1}^{\infty} k\rho^k \cos kx = \frac{1 + 2\rho^3 \cos x - 3\rho^2}{2(1 - 2\rho \cos x + \rho^2)^2}.$$

8. Просумувати числові ряди методом степеневих множників і знайти узагальнені суми:

а) $1 - 1 + 1 - 1 + \ldots$ ($\sigma = \frac{1}{2}$);  б) $1 - 2 + 3 - 4 + \ldots$ ($\sigma = \frac{1}{4}$).

9. Показати, що для $0 \leq \rho < 1$ ряди

$$\frac{1}{2} + \sum_{k=1}^{\infty} \rho^k \cos kx, \quad \sum_{k=1}^{\infty} \rho^k \sin kx$$

можна почленно диференціювати по $\rho$ і по $x$ скільки завгодно разів.

# Р О З Д І Л  V

# УЗАГАЛЬНЕНІ РОЗВ'ЯЗКИ КРАЙОВИХ ЗАДАЧ

___

### 5.1. Крайові задачі для рівняння коливань струни

Ряди Фур'є найчастіше використовуються для розв'язання задач математичної фізики $[4, 13-19]$. Розглянемо задачу про малі коливання однорідної струни довжини $l$, закріпленої на кінцях. Рівняння коливань струни, межові та початкові умови наступні:

$$a^2 \frac{\partial u}{\partial x^2} = \frac{\partial u}{\partial t^2} + f(x,t), \quad 0 < x < l, \ t > 0; \quad (5.1)$$

$$u(x,t)\big|_{x=0} = 0, \ u(x,t)\big|_{x=l} = 0, \quad t \geq 0;$$

$$u(x,t)\big|_{t=0} = \chi(x), \ \frac{\partial u}{\partial t}\bigg|_{t=0} = \psi(x), \quad 0 \leq x \leq l, \quad (5.2)$$

де $u(x,t)$ – переміщення струни; $f(x,t)$ – густина зведеної (до лінії) сили, що діє на струну; $\chi(x)$ і $\psi(x)$ – форма і швидкість струни в початковий момент часу; $a^2 = \dfrac{T_0}{\rho_0}$ – величина, що залежить від зведеної (до лінії) густини матеріалу $\rho_0$ і сили натягу струни $T_0$.

Розв'язок цієї задачі (внаслідок її лінійності) можна записати у вигляді суми розв'язків двох задач.

Перша задача про вільні коливання струни полягає у розв'язанні однорідного рівняння за неоднорідних межових умов

$$a^2 \frac{\partial u}{\partial x^2} = \frac{\partial u}{\partial t^2}, \quad 0 < x < l, \ t > 0; \quad (5.3)$$

$$u(x,t)\big|_{x=0} = 0, \ u(x,t)\big|_{x=l} = 0, \quad t \geq 0; \quad (5.4)$$

$$u(x,t)\big|_{t=0} = \chi(x), \ \frac{\partial u}{\partial t}\bigg|_{t=0} = \psi(x), \quad 0 \leq x \leq l. \quad (5.5)$$

У другій задачі про вимушені коливання струни розглядається неоднорідне рівняння за однорідних межових і початкових умов

$$a^2 \frac{\partial u}{\partial x^2} = \frac{\partial u}{\partial t^2} + f(x,t), \quad 0 < x < l, \ t > 0; \qquad (5.6)$$

$$u(x,t)\big|_{x=0} = 0, \ u(x,t)\big|_{x=l} = 0, \quad t \geq 0; \qquad (5.7)$$

$$u(x,t)\big|_{t=0} 0, \quad \frac{\partial u}{\partial t}\bigg|_{t=0} = 0, \quad 0 \leq x \leq l. \qquad (5.8)$$

**5.1.1. Вільні коливання струни.** Спочатку, наведемо схему побудови формального розв'язку задачі (5.3) – (5.5). Шукаємо його у вигляді суми ряду, що задовольняє межову умову (5.4),

$$u(x,t) = \sum_{k=1}^{\infty} u_k(t) \sin \lambda_k x, \qquad (5.9)$$

де $\lambda_k = \dfrac{k\pi}{l}$.

Підставивши вираз (5.9) у рівняння (5.3), одержимо

$$\sum_{k=1}^{\infty} \left[ u_k''(t) + a^2 \lambda_k^2 u_k(t) \right] \sin \lambda_k x = 0.$$

Звідси, враховуючи незалежність системи тригонометричних функцій $\{\sin \lambda_k x\}$ в області $0 \leq x \leq l$, знайдемо рівняння

$$u_k''(t) + a^2 \lambda_k^2 u_k(t) = 0.$$

Розв'язок цього рівняння наступний

$$u_k(t) = a_k \cos a\lambda_k t + b_k \sin a\lambda_k t.$$

Підставляючи його в ряд (5.9), одержимо

$$u(x,t) = \sum_{k=1}^{\infty} (a_k \cos a\lambda_k t + b_k \sin a\lambda_k t) \sin \lambda_k x. \qquad (5.10)$$

Припустимо, що функції розглянуті в умовах (5.5), розвиваються в ряди Фур'є

$$\chi(x) = \sum_{k=1}^{\infty} \chi_k \sin \lambda_k x, \quad \psi(x) = \sum_{k=1}^{\infty} \psi_k \sin \lambda_k x, \qquad (5.11)$$

де $\chi_k = \dfrac{2}{l}\int\limits_0^l \chi(x) \sin \lambda_k x\, dx$, $\psi_k = \dfrac{2}{l}\int\limits_0^l \psi(x) \sin \lambda_k x\, dx$.

Підставляючи ряд (5.10) в умови (5.5), одержимо з урахуванням розвинень (5.11) вирази для невідомих коефіцієнтів

$$a_k = \chi_k, \quad b_k = \frac{1}{a\lambda_k}\psi_k.$$

Враховуючи їх у (5.10), розв'язок задачі (5.3) – (5.5) запишемо у вигляді

$$u(x,t) = \sum_{k=1}^{\infty}\left(\chi_k \cos a\lambda_k t + \frac{\psi_k}{a\lambda_k}\sin a\lambda_k t\right)\sin\lambda_k x. \qquad (5.12)$$

Розглянемо умови, за яких ряд (5.12) збігається рівномірно і його можна диференціювати два рази по $x$ і $t$.

***Т е о р е м а   1 .*** *Нехай:*

*а) функція $\chi(x)$ два рази неперервно диференційовна на відрізку $[0,l]$ і має кусково-неперервну третю похідну (неперервну кусково-гладку другу похідну), а також задовольняє умови*

$$\chi(0) = 0, \quad \chi(l) = 0, \quad \chi''(0) = 0, \quad \chi''(l) = 0; \qquad (5.13)$$

*б) функція $\psi(x)$ неперервно диференційовна на $[0,l]$, має кусково-неперервну другу похідну (неперервну кусково-гладку першу похідну) і задовольняє умови*

$$\psi(0) = 0, \quad \psi(l) = 0. \qquad (5.14)$$

*Тоді функція $u(x,t)$ – сума ряду (5.12) має неперервні другі похідні і задовольняє рівняння (5.3), умови (5.4) та (5.5).*

***Д о в е д е н н я .*** Інтегруючи частинами інтеграли у виразах для коефіцієнтів $a_k$ і $b_k$ з урахуванням умов (5.13) і (5.14), одержимо

$$a_k = -\frac{b_k^{(3)}}{\lambda_k^3}, \quad b_k = -\frac{a_k^{(2)}}{a\lambda_k^3},$$

де $b_k^{(3)} = \frac{2}{l}\int_0^l \chi'''(\xi)\cos\lambda_k\xi\, d\xi;\quad a_k^{(2)} = \frac{2}{l}\int_0^l \psi''(\xi)\sin\lambda_k\xi\, d\xi.$

Підставляючи одержані вирази у ряд (5.12), маємо

$$u(x,t) = -\sum_{k=1}^{\infty} \frac{1}{\lambda_k^3}\left(b_k^{(3)}\cos a\lambda_k t + \frac{a_k^{(2)}}{a}\sin a\lambda_k t\right)\sin\lambda_k x. \quad (5.15)$$

Цей ряд мажорується збіжним числовим рядом

$$\sum_{k=1}^{\infty}\frac{1}{\lambda_k^3}\left(\left|b_k^{(3)}\right| + \frac{1}{a}\left|a_k^{(2)}\right|\right)$$

і тому збігається рівномірно і абсолютно в області $[0, l]\times[0, \infty)$.

Продиференціюємо почленно ряд (5.15) два рази по $x$ і $t$,

$$\frac{\partial^2 u}{\partial x^2} = \sum_{k=1}^{\infty}\frac{1}{\lambda_k}\left(b_k^{(3)}\cos a\lambda_k t + \frac{a_k^{(2)}}{a}\sin a\lambda_k t\right)\sin\lambda_k x,$$

$$\frac{\partial^2 u}{\partial t^2} = a^2\sum_{k=1}^{\infty}\frac{1}{\lambda_k}\left(b_k^{(3)}\cos a\lambda_k t + \frac{a_k^{(2)}}{a}\sin a\lambda_k t\right)\sin\lambda_k x. \quad (5.16)$$

Ці ряди мажоруються числовим рядом

$$\left(1+a^2\right)\sum_{k=1}^{\infty}\frac{1}{\lambda_k}\left(\left|b_k^{(3)}\right| + \frac{1}{a}\left|a_k^{(2)}\right|\right). \quad (5.17)$$

З нерівностей $\left(\left|b_k^{(3)}\right| + \frac{1}{\lambda_k}\right)^2 > 0$, $\left(\left|a_k^{(2)}\right| + \frac{1}{\lambda_k}\right)^2 > 0$ випливають такі нерівності:

$$\frac{\left|b_k^{(3)}\right|}{\lambda_k} < \frac{1}{2}\left(\left|b_k^{(3)}\right|^2 + \frac{1}{\lambda_k^2}\right), \quad \frac{\left|a_k^{(2)}\right|}{\lambda_k} < \frac{1}{2}\left(\left|a_k^{(2)}\right|^2 + \frac{1}{\lambda_k^2}\right). \quad (5.18)$$

За теоремою 2 (п. 3.2) третя похідна від функції $\varphi(x)$ і друга похідна від функції $\psi(x)$ інтегровані з квадратом на проміжку $[0, l]$ і тому збіжні ряди

$$\sum_{k=1}^{\infty}\left|b_k^{(3)}\right|^2, \quad \sum_{k=1}^{\infty}\left|a_k^{(2)}\right|^2.$$

Тоді, ґрунтуючись на нерівностях (5.18), можна стверджувати, що збігається ряд (5.17). Збіжність ряду (5.17) забезпечує рівномірну збіжність рядів (5.16) в області $[0, l]\times[0, \infty)$.

Внаслідок рівномірної збіжності рядів (5.16) і (5.11), а також виконання умов (5.13) і (5.14), можна ці ряди перетворювати і, відповідно, сума ряду (5.15) є розв'язком задачі (5.3) – (5.5).

Теорему доведено.

*З а у в а ж е н н я*. Умови (5.13) і (5.14) можна замінити умовами періодичності з періодом $2l$ і непарності відповідних функцій.

**5.1.2. Вимушені коливання струни.** Розглянемо задачу (5.6) – (5.8) про вимушені коливання однорідної струни, нерухомо закріпленої у кінцевих точках. Розв'язок задачі шукаємо у вигляді суми ряду (5.9), який задовольняє умови (5.7). Вважаємо, що ряд (5.9) збігається рівномірно і його можна диференціювати два рази по $x$ і $t$. Вважаємо також, що функцію $f(x,t)$ можна розкласти в ряд Фур'є за системою функцій $\{\sin\lambda_k x\}$

$$f(x,t) = \sum_{k=1}^{\infty} f_k(t)\sin\lambda_k x, \qquad (5.19)$$

де $f_k(t) = \dfrac{2}{l}\int_0^l f(x,t)\sin\lambda_k x\,dx$.

Підставляючи вирази (5.9), (5.19) в рівняння (5.6), одержимо

$$\sum_{k=1}^{\infty}\left[u_k''(t) + a^2\lambda_k^2 u_k(t) - f_k(t)\right]\sin\lambda_k x = 0.$$

Остання рівність можлива тоді, коли

$$u_k''(t) + a^2\lambda_k^2 u_k(t) = f_k(t). \qquad (5.20)$$

Початкові умови для цього рівняння одержимо підстановкою ряду (5.9) в умови (5.8) у вигляді

$$u_k(0) = 0, \;\; u_k'(0) = 0.$$

Інтегруючи рівняння (5.20) з урахуванням цих умов, знайдемо

$$u_k(t) = \frac{1}{\lambda_k a}\int_0^t \sin a\lambda_k(t-\tau)f_k(\tau)d\tau.$$

Підставляючи цей вираз у ряд (5.9), знайдемо розв'язок сформульованої задачі

$$u(x,t) = \frac{1}{a}\sum_{k=1}^{\infty}\frac{1}{\lambda_k}\int_0^t \sin a\lambda_k(t-\tau)f_k(\tau)d\tau\,\sin\lambda_k x. \qquad (5.21)$$

Знайдемо достатні умови, за яких функція (5.21) – розв'язок задачі (5.6) – (5.8).

***Теорема 2.*** *Нехай функція $f(x,t)$ справджує умови:*

*а) неперервна за змінною $t$ в області $t \geq 0$;*

*б) неперервна за змінною $x$, має неперервну першу похідну і кусково-неперервну другу похідну (неперервну кусково-гладку першу похідну) за змінною $x$ при кожному фіксованому $t$;*

*в) $f(0,t) = 0$, $f(l,t) = 0$.* (5.22)

*Тоді функція $u(x,t)$ – сума ряду (5.21) має неперервні другі похідні в області $[0,l] \times [0,\infty)$ і задовольняє рівняння (5.6), умови (5.7) та (5.8).*

Д о в е д е н н я. Знайдемо другі похідні від функції, що задається рядом (5.21),

$$\frac{\partial^2 u}{\partial t^2} = -\frac{1}{a}\sum_{k=1}^{\infty} \lambda_k \sin\lambda_k x \int_0^t \sin a\lambda_k(t-\tau) f_k(\tau)\, d\tau + \sum_{k=1}^{\infty} f_k(t)\sin\lambda_k x,$$

$$\frac{\partial^2 u}{\partial x^2} = -a\sum_{k=1}^{\infty} \lambda_k \sin\frac{k\pi}{l}x \int_0^t \sin a\lambda_k(t-\tau) f_k(\tau)\, d\tau. \quad (5.23)$$

Розглянемо ряд

$$\frac{1+a^2}{a}\sum_{k=1}^{\infty} \lambda_k f_k(\tau)\sin\lambda_k x \,\sin a\lambda_k \tau, \quad 0 \leq \tau < T, \quad (5.24)$$

для довільного $T > 0$. Інтегруючи два рази частинами інтеграл у формулі для $f_k(\tau)$ з урахуванням умови (5.22), одержимо

$$f_k(\tau) = -\frac{f_k^{(2)}(\tau)}{\lambda_k^2}, \quad (5.25)$$

де $f_k^{(2)}(\tau) = \dfrac{2}{l}\int_0^l \dfrac{\partial^2 f(\xi,\tau)}{\partial \xi^2} \cos\lambda_k \xi\, d\xi$.

Ряд (5.24) мажорується рядом $\dfrac{1+a^2}{a}\sum_{k=1}^{\infty} \dfrac{\left|f_k^{(2)}(\tau)\right|}{\lambda_k}$, $0 \leq \tau \leq T$.

Цей ряд збігається, внаслідок виконання умови типу (5.18),

$$\frac{\left|f_k^{(2)}(\tau)\right|}{\lambda_k} < \frac{1}{2}\left(\left|f_k^{(2)}(\tau)\right|^2 + \frac{1}{\lambda_k^2}\right),$$

і збіжності ряду $\sum_{k=1}^{\infty} \left| f_k^{(2)}(\tau) \right|^2$, що випливає з рівності Парсеваля, якщо врахувати обмеженість (за умовою) функції $\dfrac{\partial^2 f(x,t)}{\partial x^2}$, $t \in [0, T]$, і, відповідно, інтегровність її з квадратом за змінною $x$.

Внаслідок цього, ряд (5.24) збігається рівномірно для всіх $x \in [0, l]$, $\tau \in [0, T]$ і його можна інтегрувати почленно. Тому ряди (5.23) рівномірно збігаються в області $[0, l] \times [0, T]$. Другий ряд у першій формулі (5.23) також рівномірно збігається, оскільки його сума дорівнює $f(x, t)$.

Отже, сума ряду (5.21) є неперервною функцією зі своїми другими похідними в області $[0, l] \times [0, T]$ і внаслідок довільності $T$ – в області $[0, l] \times [0, \infty)$.

Теорему доведено.

**5.1.3. Узагальнені розв'язки крайових задач.** У задачах прикладних наук не завжди функції, що задають праві частини диференціальних рівнянь чи крайові умови задач (5.3) – (5.5) і (5.6) – (5.8), задовольняють відповідні умови гладкості. Такі задачі називають некоректно поставленими. Для окремого класу некоректно поставлених задач, в яких умови стосовно функцій не виконуються лише в окремих точках і на кусково-гладких лініях, вдається також дати означення розв'язків. Ці розв'язки називають узагальненими розв'язками $[15, 17 - 19]$.

Розглянемо послідовнісний підхід до побудови узагальнених розв'язків некоректно поставлених крайових задач. Він ґрунтується на згладжуванні функцій з використанням операторів згладжування (4.22). За теоремою 3 (п. 4.2), якщо $f(x)$ – кусково-неперервна функція на проміжку $[-\pi, \pi]$ і

$$f(x) \sim \frac{a_0}{2} + \sum_{k=1}^{\infty} \left( a_k \cos kx + b_k \sin kx \right) \qquad (5.26)$$

– її ряд Фур'є, то періодичне усереднення цієї функції

$$S_r^*(f; x) = \frac{a_0}{2} + \sum_{k=1}^{\infty} \varphi(kr) \left( a_k \cos kx + b_k \sin kx \right)$$

– неперервна $2\pi$-періодична функція і її можна диференціювати $m$ разів, $1 \leq m \leq p-1$. Більше того, за теоремою 3 (п. 4.3), якщо функція $f(x)$ має на деякому проміжку неперервну похідну $m$-го порядку, $1 \leq m \leq p-1$, то її ряд Фур'є, почленно продиференційований $m$ разів, рівномірно підсумовується методом $\{\varphi(kr)\}$ до $f^{(m)}(x)$ на будь-якому відрізку, строго внутрішньому до проміжку неперервності цієї функції.

Зауважимо, що $m$ разів почленно продиференційований ряд Фур'є (5.26), вже не є рядом Фур'є для функції $f^{(m)}(x)$.

**Узагальнений розв'язок крайової задачі (5.3) – (5.5).** Нехай функції $\chi(x)$ і $\psi(x)$ крайової задачі (5.3) – (5.5) є кусково-неперервними функціями на відрізку $[0, l]$. Тоді справедливі розвинення

$$\chi(x) \sim \sum_{k=1}^{\infty} \chi_k \sin\lambda_k x, \quad \psi(x) \sim \sum_{k=1}^{\infty} \psi_k \sin\lambda_k x, \qquad (5.27)$$

де $\lambda_k = \dfrac{k\pi}{l}$.

Розглянемо періодичні усереднення цих функцій

$$\chi_r(x) = \sum_{k=1}^{\infty} \varphi(\lambda_k r) \chi_k \sin\lambda_k x, \quad \psi_r(x) = \sum_{k=1}^{\infty} \varphi(\lambda_k r) \psi_k \sin\lambda_k x. \qquad (5.28)$$

Якщо члени послідовності $\{\varphi(\lambda_k r)\}$ справджують оцінку

$$\varphi(\lambda_k r) = \mathrm{O}\left(\frac{1}{k^4}\right), \quad r > 0, \qquad (5.29)$$

(теорема 5, п. 4.1), то функції $\chi_r(x)$ і $\psi_r(x)$ задовольняють умови теореми 1.

Розглянемо, наприклад, перший ряд (5.28). Безпосередньо переконуємося у виконанні умов (5.13). Покажемо, що третя похідна від функції $\chi_r(x)$ неперервна функція,

$$\frac{\partial^3 \chi_r}{\partial x^3} = -\sum_{k=1}^{\infty} \lambda_k^3 \, \varphi(\lambda_k r) \, \chi_k \cos\lambda_k x.$$

Для цього ряду справедлива оцінка

$$\left|\frac{\partial^3 \chi_r}{\partial x^3}\right| \le A \sum_{k=1}^{\infty} \frac{|\chi_k|}{\lambda_k}.$$

Враховуючи тут умову $\dfrac{|\chi_r|}{\lambda_k} < \dfrac{1}{2}\left(|\chi_r|^2 + \dfrac{1}{\lambda_k^2}\right)$ і рівність Парсеваля для інтегрованої з квадратом функції $\chi(x)$, встановимо збіжність цього ряду, а отже, рівномірну збіжність ряду для третьої похідної $\dfrac{\partial^3 \chi_r}{\partial x^3}$ і, відповідно, її неперервність.

Тому розв'язок рівняння (5.3) з однорідними межовими умовами (5.4) і початковими умовами

$$u(x,t)\big|_{t=0} = \chi_r(x), \quad \frac{\partial u}{\partial t}\bigg|_{t=0} = \psi_r(x), \qquad 0 \le x \le l,$$

існує і записується у вигляді (5.12)

$$u_r(x,t) = \sum_{k=1}^{\infty} \varphi(\lambda_k r)\left(\chi_k \cos a\lambda_k t + \frac{\psi_k}{a\lambda_k}\sin a\lambda_k t\right)\sin \lambda_k x. \quad (5.30)$$

**О з н а ч е н н я 1.** *Узагальненим розв'язком задачі* (5.3) – (5.5) *з кусково-неперервними на відрізку* $[0,l]$ *функціями* $\chi(x)$ *і* $\psi(x)$ *називається границя при* $r \to +0$ *суми ряду* (5.30),

$$u(x,t) = \lim_{r \to +0} u_r(x,t).$$

Важливо, що за теоремою 4 (п. 4.3) ряд (5.30) рівномірно підсумовується узагальненим методом, що справджує умову (5.29), на будь-якому відрізку, строго внутрішньому до проміжку неперервності функцій $\chi(x)$ і $\psi(x)$, тобто узагальнений розв'язок задачі справджує рівняння і умови задачі всюди, хіба-що за виключенням скінченного числа точок розриву цих функцій.

*П р и к л а д 1*. Знайти узагальнений розв'язок задачі про вільні коливання однорідної струни, закріпленої на кінцях $x = 0$, $x = l$, якщо початкова швидкість струни дорівнює нулю, а початкова форма наступна

$$\chi(x) = h_0 \begin{cases} \left(1 - \dfrac{\xi}{l}\right) x, & 0 \leq x \leq \zeta, \\ \left(1 - \dfrac{x}{l}\right) \xi, & \xi \leq x \leq l, \end{cases}$$

де $h_0 = \dfrac{h}{\xi(1 - \xi/l)}$, $0 < \xi < l$, $h$ – переміщення струни у точці $x = \xi$ в початковий момент часу, рис. 5.1.

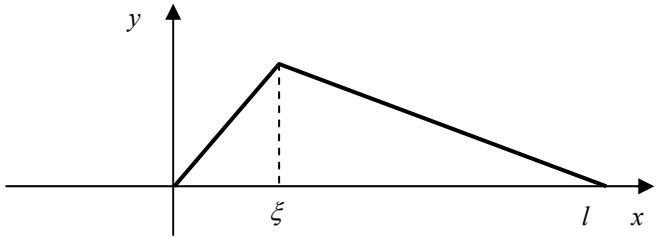

Рис. 5.1.

Функція $\chi(x)$ неперервна на проміжку $[0, l]$, однак вона має розривну першу похідну і тому шукаємо узагальнений розв'язок цієї задачі.

Знайдемо коефіцієнти першого ряду (5.27)

$$\chi_k = \frac{2}{l} \left[ \left(1 - \frac{\xi}{l}\right) \int\limits_0^{\xi} x \sin\lambda_k x\, dx + \xi \int\limits_{\xi}^{l} \left(1 - \frac{x}{l}\right) \sin\lambda_k x\, dx \right] = \frac{2h_0}{l} \frac{\sin\lambda_k \xi}{\lambda_k^2}$$

і, відповідно,

$$\chi(x) = \frac{2h_0}{l} \sum_{k=1}^{\infty} \frac{\sin\lambda_k \xi}{\lambda_k^2} \sin\lambda_k x .$$

Підставимо цей вираз у формулу (5.30), одержимо

$$u_r(x, t) = \sum_{k=1}^{\infty} \varphi(\lambda_k r)\, \chi_k \cos a\lambda_k t\, \sin\lambda_k x =$$

$$= \frac{2h_0}{l} \sum_{k=1}^{\infty} \varphi(\lambda_k r) \frac{\sin\lambda_k \xi}{\lambda_k^2} \cos a\lambda_k t\, \sin\lambda_k x .$$

Звідси знайдемо узагальнений розв'язок задачі

$$u(x,t) = \frac{2h_0}{l} \lim_{r \to 0} \sum_{k=1}^{\infty} \varphi(\lambda_k r) \frac{\sin \lambda_k \xi}{\lambda_k^2} \cos a\lambda_k t \, \sin \lambda_k x \, .$$

Одержаний тут ряд має оцінку

$$\frac{2h_0}{l} \left| \sum_{k=1}^{\infty} \varphi(\lambda_k r) \frac{\sin \lambda_k \xi}{\lambda_k^2} \cos a\lambda_k t \, \sin \lambda_k x \right| \le \frac{2h_0}{l} \sum_{k=1}^{\infty} |\varphi(\lambda_k r)| \frac{1}{\lambda_k^2} < \infty \, .$$

Тому можна перейти під знаком суми до границі при $r \to +0$ і, врахувавши рівність $\lim_{r \to 0} \varphi(\lambda_k r) = 1$, одержати узагальнений розв'язок задачі у вигляді рівномірно збіжного ряду

$$u(x,t) = \frac{2h_0}{l} \sum_{k=1}^{\infty} \frac{\sin \lambda_k \xi}{\lambda_k^2} \cos a\lambda_k t \, \sin \lambda_k x \, .$$

Цей приклад ілюструє наступне твердження.

**Т е о р е м а  3 .** *Якщо функція $\chi(x)$ неперервна, має кусково-неперервну першу похідну на відрізку $[0, l]$ і задовольняє умови*

$$\chi(0) = \chi(l) = 0,$$

*а функція $\psi(x)$ кусково-неперервна на $[0, l]$, то узагальнений розв'язок задачі* (5.3) – (5.5) *зображається рівномірно збіжним рядом*

$$u(x,t) = \sum_{k=1}^{\infty} \left( \chi_k \cos a\lambda_k t + \frac{\psi_k}{a\lambda_k} \sin a\lambda_k t \right) \sin \lambda_k x \, . \quad (5.31)$$

*Д о в е д е н н я .* Покажемо, що ряд (5.31) збігається рівномірно, тобто можна перейти почленно до границі при $r \to +0$ у ряді (5.30). Інтегруючи частинами інтеграл у виразі для коефіцієнтів $\chi_k$, одержимо

$$\chi_k = \frac{2}{l\lambda_k} \int_0^l \chi'(x) \cos \lambda_k x \, dx = \frac{\chi_k^{(1)}}{\lambda_k}$$

і, відповідно, знайдемо оцінку для ряду (5.31)

$$\left| \sum_{k=1}^{\infty} \left( \chi_k \cos a\lambda_k t + \frac{\psi_k}{a\lambda_k} \sin a\lambda_k t \right) \sin \lambda_k x \right| \le \sum_{k=1}^{\infty} \frac{1}{\lambda_k} \left( |\chi_k^{(1)}| + \frac{|\psi_k|}{a} \right).$$

Враховуючи тут інтегровність з квадратом функцій $\chi'(x)$, $\psi(x)$ і аналогічні до (5.18) нерівності

$$\frac{|\chi'_k|}{\lambda_k} < \frac{1}{2}\left(\left|\chi_k^{(1)}\right|^2 + \frac{1}{\lambda_k^2}\right), \quad \frac{|\psi_k|}{\lambda_k} < \frac{1}{2}\left(|\psi_k|^2 + \frac{1}{\lambda_k^2}\right),$$

прийдемо до такої оцінки:

$$\sum_{k=1}^{\infty}\frac{1}{\lambda_k}\left(\left|\chi_k^{(1)}\right| + \frac{|\psi_k|}{a}\right) \leq \sum_{k=1}^{\infty}\left(\left|\chi_k^{(1)}\right|^2 + \frac{|\psi_k|^2}{a}\right) + \left(1 + \frac{1}{a}\right)\sum_{k=1}^{\infty}\frac{1}{\lambda_k^2}.$$

Оскільки одержані мажорантні ряди збігаються, ряд (5.31) збігається рівномірно.

Теорему доведено.

**Узагальнений розв'язок крайової задачі (5.6) – (5.8).** Нехай функція $f(x,t)$ неперервна за змінною $t$ в області $t \geq 0$ і кусково-неперервна за змінною $x$, $0 \leq x \leq l$, при кожному фіксованому значенні $t$. Тоді її можна розкласти в ряд Фур'є

$$f(x,t) \sim \sum_{k=1}^{\infty} f_k(t)\sin\lambda_k x,$$

де $f_k(t) = \dfrac{2}{l}\int\limits_0^l f(x,t)\sin\lambda_k x\, dx$.

Розглянемо періодичне усереднення цієї функції

$$f_r(x,t) = \sum_{k=1}^{\infty} \varphi(\lambda_k r)\, f_k(t)\,\sin\lambda_k x.$$

Якщо члени послідовності $\{\varphi(\lambda_k r)\}$ справджують оцінку

$$\varphi(\lambda_k r) = \mathrm{O}\!\left(\frac{1}{k^3}\right), \quad r > 0, \qquad (5.32)$$

то функція $f_r(x,t)$ задовольняє умови теореми 2. Дійсно, умова (5.22) виконується

$$f_r(0,t) = f_r(l,t) = 0.$$

Покажемо, що неперервною є функція

$$\frac{\partial^2 f_r}{\partial x^2} = -\sum_{k=1}^{\infty}\lambda_k^2\, \varphi(\lambda_k r) f_k(t)\sin\lambda_k x.$$

Для кожного фіксованого значення $t$ маємо оцінку

$$\left|\frac{\partial^2 f_r}{\partial x^2}\right| = A\sum_{k=1}^{\infty}\frac{|f_k(t)|}{\lambda_k}.$$

За умовою функція $f(x,t)$ інтегрована з квадратом і, відповідно, для коефіцієнтів цього ряду справедлива оцінка

$$\frac{|f_k(t)|}{\lambda_k} < \frac{1}{2}\left(|f_k(t)|^2 + \frac{1}{\lambda_k^2}\right),$$

а отже, ряд збігається. Тоді ряд для функції $\frac{\partial^2 f_r}{\partial x^2}$ збігається рівномірно і ця функція неперервна.

Тому розв'язок задачі (5.6) – (5.8) з правою частиною $f_r(x,t)$ першого рівняння існує і записується у вигляді (5.21)

$$u_r(x,t) = \frac{1}{a}\sum_{k=1}^{\infty}\frac{\varphi(\lambda_k r)}{\lambda_k}\int_0^t \sin a\lambda_k(t-\tau)f_k(\tau)\,d\tau \sin\lambda_k x \quad (5.33)$$

або з урахуванням виразу для функції $f_k(\tau)$ у вигляді
$$u_r(x,t) =$$
$$= \frac{2}{la}\int_0^t\int_0^l\sum_{k=1}^{\infty}\frac{\varphi(\lambda_k r)}{\lambda_k}\sin\lambda_k\xi\,\sin\lambda_k x\,\sin a\lambda_k(t-\tau)f(\xi,\tau)\,d\xi\,d\tau. \quad (5.34)$$

*О з н а ч е н н я  2.* *Узагальненим розв'язком задачі (5.6) – (5.8) з правою частиною рівняння (5.6) – функцією $f(x,t)$, неперервною за змінною $t \geq 0$ і для кожного фіксованого $t$ кусково-неперервною за змінною $x$, $0 \leq x \leq l$, називається границя при $r \to +0$ суми ряду* (5.33),
$$u(x,t) = \lim_{r \to +0} u_r(x,t).$$

*П р и к л а д  2*. Знайти вимушені коливання однорідної струни, закріпленої на кінцях $x=0$, $x=l$. Початкове відхилення і початкова швидкість струни дорівнюють нулеві. На струну діє сила густини
$$f(x,t) = f_1(x)f_0(t),$$
де

$$f_1(x) = \begin{cases} \dfrac{1}{2h}, & \xi - h \leq x \leq \zeta + h, \\ 0, & 0 < x < \xi - h \cup \xi + h \leq x \leq l; \end{cases}$$

$f_0(t)$ – неперервна функція в області $t \geq 0$; графік функції $y = f_1(x)$ наведено на рис. 5.2.

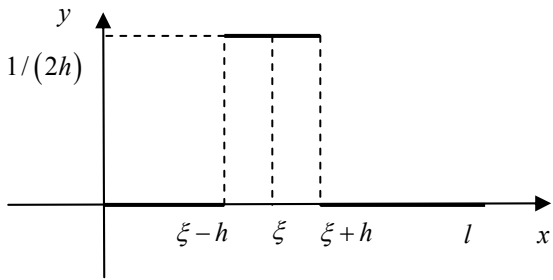

Рис. 5.2.

Функції сформульованої задачі не задовольняють умови теореми 2. Знайдемо узагальнений розв'язок цієї задачі.

Знайдемо коефіцієнти ряду Фур'є (5.19) функції $f(x, t) = f_1(x) f_0(t)$,

$$f_k(t) = \frac{2}{l} \int_{\zeta-h}^{\xi+h} \frac{1}{2h} \sin \lambda_k x \, dx \, f_0(t) = \frac{2}{l} \frac{\sin \lambda_k h}{\lambda_k h} f_0(t).$$

Підставимо цей вираз у формулу (5.33), одержимо

$$u_r(x, t) = \frac{2}{lah} \sum_{k=1}^{\infty} \frac{\varphi(\lambda_k r) \sin \lambda_k h \, \sin \lambda_k x}{\lambda_k^2} \int_0^t \sin a\lambda_k (t - \tau) f_0(\tau) \, d\tau.$$

Тоді узагальнений розв'язок задачі запишемо у вигляді

$$u(x, t) = \frac{2}{lah} \lim_{r \to +0} \sum_{k=1}^{\infty} \frac{\varphi(\lambda_k r) \sin \lambda_k h \, \sin \lambda_k x}{\lambda_k^2} \int_0^t \sin a\lambda_k (t - \tau) f_0(\tau) \, d\tau.$$

Внаслідок обмеженості функції $f_0(\tau)$ на відрізку $[0, T]$ для довільного $T > 0$ і виконання рівності $\lim\limits_{r \to +0} \varphi(\lambda_k r) = 1$, одержаний ряд після граничного переходу при $r \to +0$, збігається рівномірно.

Тому матимемо узагальнений розв'язок задачі у вигляді рівномірно збіжного ряду

$$u(x,t) = \frac{2}{lah} \sum_{k=1}^{\infty} \frac{\sin\lambda_k h \ \sin\lambda_k x}{\lambda_k^2} \int_0^t \sin a\lambda_k (t-\tau) f_0(\tau) d\tau.$$

У частинному випадку, коли зовнішня сила змінюється за гармонічним законом $f_0(t) = A\sin\omega t$, обчислюючи відповідний інтеграл, одержимо вираз для відхилень точок струни

$$u(x,t) = \frac{2A}{lah} \sum_{k=1}^{\infty} \frac{\sin\lambda_k h \ \sin\lambda_k x}{\lambda_k^2 \left(a^2\lambda_k^2 - \omega^2\right)} (a\lambda_k \sin\omega t - \omega\sin a\lambda_k t) =$$

$$= \frac{2A\sin\omega t}{lh} \sum_{k=1}^{\infty} \frac{\sin\lambda_k \xi \sin\lambda_k x}{\lambda_k \left(a^2\lambda_k^2 - \omega^2\right)} - \frac{2A\omega}{lah} \sum_{k=1}^{\infty} \frac{\sin\lambda_k \xi \ \sin\lambda_k x \ \sin a\lambda_k t}{\lambda_k^2 \left(a^2\lambda_k^2 - \omega^2\right)}.$$

Ця формула справедлива за умови $a^2\lambda_k^2 - \omega^2 \neq 0$. Якщо частота зовнішньої сили збігається з однією з власних частот коливань струни $\theta_k = a\lambda_k$, то відповідний доданок ряду зображується іншою формулою [3].

Зауважимо, що перший доданок у цій формулі визначає усталені коливання струни з частотою зовнішньої сили, другий доданок забезпечує виконання другої початкової умови мало впливає на переміщення мембрани.

***Т е о р е м а  4 .*** *Якщо функція* $f(x,t)$ *неперервна за змінною* $t \geq 0$ *і кусково-неперервна за змінною* $x$, $0 \leq x \leq l$, *для кожного фіксованого* $t$, *то узагальнений розв'язок задачі* (5.6) – (5.8) *зображується рівномірно збіжним рядом* (5.21),

$$u(x,t) = \frac{1}{a} \sum_{k=1}^{\infty} \frac{1}{\lambda_k} \int_0^t \sin a\lambda_k (t-\tau) f_k(\tau) d\tau \ \sin\lambda_k x. \qquad (5.35)$$

*Д о в е д е н н я .* Покажемо що у виразі узагальненого розв'язку (5.33) задачі (5.6) – (5.8) можна перейти до границі при $r \to +0$ і одержаний ряд (5.35) збігається рівномірно.

Розглянемо ряд

$$\sum_{k=1}^{\infty} \frac{1}{\lambda_k} \sin a\lambda_k \tau \ f_k(\tau) \sin\lambda_k x$$

для будь-яких фіксованих значень $\tau \in [0, T]$ для довільного $T$, $0 < T < \infty$. Оскільки функція $f(x,t)$ неперервна і, відповідно, інтегрована з квадратом за змінною $x \in [0, l]$, мажорантний ряд $\sum_{k=1}^{\infty} \dfrac{|f_k(\tau)|}{\lambda_k}$ збігається, а отже розглянутий ряд збігається рівномірно за змінними $x$ і $\tau$. Тоді його можна інтегрувати почленно. Одержаний при цьому ряд (5.35) рівномірно збігається в області $[0, l] \times [0, T]$, а отже в області $[0, l] \times [0, \infty)$.

Теорему доведено.

*З а у в а ж е н н я  1*. Узагальнені розв'язки (5.31), (5.35) ще називають узагальненими (в розумінні рівномірної збіжності) рівномірно збіжними розв'язками.

Розглядають також [10, 17, 19] узагальнені розв'язки крайових задач збіжні у середньому, коли граничні функції і вільні члени диференціальних рівнянь задаються у вигляді інтегрованих з квадратом функцій, або послідовностей функцій, збіжних в середньому.

Введемо функцію
$$G_r(x, \xi, t - \tau) = \frac{2}{la} \sum_{k=1}^{\infty} \frac{\varphi(\lambda_k r)}{\lambda_k} \sin \lambda_k \xi \, \sin \lambda_k x \, \sin a\lambda_k (t - \tau),$$

яка за умови $r > 0$ неперервна за усіма змінними, оскільки вважаємо, що справедлива оцінка $|\varphi(\lambda_k r)| = \mathrm{O}\left(\dfrac{1}{k^p}\right)$, $p \geq 2$. Розглянемо також граничний при $r \to +0$ елемент цієї функції (сім'ї функцій)
$$G(x, \xi, t - \tau) = \lim_{r \to +0} G_r(x, \xi, t - \tau) =$$
$$= \frac{2}{la} \lim_{r \to +0} \sum_{k=1}^{\infty} \frac{\varphi(\lambda_k r)}{\lambda_k} \sin \lambda_k \xi \, \sin \lambda_k x \, \sin a\lambda_k (t - \tau).$$

Тоді розв'язок (5.34) задачі (5.6) – (5.8) запишемо у вигляді
$$u(x, t) = \int_0^t \int_0^l G(x, \xi, t - \tau) f(\xi, \tau) \, d\tau \, d\xi. \tag{5.36}$$

***О з н а ч е н н я  3 .*** *Функція $G(x, \xi, t - \tau)$ називається функцією впливу або функцією Гріна для задачі (5.6) – (5.8).*

Функція $G(x, \xi, t - \tau)$ визначає відхилення від положення рівноваги точок, закріпленої на краях струни, внаслідок дії одиничного імпульса, прикладеного в точці $\xi$ в момент часу $\tau$.

*З а у в а ж е н н я  2 .* За теоремою 4 (п. 4.3), оскільки дія зосередженої в точці $\xi$ сили може бути описана з використанням послідовності неперервних функцій, ряд, що зображає функцію $G(x, \xi, t - \tau)$, рівномірно збігається на будь-якому відрізку з інтервалу $(0, l)$, що не містить точки $\xi$.

*З а у в а ж е н н я  3 .* В означеннях узагальнених розв'язків некоректно поставлених задач умови стосовно функцій $\chi(x), \psi(x)$ і $f(x, t)$ можуть бути замінені жорсткішими умовами: функції $\chi(x), \psi(x)$ є похідними порядку $N$ від неперервних функцій, а $f(x, t)$ є похідною порядку $N$ за змінною $x$ від неперервної функції. При цьому порядки оцінок (5.29) і (5.32) мають бути збільшені на величину порядку введених похідних.

Згідно з цим зауваженням функція

$$u(x, t) = \int_0^t G(x, \xi, t - \tau) f_0(\tau) d\tau \qquad (5.37)$$

є узагальненим розв'язком задачі (5.6) – (5.8) з вільним членом рівняння (5.6), що є другою похідною по $x$ від функції $f(x, t)$, розглянутої у прикладі 1.

Дійсно, якщо функція $f(x, t)$ має вигляд
$$f(x, t) = \chi(x) f_0(t),$$
де $f_0(t)$ – неперервна функція на проміжку $(0, T)$;

$$\chi(x) = \begin{cases} \left(1 - \dfrac{\xi}{l}\right) x, & 0 \leq x \leq \zeta, \\ \left(1 - \dfrac{x}{l}\right) \xi, & \xi \leq x \leq l. \end{cases}$$

Розвинення цієї функції у тригонометричний ряд наступне

$$f(x,t) = \frac{2 f_0(t)}{l} \sum_{k=1}^{\infty} \frac{\sin \lambda_k \xi}{\lambda_k^2} \sin \lambda_k x .$$

Друга похідна по $x$ від ряду функції $\chi(x)$ є формальним розвиненням дельта-функції у тригонометричний ряд

$$\frac{\partial^2 \chi}{\partial x^2} \sim -\frac{2}{l} \sum_{k=1}^{\infty} \sin \lambda_k \xi \sin \lambda_k x$$

Тоді узагальнений розв'язок задачі (5.6) – (5.8) з правою частиною рівняння (5.6) $f(x,t) = -\dfrac{\partial^2 \chi}{\partial x^2} f_0(t)$ запишеться у вигляді (5.37).

### 5.2. Крайові задачі для рівняння коливань мембрани

Дослідимо коливання однорідної прямокутної мембрани зі сторонами $l_1$ і $l_2$, які відбуваються внаслідок початкового відхилення, початкової швидкості і дії масової сили. Край мембрани нерухомо закріплений.

Для визначення функції $u = u(x, y, t)$ – відхилення мембрани від положення рівноваги маємо рівняння, межові та початкові умови.

Розглянемо дві задачі. Перша задача про вільні коливання мембрани. Для відшукання функції $u = u(x, y, t)$ маємо однорідне рівняння руху

$$\frac{\partial^2 u}{\partial t^2} = a^2 \left( \frac{\partial^2 u}{\partial x^2} + \frac{\partial^2 u}{\partial y^2} \right), \ \ 0 < x < l_1, \ \ 0 < y < l_2, \ \ t > 0, \quad (5.38)$$

однорідні межові умови

$$u(0, y, t) = u(l_1, y, t) = 0,$$
$$u(x, 0, t) = u(x, l_2, t) = 0, \quad t \geq 0, \tag{5.39}$$

та неоднорідні початкові умови

$$u(x, y, 0) = \chi(x, y),$$
$$\frac{\partial u(x, y, 0)}{\partial t} = \psi(x, y), \quad 0 \leq x \leq l_1, \ \ 0 \leq y \leq l_2. \tag{5.40}$$

Тут $a^2 = \dfrac{T_0}{\rho_0}$, де $T_0$ – сила натягу, зведена до одиниці довжини межі мембрани; $\rho_0$ – зведена до серединної поверхні густин матеріалу.

Друга задача про вимушені коливання мембрани. Маємо неоднорідне рівняння руху

$$\frac{\partial^2 u}{\partial t^2} = a^2 \left( \frac{\partial^2 u}{\partial x^2} + \frac{\partial^2 u}{\partial y^2} \right) + f(x, y, t), \qquad (5.41)$$

$$0 < x < l_1, \ \ 0 < y < l_2, \ \ t > 0,$$

і однорідні межові та початкові умови

$$u(0, y, t) = u(l_1, y, t) = 0, \ \ u(x, 0, t) = u(x, l_2, t) = 0, \ \ t \geq 0, \qquad (5.42)$$

$$u(x, y, 0) = 0, \ \ \frac{\partial u(x, y, 0)}{\partial t} = 0, \ \ 0 \leq x \leq l_1, \ \ 0 \leq y \leq l_2. \qquad (5.43)$$

**5.2.1. Вільні коливання мембрани.** Розглянемо схему побудови формального розв'язку задачі (5.38) – (5.40). Задовольняючи межові умови (5.42), розв'язок задачі шукаємо у вигляді подвійного ряду

$$u(x, y, t) = \sum_{k=1}^{\infty} \sum_{m=1}^{\infty} u_{km}(t) \, \Phi_{km}(x, y), \qquad (5.44)$$

де $\lambda_{1k} = \dfrac{k\pi}{l_1}$, $\lambda_{2m} = \dfrac{m\pi}{l_2}$; $\Phi_{km}(x, y) = \sin \lambda_{1k} x \sin \lambda_{2m} y$.

Підставляючи цей вираз в рівняння (5.38) з урахуванням незалежності тригонометричних систем функцій, одержимо звичайне диференціальне рівняння

$$u''_{km}(t) + a^2 \lambda_{km}^2 u_{km}(t) = 0,$$

де $\lambda_{km}^2 = \left( \dfrac{k\pi}{l_1} \right)^2 + \left( \dfrac{m\pi}{l_2} \right)^2$. Розв'язок цього рівняння запишемо у вигляді

$$u_{km}(t) = c_{1km} \cos a \lambda_{km} t + c_{2km} \sin a \lambda_{km} t,$$

де $\lambda_{km} = \left[ \left( \dfrac{k\pi}{l_1} \right)^2 + \left( \dfrac{m\pi}{l_2} \right)^2 \right]^{1/2}$.

Підставляючи цей вираз у ряд (5.44), одержимо

$$u(x, y, t) = \sum_{k=1}^{\infty}\sum_{m=1}^{\infty}\left(c_{1km}\cos a\lambda_{km}t + c_{2km}\sin a\lambda_{km}t\right)\Phi_{km}(x, y). \quad (5.45)$$

Вважаємо, що функції (5.40) розвиваються в ряд Фур'є

$$\chi(x, y) = \sum_{k=1}^{\infty}\sum_{m=1}^{\infty}\chi_{km}\Phi_{km}(x, y),\ \psi(x, y) = \sum_{k=1}^{\infty}\sum_{m=1}^{\infty}\psi_{km}\Phi_{km}(x, y),$$

де $\chi_{km} = \dfrac{4}{l_1 l_2}\int\limits_0^{l_1}\int\limits_0^{l_2}\chi(x, y)\Phi_{km}(x, y)\,dxdy$,

$$\psi_{km} = \dfrac{4}{l_1 l_2}\int\limits_0^{l_1}\int\limits_0^{l_2}\psi(x, y)\Phi_{km}(x, y)\,dxdy.$$

Підставляючи розвинення функцій $\chi(x, y)$, $\psi(x, y)$, вираз ряду (5.45) і його похідної

$$\frac{\partial u}{\partial t} = \sum_{k=1}^{\infty}\sum_{m=1}^{\infty}a\lambda_{km}(-c_{1km}\sin a\lambda_{km}t + c_{2km}\cos a\lambda_{km}t)\Phi_{km}(x, y)$$

у початкові умови (5.40), одержимо значення коефіцієнтів

$$c_{1km} = \chi_{km},\quad c_{2km} = \frac{\psi_{km}}{a\lambda_{km}}.$$

Враховуючи їх в розвиненні (5.45), прийдемо до такого виразу формального розв'язку

$$u(x, y, t) = \sum_{k=1}^{\infty}\sum_{m=1}^{\infty}\left(\chi_{km}\cos a\lambda_{km}t + \frac{\psi_{km}}{a\lambda_{km}}\sin a\lambda_{km}t\right)\Phi_{km}(x, y). \quad (5.46)$$

Знайдемо достатні умови рівномірної збіжності ряду (5.46) і його других частинних похідних.

***Т е о р е м а 1*** . Якщо функція $\chi(x, y)$ неперервна разом з похідними до четвертого порядку включно, а функція $\psi(x, y)$ неперервна разом з похідними до третього порядку включно в прямокутнику $0 \le x \le l_1,\ 0 \le y \le l_2$ і виконуються умови

$$\chi(0, y) = \chi(x, 0) = \chi(l_1, y) = \chi(x, l_2) = 0,$$

$$\frac{\partial^2 \chi}{\partial x^2}(0, y) = \frac{\partial^2 \chi}{\partial x^2}(l_1, y) = \frac{\partial^2 \chi}{\partial y^2}(x, 0) = \frac{\partial^2 \chi}{\partial y^2}(x, l_2) = 0,$$

$$\psi(0, y) = \psi(x, 0) = \psi(l_1, y) = \psi(x, l_2) = 0,$$
$$\frac{\partial^2 \psi}{\partial x^2}(0, y) = \frac{\partial^2 \psi}{\partial x^2}(l_1, y) = \frac{\partial^2 \psi}{\partial y^2}(x, 0) = \frac{\partial^2 \psi}{\partial y^2}(x, l_2) = 0,$$

*то ряд (5.46) збігається рівномірно в області* $0 \le x \le l_1$, $0 \le y \le l_2$, $t \ge 0$ *і його можна почленно диференціювати два рази по* $x$, $y$ *і* $t$.

*Д о в е д е н н я .* Позначимо через $\left(\dfrac{\partial^s \chi}{\partial x^l \partial y^{s-l}}\right)_{km}$ коефіцієнти Фур'є відповідних похідних функції $\chi(x, y)$. Інтегруємо частинами інтеграли у виразах коефіцієнтів з урахуванням умов (узгодження) теореми. Одержимо

$$|\chi_{km}| = \frac{1}{\lambda_{1k}}\left|\left(\frac{\partial \chi}{\partial x}\right)_{km}\right|, \quad |\chi_{km}| = \frac{1}{\lambda_{2m}}\left|\left(\frac{\partial \chi}{\partial y}\right)_{km}\right|,$$

$$|\chi_{km}| = \frac{1}{(\lambda_{1k})^2}\left|\left(\frac{\partial^2 \chi}{\partial x^2}\right)_{km}\right|,$$

$$|\chi_{km}| = \frac{1}{(\lambda_{2m})^2}\left|\left(\frac{\partial^2 \chi}{\partial y^2}\right)_{km}\right|, \quad |\chi_{km}| = \frac{1}{\lambda_{1k}\lambda_{2m}}\left|\left(\frac{\partial^2 \chi}{\partial y \partial x}\right)_{km}\right|.$$

Звідси

$$|\chi_{km}| = (\lambda_{1k} + \lambda_{2m})^{-1}\left(\left|\left(\frac{\partial \chi}{\partial x}\right)_{km}\right| + \left|\left(\frac{\partial \chi}{\partial y}\right)_{km}\right|\right).$$

Подібно до цієї формули знайдемо

$$\left|\left(\frac{\partial \chi}{\partial x}\right)_{km}\right| = (\lambda_{1k} + \lambda_{2m})^{-1}\left(\left|\left(\frac{\partial^2 \chi}{\partial x^2}\right)_{km}\right| + \left|\left(\frac{\partial^2 \chi}{\partial x \partial y}\right)_{km}\right|\right),$$

$$\left|\left(\frac{\partial \chi}{\partial y}\right)_{km}\right| = (\lambda_{1k} + \lambda_{2m})^{-1}\left(\left|\left(\frac{\partial^2 \chi}{\partial y \partial x}\right)_{km}\right| + \left|\left(\frac{\partial^2 \chi}{\partial y^2}\right)_{km}\right|\right)$$

і, відповідно,

$$|\chi_{km}| = (\lambda_{1k} + \lambda_{2m})^{-2}\left(\left|\left(\frac{\partial^2\chi}{\partial x^2}\right)_{km}\right| + 2\left|\left(\frac{\partial^2\chi}{\partial y\partial x}\right)_{km}\right| + \left|\left(\frac{\partial^2\chi}{\partial y^2}\right)_{km}\right|\right)$$

або

$$|\chi_{km}| \leq \frac{2}{(\lambda_{1k}+\lambda_{2m})^2}\sum_{i=0}^{2}\left|\left(\frac{\partial^2\chi}{\partial y^{2-i}\partial x^i}\right)_{km}\right|.$$

Оскільки існують четверті похідні від функції $\chi(x, y)$, виконується нерівність

$$|\chi_{km}| \leq \frac{6}{(\lambda_{1k}+\lambda_{2m})^4}\sum_{i=0}^{4}\left|\left(\frac{\partial^4\chi}{\partial y^{4-i}\partial x^i}\right)_{km}\right|. \qquad (5.47)$$

Для коефіцієнтів функції $\psi(x, y)$, оскільки існують треті частинні похідні, знайдемо з урахуванням умов теореми таку нерівність:

$$|\psi_{km}| \leq \frac{3}{(\lambda_{1k}+\lambda_{2m})^3}\sum_{i=0}^{3}\left|\left(\frac{\partial^3\psi}{\partial y^{3-i}\partial x^i}\right)_{km}\right|. \qquad (5.48)$$

Продиференціюємо почленно ряд (5.46) два рази по $x$, $y$ і $t$. Маємо

$$\frac{\partial^2 u}{\partial x^2} = -\sum_{k=1}^{\infty}\sum_{m=1}^{\infty}(\lambda_{1k})^2\left(\chi_{km}\cos a\lambda_{km}t + \frac{\psi_{km}}{a\lambda_{km}}\sin a\lambda_{km}t\right)\Phi_{km}(x, y),$$

$$\frac{\partial^2 u}{\partial y^2} = -\sum_{k=1}^{\infty}\sum_{m=1}^{\infty}(\lambda_{2m})^2\left(\chi_{km}\cos a\lambda_{km}t + \frac{\psi_{km}}{a\lambda_{km}}\sin a\lambda_{km}t\right)\Phi_{km}(x, y),$$

$$\frac{\partial^2 u}{\partial t^2} = -\sum_{k=1}^{\infty}\sum_{m=1}^{\infty}(a\lambda_{km})^2\left(\chi_{km}\cos a\lambda_{km}t + \frac{\psi_{km}}{a\lambda_{km}}\sin a\lambda_{km}t\right)\Phi_{km}(x, y).$$

Оцінимо перший ряд з урахуванням нерівностей (5.47), (5.48)

$$\left|\frac{\partial^2 u}{\partial x^2}\right| \leq \sum_{k=1}^{\infty}\sum_{m=1}^{\infty}(\lambda_{1k})^2\left(|\chi_{km}| + \frac{|\psi_{km}|}{a\lambda_{km}}\right) \leq$$

$$\leq \sum_{k=1}^{\infty}\sum_{m=1}^{\infty}\left[\frac{6(\lambda_{1k})^2}{(\lambda_{1k}+\lambda_{2m})^4}\sum_{i=0}^{4}\left|\left(\frac{\partial^4\chi}{\partial y^{4-i}\partial x^i}\right)_{km}\right| + \right.$$

$$+ \frac{3(\lambda_{1k})^2}{a\lambda_{km}(\lambda_{1k}+\lambda_{2m})^3}\sum_{i=0}^{3}\left|\left(\frac{\partial^3\psi}{\partial y^{3-i}\partial x^i}\right)_{km}\right|\right] \le$$

$$\le 6\sum_{k=1}^{\infty}\sum_{m=1}^{\infty}\left[\frac{1}{(\lambda_{1k}+\lambda_{2m})^2}\sum_{i=0}^{4}\left|\left(\frac{\partial^4\chi}{\partial y^{4-i}\partial x^i}\right)_{km}\right|\right]+$$

$$+\frac{3}{a}\sum_{k=1}^{\infty}\sum_{m=1}^{\infty}\left[\frac{1}{(\lambda_{1k}+\lambda_{2m})^2}\sum_{i=0}^{3}\left|\left(\frac{\partial^3\psi}{\partial y^{3-i}\partial x^i}\right)_{km}\right|\right].$$

Тут враховано нерівності $\lambda_{1k}+\lambda_{2m}>\lambda_{1k}$ і $\lambda_{km}>\lambda_{1k}$.

Застосовуючи до членів одержаних рядів нерівності $|a\|b|\le\frac{1}{2}\left(a^2+b^2\right)$ і $\left(|a_1|+...+|a_n|\right)^2\le n\left(a_1^2+...+a_n^2\right)$, знайдемо

$$\left|\frac{\partial^2 u}{\partial x^2}\right|\le 6\sum_{k=1}^{\infty}\sum_{m=1}^{\infty}\left[\frac{1}{2(\lambda_{1k}+\lambda_{2m})^4}+\frac{5}{2}\sum_{i=0}^{4}\left|\left(\frac{\partial^4\chi}{\partial y^{4-i}\partial x^i}\right)_{km}\right|^2\right]+$$

$$+3\sum_{k=1}^{\infty}\sum_{m=1}^{\infty}\left[\frac{1}{2(\lambda_{1k}+\lambda_{2m})^4}+2\sum_{i=0}^{3}\left|\left(\frac{\partial^3\psi}{\partial y^{3-i}\partial x^i}\right)_{km}\right|^2\right].$$

Перші доданки цих рядів за ознакою Вейєрштрасса утворюють збіжні ряди. Збіжність рядів з членами, якими є другі доданки одержаних рядів, випливає з рівності Парсеваля (3.68).

Отже, почленно продиференційований два рази по $x$ ряд (4.92) мажорується збіжним числовим рядом і, відповідно, рівномірно збігається в області $0\le x\le l_1$, $0\le y\le l_2$, $t\ge 0$.

Аналогічно доводиться рівномірна збіжність ряду (5.46) і рядів, одержаних від почленного диференціювання ряду (5.46) два рази по $y$ і $t$.

Теорему доведено.

**5.2.2. Вимушені коливання мембрани.** Розв'язок задачі (5.41) – (5.43) шукаємо у вигляді суми ряду (5.44). При цьому межові умови (5.42) задовольняються тотожньо. Вважаємо, що функція $f(x,y,t)$ розвивається в тригонометричний ряд

$$f(x, y, t) = \sum_{k=1}^{\infty} \sum_{m=1}^{\infty} f_{km}(t) \Phi_{km}(x, y), \ t \geq 0,$$

де $f_{km}(t) = \dfrac{4}{l_1 l_2} \int\limits_0^{l_1} \int\limits_0^{l_2} f(x, y, t) \Phi_{km}(x, y) \, dx dy$.

Якщо підставити цей ряд і ряд (5.44) у рівняння (5.41), то матимемо

$$\sum_{k=1}^{\infty} \sum_{m=1}^{\infty} \left[ -u''_{km}(t) - a^2 \lambda_{km}^2 u_{km}(t) + f_{km}(t) \right] \Phi_{km}(x, y) = 0.$$

Звідси, внаслідок лінійної незалежності системи тригонометричних функцій, одержимо звичайне диференціальне рівняння для визначення $u_{km}(t)$

$$u''_{km}(t) + a^2 \lambda_{km}^2 u_{km}(t) = f_{km}(t).$$

Підстановкою ряду (5.44) у початкові умови (5.43) прийдемо до таких початкових умов для функції $u_{km}(t)$

$$u_{km}(0) = 0, \ \dfrac{du_{km}(0)}{dt} = 0.$$

Розв'язок цієї задачі запишемо в інтегральній формі

$$u_{km}(t) = \dfrac{1}{a \lambda_{km}} \int\limits_0^t \sin a \lambda_{km}(t - \tau) f_{km}(\tau) \, d\tau.$$

Підставивши його у вираз ряду (5.44), запишемо формальний розв'язок

$$u(x, y, t) = \int\limits_0^t \sum_{k=1}^{\infty} \sum_{m=1}^{\infty} \dfrac{f_{km}(\tau)}{a \lambda_{km}} \Phi_{km}(x, y) \sin a \lambda_{km}(t - \tau) \, d\tau. \quad (5.49)$$

Достатні умови рівномірної збіжності ряду (5.49) і його похідних, що визначаються рівнянням (5.41), встановлюється наступною теоремою.

**Т е о р е м а  2.** *Якщо функція $f(x, y, t)$ неперервна в області $0 \leq x \leq l_1, \ 0 \leq y \leq l_2, \ t \geq 0$ має неперервні частинні похідні до третього порядку включно за змінними $x$ та $y$ в цій області, а також справджує умови*

$$f(0, y, t) = 0, \ f(l_1, y, t) = 0,$$

$$f(x, 0, t) = 0, \quad f(x, l_2, t) = 0,$$
$$\frac{\partial^2 f}{\partial x^2}(0, y, t) = 0, \quad \frac{\partial^2 f}{\partial x^2}(l_1, y, t) = 0,$$
$$\frac{\partial^2 f}{\partial y^2}(x, 0, t) = 0, \quad \frac{\partial^2 f}{\partial y^2}(x, l_2, t) = 0,$$

*то функція* $u(x, y, t)$ – *сума ряду* (5.49) *має неперервні другі похідні в області* $0 \leq x \leq l_1$, $0 \leq y \leq l_2$, $t \geq 0$ *і задовольняє рівняння* (5.41), *умови* (5.42) *та* (5.43).

*Д о в е д е н н я*. Позначимо через $\left(\dfrac{\partial^s f}{\partial x^l \partial y^{s-l}}\right)_{km}$ коефіцієнти Фур'є відповідних похідних від функції $f(x, y, t)$. Інтегруємо частинами інтеграли у виразах коефіцієнтів з урахуванням умов теореми. Одержимо

$$\left|\left(\frac{\partial f}{\partial x}\right)_{km}\right| = \lambda_{1k}|\chi_{km}|, \quad \left|\left(\frac{\partial f}{\partial y}\right)_{km}\right| = \lambda_{2m}|\chi_{km}|, \quad \left|\left(\frac{\partial^2 f}{\partial x^2}\right)_{km}\right| = (\lambda_{1k})^2|\chi_{km}|,$$

$$\left|\left(\frac{\partial^2 f}{\partial x \partial y}\right)_{km}\right| = \lambda_{1k}\lambda_{2m}|\chi_{km}|, \quad \left|\left(\frac{\partial^3 f}{\partial y^3}\right)_{km}\right| = (\lambda_{2m})^3|\chi_{km}|.$$

Звідси знайдемо рівність

$$(\lambda_{1k} + \lambda_{2m})^3 |f_{km}| = \left|\left(\frac{\partial^3 f}{\partial x^3}\right)_{km}\right| + ... + \left|\left(\frac{\partial^3 f}{\partial y^3}\right)_{km}\right|$$

і встановимо нерівність

$$|f_{km}| \leq \frac{3}{(\lambda_{1k} + \lambda_{2m})^3} \sum_{i=0}^{3} \left|\left(\frac{\partial^3 f}{\partial y^{3-i} \partial x^i}\right)_{km}\right|. \qquad (5.50)$$

Знайдемо другі частинні похідні від розв'язку (5.49)

$$\frac{\partial^2 u}{\partial x^2} = -\int_0^t \sum_{k=1}^{\infty} \sum_{m=1}^{\infty} \frac{(\lambda_{1k})^2 f_{km}(\tau)}{a\lambda_{km}} \Phi_{km}(x, y) \sin a\lambda_{km}(t - \tau)\, d\tau,$$

$$\frac{\partial^2 u}{\partial y^2} = -\int_0^t \sum_{k=1}^{\infty} \sum_{m=1}^{\infty} \frac{(\lambda_{2m})^2 f_{km}(\tau)}{a\lambda_{km}} \Phi_{km}(x, y) \sin a\lambda_{km}(t - \tau)\, d\tau,$$

$$\frac{\partial^2 u}{\partial t^2} = -a\int\limits_0^t \sum_{k=1}^\infty \sum_{m=1}^\infty \lambda_{km} f_{km}(\tau) \Phi_{km}(x,y) \sin a\lambda_{km}(t-\tau)\,d\tau +$$

$$+ \sum_{k=1}^\infty \sum_{m=1}^\infty f_{km}(\tau)\Phi_{km}(x,y) =$$

$$= -a\int\limits_0^t \sum_{k=1}^\infty \sum_{m=1}^\infty \lambda_{km} f_{km}(\tau)\Phi_{km}(x,y)\sin a\lambda_{km}(t-\tau)\,d\tau + f(x,y,t).$$

Оцінимо ряди в цих формулах для будь-якого значення $t < \infty$ з урахуванням нерівностей $\lambda_{1k} + \lambda_{2m} > \lambda_{1k}$, $\lambda_{km} > \lambda_{1k}$ і оцінки (5.50). Для першого ряду маємо

$$I_1 = \left| \sum_{k=1}^\infty \sum_{m=1}^\infty \frac{(\lambda_{1k})^2 f_{km}(\tau)}{a\lambda_{km}} \Phi_{km}(x,y) \sin a\lambda_{km}(t-\tau) \right| \leq$$

$$\leq \sum_{k=1}^\infty \sum_{m=1}^\infty \frac{(\lambda_{1k})^2 |f_{km}(\tau)|}{a\lambda_{km}} \leq \frac{1}{a}\sum_{k=1}^\infty \sum_{m=1}^\infty \lambda_{1k}|f_{km}(\tau)| \leq$$

$$\leq \frac{3}{a}\sum_{k=1}^\infty \sum_{m=1}^\infty \frac{\lambda_{1k}}{(\lambda_{1k}+\lambda_{2m})^3} \sum_{i=0}^3 \left|\left(\frac{\partial^3 f}{\partial y^{3-i}\partial x^i}\right)_{km}\right| \leq$$

$$\leq \frac{3}{a}\sum_{k=1}^\infty \sum_{m=1}^\infty \frac{1}{(\lambda_{1k}+\lambda_{2m})^2} \sum_{i=0}^3 \left|\left(\frac{\partial^3 f}{\partial y^{3-i}\partial x^i}\right)_{km}\right|.$$

Застосовуючи до членів одержаного ряду нерівності $|a||b| \leq \frac{1}{2}(a^2+b^2)$ і $(|a_1|+...+|a_n|)^2 \leq n(a_1^2+...+a_n^2)$, знайдемо

$$I_1 \leq \frac{3}{a}\sum_{k=1}^\infty \sum_{m=1}^\infty \left[ \frac{1}{2(\lambda_{1k}+\lambda_{2m})^4} + 2\sum_{i=0}^3 \left|\left(\frac{\partial^3 f}{\partial y^{3-i}\partial x^i}\right)_{km}\right|^2 \right].$$

Перший доданок цього ряду утворює за ознакою Вейєрштрасса збіжний ряд. Збіжність рядів, членами яких є наступні доданки членів одержаного ряду, випливає з рівності Парсеваля (3.68).

Отже, ряд у виразі другої частинної похідної по $x$ від функції $u(x,y,t)$ мажорується збіжним рядом. Тому він збігається рівномірно в області $0 \leq x \leq l_1$, $0 \leq y \leq l_2$, $t \geq 0$, а також рівномірно збігається в цій області ряд, одержаний після почленного інтегрування даного ряду.

Аналогічно доводиться рівномірна збіжність ряду, що задає розв'язок задачі, а також рядів, що є другими похідними від розв'язку за змінними $y$ та $t$.

Якщо ряд (5.49) справджує умови теореми, то його сума справджує (за побудовою) рівняння і умови задачі.

Теорему доведено.

**5.2.3. Узагальнені розв'язки задач для рівняння коливань мембрани. Узагальнений розв'язок задачі (5.38) – (5.40).** Вважаємо, що функції $\chi(x, y)$, $\psi(x, y)$ неперервні в області $0 \leq x \leq l_1$, $0 \leq y \leq l_2$. Розглянемо періодичні продовження усереднень цих функцій

$$\chi_r(x, y) = \sum_{k=1}^{\infty} \sum_{m=1}^{\infty} \varphi(\lambda_{1k} r) \varphi(\lambda_{2m} r) \chi_{km} \Phi_{km}(x, y),$$

$$\psi_r(x, y) = \sum_{k=1}^{\infty} \sum_{m=1}^{\infty} \varphi(\lambda_{1k} r) \varphi(\lambda_{2m} r) \psi_{km} \Phi_{km}(x, y), \quad (5.51)$$

де $\chi_{km} = \dfrac{4}{l_1 l_2} \int\limits_0^{l_1} \int\limits_0^{l_2} \chi(x, y) \Phi_{km}(x, y)\, dxdy$,

$\psi_{km} = \dfrac{4}{l_1 l_2} \int\limits_0^{l_1} \int\limits_0^{l_2} \psi(x, y) \Phi_{km}(x, y)\, dxdy$.

За умови $\varphi(\lambda_k r) = \mathrm{O}\left(\dfrac{1}{k^5}\right)$, $r > 0$, ряди (5.51) рівномірно збігаються, а також рівномірно збіжними є четверті частинні похідні від першого ряду і треті частинні похідні від другого ряду, тобто виконуються умови теореми 1. Тоді за формулою (5.46) знайдемо розв'язок задачі (5.38) – (5.40) з правими частинами межових умов у вигляді (5.51)

$$u_r(x, y, t) = \sum_{k=1}^{\infty} \sum_{m=1}^{\infty} \varphi(\lambda_{1k} r) \varphi(\lambda_{2m} r) \Bigg( \chi_{km} \cos a\lambda_{km} t +$$

$$+ \dfrac{\psi_{km}}{a \lambda_{km}} \sin a\lambda_{km} t \Bigg) \Phi_{km}(x, y). \quad (5.52)$$

***О з н а ч е н н я  1.*** *Узагальненим розв'язком задачі* (5.38) – (5.40), *праві частини рівнянь яких* $\chi(x, y)$ *і* $\psi(x, y)$ – *неперервні функції в області* $0 \le x \le l_1$, $0 \le y \le l_2$, *називається границя при* $r \to +0$ *суми ряду* (5.52)
$$u(x, y, t) = \lim_{r \to +0} u_r(x, y, t).$$

**Узагальнений розв'язок задачі (5.41) – (5.43).** Нехай права частина рівняння (5.41) – функція $f(x, y\,t)$ неперервна в області $0 \le x \le l_1$, $0 \le y \le l_2$, $t \ge 0$. Тоді її можна розкласти в ряд Фур'є

$$f(x, y, t) \sim \sum_{k=1}^{\infty} \sum_{m=1}^{\infty} f_{km}(t) \Phi_{km}(x, y),$$

де $f_{km}(t) = \dfrac{4}{l_1 l_2} \int_0^{l_1}\!\!\int_0^{l_2} f(x, y, t) \Phi_{km}(x, y)\, dx dy$.

Розглянемо періодичне усереднення цієї функції

$$f_r(x, y, t) = \sum_{k=1}^{\infty} \sum_{m=1}^{\infty} \varphi(\lambda_{1k} r)\, \varphi(\lambda_{2m} r)\, f_{km}(t)\, \Phi_{km}(x, y).$$

Якщо члени послідовності $\{\varphi(\lambda_k r)\}$ справджують оцінку

$$\varphi(\lambda_k r) = \mathrm{O}\!\left(\frac{1}{k^3}\right), \quad r > 0,$$

то функція $f_r(x, t)$ задовольняє умови теореми 2. Тому розв'язок задачі (5.41) – (5.43) з правою частиною $f_r(x, y, t)$ першого рівняння існує і записується за аналогією (5.49) у вигляді
$$u_r(x, y, t) = \qquad\qquad (5.53)$$
$$= \int_0^t \sum_{k=1}^{\infty} \sum_{m=1}^{\infty} \frac{\varphi(\lambda_{1k} r)\varphi(\lambda_{2m} r) f_{km}(\tau)}{a \lambda_{km}} \Phi_{km}(x, y) \sin a\lambda_{km}(t - \tau)\, d\tau.$$

***О з н а ч е н н я  1.*** *Узагальненим розв'язком задачі* (5.41) – (5.43) *з правою частиною рівняння* (5.41) – *функцією* $f(x, y, t)$ *неперервною в області* $0 \le x \le l_1$, $0 \le y \le l_2$, $t \ge 0$ *називається границя при* $r \to +0$ *суми ряду* (5.53)
$$u(x, y, t) = \lim_{r \to +0} u_r(x, y, t).$$

Перетворимо інтеграл у формулі (5.53) з урахуванням формули для коефіцієнтів Фур'є функції $f(x, y, t)$ і запишемо вираз узагальненого розв'язку у вигляді

$$u(x, y, t) = \frac{4}{l_1 l_2} \lim_{r \to +0} \int_0^t \int_0^{l_1} \int_0^{l_2} \sum_{k=1}^{\infty} \sum_{m=1}^{\infty} \frac{\varphi(\lambda_{1k} r) \varphi(\lambda_{2m} r)}{a \lambda_{km}} \cdot$$
$$\cdot \Phi_{km}(x, y) \Phi_{km}(x_0, y_0) \sin a\lambda_{km}(t - \tau) f(x_0, y_0, \tau) \, dx_0 \, dy_0 \, d\tau .$$

Ввівши позначення

$$G(x, y, x_0, y_0, t - \tau) = \frac{4}{l_1 l_2} \lim_{r \to +0} \sum_{k=1}^{\infty} \sum_{m=1}^{\infty} \frac{\varphi(\lambda_{1k} r) \varphi(\lambda_{2m} r)}{a \lambda_{km}} \cdot$$
$$\cdot \Phi_{km}(x, y) \Phi_{km}(x_0, y_0) \sin a\lambda_{km}(t - \tau), \qquad (5.54)$$

запишемо узагальнений розв'язок задачі (5.41) – (5.43) у вигляді

$$u(x, y, t) = \int_0^t \int_0^{l_1} \int_0^{l_2} G(x, y, x_0, y_0, t - \tau) f(x_0, y_0, \tau) \, dx_0 \, dy_0 \, d\tau .$$

*О з н а ч е н н я  2 . Функція* $G(x, y, x_0, y_0, t - \tau)$ *називається функцією впливу або функцією Гріна для задачі* (5.41) – (5.43).

Функція $G(x, y, x_0, y_0, t - \tau)$ визначає відхилення точок, закріпленої на краях прямокутної мембрани, від положення рівноваги внаслідок дії одиничного імпульса, прикладеного в точці $(x_0, y_0)$ в момент часу $\tau$.

*З а у в а ж е н н я  4* . За теоремою 2 (п. 4.4), оскільки дія зосередженої в точці $(x_0, y_0)$ сили може бути описана з використанням послідовності неперервних функцій, ряд (5.54) рівномірно підсумовується в будь-якій області, що не містить точки $(x_0, y_0)$.

## 5.3. Узагальнені розв'язки граничних задач для рівняння Гельмгольца

**5.3.1. Формулювання задач для рівняння коливань без початкових умов.** В електростатиці, механіці та інших прикладних науках для опису усталених динамічних процесів формулюють крайові задачі без початкових умов, коли коливний процес відбувається за гармонічним законом.

Розглянемо задачу

$$\frac{\partial^2 u}{\partial t^2} = a^2\left(\frac{\partial^2 u}{\partial x^2} + \frac{\partial^2 u}{\partial y^2}\right) + f(x, y, t), \qquad (5.55)$$

$$(x, y) \in D, \quad -\infty < t < +\infty$$

$$Au(x, y, t) = \psi(x, y, t), \qquad (5.56)$$

$$(x, y) \in L, \quad -\infty < t < +\infty,$$

де

$$f(x, y, t) = f_1(x, y)\cos\omega t - f_2(x, y)\sin\omega t,$$
$$\psi(x, y, t) = \psi_1(x, y)\cos\omega t - \psi_2(x, y)\sin\omega t; \qquad (5.57)$$

$\omega$ – частота коливань; $L$ – границя (гладка крива) області $D$; $A$ – лінійний оператор, вирази $A = 1$, $A = \dfrac{\partial}{\partial n}$ або $A = \dfrac{\partial}{\partial n} - \alpha(x, y)$ відповідають першій, другій або третій граничним задачам, $\dfrac{\partial}{\partial n}$ – похідна за напрямком нормалі до $L$, $\alpha(x, y)$ – функція точки кривої $L$.

Розв'язок задачі (5.55), (5.56) шукаємо у вигляді

$$u(x, y, t) = u_1(x, y)\cos\omega t - u_2(x, y)\sin\omega t. \qquad (5.58)$$

Тоді для визначення функцій $u_1(x, y)$, $u_2(x, y)$ одержимо з урахуванням незалежності тригонометричних функцій такі граничні задачі:

$$\frac{\partial^2 u_1}{\partial x^2} + \frac{\partial^2 u_1}{\partial y^2} + \theta^2 u_1 = -\frac{1}{a^2}f_1(x, y), \quad (x, y) \in D,$$
$$Au_1 = \psi_1(x, y), \qquad (x, y) \in L;$$

$$\frac{\partial^2 u_2}{\partial x^2} + \frac{\partial^2 u_2}{\partial y^2} + \theta^2 u_2 = -\frac{1}{a^2}f_2(x, y), \quad (x, y) \in D,$$
$$Au_2 = \psi_2(x, y), \qquad (x, y) \in L,$$

де $\theta^2 = \dfrac{\omega^2}{a^2}$.

Якщо тут ввести комплексні функції $U = u_1 + iu_2$, $F = \dfrac{f_1}{a^2} + i\dfrac{f_2}{a^2}$ і $\Psi = u_1 + iu_2$, то одержимо

$$\dfrac{\partial^2 U}{\partial x^2} + \dfrac{\partial^2 U}{\partial y^2} + \theta^2 U = -F(x, y), \quad (x, y) \in D, \qquad (5.59)$$
$$AU = \Psi(x, y), \qquad (x, y) \in L.$$

Диференціальне рівняння (5.59) називається рівнянням Гельмгольца, а функція $U = u_1 + iu_2$ називається *комплексною амплітудою*.

Функція (5.58) виражається через комплексну амплітуду наступним чином $u = \operatorname{Re}\!\left(Ue^{i\omega t}\right)$.

**5.3.2. Функція Гріна для рівняння Гельмгольца у прямокутнику.** Розглянемо задачу про відшукання розв'язку неоднорідного рівняння Гельмгольца у прямокутнику $\Pi = \{(x, y): 0 < x < l_1; 0 < y < l_1\}$ за однорідних умов

$$\dfrac{\partial^2 u}{\partial x^2} + \dfrac{\partial^2 u}{\partial y^2} + \theta^2 u = -F(x, y), \quad (x, y) \in \Pi,$$
$$u(x, y) = 0, \qquad (x, y) \in \partial\Pi, \qquad (5.60)$$

де $\partial\Pi$ – границя прямокутника.

За теоремою 2 (п. 5.2), якщо непарна періодична з періодом $2l_1$ функція $F(x, y)$ – неперервна і має неперервні частинні похідні до третього порядку включно, то розв'язок задачі (5.60) можна знайти методом Фур'є у вигляді суми тригонометричного ряду

Вважаємо, що вільний член рівняння (5.60) - періодичне продовження достатньо гладкої дельтоподібної функції за двома змінними $F(x, y) = \delta_r^*(x, x_0)\delta_r^*(y, y_0)$. Розвинення дельтоподібної функції за системою $\{\sin \lambda_{1k} x\}$ одержимо з використанням $2\pi$-періодичного розвинення (4.19)

$$\delta_r^*(x, x_0) = \delta_r^*(x - x_0) - \delta_r^*(x + x_0) = \dfrac{2}{l_1} \sum_{k=1}^{\infty} \varphi(\lambda_{1k} r) \sin \lambda_{1k} x_0 \sin \lambda_{1k} x.$$

Аналогічно одержимо вираз дельтоподібної функції за системою $\{\sin\lambda_{2m}y\}$

$$\delta_r^*(y, y_0) = \frac{2}{l_2}\sum_{m=1}^{\infty}\varphi(\lambda_{2m}r)\sin\lambda_{2m}y_0 \sin\lambda_{2m}y.$$

Тоді вираз дельтоподібної функції за двома змінними запишемо у вигляді

$$\delta_r^*(x, x_0)\delta_r^*(y, y_0) = \frac{4}{l_1 l_2}\sum_{k=1}^{\infty}\sum_{m=1}^{\infty}\varphi_{km}(r)\Phi_{km}(x_0, y_0)\Phi_{km}(x, y), \quad (5.61)$$

де $\lambda_{1k} = \dfrac{k\pi}{l_1}$; $\lambda_{2m} = \dfrac{m\pi}{l_2}$; $\varphi_{km}(r) = \varphi(\lambda_{1k}r)\,\varphi(\lambda_{2m}r)$;

$\Phi_{km}(x, y) = \sin\lambda_{1k}x\,\sin\lambda_{2m}y$.

Розв'язок задачі (5.60) шукаємо у вигляду суми ряду за системою функцій $\{\Phi_{km}(x, y)\}$

$$u_r(x, y) = \sum_{k=1}^{\infty}\sum_{m=1}^{\infty}u_{km}\,\Phi_{km}(x, y),$$

який задовольняє граничні умови цієї задачі. Підставляючи його разом з розвиненням (5.61) у рівняння (5.60), знайдемо формулу для визначення коефіцієнтів

$$u_{km} = \frac{4}{l_1 l_2}\frac{\varphi_{km}(r)\,\Phi_{km}(x_0, y_0)}{\lambda_{km}^2 - \theta^2}$$

за умови $\theta \neq \lambda_{km}$. Тоді розв'язок задачі наступний

$$u_r(x, y) = \frac{4}{l_1 l_2}\sum_{k=1}^{\infty}\sum_{m=1}^{\infty}\frac{\varphi_{km}(r)\Phi_{km}(x_0, y_0)}{\lambda_{km}^2 - \theta^2}\,\Phi_{km}(x, y). \quad (5.62)$$

***О з н а ч е н н я  1 .*** *Границя при $r \to +0$ суми ряду (5.62) називається функцією Гріна для задачі* (5.60)

$$G(x, y, x_0, y_0) = \frac{4}{l_1 l_2}\lim_{r\to +0}\sum_{k=1}^{\infty}\sum_{m=1}^{\infty}\frac{\varphi_{km}(r)\,\Phi_{km}(x_0, y_0)}{\lambda_{km}^2 - \theta^2}\Phi_{km}(x, y). \,(5.63)$$

Зауважимо, що узагальнена сума (5.63) неперервна функція в будь-якій області, що не містить точки $(x_0, y_0)$.

Якщо функція $F(x, y)$ приймає ненульові значення тільки в області $D \subset \Pi$

$$F(x, y) = \begin{cases} F_0(x, y), & (x, y) \subset D, \\ 0, & (x, y) \neq D, \end{cases}$$

де $F_0(x, y)$ – кусково-неперервна функція в області $D$, то розв'язок відповідної задачі (5.60) запишеться у вигляді

$$u(x, y) = \iint\limits_D G(x, y, x_0, y_0) F(x_0, y_0) \, dx_0 dy_0 . \qquad (5.64)$$

Якщо ж функція $F(x, y)$ приймає ненульові значення тільки на кусково-гладкій кривій $L \subset \Pi$

$$F(x, y) = \begin{cases} F_1(x, y), & (x, y) \subset L, \\ 0, & (x, y) \notin L, \end{cases}$$

де $F_1(x, y)$ – кусково-неперервна функція на $L$, то розв'язок відповідної задачі (5.60) запишеться у вигляді

$$u(x, y) = \int\limits_L G(x, y, x_0, y_0) F_1(x_0, y_0) dl(x_0, y_0).$$

*П р и к л а д  1*. Знайти розв'язок задачі (5.60), якщо $F(x, y) = 1$.

Оскільки не виконуються умови теореми 2 (п. 5.2), $F(0, y) = F(l_1, y) = F(x, 0) = F(x, l_2) = 1 \neq 0$, шукаємо узагальнений розв'язок задачі за формулою

$$u(x, y) = \int\limits_0^{l_1} \int\limits_0^{l_2} G(x, y, x_0, y_0) \, dx_0 dy_0 =$$

$$= \frac{4}{l_1 l_2} \lim_{r \to +0} \sum_{k=1}^{\infty} \sum_{m=1}^{\infty} \frac{\varphi_{km}(r) \Phi_{km}(x, y)}{\lambda_{km}^2 - \theta^2} \int\limits_0^{l_1} \sin \lambda_{1k} x_0 \, dx_0 \int\limits_0^{l_2} \sin \lambda_{2m} y_0 \, dy_0 =$$

$$= \frac{16}{l_1 l_2} \lim_{r \to +0} \sum_{k=1,3,\ldots}^{\infty} \sum_{m=1,3,\ldots}^{\infty} \frac{\varphi_{km}(r)}{\lambda_{1k} \lambda_{2m} \left( \lambda_{km}^2 - \theta^2 \right)} \Phi_{km}(x, y), \quad \theta \neq \lambda_{km}.$$

Внаслідок рівномірної збіжності одержаного ряду за умови $\lambda_{km} \neq \theta$, можливий почленний граничний перехід при $r \to +0$. Тому з урахуванням рівності $\lim\limits_{r \to +0} \varphi_{km}(r) = 1$ узагальнений розв'язок задачі запишемо у вигляді суми рівномірно збіжного ряду

$$u(x, y) = \frac{16}{l_1 l_2} \sum_{k=1,3,\ldots}^{\infty} \sum_{m=1,3,\ldots}^{\infty} \frac{1}{\lambda_{1k} \lambda_{2m} \left(\lambda_{km}^2 - \theta^2\right)} \Phi_{km}(x, y).$$

Очевидно цей узагальнений розв'язок не є класичним розв'язком сформульованої задачі, оскільки другі частинні похідні не зображуються рівномірно збіжними рядами, тобто не є неперервними функціями.

### 5.4. Завдання до п'ятого розділу

1. Знайти закон вільних коливання однорідної струни, кінцеві точки якої $x = 0$ і $x = l$ нерухомо закріплені, початкова форма якої $u(x,0) = 3x(l-x)$ і початкова швидкість її точок дорівнює нулю.

2. Знайти закон вільних коливання однорідної струни без початкових зміщень. Кінцеві точки $x = 0$ і $x = l$ струни нерухомо закріплені, а початкова швидкість її точок задана $\dfrac{\partial u(x,0)}{\partial t} = 4\sin\dfrac{\pi}{l}x$.

3. У скінченному однорідному стержні довжини $l$ з теплоізольованою бічною поверхнею початкова температура дорівнює $\psi(x)$. На кінцях стержня підтримується нульова температура. Визначити температуру стержня $u(x,t)$.

а) Методом Фур'є знайти формальний розв'язок задачі

$\dfrac{\partial u}{\partial t} = a^2 \dfrac{\partial^2 u}{\partial x^2}$, $0 < x < l$, $t > 0$;

$u(0,t) = u(l,t) = 0$, $t \geq 0$;

$u(x,0) = \psi(x)$, $0 \leq x \leq l$.

б) Довести: якщо $\psi(x)$ – неперервна кусково-гладка функція на відрізку $[0, l]$ і $\psi(0) = \psi(l) = 0$, то

$$u(x,t) = \sum_{k=1}^{\infty} a_k \exp\left\{-\left(\frac{k\pi}{l}\right)t\right\} \sin\frac{k\pi}{l}x,$$

де $a_k = \dfrac{2}{l} \displaystyle\int_0^l \psi(x) \sin\dfrac{k\pi}{l}x\, dx$.

в) Знайти узагальнений розв'язок задачі а), якщо $\psi(x)$ – кусково-неперервна функція на відрізку $[0, l]$

4. Скінченний однорідний стержень довжини $l$ з теплоізольованою бічною поверхнею нагрівається джерелами тепла інтенсивності $f(x,t)$. На кінцях стержня підтримується нульова температура. Початкова температура дорівнює нулеві. Визначити температуру стержня $u(x,t)$.

а) Методом Фур'є знайти формальний розв'язок задачі

$$\frac{\partial u}{\partial t} = a^2 \frac{\partial^2 u}{\partial x^2} + f(x,t), \quad 0 < x < l, \quad t > 0;$$
$$u(0,t) = u(l,t) = 0, \quad t \geq 0;$$
$$u(x,0) = 0, \quad 0 \leq x \leq l.$$

б) Довести: якщо функція $f(x,t)$ неперервна в області $0 \leq x \leq l$, $t \geq 0$, має неперервну кусково-гладку другу похідну і $f(0,t) = f(l,t) = 0$, $\frac{\partial^2 f(0,t)}{\partial x^2} = \frac{\partial^2 f(l,t)}{\partial x^2} = 0$, то

$$u(x,t) = \sum_{k=1}^{\infty} \sin\frac{k\pi}{l}x \int_0^t \exp\left\{-\left(\frac{k\pi}{l}\right)^2 (t-\tau)\right\} f_n(\tau)\, d\tau,$$

де $f_k(\tau) = \frac{2}{l} \int_0^l f(x,\tau) \sin\frac{k\pi}{l}x\, dx$.

в) Знайти узагальнений розв'язок задачі а), якщо $f(x,t)$ неперервна в області $0 \leq x \leq l$, $t \geq 0$ функція.

5. Знайти узагальнений розв'язок задачі:

$$\frac{\partial^2 u}{\partial x^2} + \frac{\partial^2 u}{\partial y^2} + \theta^2 u = -F(x,y), \quad (x,y) \in \Pi,$$
$$u(x,y) = 0, \quad (x,y) \in \partial\Pi,$$

де $F(x,y) = xy(l_1 - x)(l_2 - y); \quad \Pi = \{(x,y): 0 < x < l_1; 0 < y < l_1\}$.

6. Знайти закон вільних коливання однорідної струни, кінцеві точки якої $x = 0$ і $x = l$ нерухомо закріплені, початкова форма якої $u(x,0) = 3x(l-x)$ і початкова швидкість її точок дорівнює нулю.

# СПИСОК ЛІТЕРАТУРИ